\documentclass[12pt]{report}
\usepackage{manthesis,axodraw,epsfig,amssymb}
\usepackage[]{graphicx}
\usepackage{bm}
\usepackage{amsmath}
\usepackage{bbm}
\usepackage{ulem}
\usepackage{rotating}
\usepackage{cite}

\begin{document}
\title{Rapidity gap physics at contemporary colliders}
\author{Robert Appleby}
\supervisor{Dr. Mike Seymour}
\department{Department of Physics and Astronomy}
\submissiondate{September 2003}
\degree{Doctor of Philosophy}
\faculty{Faculty of Science and Engineering}
\institution{University of Manchester}
\maketitle{yes}{yes}{%abs
This thesis is concerned with the theory and the phenomenology of rapidity gap processes. We perform perturbative 
calculations of energy flow observables in jet-gap-jet processes, 
which consist of resummed primary emission calculations specific to the soft 
gluon geometry at HERA and an estimate of non-global (secondary emission) effects in clustered energy flow 
observables. The resulting predictions agree well with H1 data. We also study
hard diffraction and use a factorised model, with a Monte Carlo event generator, to make detailed 
predictions for gap-jet-gap events at the Tevatron. We find that we can describe the data in a natural way by using
HERA parton densities and a gap surivial factor consistent with theoretical estimates. 

}
\preface{Declaration}
% My declaration
No portion of the work referred to in this thesis has been submitted in support of an 
application for another degree or qualification at this or any other university or other
institution of learning.
\cnotice

\preface{Acknowledgements}
% acknowledgements
First and foremost, I would like to acknowledge Mike Seymour for 
the advice, encouragement and superb supervision he has offered me 
over the course of my PhD. Without Mike carefully guiding me in the 
right direction and helping solve my myriad of problems, this thesis would
never exist. 
Many thanks and acknowledgement also go to Jeff Forshaw, for excellently
supervising me in the first year of my PhD and for providing encouragement 
throughout. \\
\\
Thanks also to all the people who have offered me help, either face-to-face or
by email. These include (in no particular order) George Sterman, Carola Berger, 
Mrinal Dasgupta, Gavin Salam,
Nikolaos Kidonakis, Brian Cox, Arjun Berera and Dino Goulianos. Thanks also go  
to all the people I have accidently omitted. \\
I would also like to thank the theoretical physics group of Manchester University, for 
providing a stimulating, enjoyable place to work and for agreeing to fund my many foreign
trips, including a chance to enjoy the Rio de Janeiro carnival (all in the name of physics, of
course!)\\
\\
Three years is a long time to spend working for one goal, and this has been made bearable 
by the friendship of my office-mates Gavin Poludniowski and Tom Barford. Many thanks go to them for
making my PhD a lot of fun, and for making me sad to complete it and finally leave the office.\\
\\
Finally, I would like to thank Naomi Baker for her love and support.

\preface{Autobiographical Note}
% me
Robert Barrie Appleby was born in Gateshead, in the north of England. He then moved
to the town of Alnwick in Northumberland, where he stayed until passing his A-levels
in 1996. He then spent
four years in the city of York, gaining his masters degree in 
theoretical physics in 2000 and working on problems in theoretical condensed
matter physics for his final year project, before moving across the Pennines 
to Manchester. He then started his PhD in theoretical particle physics 
with Jeff Forshaw and Mike Seymour, which was completed in September 2003.

\preface{Publications}
% publications

R.~B.~Appleby and J.~R.~Forshaw,\\
``Diffractive dijet production,'' \\
Phys.\ Lett.\ B {\bf 541} (2002) 108 
[arXiv:hep-ph/0111077]. \\
%%CITATION = HEP-PH 0111077;%%
\\
R.~B.~Appleby and M.~H.~Seymour,\\
``Non-global logarithms in inter-jet energy flow with kt clustering  requirement,'' \\
JHEP {\bf 0212} (2002) 063 
[arXiv:hep-ph/0211426]. \\
%%CITATION = HEP-PH 0211426;%%
\\
R.~B.~Appleby and G.~P.~Salam,\\
``Theory and phenomenology of non-global logarithms,'' \\
The proceedings of ``QCD and hadronic interactions'', Moriond 2003, Les Arcs, France
[arXiv:hep-ph/0305232]. \\
%%CITATION = HEP-PH 0305232;%%
\\
R.~B.~Appleby and M.~H.~Seymour,\\
``The resummation of inter-jet energy flow for gaps-between-jets processes at HERA,'' \\
JHEP {\bf 0309} (2003) 056 
[arXiv:hep-ph/0308086].

\preface{Dedication}
% ded
This thesis is dedicated to my parents

\chapter{Introduction}
\label{ch0}
% Introduction

In this thesis we will use the theory of the strong force, Quantum Chromodynamics or QCD, to calculate detailed 
predictions for rapidity gap processes and perform comparisons to experimental observation. 
%First of all we will give a
%summary of each of the following chapters, in order to get an overview of the thesis.

Chapter \ref{ch1} is devoted to a survey of the fundamental ideas of QCD, and equips the reader with
some of the tools that are used in the rest of the thesis. We start with an introduction to QCD, 
explain the crucial idea of asymptotic freedom, and proceed with a discussion of factorisation. The colour 
mixing matrices of QCD are then described, with an example calculation for a quark process 
and a gluon process, and then
we outline the basic ideas of Regge theory, in preparation for our later studies of diffractive processes.
We continue with brief sections on Monte Carlo event generators and rapidity gaps, and finish with 
a summary.

In chapter \ref{ch2} we study the diffraction of hadrons at the Tevatron. In these processes, diffracting 
hadrons produce a central dijet system which is separated from the intact hadrons by rapidity gaps. These 
gaps are attributed to pomeron exchange. We describe the factorised model of Ingelman and Schlein, which views 
the process by double pomeron exchange, and use the diffractive Monte Carlo event generator POMWIG, coupled 
with pomeron parton densities from HERA, to produce a set of diffractive predictions. We test these 
predictions against Tevatron data and find, by using a gap survival factor consistent with 
theoretical estimates, we can naturally describe the experimental observations.

Chapter \ref{ch3} is a survey chapter examining resummation, which is a consequence of the factorisation 
properties obeyed by QCD cross sections, and applies the idea to rapidity gap processes. We start by 
writing down the factorisation
properties of the cross section in a specific region of phase space, using jet functions to describe 
approximately collinear quanta and a soft function to describe soft gluon emission, 
and develop the resummation 
formalism. The result
is an expression for the cross section in which the large logarithms in the jet and soft functions are resummed. 
We develop the application of resummation 
to rapidity gap processes, focusing our attention on the soft function, which describes the soft, 
wide angle emission of gluons into a restricted region of phase space.

The theme of chapter \ref{ch4} is non-global logarithms in interjet energy flow observables. The resummation formalism 
to describe interjet energy flow, discussed in chapter~\ref{ch3}, fails to include the effects of 
secondary gluons, which are radiated outside of the 
rapidity gap and subsequently radiate into it. We study the effect of these secondary gluons on~2 jet cross sections 
in the presence of a clustering algorithm, at leading order and at all-orders, and make predictions 
for the impact of non-global logarithms based on an overall, gap dependent suppressive constant. We find that, compared 
to the non-clustered case, the 
use of a clustering algorithm reduces, but does not remove, the suppressive effect.

In Chapter \ref{ch5} we draw on the analyses of the previous chapters and apply our ideas of the 
resummation of interjet energy flow and non-global 
observables to gaps-between-jets measurements at HERA. We include primary interjet logarithms using the resummation 
formalism of chapter \ref{ch3} by making detailed 
soft gluon calculations for the specific gap geometry at HERA. Non-global logarithms 
are approximately included by an extension of the work in chapter \ref{ch4}. 
We find that our calculations are consistent with 
H1 data, and we make predictions for the ZEUS gaps-between-jets analysis.

Finally, in chapter \ref{conc} we draw our conclusions.

\chapter{QCD at the frontier}
\label{ch1}
% Chapter 1

\section{The theory of the strong force}

We are concerned with the theory of the strong force - Quantum Chromodynamics, or QCD. This theory attempts to
describe the fundamental 
constituents of hadrons using point-like quarks and gluons; the former making the matter content of the hadrons, and 
the latter mediating the colour force which binds the quarks together.

Baryons and mesons were suggested to have a composite nature in the early sixties, resulting in 
the colour degree of freedom being
introduced to maintain the fundamental link between spin and statistics, and providing the baryons with an 
antisymmetric wavefunction. Feynman then continued the theme of hadron constituents with his high energy parton 
model, which he used to explain scaling properties of DIS and Drell-Yan cross sections. Today the theory of quarks and 
gluons, enjoying the status of a Yang-Mills non-abelian gauge theory, is a cornerstone of the ``standard'' model 
of particle physics. There are six quarks, each of which is an SU(3) triplet, interacting through the gluons, 
which are the SU(3) gauge bosons. 

We shall now give an overview of some of the tools of QCD we will be using in the rest of this
thesis; we shall resist the impulse to list
the QCD Lagrangian and the other trappings of modern gauge theories and assume the reader is familiar with the 
basic ideas. For further details of this, and the rest of the material in this chapter 
see \cite{Ellis:qj,Brock:1993sz,Peskin:ev,Sterman:ce}.
In all of this work, we will use dimensional regularisation with $d=4-2\epsilon$ and use the 
Feynman gauge, unless otherwise stated. Some of the figures in this thesis were produced using the 
package {\it axodraw} \cite{Vermaseren:1994je}. 

The main property that allows a perturbative approach to QCD is the feature of asymptotic freedom. 
This means that the coupling constant, $\alpha_s$, is a function 
of the scale $Q$ and decreases as the scale increases. For a one-scale problem we write our (dimensionless) 
observable as a
perturbative expansion in $\alpha_s$,
\begin{equation}
\sigma(Q^2)=\sum_{n} \alpha_s^n(\mu^2) \sigma^{(n)}\left(\frac{Q^2}{\mu^2}\right),
\label{secintro:eqsigma}
\end{equation}
where $\mu$ is the scale used to renormalise the theory. We now note that the observable~$\sigma$~cannot possibly depend 
on the precise choice of $\mu$, for we can (in principle) measure~$\sigma$~in our laboratory and a change in 
a theoretical scale cannot affect such a measurement. Hence 
\begin{equation}
\mu \frac{\partial}{\partial \mu} \sigma(Q^2)=0.
\end{equation}
This equation, and many like it, are a vital part of a particle physicist's toolkit and will play a central role 
in the analysis of this thesis. The solution of this equation indicates that the scale dependence, or running, of 
$\alpha_s$ is given by the solution of the so-called renormalisation group 
equation (RGE),
\begin{equation}
Q^2 \frac{\partial\alpha_s}{\partial Q^2}=\beta(\alpha_s),
\end{equation}
where the QCD $\beta$-function is calculable using QCD. By restricting ourself to a one-loop solution we
find,
\begin{equation}
\alpha_s(Q^2)=\frac{1}{b \log(Q^2/\Lambda_{\mathrm{QCD}}^2)},
\end{equation}
where $\beta(\alpha_s)=-b\alpha_s^2$ and\footnote{$n_f$ is the number of flavours.}
\begin{equation}
b=\frac{33-2n_f}{12\pi}.
\end{equation}
We have written this equation in terms of the experimentally determined parameter~$\Lambda_{\mathrm{QCD}}$, at which 
the coupling diverges and perturbative calculation breaks down. The most commonly used value of $\Lambda_{\mathrm{QCD}}$ 
is the five-flavour QCD scale $\Lambda_{\mathrm{QCD}}^{(5)\overline{MS}}$, where we calculate the contributing one-loop 
Feynman diagrams in the modified minimal subtraction scheme.  The world average value for the coupling at 
the $Z^0$ mass is
\begin{equation}
\alpha_s(M_Z)=0.118 \pm 0.002,
\end{equation}
for which $\Lambda_{\mathrm{QCD}}^{(5)\overline{MS}}$ is deduced to be
\begin{equation}
\Lambda_{\mathrm{QCD}}^{(5)\overline{MS}}=208^{+25}_{-23}\,\mathrm{MeV}.
\end{equation}
The non-abelian gauge group of QCD ensures a negative $\beta$-function (assuming~$n_f~\le~16$) 
and hence a reduction\footnote{QED, conversely, 
has an abelian gauge group and a positive $\beta$-function. The result is that $\alpha_{\mathrm{em}}$ weakly rises
with scale.} of the coupling with scale. The strategy of perturbation theory is to expand the observable in 
powers of $\alpha_s$, equation (\ref{secintro:eqsigma}), and hope that, by the smallness of $\alpha_s$, it is sufficient
to calculate just the first one or two terms of the expansion. 

In general, the perturbative expansions of QCD two-scale 
observables, generally denoted $R$, are littered by large logarithmic enhancements of the form
\begin{equation}
R(V)=\sum_n \sum_{m} R_{mn} \alpha_s^n \log^m(1/V),
\end{equation}
where $V$ is the ratio of the two scales (normally the hard scale and a softer scale). The leading logarithmic (LL) set 
is the set of terms with the most number of logarithms for a given~$\alpha_s$; 
note that the definition of the LL set is 
observable dependent, and can be up to two logarithms per~$\alpha_s$. 
If these large logarithms overcome the smallness of~$\alpha_s$, then it is insufficient to 
calculate the first one or two terms in the perturbation series, because all the terms are potentially 
large, and all orders must be considered. 

The physical origin of these large logarithms is the soft and/or collinear limit of Feynman diagrams. In these 
regions of phase space, the denominators of some internal propagators vanish and these regions are
logarithmically enhanced. Examples 
of observables which include large logarithms include the thrust~($T$) distribution as~$T\rightarrow 1$ in
electron-positron cross sections and cross sections in the vicinity of partonic threshold, where the partonic 
system has just enough energy~$\sqrt{\hat{s}}$ to produce the observed final state of mass~$Q$. In the latter example,
large logarithms occur in the limit~$z\rightarrow 1$, where~$z=Q^2/\hat{s}$. The 
theoretical machinery to include the effect of terms of all-orders is known as resummation. In this thesis we 
will be primarily concerned (beginning with chapter~\ref{ch3}, once we have concluded our study of diffractive 
processes in chapter~\ref{ch2}) with the effect of large logarithms which arise from 
restricting soft particle emission in restricted regions of phase space, where the large terms arise from an 
incomplete real and virtual diagram cancellation. An important tool for such an analysis is the use of quantum 
mechanical incoherence, to which we now turn.

\section{Factorisation and refactorisation}

Factorisation is a statement of the quantum mechanical incoherence of short and long distance physics, and plays a
central role in the predictive power of QCD. In this section we will outline the statements of factorisation that are
of most use to us, namely the ``standard'' factorisation theorems, which write cross sections as 
convolutions of long and short distance functions, and the refactorisation theorems of the short distance function.  

Factorisation is the QCD generalisation of Feynman's parton model. Colliding hadrons, in the centre-of-mass frame, are
highly Lorentz contracted and time-dilated and the interaction probes a frozen configuration of 
partons.  The interaction thus proceeds by one parton from 
each hadron undergoing a QCD hard scattering event; the spectator partons cannot 
interfere with this process as the interactions 
between these take place at longer, time-dilated scales. The unscattered partons go on to form the hadron
remnants. The inclusive cross section is written as a convolution of a long distance function describing the dynamics 
of partons in the hadrons with a short distance function describing the hard event,
\begin{equation}
\sigma^{\mathrm{inc}}(Q)=\sum_{i,j} \int\,dx_1 \int\,dx_2 f_{i/h_A}(x_1,\mu) \, f_{j/h_B}(x_2,\mu) \, 
\sigma^{\mathrm{hard}}_{ij}(Q,x_1,x_2,\mu),
\end{equation}
where $\mu$ is a factorisation scale separating the long distance dynamics from the short distance dynamics, 
$f_{i/h_A}(x,\mu)$ denotes the distribution of parton $i$ in hadron $A$ with momentum fraction $x$ (known as 
a parton distribution function, a parton density or a PDF) and $Q$ is the scale of the process. 
The parton densities are non-perturbative and are required to be determined experimentally, whilst the 
short distance function 
is calculable using perturbation theory.

A remarkable consequence of factorisation is that measuring a parton density at one scale $\mu$ allows us to predict
the parton density at another scale~$\mu'$, provided that~$\mu,\mu'\gg \Lambda$. This result, known as the 
evolution of parton densities and structure functions, is a powerful predictive tool in perturbative QCD (pQCD). This 
evolution is most transparently expressed using a set of integro-differential equations,
\begin{equation}
\mu^2 \frac{d}{d\mu^2} f_{i/h_A}(x,\mu^2)=\sum_{j=f,\bar{f},g} 
\int_x^1 d\xi P_{ij}\left(\frac{x}{\xi},\alpha_s(\mu^2)\right)  f_{j/h_A}(\xi,\mu^2),
\end{equation}
which are known as the Dokshitzer-Gribov-Lipatov-Altarelli-Parisi (DGLAP) equations and are one of the most
important sets of equations in pQCD. The evolution kernels (or splitting functions)~$P_{ij}$ give 
the probability of finding species~$i$ in species~$j$, and are calculable as a power series in~$\alpha_s$. 
The combination of factorisation of observables 
into short distance functions and non-perturbative parton densities, and the subsequent evolution of the 
parton densities using the DGLAP equations is central in the impressive success of QCD as the 
gauge theory of the strong force.

We can now perform a further refactorisation on the short distance function, for a specific class of observables 
in a particular limit of their final state phase space. In this region, we are interested in 
a QCD hard process at scale $Q$, which is only accompanied by soft radiation up to the soft scale~$Q_s$. This region of
phase space is known as the threshold region, and is relevant if we make a specific 
restriction on the energy of gluon emission or, for
example, in the production of heavy quarks near threshold. 
The soft radiation is described by a function $S$ and we 
write the cross section as the product of a hard and a soft matrix,  
\begin{equation}
\sigma^{\mathrm{hard}}_{ij}(Q,Q_s,\alpha_s)=\sum_{I,L} 
H_{IL}(Q,\alpha_s(\mu_f))
S_{LI}(Q_s,\alpha_s(\mu_f)),
\end{equation}
where~$\mu_f$ is a new factorisation scale. The proof of this statement follows standard 
factorisation arguments~\cite{Collins:ig,Kidonakis:1998bk}.
The soft gluon emission is sensitive to the colour state of the hard event and hence we have written 
the soft and hard functions as matrices in the space of possible colour flows of the process. 
Now all the dynamics at the softer 
scale~$Q_s$ are described by the soft matrix, with all higher energy dynamics described by the hard matrix. 
For example, we can apply this factorisation to a dijet process, and restrict the interjet radiation to the
scale~$Q_s$; in these rapidity gap processes, the soft function describes soft radiation into the interjet gap.
 We will use this factorisation in chapter~\ref{ch3} and in chapter~\ref{ch5}, where we will apply these ideas
to rapidity gap processes and exploit the factorisation to resum large interjet logarithms in energy flow processes
at HERA.

\section{Colour mixing in QCD}

\label{secintro:mixing}

The refactorisation properties discussed in the last section will be exploited in later chapters to 
resum large QCD logarithms. An important tool in these calculations is the mixing of the basis of colour 
tensors\footnote{The full set of colour bases (or tensors) used in this work is in appendix \ref{secef:appbases}.}, 
over which the hard and soft matrices of the previous section are expressed, by quantum corrections. In this section
 we will calculate these mixing matrices for a quark process and a gluon process at one loop. The full set of mixing 
matrices for all processes is in appendix \ref{secef:appdecomp} and appear in \cite{Kidonakis:2000gi,Berger:2003zh}.

The physical importance of the decomposition of an observable into its possible colour flows becomes clear 
when we note that 
the emission of soft radiation in the QCD process is sensitive to the colour state of the hard interaction. This colour
coherence effect means that the soft gluon emission pattern depends on the overall colour charge of the parent system. 
This dependence of the radiation on the colour state can be understood by consideration of a 
QCD quark-antiquark scattering process, for example, in the large~$N_c$ limit. 
If the quarks interact by exchanging colour, then the outgoing quarks will be colour 
connected, and a colour dipole will be stretched between the outgoing partons.  Hence the region between the quarks
will be filled with gluonic radiation. However, if the outgoing quarks are not colour connected then there will be no
dipole stretched between them. Therefore the soft gluon radiation pattern is sensitive to the colour state of the
hard scattering, and we are required to consider the decomposition of an observable into its different colour flows.

We shall consider the colour flow of QCD $2\rightarrow 2$ scattering, which is described by~4 colour indices. The initial
state particles will be labelled~$A$ and~$B$, the final state particles will be labelled~$1$ and~$2$ and we will use lower
case roman indices for internal lines. In this notation, the standard colour algebra for $q\bar{q}$ 
scattering with a~t-channel gluon
would then be written as $t^a_{1A}t^a_{B2}$. In order to decompose the colour flow of a matrix element, we need to specify a
basis of colour tensors, linking the~4~indices, which describe the possible underlying colour flows.
For example, the process $q\bar{q}\rightarrow q\bar{q}$ will have a two-element basis consisting of elements which have the
physical interpretation of singlet or octet colour exchange exchange,
\begin{eqnarray}
c_1&=&\delta_{A1}\delta_{B2}, \nonumber \\
c_2&=&-\frac{1}{2N_c}\delta_{A1}\delta_{B2}+\frac12\delta_{AB}\delta_{12},
\label{secintro:qqbarbasis}
\end{eqnarray}
where we denote elements of the basis as $c_i$. Note that $c_2$ in this basis is interpreted as the colour flow
for a~t-channel gluon.
The $2\rightarrow 2$ scattering amplitude
can then be decomposed over this basis,
\begin{equation}
\mathcal{M}=\sum_i c_i \mathcal{M}_i,
\end{equation}
where the~$\mathcal{M}_i$ coefficients encode the amount of colour tensor~$c_i$ in~$\mathcal{M}$. The calculation of
these coefficients will allow the successful decomposition of an arbitrary QCD amplitude over an 
appropriate basis.

The colour tensor basis set will be mixed into itself by higher order diagrams. For example, a diagram with 
singlet colour flow will become a diagram with octet colour flow by the addition of a virtual~t-channel gluon. 
To compute this effect, which is the aim of this section, we dress a colour tensor with a virtual gluon connecting
two external legs and, by considering the colour content of the resulting diagram, express the result in terms
of elements of the basis set. A diagram illustrating the addition of a virtual gluon to the 
tensor $c_i$ is shown in diagram \ref{secintro:eikonal}. Therefore the virtual gluons will cause
the bare colour tensors to mix into each other, with a matrix describing this mixing; we refer to this matrix 
as the colour mixing matrix. Therefore if the undressed basis set is denoted $(c_1,c_2)^T$ and the basis set 
which has mixed under quantum corrections is denoted $(c_1',c_2')^T$, then
\begin{equation}
\left(\begin{array}[c]{c}
c_1'\\
c_2'
\end{array}\right)
=
\left(\begin{array}[c]{cc}
\mathcal{C}_{11} &  \mathcal{C}_{12} \\
\mathcal{C}_{21} &  \mathcal{C}_{22}
\end{array}\right)
\left(\begin{array}[c]{c}
c_1\\
c_2
\end{array}\right).
\end{equation}
This is just standard operator mixing under quantum corrections and we produce a process and basis 
dependent matrix describing how the colour tensors mix. 
%To summarise, in this section we rewrite the colour 
%factors of QCD diagrams in terms of the set of colour tensors, and the higher order dynamics, which appear through
%the additional colour structure of an extra virtual gluon, then mix
%up this set of colour tensors. This colour mixing is controlled by the colour mixing matrix.

Note that although it is the dynamics which cause the colour tensor mixing, we are only interested in
the resulting colour structure in this section. 
We will start with a detailed example for 
a quark-only process and then describe the complications in the presence of external gluons. 
Appendix \ref{secintro:app1} contains a set of SU(3) group identities, which are used in this section.

\subsection{Quark-only processes}

\begin{figure}
\begin{center}
\epsfig{figure=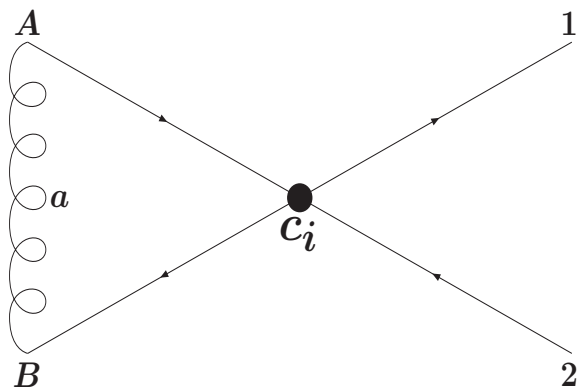,width=3in,height=2in}
\caption{The diagram obtained by dressing the colour tensor $c_i$ with a virtual gluon $a$ which connects external 
legs $A$ and $B$, for the process $q(A)\,\bar{q}(B)\rightarrow q(1)\,\bar{q}(2)$. In the notation in the text, this 
diagram is denoted $\mathcal{C}_i^{(AB)}$. Note that in this section we are only concerned with the 
colour flow, and not the 
dynamics, of this diagram.
\label{secintro:eikonal}}
\end{center}
\end{figure}

The colour mixing matrix for quark-only processes is found using the fundamental identity
\begin{equation}
t^a_{ij}t^a_{kl}=\frac12\delta_{il}\delta_{kj}-\frac{1}{2N_c}\delta_{ij}\delta_{kl},
\label{secintro:eqtexpand}
\end{equation}
where $t^a$ denotes a SU(3) matrix in the fundamental representation. 
For the process $q(A)\,\bar{q}(B)\rightarrow q(1)\,\bar{q}(2)$ we choose the basis 
of equation (\ref{secintro:qqbarbasis}), which encodes singlet and octet exchange in the t-channel. 
Note that any other choice that completely spans 
the colour space is acceptable; the benefit of a singlet-octet choice is that the basis is orthogonal, and 
the tensor product of different elements of the basis is zero. 
We can attach the virtual, soft gluon to any two of the external legs, which gives six possible attachments with
three classes of colour structure. Let $\mathcal{C}^{(ij)}_{k}$ represent the colour decomposition obtained from 
dressing the colour tensor $c_k$ with a soft gluon connecting the $i$ and $j$ external legs. 
Each class of diagram has an associated dynamical piece, but we are only interested in the colour structure at 
this stage. It is
these dynamical pieces that will ultimately undergo the mixing, through their associated colour structure.
We have illustrated the dressing
of a colour tensor in figure~\ref{secintro:eikonal}, where we show~$\mathcal{C}_i^{(AB)}$. We can now work out the colour
 decomposition. Starting with~$\mathcal{C}_1^{(AB)}$ we get the following,
\begin{eqnarray}
\mathcal{C}_1^{(AB)}&=&t^a_{Bn}\,c_1\,t^a_{mA}, \nonumber \\
&=&t^a_{Bn}\,\delta_{m1}\delta_{n2}\,t^a_{mA},
\end{eqnarray}
where~$m$ and~$n$ are dummy indices. Note that we have suppressed the colour indices on the colour tensors~$c_i$, in 
accordance with equation~(\ref{secintro:qqbarbasis}).
By contracting the indices with the~$\delta$-functions we find 
that~$\mathcal{C}_1^{(AB)}=c_2$, which also means that~$\mathcal{C}_1^{(12)}=c_2$.
By similar manipulations we find that
\begin{eqnarray}
\mathcal{C}^{(A1)}&=&\mathcal{C}^{(B2)}_1=C_F c_1, \\
\mathcal{C}^{(B1)}_1&=&\mathcal{C}^{(A2)}=c_2. 
\end{eqnarray}
For the first of these results we have 
used the fact that
\begin{equation}
\left( t^a t^a \right)_{ij}=C_F \delta_{ij}.
\label{secintro:eqttrace}
\end{equation}
Now we turn to the cases of diagrams involving $c_2$. Starting with $\mathcal{C}_2^{(12)}$ we have
\begin{eqnarray}
\mathcal{C}_2^{(12)}&=&t^a_{Bn}\,c_2\,t^a_{mA}, \nonumber \\
&=&t^a_{Bn}\,\frac12\left(\delta_{mn}\delta_{12}-\frac{1}{N_c}\delta_{m1}\delta_{n2}\right)\,t^a_{mA}.
\end{eqnarray}
If we expand the terms, then we can use equation~(\ref{secintro:eqttrace}) on the first term and 
equation~(\ref{secintro:eqtexpand}) on the second to obtain
\begin{equation}
\mathcal{C}_2^{(12)}=\frac{C_F}{2N_c} c_1 +\left(C_F-\frac{1}{2N_c}\right)c_2.
\end{equation}
A similar calculation for the other classes of diagram yields,
\begin{eqnarray}
\mathcal{C}^{(A1)}_2&=&\mathcal{C}^{(B2)}_2=-\frac{1}{2N_c}c_2, \nonumber \\
\mathcal{C}^{(B1)}_2&=&\mathcal{C}^{(A2)}_2=-\frac{1}{N_c}c_2 + \frac{C_F}{2N_c} c_1.
\end{eqnarray}
We can now write down the mixing matrix for this process in terms of the three classes of diagram, and we 
use~$\alpha$, $\beta$ and $\gamma$ to denote the associated dynamical piece,~$\zeta^{(ij)}$, of the relevant diagram,
\begin{eqnarray}
\alpha=\zeta^{(AB)}+\zeta^{(12)}, \nonumber \\
\beta=\zeta^{(A1)}+\zeta^{(B2)}, \nonumber \\
\gamma=\zeta^{(B1)}+\zeta^{(A2)}.
\end{eqnarray}
We will use this notation, and discuss the origin of the dynamical pieces, in chapter~\ref{ch5}. 
Recall that if the undressed set is denoted $(c_1,c_2)^T$ and the set which has mixed under quantum corrections 
is denoted $(c_1',c_2')^T$, then
\begin{equation}
\left(\begin{array}[c]{c}
c_1'\\
c_2'
\end{array}\right)
=
C^{q\bar{q}\rightarrow q\bar{q}}
\left(\begin{array}[c]{c}
c_1\\
c_2
\end{array}\right),
\end{equation}
where the mixing matrix, $C^{q\bar{q}\rightarrow q\bar{q}}$, is given by
\begin{equation}
\mathcal{C}^{q\bar{q}\rightarrow q\bar{q}}=
\left(\begin{array}[c]{cc}
C_F \beta \,\,\,& \frac{C_F}{2N_c}(\alpha+\gamma) \\
\alpha+\gamma \,\,\,&  C_F\alpha-\frac{1}{2N_c}(\alpha+\beta+2\gamma)
\end{array}\right).
\end{equation}
This matrix describes how the colour tensors, and the associated dynamical pieces, mix under 
quantum corrections.
The mixing matrix for the other quark process,~$qq\rightarrow qq$, can be calculated in a similar way and 
appears in appendix~\ref{secef:appdecomp}.

\subsection{The addition of gluons}

The presence of external gluons increases the complexity of the colour mixing calculations, due to 
more involved group theory.
In this section we will describe the calculation of the mixing matrix for the process 
$q(A)\,g(B)\rightarrow q(1)\,g(2)$, which will 
introduce the majority of the tools needed for the gluon processes. The basis for this process is
\begin{eqnarray}
c_1&=&\delta_{A1}\delta_{B2}, \nonumber \\
c_2&=&d_{B2c}(t_F^c)_{1A}, \nonumber \\
c_3&=&if_{B2c}(t^c_F)_{1A},
\label{secintro:eqqgbasis}
\end{eqnarray}
where the first element describes singlet exchange, and the second and third members describe 
symmetric and antisymmetric colour exchange respectively. The constants~$f^{ABC}\,(A,B,C=1,\dots,8)$ are the 
antisymmetric structure constants of SU(3). They are antisymmetric under the exchange of any two indices and 
satisfy the Jacobi identity,
\begin{equation}
f_{ABE}f_{ECD}+f_{CBE}f_{AED}+f_{DBE}f_{ACE}=0.
\end{equation}
The constants $d^{ABC}\,(A,B,C=1,\dots,8)$ are the symmetric structure constants of~SU(3). They are symmetric 
under the exchange of any two indices, obey $d_{ABB}=0$ and satisfy
\begin{equation}
f_{ABE}d_{ECD}+f_{CBE}d_{AED}+f_{DBE}d_{ACE}=0.
\end{equation}
From these, and the group theory identities in 
appendix \ref{secintro:app1}, from which we mainly use
\begin{equation}
t^a_{ij} t^b_{jk}=\frac{1}{2}\left[\frac{1}{N_c}\delta_{ab}\delta_{ik}+\left(d_{abc}+i f_{abc}\right)t^c_{ik}\right],
\label{secintro:eqtprod}
\end{equation}
we can deduce the following basis relationships,
\begin{eqnarray}
c_3 \times c_1&=& c_3, \\
c_2 \times c_1&=& c_2, \\
c_3 \times c_2 &=& -\frac{N_c}{4} c_2 + \frac{N_c^2-4}{4N_c}c_3, \\
c_3 \times c_3 &=& \frac12 c_1 + \frac{N_c}{4}c_2 - \frac{N_c}{4} c_3, \label{secintro:eqc3tc2}
\end{eqnarray}
which are useful when we see that adding a virtual t-channel soft gluon to the colour tensor $c_i$ is 
equivalent to taking the tensor product of this colour tensor with $c_3$\footnote{This is true for the t-channel 
basis set we are using for this process.}. These identities are straightforward to
prove; for example, if we label the two exchanged gluons as $c$ and $d$, the virtual quark as $i$ and the
virtual gluon as $j$, we get the following group algebra 
for $c_3 \times c_3$,
\begin{equation}
c_3 \times c_3 = i f^{Bjc}\,t^{c}_{iA} \times if^{j2d}\,t^d_{1i},
\end{equation}
where we can see the meaning of the $\times$ notation.
By using equation (\ref{secintro:eqtprod}) 
and equations (\ref{secintroappen:eqdfstart}-\ref{secintroappen:eqdfend}) we can prove 
equation~(\ref{secintro:eqc3tc2}). 
We can now use these identities and deduce that
\begin{eqnarray}
\mathcal{C}_1^{(AB)}&=&c_3, \\
\mathcal{C}_2^{(AB)}&=&c_3\times c_2 =  -\frac{N_c}{4} c_2 + \frac{N_c^2-4}{4N_c}c_3, \\
\mathcal{C}_3^{(AB)}&=&c_3\times c_3 =  \frac12 c_1 + \frac{N_c}{4}c_2 - \frac{N_c}{4} c_3,
\end{eqnarray}
and $\mathcal{C}_i^{(12)}=\mathcal{C}_i^{(AB)}$. Hence 
\begin{equation}
\mathcal{C}_i^{(12)}=\mathcal{C}_i^{(AB)}
=
\left(\begin{array}[c]{ccc}
0 & 0 & \frac12 \\
0 & -\frac{N_c}{4} & \frac{N_c}{4} \\
1 & \frac{N_c-4}{4N_c} & -\frac{N_c}{4}
\end{array}\right).
\end{equation}
Turning now to $\mathcal{C}^{(A1)}$, we see that
\begin{equation}
\mathcal{C}_1^{(A1)}=C_F c_1,
\end{equation}
and that 
\begin{eqnarray}
\mathcal{C}_2^{(A1)}&=&d_{B2c}\,t^c_{ji}\,t^d_{iA}\,t^d_{1j}, \nonumber \\
&=&d_{B2c}\,t^c_{ji}\,\frac12\left(\delta_{ij}\delta_{A1}-\frac{1}{N_c}\delta_{iA}\delta_{1j}\right), \nonumber \\
&=&-\frac{1}{2N_c} d_{B2c}\,t^c_{1A}, \nonumber \\
&=&-\frac{1}{2N_c} c_2,
\end{eqnarray}
where we have expanded the pair of SU(3) matrices $t^d_{iA}\,t^d_{1j}$ 
and used the fact that~$t^a_{ii}=0$. Finally, working in 
a similar way,
\begin{eqnarray}
\mathcal{C}_3^{(A1)}&=&if_{B2c}\,t^c_{ji}\,t^d_{1j}\,t^d_{iA}, \nonumber \\
&=&-\frac{1}{2N_c}c_3.
\end{eqnarray}
Note that these last three results are not the same for $\mathcal{C}^{(B2)}$. For this case the group algebra 
gives (where we note that $N_c=C_A$),
\begin{eqnarray}
\mathcal{C}_1^{(B2)}&=&i f_{Bdc}\,if_{d2c}\,\delta_{A1}=C_A c_1, \\
\mathcal{C}_2^{(B2)}&=&t^d_{1A}\,d_{ijd}\,i f_{Bic}\,i f_{j2C}=\frac{N_c}{2}c_2, \\
\mathcal{C}_3^{(B2)}&=&-t^d_{1A}\,f_{ijd}\,f_{Bic}\,f_{j2C}=\frac{N_c}{2}c_3.
\end{eqnarray}
For the first of these results we have used the fact that $f_{Bcd}\,f_{2cd}=N_c\delta_{B2}$, for the second of
these results we have used the fact that $d_{jdi}\,f_{iBc}\,f_{c2j}=-d_{B2d}$ and for the third of these results
we have used the fact that $f_{jdi}\,f_{iBc}\,f_{c2j}=-\frac{N_c}{2} f_{db2}$. Therefore we find that
\begin{eqnarray}
\mathcal{C}_i^{(A1)}
&=&
\left(\begin{array}[c]{ccc}
C_F & 0 & 0 \\
0 & -\frac{1}{2N_c} & 0 \\
0 & 0 & -\frac{1}{2N_c}
\end{array}\right), \\
\mathcal{C}_i^{(B2)}
&=&
\left(\begin{array}[c]{ccc}
C_A & 0 & 0 \\
0 & \frac{N_c}{2} & 0 \\
0 & 0 & \frac{N_c}{2}
\end{array}\right).
\end{eqnarray}
After similar, albeit longer, calculations we find 
\begin{eqnarray}
c_1^{(A2)}&=&c_3, \\
c_2^{(A2)}&=&i d_{j2c}\,f_{Bjd}\,t^d_{1i}\,t^c_{iA}=\frac{N_c^2-4}{4N_c}c_3+\frac{N_c}{4}c_2, \\
c_3^{(A2)}&=&-f_{j2c}\,f_{Bjd}\,t^d_{1i}\,t^c_{iA}=\frac{1}{2}c_1+\frac{N_c}{4}c_2+\frac{N_c}{4}c_3,
\end{eqnarray}
and hence the result
\begin{equation}
\mathcal{C}_i^{(A2)}
=
\mathcal{C}_i^{(B1)}
=
\left(\begin{array}[c]{ccc}
0 & 0 & \frac12 \\
0 & \frac{N_c}{4} & \frac{N_c}{4} \\
1 & \frac{N_c^2-4}{4N_c} & \frac{N_c}{4}
\end{array}\right),
\end{equation}
and so we find the colour mixing matrix for the process $qg\rightarrow qg$, in basis \ref{secintro:eqqgbasis},
is 
\begin{equation}
\mathcal{C}^{qg\rightarrow qg}=
\left(\begin{array}[c]{ccc}
C_F \mathcal{C}^{(A1)}+C_A\mathcal{C}^{(B2)} & 0 &  -\frac12(\alpha+\gamma) \\
0 & \chi & -\frac{N_c}{4}(\alpha+\gamma) \\
-(\alpha+\gamma) & -\frac{N_c^2-4}{4N_c}(\alpha+\gamma) & \chi
\end{array}\right),
\end{equation}
where
\begin{equation}
\chi=\frac{N_c}{4}(\alpha-\gamma)-\frac{1}{2N_c}\zeta^{(A1)}+\frac{N_c}{2}\zeta^{(B2)}.
\end{equation}
In this equation, as in the previous section, $\zeta^{(ij)}$ denotes the dynamical piece of the 
relevant diagram. We can proceed in a similar way and calculate the mixing matrices for any partonic process that may 
interest us. The bases and the mixing matrices for all the processes used in this thesis can be found in
appendices \ref{secef:appbases} and \ref{secef:appdecomp} and we will use these results when we study
the resummation of rapidity gap events in chapters~\ref{ch3} and~\ref{ch5}.

\section{Regge theory and the pomeron}

Regge theory \cite{regge} provides a framework to describe scattering amplitudes in the Regge limit $s \gg -t$. Whilst in this thesis 
we are primarily concerned with the perturbative description of rapidity gap processes, we will use some of the ideas 
of Regge theory in chapter \ref{ch2} to examine diffractive processes and so in this section we will outline some of the
basic concepts. Further details can be found in \cite{Forshaw:dc,Donnachie:en,Collins:jy}. 

The whole idea of Regge theory, starting from the analytic properties of the S-matrix, is to extract the high 
energy behaviour of scattering amplitudes in a model 
independent way. The fundamental result is that the dominant contribution to the high energy scattering amplitude has a 
general form, which has the interpretation of the exchange of objects known as reggeons in the t-channel. It can 
be shown that if the high energy cross section in the Regge region can be described only in terms of reggeon exchanges,
then the total cross section will behave like
\begin{equation}
\sigma_{\mathrm{tot}}\propto s^{\alpha(0)-1},
\end{equation}
where $\alpha(0)$ is known as the intercept of the Regge trajectory (the Regge trajectory is a straight line in 
the spin-mass squared plane of the exchanged mesons). The fits of Chew and 
Frautschi \cite{chewandf} to meson data showed that 
$\alpha(0)=0.55$, and this observation of~$\alpha(0) < 1$ was found to be true for other exchange particles. 
Hence the total cross section should decrease with $s$, and vanish asymptotically. 
However when the total cross section was measured experimentally, it was found that it slowly increased 
at high $s$. Therefore reggeon exchange is not the whole story, and a new trajectory is needed to describe the data, 
with an intercept greater than unity, $\alpha(0) > 1$. This new trajectory is known as the pomeron 
trajectory, or the pomeron.

From fits to 
total $pp$ and $p\bar{p}$ data, Donnachie and Landshoff \cite{Donnachie:1992ny} extracted the pomeron intercept and found
\begin{equation}
\alpha(0)\simeq 1.08.
\end{equation}
This is normally known as the ``soft'' pomeron intercept. We will use the ideas of Regge theory in
chapter~\ref{ch2}, when we study diffractive processes at the Tevatron and we will describe
the diffractive interaction of two hadrons by the exchange of both pomerons and reggeons.

\section{Event generators}

Monte Carlo event generators have many uses in particle physics, for example estimating the 
backgrounds to measured processes and calculating production rates of exotic particles. 
In this thesis we will use the Monte Carlo event generator 
HERWIG \cite{Corcella:2000bw,Corcella:2002jc} on two occasions as a calculation tool and in this section we will give an overview of the theoretical methods 
used. Our discussion in this section on Monte Carlo event generators will hence refer to the operation of HERWIG.

The idea of a Monte Carlo event generator is to give a simulation of a particle physics event, starting from the
initial interaction of the colliding particles and culminating in the
angular distribution and energy of (colourless) final state particles. This is still a level removed from 
what is seen in particle detectors, and the inclusion of detector effects is possible, but it is largely 
sufficient for the research work in the following chapters. To simulate the event, HERWIG starts with a hard subprocess 
and evolves the resulting off-shell partons into colourless final state particles; to do this 
HERWIG is separated into a number of distinct phases.  

For the process of two hadrons interacting to produce a number of hadronic jets, HERWIG starts by 
calculating the hard subprocess, producing
two final state partons using the QCD factorisation theorems with a leading order calculation and the 
parton distribution functions of the hadrons. At this point we have selected a 
QCD $2\rightarrow 2$ subprocess, but any perturbatively calculable process of the 
standard model or any of its extensions is permissible. The result of the calculation is two initial state 
and two final state partons, with their associated 4-momenta, and the total cross section for the process, which 
is calculated by evaluating the integral,
\begin{equation}
\sigma^{\mathrm{hard}}(s,p_{t,\mathrm{min}})=\int_{p^2_{t,\mathrm{min}}}^{s/4} dp_t^2 
\int \frac{d\sigma^{\mathrm{hard}}(s)}{dx_1\,dx_2\,dp_t^2} dx_1\,dx_2,
\end{equation}
where~$p_t$ is the transverse momentum of the outgoing partons. The cut~$p_{t,\mathrm{min}}$ is necessary to 
prevent the integrand approaching divergences in the matrix element and the Landau pole in
the coupling. 

Once the momenta of the final state partons produced by the hard event have been generated, the initial and
final state partons are evolved through perturbative branching, and all emitted partons are themselves evolved, until all
partons reach an infrared hadronisation scale which characterises the onset of non-perturbative physics. 

\subsection*{Final state parton shower}
The off-shell partons produced by the hard subprocess subsequently evolve with the emission of
QCD radiation. The exact calculation of the matrix element for processes with many final state particles is not
possible, and we must use a branching algorithm which correctly takes into account the regions of phase space 
in which QCD radiation is enhanced.  These regions are associated with kinematical configurations in which the
relevant matrix elements are enhanced, and correspond to emission of a soft gluon, or when a gluon or a quark
splits into two collinear partons. The aim of a parton shower algorithm is to identify and sum up the leading 
behaviour in these regions.

The parton showering is controlled by a set of Sudakov form factors, which encode the probability that 
a parton with a virtual mass scale~$t$ will evolve without resolvable branching to the lower 
virtual mass scale~$t_1$. 
If branching does occur, splitting functions are used to compute the
momentum fraction and virtual mass scale of the products. The Sudakov form factors take into account collinear 
enhancements and also soft enhancements using the {\it coherent branching formalism} and the parton shower
is terminated when all partons reach the cut-off scale $t_0$. 
The virtual mass scale variable, $t$, which is evolved to $t_0$,
is a combination of parton energy and
branching opening angle; this choice of the evolution variable is the correct one to include both soft and 
collinear enhancements. There is an excellent explanation of 
the parton shower formalism, and the application to Monte Carlo event generators, in~\cite{Ellis:qj,Webber:mc}.

\subsection*{Initial state parton shower}

The final state parton shower is called a forward evolution algorithm, as in every step the partons move
to a lower virtual mass scale. For the initial state parton shower, it is more convenient to 
%work backwards 
%and evolve partons to higher virtual mass scales, starting 
%from the most virtual partons which participate in the
start with the (most negative virtual) partons which particpate in the 
hard event and evolve backward to the lower virtual mass scales of the partons in the interacting hadrons. 
In this backward evolution 
formalism, the parton distribution functions are used as part 
of the input, to ``guide'' the evolution to the correct distribution of partons.

\subsection*{Hadronisation}

Using the parton shower formalism, all partons are evolved until they reach the cut-off scale $t_0$. It is
at this point that we enter the long-wavelength, non-perturbative regime where the partons group 
themselves into the colourless hadrons we observe; this process is included in HERWIG using a phenomenological 
model.

A property of the parton showering is that the flow of momentum and quantum numbers at the hadron level tends to follow
the flow at the parton level. This hypothesis is known as local parton-hadron duality (LPHD), and underlies 
the {\it cluster hadronisation} model \cite{Webber:1983if,Marchesini:1987cf} used by HERWIG. 
In this model, clusters of colour singlet partons form after the parton shower, inheriting the 
momentum structure of their constituent partons. The process occurs by the splitting of gluons into 
quark/antiquark pairs, and then neighbouring quarks and antiquarks form colour singlet objects. Unstable 
hadrons are then allowed to decay according to experimentally determined 
branching ratios and this process, which tends to disfavour heavy mesons and baryons, provides a good model
of observed final states in $e^+ e^-$ collisions.

\vspace{0.5cm}

HERWIG also includes a model of the underlying soft event, in which the hadron remnants undergo secondary 
soft interactions. 
Once the final state has been determined, the experimental cuts for the analysis of 
interest can be applied  and observables calculated. It is also possible, as we briefly mentioned earlier, to add a further
phase of detector simulation, where the experimental signature of the produced final state is 
simulated. Such a process allows for detailed comparisons to be made with experimental observation.

We will use HERWIG in chapter 
\ref{ch2}, where we are especially interested in the effects of parton showering and hadronisation and again in 
chapter \ref{ch5}, where we will be interested in the computation of the hard event.

\section{Rapidity gaps}

The aim of this thesis is to examine the physics of rapidity gap processes, and make predictions to compare 
to recent experimental data; in this section we will give a brief overview of rapidity and rapidity gaps in
particle collisions. The scattering of two hadrons provides two beams of incoming partons, with a spectrum of 
longitudinal momenta described by hadronic parton densities. The parton centre-of-mass frame is boosted 
with respect to the incoming hadron frame, and it is natural to describe the final state in terms of variables which 
have simple transformation properties under longitudinal boosts. 
To do this we describe the final state in terms of transverse
momentum $p_t$, rapidity $y$ and azimuthal angle $\phi$, and a general 4-vector is written
\begin{eqnarray}
p^{\mu}&=&(E,p_1,p_2,p_3), \nonumber \\
&=&\left(m_t \cosh y,p_t \sin\phi,p_t \cos\phi,m_t \sinh y\right),
\end{eqnarray}
where we define the transverse mass for a particle of mass $m$ by $m_t=\sqrt{p_t^2+m^2}$. Rapidity is defined as
\begin{eqnarray}
y&=&\frac12 \log\left(\frac{E+p_3}{E-p_3}\right), \\
&=&\frac12 \log\left(\frac{1+v\cos\theta}{1-v\cos\theta}\right),
\end{eqnarray}
and is a measure of polar angle and speed.

Pseudorapidity is often more useful in practice, which is defined
\begin{equation}
\eta=-\log\tan\left(\frac{\theta}{2}\right).
\end{equation}
Rapidity and pseudorapidity coincide in the massless limit, but pseudorapidity is more convenient in that the
polar angle $\theta$ can be measured directly from the detector and we require no knowledge of the particle mass.
Rapidities (and approximately pseudorapidities) are additive under z-axis Lorentz boosts, and (pseudo)rapidity differences,
\begin{equation}
\Delta\eta=|\eta_1-\eta_2|,
\end{equation}
are boost invariant.

\begin{figure}
\begin{center}
\epsfig{figure=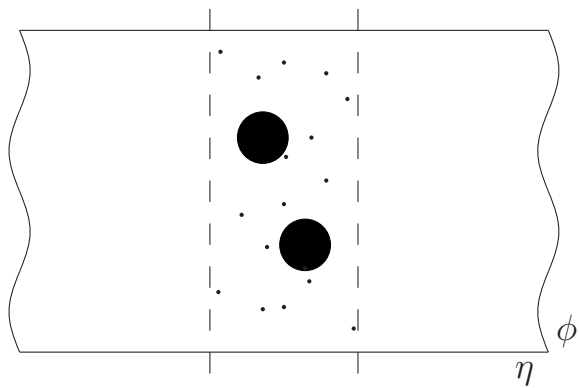,width=3in,height=2.0in}
\caption{The experimental signature of a ``gap-jet-gap'' event. The dark circles denote two jets, and the gap regions 
are separated from the central dijet system by the dashed lines.
\label{secintro:picgjg}}
\end{center}
\end{figure}

\begin{figure}
\begin{center}
\epsfig{figure=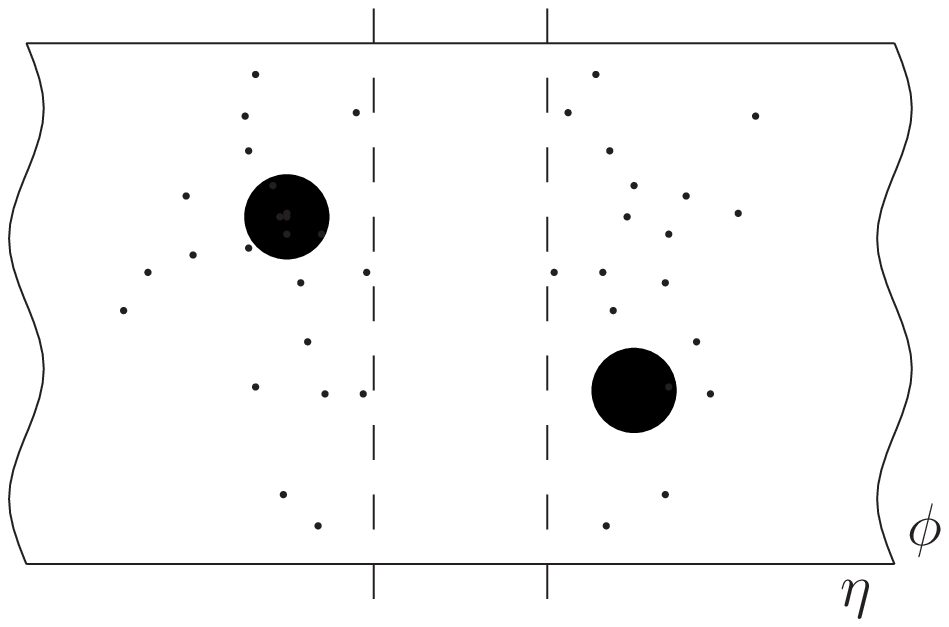,width=3in,height=2.0in}
\caption{The experimental signature of a ``jet-gap-jet'' event. The dark circles denote two jets, and the gap region 
is bounded by the dashed lines.
\label{secintro:picjgj}}
\end{center}
\end{figure}

%The required boost to change frames is simple to deduce; consider 
%electron-proton scattering with electron energy $E_e$, proton energy
%$E_p$, with the proton travelling down the +z direction. In the centre-of-mass frame, the total rapidity
%\begin{equation}
%\eta_{t}=\frac12(\eta_1+\eta_2),
%\end{equation}
%of the electron-positron system is 0, and the rapidity difference,
%\begin{equation}
%\Delta\eta=|\eta_1-\eta_2|,
%\end{equation}
%equals $2\hat{\eta}$. This is because the two particles are back-to-back. In the lab frame, the total jet 
%rapidity is seen to be (using the definition of rapidity),
%\begin{equation}
%\eta_{t}^{\mathrm{lab}}=\frac12 \log \left( \frac{E_p}{E_e} \right),
%\end{equation}
%and hence, using the boost invariance of $\Delta\eta$,
%\begin{equation}
%\eta^{\mathrm{lab}}=\eta+\frac12 \log \left( \frac{E_p}{E_e}\right).
%\end{equation}
%This can be used to boost to any frame. 
%For example, we can boost from the centre-of-mass frame of $p\bar{p}$ 
%to the centre-of-parton frame by the additive boost
%\begin{equation}
%\hat{\eta}=\eta-\frac12 \log \left( \frac{x_1}{x_2} \right).
%\end{equation}
A rapidity gap event is defined as an event producing jets, with 
the region in pseudorapidity between any two jets, $\Delta\eta$, being devoid of 
particle activity. This interjet region, 
known as a gap, is often symmetrical in azimuthal angle $\phi$. The precise definition of the gap depends on the 
experimental geometry, and is experimentally defined by a cut on the total transverse energy flowing in the 
gap region. In this thesis we will consider two kinds of rapidity gap processes:
\begin{itemize}
\item
``gap-jet-gap'' processes in diffractive hadron collisions. In this class of process, the final state consists of
a central (in rapidity) dijet system, separated from the intact colliding hadrons by a rapidity gap on each side. 
Hence the ``gap-jet-gap'' experimental signature, which is illustrated in figure \ref{secintro:picgjg}. 
We shall study these processes in chapter \ref{ch2}, where the 
rapidity gap is produced by the exchange of a pomeron between the hadron and the central system.
\item
``jet-gap-jet'' processes in photoproduction. In this class of process we see a rapidity gap between the two dijets.
We shall study these gaps-between-jets processes in chapters \ref{ch3}, \ref{ch4} and \ref{ch5}, where we will
make perturbative calculations of the rate of gap-event production and make comparisons to experimental data.
The experimental signature is illustrated in figure \ref{secintro:picjgj}. 
\end{itemize}
For both classes of rapidity gap process, we will find that the predictions of the methods that we use 
will give a good description 
of the data.

\section{Summary}

In this chapter we have outlined some of the tools we shall be using in this thesis. The features of asymptotic 
freedom and factorisation give QCD immense predictive power, and one of the central aims of our work is to 
examine the consequences of factorisation for the perturbative calculation of rapidity gap processes at HERA.
 An important element of our calculations will be colour mixing, and in section \ref{secintro:mixing} we derived a
set of process-dependent QCD mixing matrices for a fixed basis. We will use these matrices in 
chapter~\ref{ch5}. We then described the basic ideas of Regge theory, event generators and 
rapidity gaps, in preparation 
for the use of these ideas in our calculations. 
So in conclusion, the research work in this thesis deals with the calculation of rapidity gap processes in QCD
and builds on the fundamental tools presented in this chapter.

\chapter{Diffractive dijet production}
\label{ch2}
% Chapter 2

\section{Introduction}

\label{secdd:intro}

In this chapter we will study hard diffractive processes in hadron-hadron collisions. To be specific we
will make detailed calculations for the process $p\bar{p} \rightarrow p + JJX + \bar{p}$, where $JJX$ denotes
a centrally produced cluster of hadrons, containing at least two jets. The process is theoretically 
diffractive in the sense that both 
the initiating hadrons remain intact in the collision and experimentally diffractive in the sense that the initiating 
proton remains intact, whilst the antiproton either remains intact or dissociates to a low-mass system. 
In both cases the initiating hadrons suffer only a small loss of longitudinal
momentum, and the process is hard in the sense that the central subprocess 
takes place at high momentum transfer. The central (in rapidity) jet-producing system is separated by a rapidity
gap from each of the interacting hadrons, giving the experimental signature of ``gap-jet-gap'' events and the
term hard double diffraction. 

Space-time arguments \cite{collins98} suggest that hard events are well localised in space and time, 
and therefore the effect
of the incoming particles is to act independently of the hard event. This has led, in analogy to 
conventional QCD factorisation theorems, to the concept of a diffractive parton density and the hard diffractive
factorisation theorems. Diffraction factorisation has been proven for diffractive DIS, but for hard diffraction
in pure hadron collisions counterarguments exist which predict that the factorisation will be violated at the
Tevatron \cite{berera2000,collins98,collins1993,martin1997}. The manifestation of this violation is the invalidity of using diffractive parton densities, 
obtained
 in DIS experiments at HERA, in hadron-hadron collisions at the Tevatron and we should expect to see an
overestimation of predicted cross sections. The extent of the breakdown of factorisation is still under investigation
and has been embodied in so-called gap survival factors, which will have direct relevance in this work.

\begin{figure}
\begin{center}
\epsfig{figure=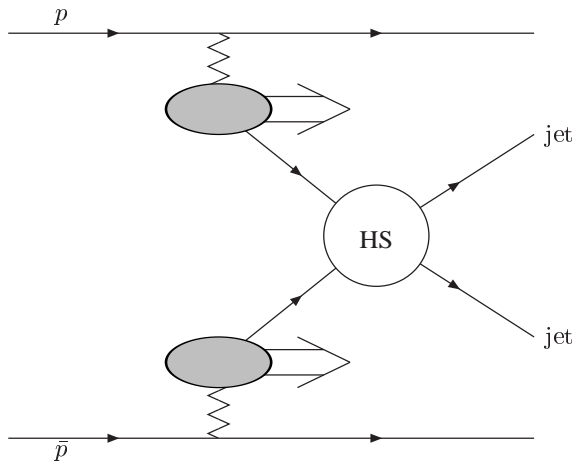,width=3in,height=2.4in}
\caption{Dijet production in the factorised model. The zig-zag lines denote
the exchanged pomerons, and the double-lined arrows represent the pomeron remnants.
\label{figdd:pic1}}
\end{center}
\end{figure}

The standard factorised model of double diffraction draws its inspiration from Regge theory. In the Ingelman-Schlein 
(IS) model \cite{ingelmann1985}, diffractive scattering is attributed to the exchange of a pomeron, which is defined as a colourless object
with vacuum quantum numbers. The model assumes that each of the diffracting hadrons ``emits'' a pomeron, which then
collide to produce the central dijet system. This is illustrated in figure \ref{figdd:pic1}. 
Implicit in the model is Regge factorisation, which writes a diffractive
parton density as the product of the pomeron parton density and a pomeron flux factor. The double diffractive
process then proceeds by the QCD interactions of the emitted pomerons. This kind of event mechanism is known
as double pomeron exchange.

In this chapter we will use the Ingelman-Schlein model of double diffraction to make predictions for gaps-between-jets
processes at the Tevatron. The CDF collaboration has recently presented analyses of such events \cite{cdf2000}. We will
include the effects of fragmentation using the diffractive event generator POMWIG \cite{pomwig}, and use the pomeron parton
densities from HERA DIS experiments \cite{adloff1997}. Previous calculations of this type of event \cite{berera2000,acw} 
have led to an overestimation of the
dijet production cross section by several orders of magnitude.
In this chapter we will show that a 
combination of the HERA parton
densities and a gap survival factor \cite{bjorken1993} consistent with theoretical estimates 
\cite{kaidalov2001,gotsman1999} 
is sufficient to describe the recent
CDF data. Central dijet production within a factorised model has been studied before 
\cite{berera2000,acw,boonekamp2001,cox2001} and the research work in this chapter is published in \cite{Appleby:2001xk}.

This chapter is organised as follows. We start by introducing the experimental analyses in section 
\ref{secdd:tt}, and section \ref{secdd:is}
is a detailed discussion of the IS model and the concept of diffractive parton densities. Section \ref{secdd:pomwig} then
discusses the diffractive event generator POMWIG, and how it is constructed from HERWIG, 
and section \ref{secdd:results} gives our results. Finally, in section \ref{secdd:conc}, we draw
our conclusions.

\section{Diffractive dijet production at the Tevatron}

\label{secdd:tt}

The first observation at the Tevatron of dijet events with a double pomeron exchange
topology was published in 2000 by the CDF collaboration \cite{cdf2000}. In this section we will describe 
the experimental signature of these
events, the observables and cuts used, and finish with a description
of the out-of-cone corrections implemented by CDF to ``undo'' the jet broadening effects
of fragmentation.

\subsection{Diffractive events and the CDF analysis}

\begin{figure}
\begin{center}
\epsfig{figure=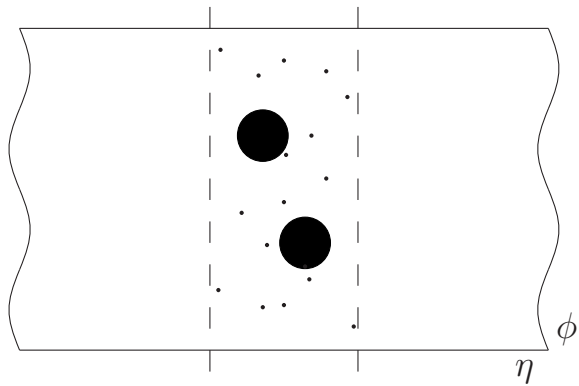,width=3in,height=2.0in}
\caption{The experimental signature of a ``gap-jet-gap'' event. The dark circles denote two jets, and the gap regions 
are separated from the central dijet system by the dashed lines.
\label{secdd:gjg}}
\end{center}
\end{figure}

The diffractive data used in this work was collected by the CDF collaboration when the Tevatron
was colliding protons and antiprotons at a centre-of-mass energy $\sqrt{s}=1800$~GeV.
The events were obtained by making a further analysis of the single diffractive (SD) data set; SD meaning that 
only one interacting hadron undergoes diffraction. In this analysis, 
events with a rapidity gap on the antiproton side were selected by requiring a fractional momentum loss
of the antiproton, $\xi_{\bar{p}}$,  to satisfy $0.035 < \xi_{\bar{p}} < 0.095$, by detecting the antiproton in a forward
Roman Pot Spectrometer (RPS). The requirement was then made of at least two centrally produced jets
of transverse energy $E_T^{\mathrm{min}}>7$~GeV and a 4-momentum transfer from the antiproton squared of~$|t|<1$~GeV$^2$.
This SD sample (of 30,439 events) was then used to search for double diffractive events by demanding
a rapidity gap on the proton side. There is no RPS on this side of the CDF detector and the events
were selected by directly demanding a region devoid of particle activity in the pseudorapidity 
region between the central dijet system and the outgoing proton. This method is not a very accurate way
to determine whether a given event is indeed a genuine double diffractive event, and consequently gives 
a large uncertainty on the the fractional momentum loss, $\xi_{p}$, of the proton. As a result, CDF found that 
the $\xi_p$ distribution was concentrated in the
region $0.01 < \xi_p < 0.03$ and used this region for the $\xi_{p}$ cut. The subset of the SD
sample satisfying this criteria is the double diffractive sample (denoted DD, in which two interacting hadrons undergo 
diffraction). The double diffractive experimental signature is illustrated in figure \ref{secdd:gjg}.
In summary, the CDF cuts used to select the 
double pomeron exchange events were
\begin{itemize}
\item The antiproton fractional energy loss \( \xi _{\bar{p}} \) satisfies \( 0.035<\xi _{\bar{p}}<0.095 \).
Such a condition corresponds to a rapidity gap on the antiproton side. This
cut is made by tagging the antiproton. 
\item The proton fractional energy loss \( \xi _{p} \) satisfies \( 0.01<\xi _{\bar{p}}<0.03 \).
Such a condition corresponds to a rapidity gap on the proton side. Note that
the proton is not tagged at CDF and the cut is made by looking for a rapidity
gap on the proton side. 
\item \( |t_{\bar{p}}|<1 \) GeV\( ^{2} \). \( |t_{\bar{p}}| \) is the four-momentum
transfer from the antiproton squared. 
\item The event must have two or more jets in the pseudorapidity region 
\begin{equation}
-4.2<\eta <2.4,
\end{equation}
and at least two jets must have a minimum transverse energy of \( E_{T}^{min}=7~\)~GeV or $10$ GeV. 
The CDF collaboration define the outgoing proton direction as the positive $\eta$ direction, also known
as ``East''.
\end{itemize}
The CDF analyses presented the total dijet cross section at $E_T^{min}>7$~GeV and $E_T^{min}>~10$~GeV. 
They also presented event distributions in $\xi_p$, $\xi_{\bar{p}}$, the mean
jet transverse energy $E_T^*$, the mean jet rapidity $\eta^*$, the azimuthal angle separation
of the dijets and the dijet mass fraction $R_{jj}$. The last quantity is defined as the mass of the dijet
system evaluated using only the energy within the cones of the two leading jets,~$M_{jj}^{\mathrm{cone}}$, 
divided by the mass of the central system,~$M_x=\sqrt{s \xi_p \xi_{\bar{p}}}$.

\subsection{Jet finding and the cone algorithm}

\begin{figure}
\begin{center}
\epsfig{figure=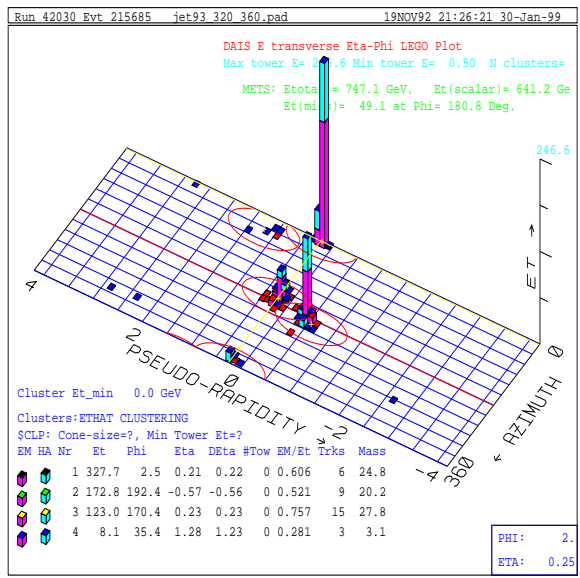,width=3in,height=3in}
\caption{A 3 jet event, defined using the cone algorithm with $R=0.7$, from the CDF experiment. 
The clusters of particles defined as a jet are ringed by the cone radius. This figure 
is taken from \cite{Blazey:1999tm} with the authors' permission.
\label{figdd:3jet}}
\end{center}
\end{figure}

The classification of a hadronic final state into one or more ``jets'' of
hadronic particles is a task performed by a jet algorithm. Modern algorithms have their roots in the
Sterman-Weinberg jet definition \cite{Sterman:1977wj}, and today several different algorithms exist. 
The method used by the CDF
collaboration in their double diffractive dijet production analyses was the cone definition. In this algorithm, 
a jet is defined as the concentration of all particles inside a cone of $(\eta,\phi)$ space, of radius $R$ in the
$(\eta,\phi)$ plane,
\begin{equation}
R=\sqrt{(\Delta\eta)^2+(\Delta\phi)^2}.
\end{equation}
The jets are reconstructed to place as much of the transverse energy as possible
inside a cone, with a minimum energy of $\epsilon^{\mathrm{cut}}$~GeV. Overlapping cones are handled by excluding jets
with more than a certain fraction, normally known as the overlap parameter, of their energy coming from other jets. 
By defining $R$ in terms of $\Delta\eta$
 (other choices could be in terms of $\Delta\theta$, for example) we obtain a jet measurement invariant under
longitudinal boosts. Note that any cross section is dependent on the choice of the cone radius $R$. In this work we will
follow CDF and set $R=0.7$. Figure \ref{figdd:3jet} shows the application of the cone algorithm to a 3 jet
event at the Tevatron. The clusters of particles defined as a jet are ringed by the cone radius in this figure.

Once a concentration of particles has been identified as a jet, the transverse energy of that jet is given
by the sum of all the transverse energies of the constituent particles,
\begin{equation}
E_T^{\mathrm{jet}}=\sum_{i\in \mathrm{jet}} E_T^i.
\end{equation}
Other kinematic variables have similar definitions, of which
\begin{eqnarray}
\eta^{\mathrm{jet}} = \frac{\sum_{i\in \mathrm{jet}} E_{T}^i\,\eta^i}{E_T^{\mathrm{jet}}}, \nonumber \\
\phi^{\mathrm{jet}} = \frac{\sum_{i\in \mathrm{jet}} E_{T}^i\,\phi^i}{E_T^{\mathrm{jet}}},
\end{eqnarray}
is one possible form.

\subsection{Out-of-cone corrections}

The CDF collaboration have applied so-called out-of-cone (oc) corrections to the diffractive
dijet data sample. These corrections are applied as the jet clustering algorithm may not include all
of the energy from the initiating partons, and some of the partons generated during fragmentation
may fall outside of the jet algorithm cones. Therefore the out-of-cone corrections add transverse energy
to the jets to account for this missing $p_t$ and ``correct'' the particle-level jet energies back to
the parton level. This procedure (which is difficult to justify) 
means that CDF no longer presents a true observable.
These corrections depend on the parton fragmentation functions, and are totally
independent of the CDF detector. The out-of-cone corrections take the form of a simple addition to the 
jet $p_t$, and were derived from Run I data. The additional energy is given by
\begin{equation}
p_t^+=A(R)\left(1-B(R)e^{-C(R)p_t}\right),
\end{equation}
where $p_t$ is the particle-level jet energy and the functions $A(R)$,$B(R)$ and $C(R)$ depend on the
cone radius ($R$) and have the values given in table \ref{tabdd:oc}.
\begin{table}
\begin{center}
\begin{tabular}{|c|c|c|c|} % all cell contents aligned center
 \hline
 & A [GeV] & B & C [GeV$^{-1}$] \\ \hline\hline
0.4 & 22.999 & 0.915 & 0.00740 \\ \hline
0.7 & 8.382 & 0.846 & 0.00728 \\ \hline
1.0 & 3.227 & 0.832 & 0.00817 \\ \hline
\end{tabular}
\end{center}
\caption{The out-of-cone correction parameters, as derived from Run I Tevatron data \cite{outofcone}. We have
shown the out-of-cone correction parameters for three different values of the radius, although only $R=0.7$ is
relevant to the current work.}
\label{tabdd:oc}
\end{table}
The impact of the out-of-cone corrections to the work in this chapter is that it is necessary to implement the 
out-of-cone corrections in the Monte Carlo simulation and compare the hadron level
Monte Carlo cross sections with the corrections to the CDF data, and not (as it may first appear) compare the
data to the hadron level
cross sections. Nevertheless, the use of Monte Carlo event generators allows the impact of the fragmentation process 
to be assessed.

\section{Diffractive factorisation and the Ingelman-Schlein model}

\label{secdd:is}

We will now examine the mechanisms for hard double pomeron exchange dijet production. In this
work we will use the factorised Ingelman-Schlein model, which takes its inspiration
from Regge theory. 
We will also describe the non-factorised model of lossless jet production 
\cite{collins1993,berera1996,khoze2001,kms}, but we will not make any 
calculations using this model in this thesis.

\subsection{Hard diffraction}

\begin{figure}
\begin{center}
\includegraphics*[width=4in,height=1.1in]{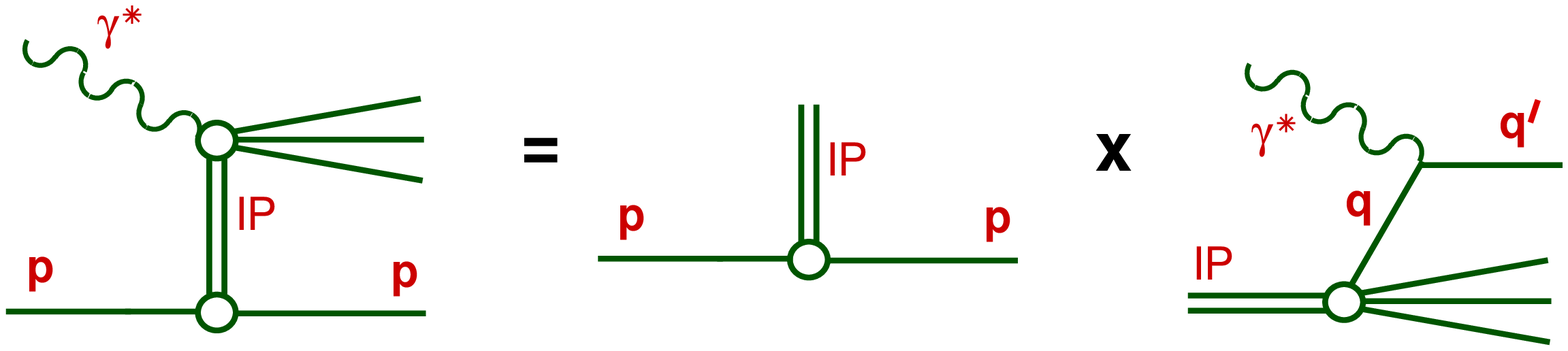}
\caption{An illustration of Regge factorisation. Figure from \cite{Thompson}. 
\label{figdd:reggefact}}
\end{center}
\end{figure}

The pomeron is the object thought to be responsible for rapidity gap processes. Defined as an exchanged object
carrying the quantum numbers of the vacuum, it is postulated to exist in many forms - from 
the soft pomeron of Regge theory
to the perturbative QCD pomeron embodied by the famous Balitsky-Fadin-Kuraev-Lipatov (BFKL) 
equation. Close to 20 years ago Ingelman and Schlein proposed that it was possible to probe 
the content of the pomeron by looking at the diffractive
production of high $p_t$ dijets. In so-called hard diffraction the momentum transfer across the rapidity gap is small,
with the gap corresponding to the exchange of a colour singlet object, but a 
high $Q^2$ process occurs between the exchanged object
and the other hadron. This is be contrasted with diffractive hard scattering, in which the colour singlet object
is exchanged with a high momentum transfer across the gap. Therefore the basic idea of Ingelman and Schlein 
\cite{ingelmann1985} is that
the exchange mechanism for the pomeron is the same for hard diffraction and soft diffraction, and the pomeron 
interacts through its partonic constituents. These ideas took shape when looking at single diffraction, 
 $\bar{p}+p\rightarrow \bar{p}+X$, and led to the following picture:
\begin{itemize}
\item
The diffracting antiproton will emit a pomeron with a small momentum transfer~$t$.  The pomeron
is sometimes said to ``float'' out of the ``parent'' antiproton.
\item
The pomeron, carrying a partonic content, then interacts with the partons in the proton at high 
$Q^2$.
\end{itemize}
The standard Regge single diffractive cross section (single diffraction with
a diffracted proton) is
\begin{equation}
\frac{d^2 \sigma_{SD}^{\bar{p}p}}{d\xi dt}=f_{\mathbb{P}/p}(\xi,t)\sigma_t^{\mathbb{P}\bar{p}}(s',t),
\end{equation}
where $s'=s\xi$ is the centre-of-mass energy of the pomeron-antiproton system, 
$\sigma_t^{\mathbb{P}\bar{p}}(s',t)$ is the total pomeron-proton cross section ($\sigma(p\mathbb{P}\rightarrow X)$)
 and $f_{\mathbb{P}/p}(\xi,t)$ is the pomeron flux factor. We have illustrated this Regge factorisation of the 
cross section in figure~\ref{figdd:reggefact}.
This equation can be treated as a definition 
of the pomeron flux, and allows us to 
write\footnote{In this section we use $\xi$ and $x_{\mathbb{P}}$ interchangably.} the 
single hard diffractive cross section for the production of dijets
as,
\begin{equation}
\frac{d^4 \sigma^{jj}_{SD}}{dx_{\mathbb{P}}dt dx_1 dx_2}=
\left[\frac{1}{\sigma(p\mathbb{P}\rightarrow X)}\frac{d^2\sigma_{SD}}{dx_{\mathbb{P}} dt}\right]
\frac{d^2 \sigma(p\mathbb{P}\rightarrow jj + X)}{dx_1 dx_2},
\end{equation}
where $x_{\mathbb{P}}$ is the momentum loss fraction of the antiproton (i.e. momentum fraction
of the pomeron to the antiproton), $t$ is the 4-momentum transfer-squared and 
$\sigma(p\mathbb{P}\rightarrow jj + X)$ is the proton-pomeron hard-scattering cross section. The pomeron flux
factor (in square brackets) depends on $x_{\mathbb{P}}$ and $t$. The pomeron-proton hard scattering cross section
depends on the momentum fractions, $x_1$ and $x_2$, of partons in the parton and the pomeron respectively. Therefore
this differential cross section is given by
\begin{equation}
\frac{d^3 \sigma(p\mathbb{P}\rightarrow jj + X)}{dx_1 dx_2 dQ^2}=
\sum_{i,j} f_{i/p}(x_1,Q^2) f_{j/\mathbb{P}}(x_2,Q^2) \frac{d\sigma(ij\rightarrow \mathrm{dijets})}{dQ^2},
\end{equation}
where $f_{i/p}(x_1,Q^2)$ and $f_{j/\mathbb{P}}(x_2,Q^2)$ are the parton densities of the proton and 
the pomeron respectively. In the original IS model it was assumed that the pomeron was a purely gluonic object
with a parton density that was $Q^2$ and $t$ independent, and two simple forms 
where proposed,
\begin{eqnarray}
x f_{g/\mathbb{P}}(x)&=&6x(1-x), \\
x f_{g/\mathbb{P}}(x)&=&6(1-x)^5,
\end{eqnarray}
the so-called hard and soft gluon densities respectively. However, these parton densities (combined with the
(soon to be described) DL flux factor) failed to describe Tevatron data. In this 
work we will use modern pomeron parton densities obtained from H1 fits to single-diffractive structure function data 
\cite{adloff1997}.

The most commonly used pomeron flux factor was suggested by A. Donnachie and P.V. Landshoff~\cite{Donnachie:1992ny}, 
now known as
the DL pomeron flux,
\begin{equation}
f_{\mathbb{P}/p}(x_{\mathbb{P}})=\frac{9 \beta_0^2}{4\pi^2}
\left[F_1(t)\right]^2 \left(\frac{1}{x_{\mathbb{P}}}\right)^{2\alpha_{\mathbb{P}}(t)-1},
\end{equation}
where $F_1(t)$ is the proton form factor, given by
\begin{equation}
F_1(t)=\frac{4m_p^2-2.8t}{4m_p^2-t}\left(\frac{1}{1-t/(0.7\,\mathrm{GeV}^2)}\right),
\end{equation}
where $m_p=0.938$~GeV is the proton mass and $\beta_0=1.8$~GeV$^{-1}$ is the pomeron-quark coupling.
The pomeron trajectory $\alpha_{\mathbb{P}}(t)$ has the form 
$\alpha_{\mathbb{P}}(t)=\alpha_{\mathbb{P}}(0)+\alpha'_{\mathbb{P}}t$, where~$\alpha_{\mathbb{P}}(0)$ is
 known as the pomeron intercept. The DL model was fitted to diffractive data, which gave
\begin{eqnarray}
\alpha'_{\mathbb{P}}&=&0.25\,\mathrm{GeV}^{-2}, \nonumber \\
\alpha_{\mathbb{P}}(0)&=&1.08.
\end{eqnarray}
This form of the flux was found to poorly describe Tevatron data and in this work we use the 
POMWIG \cite{adloff1997} parameterisation of the flux, with a harder pomeron 
intercept of $\alpha_{\mathbb{P}}(0)=1.20$, coming from
more recent HERA fits. Apart from this harder intercept (which is important when we come to
compare to the Tevatron data), this parameterised 
form of the flux is approximately the same as the DL flux.

\subsection{Diffractive structure functions}

The conventional way to write the single diffractive dijet production cross section is in terms
of diffractive structure functions. As we have discussed in the last section, this SD cross section
can be written as 
\begin{equation}
\frac{d^5 \sigma^{jj}_{SD}}{d\xi dt dx d\beta dQ^2}=
f_{\mathbb{P}/\bar{p}}(\xi,t)
\sum_{i,j} f_{i/p}(x,Q^2) f_{j/\mathbb{P}}(\beta,Q^2) \frac{d\sigma(ij\rightarrow \mathrm{dijets})}{dQ^2},
\end{equation}
where we have relabelled the momentum fraction of a parton in the proton as $x$, and the momentum fraction 
of a parton in the pomeron as $\beta=x_p/x_{\mathbb{P}}$. We have also denoted the momentum fraction of the pomeron 
with respect to the
antiproton as $\xi$. The diffractive structure function $F_{i/\bar{p}}^{D}$ of the antiproton is
\begin{equation}
F_{i/\bar{p}}^{D}(\xi,t,\beta,Q^2)=f_{\mathbb{P}/\bar{p}}(\xi,t) f_{i/\mathbb{P}}(\beta,Q^2),
\end{equation}
which is the product of the pomeron flux and the pomeron parton density (this level of 
factorisation, inspired from Regge theory, is unproven in hard diffraction and is known as 
Regge factorisation \cite{ingelmann1985}. It is illustrated in figure \ref{figdd:reggefact}). This structure function
is sometimes known as $F_{i/\bar{p}}^{D(4)}$.
The SD dijet cross section is then written
\begin{equation}
\frac{d^5 \sigma^{jj}_{SD}}{d\xi dt dx d\beta dQ^2}=
\sum_{i,j} f_{i/p}(x,Q^2) F_{i/\bar{p}}^{D}(\xi,t,\beta,Q^2) \frac{d\sigma(ij\rightarrow \mathrm{dijets})}{dQ^2}.
\end{equation}
This expression is known as the (hard) factorisation in the diffractive structure function. The $t$-integrated
diffractive structure function is known as $F_{i/\bar{p}}^{D(3)}$ with definition
\begin{eqnarray}
F_{i/\bar{p}}^{D(3)}(\xi,\beta,Q^2)&=&
\int_{t_{max}}^{t_{min}} dt f_{\mathbb{P}/\bar{p}}(\xi,t) f_{j/\mathbb{P}}(\beta,Q^2), \nonumber \\
&=&  \tilde{f}_{\mathbb{P}/\bar{p}}(\xi) f_{j/\mathbb{P}}(\beta,Q^2).
\end{eqnarray}
where $\tilde{f}_{\mathbb{P}/\bar{p}}(\xi)$ is the t-integrated pomeron flux. Performing the $\xi$ integration
gives us the definition of $F_{i/\bar{p}}^{D(2)}$,
\begin{equation}
F_{i/\bar{p}}^{D(2)}(x,Q^2)=
\int_x^{\xi_{max}} d\xi \tilde{f}_{\mathbb{P}/\bar{p}}(\xi) f_{j/\mathbb{P}}(\frac{x}{\xi},Q^2).
\end{equation}
This final quantity, which now only only depends on the momentum fraction of the parton relative to the
diffracting hadron ($x$) and $Q^2$, is often called the diffractive parton density in analogy to the
non-diffractive parton densities.

\subsection{Double pomeron exchange}

The double diffractive process in hadron-hadron collisions is formulated as doubly occurring single diffraction. 
Therefore the process can be modelled using (a copy of) the IS model for both the proton and the antiproton. 
The diffractive event therefore proceeds by both
the proton and the antiproton emitting a pomeron with a small momentum transfer, and these two pomerons 
interacting with each other through their partonic content at high $Q^2$. This QCD process produces two
centrally produced jets, with a gap on each side of rapidity (``bridged'' by the pomerons) to 
the intact parent hadrons. This type of event topology is known as double pomeron exchange or DPE (in such terminology,
single diffractive events proceed by single pomeron exchange and only a single rapidity gap is seen in the
final state). The DPE event topology is depicted in figure \ref{figdd:pic1}. 
%Figure \ref{figdd:diffrac} compares the single diffraction and double pomeron exchange 
%event topologies.
%\begin{figure}
%\begin{center}
%\includegraphics[width=.6\textwidth]{ch2/diffrac.eps}
%\caption{A comparison of single diffraction and double pomeron exchange. Figure from \cite{Forshaw:2002is}.
%\label{figdd:diffrac}}
%\end{center}
%\end{figure}

The DPE cross section can be written
\begin{eqnarray}
\frac{d\sigma_{DPE}^{dijet}}{d\eta_3 d\eta_4 dp_{\perp}^2}(p\bar{p} \rightarrow
p+JJX+\bar{p}) = \int d\xi_p \int d\xi_{\bar{p}} F_{\mathbb{P}/p}(\xi_p) F_{\mathbb{P}/\bar{p}}(\xi_{\bar{p}}) \nonumber \\
\sum_{i,j} \beta_p f_{i/\mathbb{P}}(\beta_p) \beta_{\bar{p}} f_{j/\mathbb{P}}(\beta_{\bar{p}}) \frac{d\hat{\sigma}_{HS}}{d\hat{t}}(ij \rightarrow kl),
\label{secdd:eqdpe}
\end{eqnarray}
where \( F_{\mathbb {P}/p}(\xi )=F_{\mathbb {P}/\overline{{p}}}(\xi ) \) is
the pomeron flux factor, \( \beta  \) is the fraction of the pomeron momentum
carried by the parton entering the hard scattering and \( f_{i/\mathbb {P}}(\beta ) \)
is the pomeron parton density function for partons of type \( i \). The rapidity
of the outgoing partons are denoted \( \eta _{3} \) and \( \eta _{4} \),
their transverse momentum is \( p_{\perp } \) and $\frac{d\hat{\sigma}_{HS}}{d\hat{t}}(ij \rightarrow kl)$
denotes the QCD 2-to-2 scattering amplitudes. 
The parton transverse momentum,~$p_{\perp }$ is equal to
the jet transverse energy $E_T$ at the parton level. In this model, the partonic content of the two pomerons
which does not participate in the hard event forms pomeron remnants in the final state; hence DPE event
topologies are inclusive. 
In this work we will use the $t$-integrated flux parametrisation of POMWIG (discussed in section \ref{secdd:pomwig})
with a pomeron intercept of $\alpha_{\mathbb{P}}(0)=1.20$ as found by H1 \cite{adloff1997} 
in their fits to~$F_2^{D(3)}$ data,
 instead of the softer $\alpha_{\mathbb{P}}(0)=1.08$.

\begin{figure}
\begin{center}
\epsfig{figure=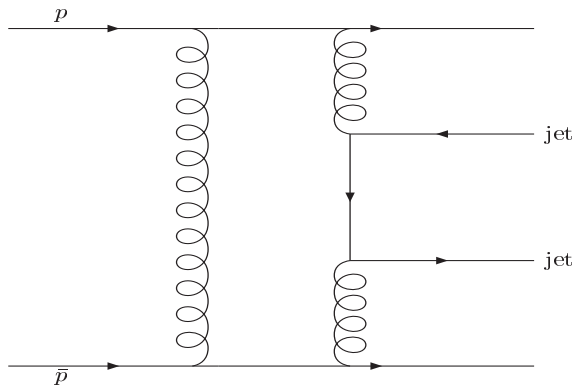,width=3in,height=2in}
\caption{Dijet production in the non-factorised model. This model is characterised by an exclusive
event topology.
\label{figdd:pic2}}
\end{center}
\end{figure}

We should also briefly mention the non-factorising models \cite{collins1993,berera1996,khoze2001,kms} 
for diffractive central dijet production. 
In these models, all the
momentum lost by the diffracting hadrons goes into the hard event, giving an exclusive event topology and
no pomeron remnants. This topology is depicted in figure 
\ref{figdd:pic2}\footnote{In \cite{kms} the possibility that there is additional radiation into the 
final state from the gluons in figure \ref{figdd:pic2} is considered.}.
A significant non-factoring contribution would manifest itself as a peak (at unity) in the distribution showing 
the available centre-of-mass energy which goes into the jets, known as the dijet mass or~$R_{jj}$; a feature which is 
absent in the CDF data we
consider in this thesis. However, this need not be the case for the Tevatron Run II due to higher
luminosity, and this ought to be a good 
place to look for such evidence of a non-factorised contribution. Consequently this thesis does
not consider such models.

\subsection{Breakdown of factorisation and gap survival}

The notion of a parton density in QCD is only useful if it has the property of
universality. This means that once a parton density has been extracted from one process, it should be able
to predict others processes with great accuracy. The current pomeron parton densities have been
extracted from HERA DIS experiments, where the structure function $F_2^{D(3)}$ was measured. However when it came
to testing HERA parton densities at the Tevatron, for hadron-hadron collisions, it was found that they
overestimated the data \cite{berera2000,acw}. 
This violation of factorisation has been understood in terms of simple models, which
indicate that the rapidity gaps will be filled with secondary interactions, from spectator partons in the beam.
 These models also indicated that the process of spoiling the gap can be approximated by a simple 
overall multiplicative factor $S^2$ 
\cite{bjorken1993,gotsman1999,kaidalov2001,Appleby:2001xk,khoze2001,kms,Forshaw:2002is}, 
with a weak process and $\sqrt{s}$ dependence. For some processes at HERA, this
factor has been estimated to be $\sim0.6$~\cite{Cox:2000dg}, and some Tevatron processes have been estimated to have a
$S^2\sim 0.1$~\cite{kaidalov2001,cox2001}. In recent times various models have been postulated 
for the calculation of the gap survival
factor \cite{bjorken1993}, and seem to support this latter figure when applied to Tevatron observables. 
Therefore, as this work
involves using HERA parton densities at the Tevatron, we expect that factorisation will be violated and 
we will include in our predictions an overall
gap survival factor,~$S^2$. We will extract this factor by fitting our overall cross sections to data.

\section{POMWIG - a Monte Carlo for Diffractive processes}

\label{secdd:pomwig}

\begin{figure}
 \begin{minipage}{0.48\textwidth}
\includegraphics*[width=7cm,height=6cm]{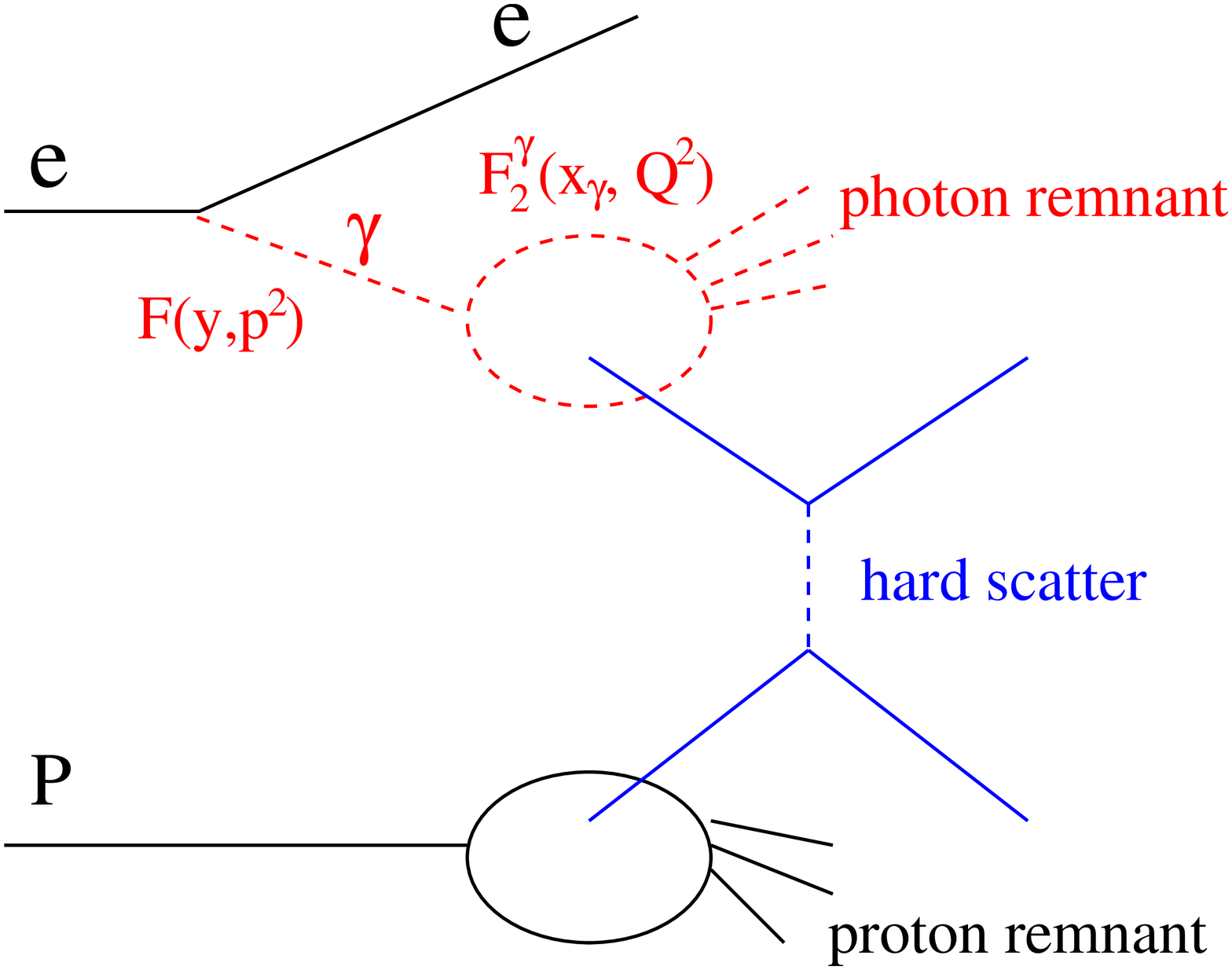}
\end{minipage}
\hfill
\begin{minipage}{0.48\textwidth}
\includegraphics*[width=7cm,height=6cm]{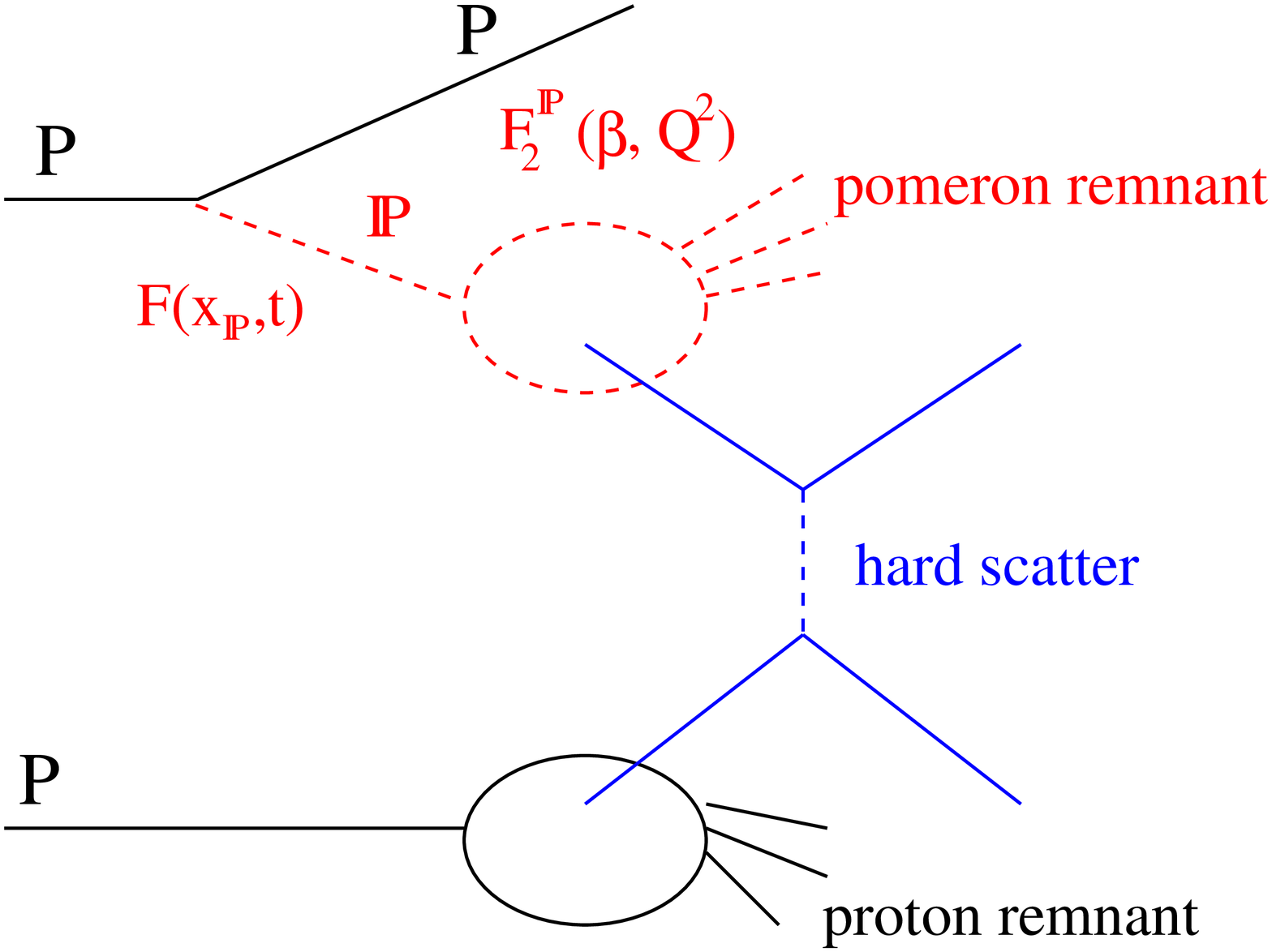}
\end{minipage}
\caption{The similarity between the resolved part of photoproduction and pomeron exchange events. 
Figure from \cite{pomwig}.}
\label{figdd:pomwig}
\end{figure}

The event generator POMWIG \cite{pomwig} is used in this work to study the hadronic 
diffractive process. The modifications
to the HERWIG Monte Carlo event generator~\cite{Corcella:2000bw,Corcella:2002jc} to study
diffractive events are very simple once it is noticed that pomeron exchange events in hadron-hadron
collisions look very much like the resolved part of photoproduction in electron/proton collisions. 
In the latter type of event, the 
electron radiates a quasi-real photon according to a flux formula. This photon is then considered to have
a partonic content which interacts, through QCD processes, with the partonic content of the proton. Therefore we
can study pomeron exchange events in POMWIG by replacing the photon flux with a pomeron flux and by replacing 
the parton density of the photon with that of the pomeron. This philosophy is illustrated in figure 
\ref{figdd:pomwig}.

POMWIG is available as a modification package to HERWIG, and adopts the philosophy of minimal changes
to the HERWIG structure. The generator uses the factorised IS model at the parton level, with pomeron
fluxes and parton densities we will discuss later in this section, and allows the final state partons
to evolve into hadrons using a parton shower and hadronisation model, as discussed in the introduction 
to this thesis. The inclusion of the fragmentation physics through a Monte Carlo simulation allows the phenomenological
impact to be assessed.

In POMWIG, the pomeron and reggeon fluxes are parameterised as 
\begin{eqnarray}
f_{\mathbb{P}/p}(x_{\mathbb{P}})&=& N \int_{t_{\mathrm{max}}}^{t_{\mathrm{min}}} \frac{e^{B_{\mathbb{P}}t}}
{x_{\mathbb{P}}^{2\alpha_{\mathbb{P}}(t)-1}}, \nonumber \\
f_{\mathbb{R}/p}(x_{\mathbb{P}})&=& \mathcal{C}_{\mathbb{R}} 
\int_{t_{\mathrm{max}}}^{t_{\mathrm{min}}} \frac{e^{B_{\mathbb{R}}t}}
{x_{\mathbb{P}}^{2\alpha_{\mathbb{R}}(t)-1}},
\end{eqnarray}
where $\alpha_{\mathbb{P}}(t)=\alpha_{\mathbb{P}}(0)+\alpha'_{\mathbb{P}}t$ and
$\alpha_{\mathbb{R}}(t)=\alpha_{\mathbb{R}}(0)+\alpha'_{\mathbb{R}}t$. The constant~$N$ is chosen so 
that~$F_2^{D(3)}$ matches the H1 data at~$x_{\mathbb{P}}=0.003$ and similarly the 
constant~$\mathcal{C}_{\mathbb{R}}$ is found from data. 
The flux parameters used are those found by the H1 collaboration, assuming there is no pomeron/reggeon interference
contribution to $F_2^{D(3)}$. The default parameters are given in table \ref{tabdd:pomwigpara}.

\begin{table}
\begin{center}
\begin{tabular}{|c|c|} % all cell contents aligned center
 \hline
Quantity & Value  \\ \hline \hline
$\alpha_{\mathbb{P}}$ & 1.20 \\ \hline
$\alpha_{\mathbb{R}}$ & 0.57 \\ \hline
$\alpha'_{\mathbb{P}}$ & 0.26 \\ \hline
$\alpha'_{\mathbb{R}}$ & 0.9 \\ \hline
$B_{\mathbb{P}}$ & 4.6 \\ \hline
$B_{\mathbb{R}}$ & 2.0 \\ \hline
$C_{\mathbb{R}}$ & 48 \\ \hline
\end{tabular}
\end{center}
\caption{The default POMWIG parameters.}
\label{tabdd:pomwigpara}
\end{table}

\begin{figure}
\begin{center}
\epsfig{figure=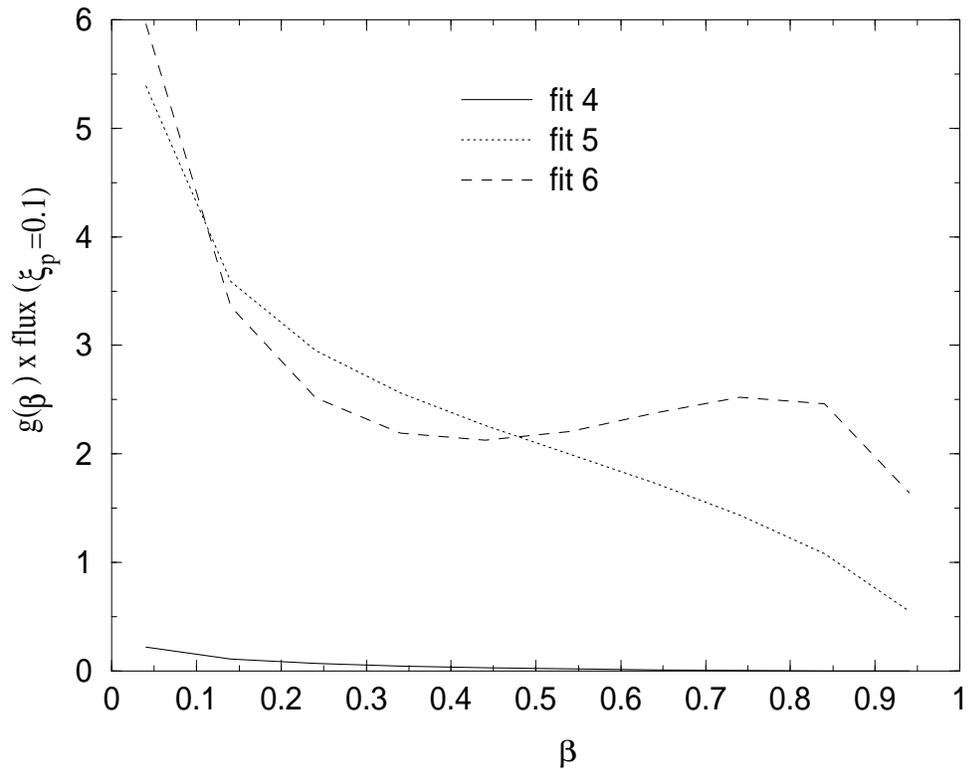,width=5in,height=4in}
\caption{The gluon densities in the pomeron as fitted by H1. The fits are evaluated at $Q^2=50$~GeV$^2$, the typical
scale of the jet transverse energy at CDF, and are multiplied by the pomeron flux factor at $\xi_p=0.1$.
\label{figdd:gluondens}}
\end{center}
\end{figure}

In this work we use the H1 leading order (LO) pomeron fits to $F_2^{D(3)}$ measurements~\cite{adloff1997}. 
The measurement of $F_2^{D(3)}$
is quark dominated and gluon sensitivity enters only through scaling violations. Hence the gluon density has a
large uncertainty of around 30\% in the relevant region. This is important for the gluon dominated DPE process. The gluon densities are
illustrated in figure \ref{figdd:gluondens}.
The mean value of $\beta$ relevant at the Tevatron is in the region of 0.3 to 0.4. Fit 4\footnote{For details of
the fits, see~\cite{adloff1997}.} has a gluon content that is
heavily suppressed relative to fits 5 and 6 (hence fit~4 is quark-dominated), and fit 6 
is peaked at high $\beta$. Fits 5 and 6 are now the
favoured fits to describe H1 data~\cite{adloff1997}.

POMWIG also allows us to include the effect of non-diffractive contributions through an additional Regge 
exchange, which we refer to as the reggeon contribution. This is expected to be important in the region 
of $\xi_{\bar{p}}$ explored at the Tevatron. Following H1, we estimate reggeon exchange by assuming that the
reggeon can be described by the pion parton densities of Owens \cite{Owens}. This contribution is added incoherently 
to the pomeron contribution. For further details of the implementation we refer the reader to \cite{pomwig}.

\section{Results}

\label{secdd:results}

In this section we compare our results using the IS model and POMWIG with the Tevatron 
DPE data. We start by testing POMWIG against an independent IS calculation, which we performed by the direct 
integration of equation~(\ref{secdd:eqdpe}) using the H1 pomeron flux and parton density. 
All results using POMWIG
are then shown at the parton level, at the hadron level and at the hadron level with out-of-cone
corrections implemented. To assist in the interpretation of the results, we then comment on results at the 
parton shower level.

\subsection{Hadronisation and parton shower effects}

\begin{figure}
\begin{center}
\epsfig{figure=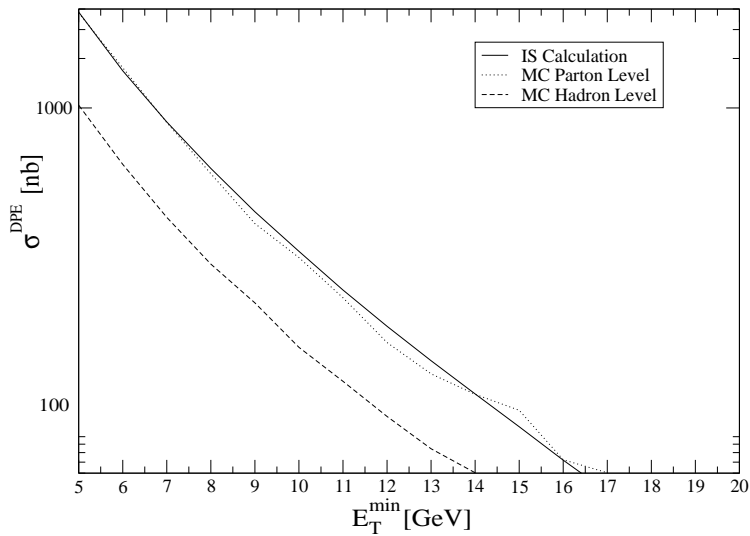,width=4.5in,height=3.5in}
\caption{The total cross-section at both parton and hadron level, plotted as
a function of the minimum transverse energy of the dijets.}
\label{figdd:isvmc}
\end{center}
\end{figure}

Figure \ref{figdd:isvmc} demonstrates the agreement between POMWIG and an independent 
calculation
for the total DPE cross-section as a function of the minimum jet transverse
energy $E^{min}_{T}$. These curves were produced using H1 fit 5 and contain
only the pomeron exchange contribution.
More interestingly, we see that the curve which describes the total 
cross section after the effects of fragmentation (hadronisation and parton showering) 
have been included shows
a significant reduction relative to the naive parton level calculation. This
suppression effect may be understood by observing that the effect is a shift
in the cross-section by $\Delta E_{T}=2$~GeV. This is a direct consequence
of the broadening of the jet profile by the parton shower and hadronisation.
The reduction is lower for the quark dominated fit (H1 fit 4) since quark jets
tend to have a narrower profile. In practice, this hadronisation suppression will reduce
the cross section by around a factor of 5.

\subsection{Total cross section}
We now turn our attention to the comparison of our theoretical predictions to
the measured total cross section. The analysis we perform includes the processes in which the 
exchange particles are either both pomerons or both reggeons. We do not include the case where one
is a reggeon and the other is a pomeron, nor do we include interference contributions.
Whilst the latter may be small, the former will not be if the pure reggeon
contribution is not negligible. This limitation arises since pomeron-reggeon interactions 
are not yet included in POMWIG. The results
are presented in table \ref{tabdd:sigma7} for $E_T^{\mathrm{min}}>7$~GeV and in table \ref{tabdd:sigma10} 
for $E_T^{\mathrm{min}}>10$~GeV at $\sqrt{s}=1.8$~TeV (for the data, the first error is statistical
and the second is systematic). The parton level result is found by considering the two partons produced
from the hard event as final state jets, the hadron level result is found by applying the cone jet algorithm 
to the produced hadrons, and the out-of-cone corrections result is found by applying the CDF out-of-cone corrections
to the hadron level result \footnote{We have noticed that there is a considerable difference between our results for the Tevatron
Run~I, at~$\sqrt{s}=1.8$~TeV, and results for the Tevatron Run~II, at~$\sqrt{s}=1.96$~TeV; for example the pomeron
only, fit~5 total cross section for the Tevatron Run II is 1036.6~nb at the parton level and 254.4~nb at the 
hadron level, compared 
to 859.3~nb at the parton level and 190.6~nb at the hadron level from table~\ref{tabdd:sigma7}.}.

%%%%%%%%%%%%%%%%%%%%%%%%%%%%%%%%%%%%%%%%%%%%%%%%%
\begin{table}
\begin{center}
\begin{tabular}{|c|c|c|c|}
\hline 
&
 Parton level {[}nb{]} &
 Hadron level {[}nb{]} &
 Hadron level + oc {[}nb{]}\\
\hline \hline
CDF Result &
&
&
 43.6 \( \pm  \) 4.4 \( \pm  \) 21.6\\
\hline 
I\( \!  \)P fit 4 &
6.4  &
2.2  &
7.3   \\
\hline 
I\( \!  \)P fit 5 &
859.3 &
190.6 &
661.8 \\
\hline 
I\( \!  \)P fit 6 &
886.7 &
230.8 &
702.1   \\
\hline 
I\( \!  \)R &
184.7     &
13.2     &
244.1      \\
\hline 
I\( \!  \)P+I\( \!  \)R fit 4 &
191.1     &
15.4     &
251.4     \\
\hline 
I\( \!  \)P+I\( \!  \)R fit 5 &
1044.0 &
203.8     &
905.9     \\
\hline 
I\( \!  \)P+I\( \!  \)R fit 6 &
1071.4     &
244.0     &
946.2      \\
\hline 
\end{tabular}
\end{center}
\caption{The total DPE cross-sections for a $E_T^{\mathrm{min}}$ cut of
$7$ GeV at $\sqrt{s}=1.8$~TeV. The contributions from pomeron and reggeon exchange are shown separately.}
\label{tabdd:sigma7}
\end{table}
%%%%%%%%%%%%%%%%%%%%%%%%%%%%%%%%%%%%%%%%%%%%%%%%

%%%%%%%%%%%%%%%%%%%%%%%%%%%%%%%%%%%%%%%%%%%%%%%%%
\begin{table}
\begin{center}
\begin{tabular}{|c|c|c|c|}
\hline 
&
 Parton level {[}nb{]} &
 Hadron level {[}nb{]} &
 Hadron level + oc {[}nb{]}\\
\hline \hline
CDF Result &
&
&
 3.4 \( \pm  \) 1.0 \( \pm  \) 2.0\\
\hline 
I\( \!  \)P fit 4 &
1.2     &
0.4     &
1.4      \\
\hline 
I\( \!  \)P fit 5 &
138.6     &
31.2     &
110.9      \\
\hline 
I\( \!  \)P fit 6 &
174.7     &
42.9     &
141.7     \\
\hline 
I\( \!  \)R &
13.2    &
0     &
13.2     \\
\hline 
I\( \!  \)P+I\( \!  \)R fit 4 &
14.4     &
0.4     &
14.6     \\
\hline 
I\( \!  \)P+I\( \!  \)R fit 5 &
151.8     &
31.2     &
124.1     \\
\hline 
I\( \!  \)P+I\( \!  \)R fit 6 &
187.9     &
42.9     &
154.9      \\
\hline 
\end{tabular}
\end{center}
\caption{The total DPE cross-sections for a $E_T^{\mathrm{min}}$ cut of
10 GeV at $\sqrt{s}=1.8$~TeV. The contributions from pomeron and reggeon exchange are shown separately.}
\label{tabdd:sigma10}
\end{table}
%%%%%%%%%%%%%%%%%%%%%%%%%%%%%%%%%%%%%%%%%%%%%%%%

Using fit 5, the overall cross section that we predict for a $E_T^{\mathrm{min}}$ cut of $7$~GeV
is~$203.8$~nb at the hadron level. When we apply out-of-cone corrections to this hadron level
result we obtain a total cross section of $905.9$~nb; this is close to the parton level result and
indicates that the out-of-cone corrections, as derived from Run I data, are approximately achieving what they were 
intended to do and sufficient 
amount of transverse energy is being added to the jets to ``correct them'' back to parton level. 
As we discussed in section 
\ref{secdd:tt}, it is the out-of-cone corrected data that we need to compare to the CDF experimental data.

This (out-of-cone corrected) predicted total cross section of 905.9 nb is in excess of the experimental value of 43.6 nb. 
A similar
excess is present with a~$E_T^{\mathrm{min}}$~cut of~$10$~GeV. However, we can match our results to
the data if we assume an overall multiplicative gap survival probability of
around 5\%\footnote{These estimates are extracted from the $E_T^{\mathrm{min}}=7$~GeV predictions but  
approximately apply to all cuts.}. The large reggeon contribution implies a non-negligible
pomeron-reggeon contribution and naively estimating this as twice the geometric mean
of the pomeron-pomeron and reggeon-reggeon contributions would push the gap
survival factor down to around 3\%. Given that the systematic error on the CDF 
cross sections is 
high, that the uncertainty in our knowledge of the gluon density
directly affects the normalisation of the cross section, and that the size of
the reggeon contribution is also uncertain it is not possible
to make a more precise statement about gap survival. In any case the value we
obtain agrees well with the expectations of \cite{kaidalov2001,gotsman1999}. 
Both fits~5 and~6 can describe the data in this way, although measurements of
diffractive dijet production at HERA suggest that fit~5 is favoured~\cite{dijets}.
The ratio of the fit~5 to the fit~4 cross sections is of the order of~100, 
which we can understand from the ratio of the gluon densities, illustrated in 
figure~\ref{figdd:gluondens}. This ratio is~10, which becomes~100 when we consider the gluons in both
pomerons. Note that the relative size of the reggeon contribution 
compared to the pomeron contribution is not small.

The suppression of the total cross section, relative to
the naive parton level result, resulting from the parton
shower phase in POMWIG has been looked at in~\cite{Appleby:2001xk}. 
Not surprisingly, the parton shower phase of POMWIG is responsible for a large
part of the suppression relative to the naive parton level prediction.

%We have performed the total cross-section calculation
%for H1 fit 5 after parton showering but before hadronisation. The results are
%presented in figure \ref{tabdd:sigmaps}, for an \( E^{min}_{T} \) of 7 GeV.

%\begin{table}
%{\centering \begin{tabular}{|c|c|c|c|}
%\hline 
%&
% Parton Level {[}nb{]} &
% Parton Shower {[}nb{]} &
% Hadron Level {[}nb{]} \\
%\hline 
%CDF result &
%&
%&
% 43.6 \( \pm  \)4.4 \( \pm  \)21.6\\
%\hline 
%I\( \!  \)P &
%     &
%     &
%     \\
%\hline 
%I\( \!  \)R &
%     &
%     &
%    \\
%\hline 
%I\( \!  \)P+I\( \!  \)R &
%      &
%     &
%      \\
%\hline 
%\end{tabular}\par}

%\caption{The total DPE cross-section at the parton shower
%and hadron levels. }
%\label{tabdd:sigmaps}
%\end{table}

%Not surprisingly, the parton shower phase of POMWIG is responsible for a large
%part of the suppression relative to the naive parton level prediction.

\subsection{Event distributions}

\begin{figure}
\begin{center}
\epsfig{figure=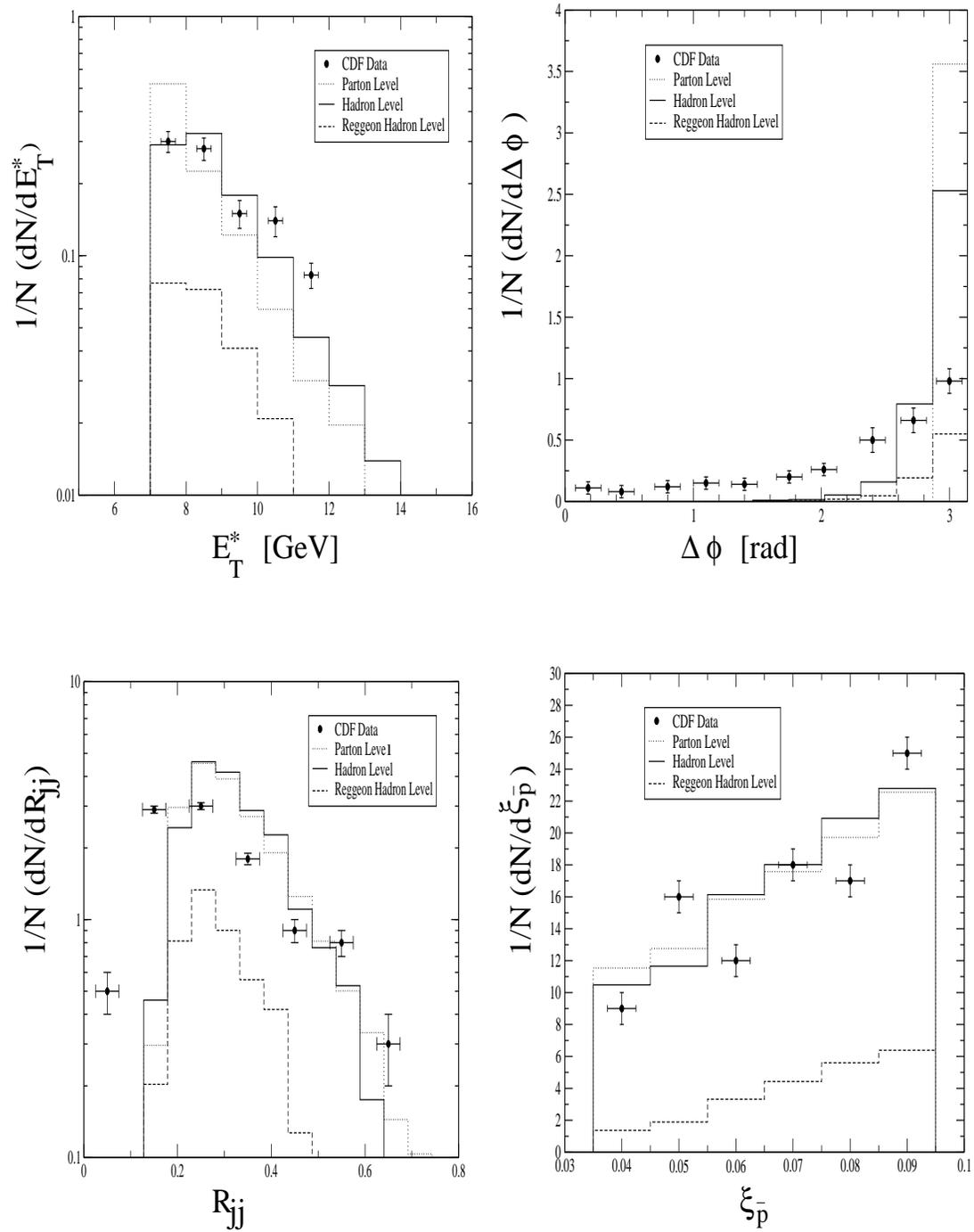,width=6in,height=8in}
\caption{Comparison of theoretical predictions at the parton and the hadron
level. Also shown is the separate contribution from the reggeon (normalised
relative to the total).}
\label{figdd:dista}
\end{center}
\end{figure}

\begin{figure}
\begin{center}
\epsfig{figure=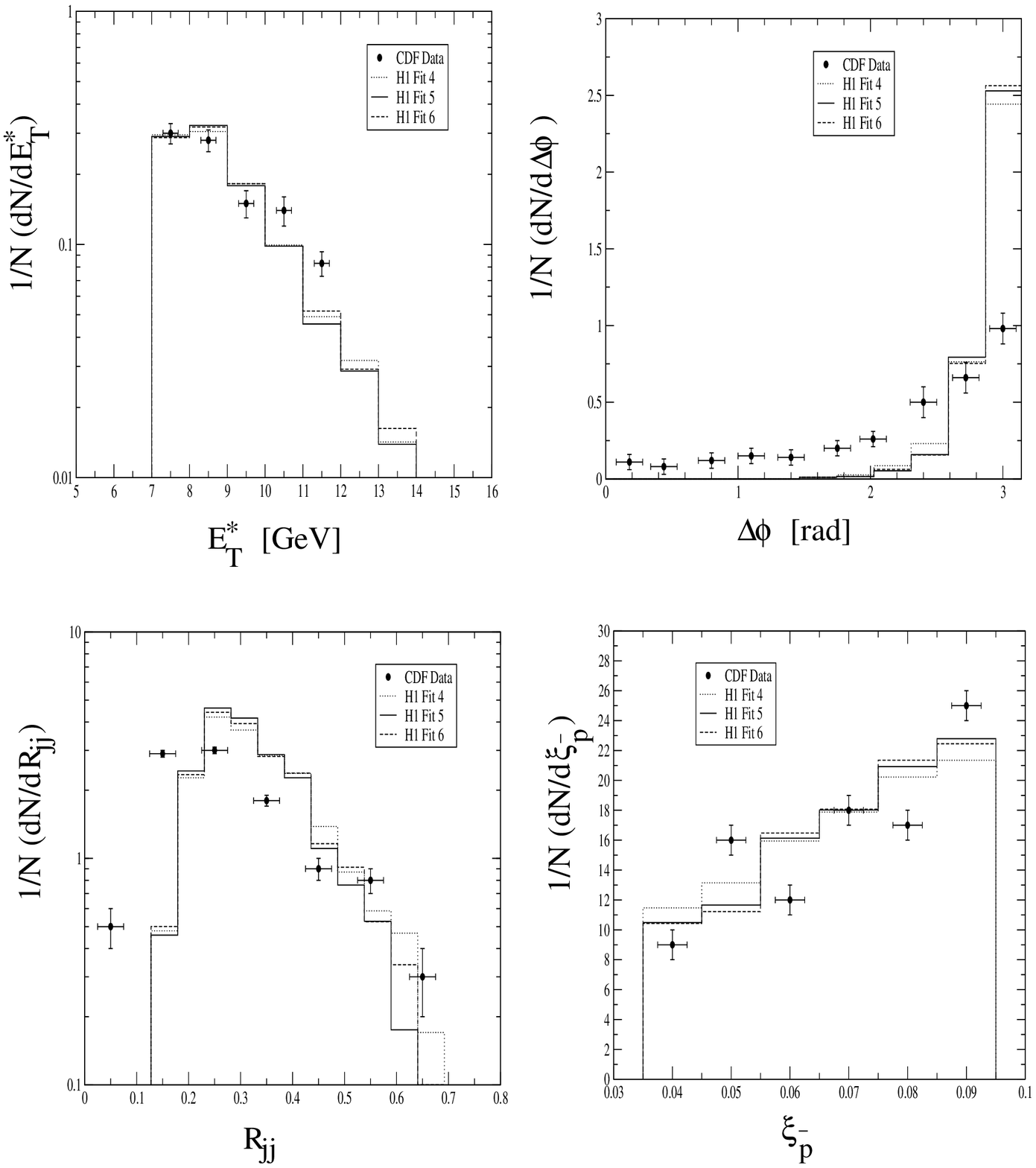,width=6in,height=8in}
\caption{Comparison of theoretical predictions for the different H1 fits to
the pomeron parton density functions at the hadron level.}
\label{figdd:distb}
\end{center}
\end{figure}

In figures \ref{figdd:dista} and \ref{figdd:distb} we show distributions in number of events of the mean jet 
transverse energy,
\( E_{T}^{*} \), the mean jet rapidity, \( \eta ^{*} \), the azimuthal separation
of the jets,~\( \Delta \phi  \), and the dijet mass fraction, \( R_{jj} \):
\begin{eqnarray}
R_{jj}=\frac{\sum_i E_i}{\xi_{p}\xi_{\bar{p}}s} \nonumber \\
\approx \beta_p\beta_{\bar{p}}.
\end{eqnarray}
The sum in the numerator is over all particles in the dijets.
Some of these distributions have also been examined in \cite{boonekamp2001}.
In figure \ref{figdd:dista} we compare the data to results at the parton and hadron level, and
we show the reggeon contribution separately\footnote{All curves except the 
reggeon are area normalised to unity. The reggeon is normalised
relative to the total.}. In figure~\ref{figdd:distb} we show results at the hadron level 
for the three different H1 pomeron parton density functions.

We urge caution when comparing the data and theory as is done in figures~\ref{figdd:dista} and~\ref{figdd:distb} 
since the data are not corrected for detector 
effects\footnote{Primarily because of the low \( E_{T} \) of the CDF jets.
}. The existence of the long tail to low angles in the \( \Delta \phi  \) 
distribution
illustrates the dangers: there is no possibility to produce such a long tail
in a hadron level Monte Carlo simulation. This tail is caused by misidentifying the two highest $E_T$ jets 
in a three-jet event and defining \( \Delta \phi  \) as the azimuthal separation between two 
neighbouring jets; this cannot occur in a Monte Carlo simulation that has not been corrected 
for detector effects. 
Hence we are unable to draw any
strong conclusions until the corrected data become available. 

\section{Conclusions}

\label{secdd:conc}

The work of this chapter has been focused on the so-called double pomeron exchange (DPE)
process recently measured at the Tevatron \cite{cdf2000}. This process is characterised by a double rapidity
gap separating the intact, diffracted interacting hadrons from a central dijet system.
We have extended previous 
calculations \cite{berera2000,acw} by including the effects of parton showering and hadronisation
and found that they lead to a suppression of the cross-section relative to 
the naive parton level by a factor of around 5. We also found that, in the 
kinematic region probed by the Tevatron, the effect of non-diffractive 
(reggeon) exchange is probably important. The out-of-cone corrections used
by CDF in their experimental analyses have been included in our cross section
predictions and we found that, if we use a gap survival probability \cite{bjorken1993} of around 5\%, we
can describe the data in a natural way. At the present time the issue of 
gap survival, which embodies a violation of diffractive factorisation,
  is not very well understood. However, it is encouraging that
we are able to describe the data with a gap survival probability which is consistent with 
previous theoretical estimates.

The Tevatron Run II has produced
around 10,000 DPE events, which are in the process of being analysed. Once this has occurred, and theoretical studies
like this one have been carried out, we will be in a far stronger position to understand double diffractive 
processes.

Finally, we can claim that one can use HERA partons at the Tevatron. When the parton level calculations of 
\cite{berera2000,acw} were carried out, the very large theory-to-experiment ratio meant that a description of
 the data was impossible, even with gap survival. However, a harder 
(and more appropriate) pomeron intercept, new parton densities in the pomeron and a gap survival factor combine 
to explain this excess.

\chapter{Resummation from factorisation}
\label{ch3}
% Chapter 3
%Insert Ch3 here!

\section{Introduction}

\label{secresum:intro}

In this survey chapter we shall discuss the factorisation of cross sections in specific regions of 
phase space \cite{Collins:ig}. We shall make
a statement of factorisation specific to colour exchange processes in QCD \cite{Contopanagos:1996nh}, 
and develop the consequences: 
the resummation of the soft and the jet functions 
\cite{Botts:nd,Botts:kf,Sotiropoulos:1993rd,Contopanagos:1996nh,Kidonakis:1999ze}. 
The final form of the cross section is derived, with 
the leading logarithms of the jet function and the next-to-leading logarithms of the soft function resummed, the 
latter being in terms of so-called soft anomalous dimension matrices 
\cite{Contopanagos:1996nh,Kidonakis:1998nf,Kidonakis:1997gm}. The application of these techniques to 
rapidity gap processes is then discussed, where we are interested in soft, wide angle radiation \cite{Kidonakis:1998nf} 
into a restricted angular
region. In these applications we are interested in colour evolution and not in the resummation of the colour-diagonal jets;
hence the soft logarithms become the leading logarithms.  
In this chapter we refer the reader to the literature for the proofs 
of the factorisation theorems we use \cite{Collins:ig}, and our aim is a survey of the consequences of factorisation and 
the phenomenological applications to rapidity gap processes.

\section{Factorisation}

\label{secresum:factor}

In this section we will describe the factorisation properties of the QCD cross sections we are interested in.
The resummation formalism developed by Collins, Soper and 
Sterman \cite{Collins:ig,Sotiropoulos:1993rd,Contopanagos:1996nh,Kidonakis:1998bk} (known as CSS) depends 
on the factorisation properties
obeyed by a cross section, with a hard scale $Q$, 
in a particular limit of its final state phase space. In such regions of phase
space, partons which are off-shell by order $Q^2$ are described by a hard scattering function $H$, and its complex
conjugate $H^*$. On-shell particles with momenta $\sim Q$ fall into ``jets'' of collinear particles, which will have
the interpretation of hadronic jets if they lie in the final state and as parton densities if they
lie in the initial state of a hadron-initiated process. We assume that the final state of this process is in 
the elastic limit and all finite energy particles are concentrated in the jet functions; hence the 
factorisation is valid at an ``edge'' of phase space for a given observable. This part of phase space, where there
is just enough energy to produce the final state jets and very little else, 
is known as the threshold region\footnote{The resulting resummation formalism, valid in this region of phase 
space, is consequently known as threshold resummation.}. In addition, the cross section involves 
the emission of soft
particles, represented by a function $S$.
In this work we are interested in hadron-initiated processes to jets, where it has been observed that 
there is no unique way of defining colour exchange in a finite amount of time. This is due to the fact that even
very soft gluons carry colour and we can exchange an arbitrary number of these long-wavelength gluons in the period
of time that is just after the short distance (short time) hard event. 
Therefore we expect the functions from which we construct the cross section
to be written as matrices in the space of possible colour flows - specifically the hard and soft functions, as the jets
themselves contain only collinear partons and are incoherent to the colour exchange. We can understand the need for a matrix
structure by noticing that as the factorisation scale changes, gluons that are part of the hard function may need 
to be moved to the soft function (or vice versa), changing the colour content of both. The 
matrix structure of the soft function 
will be discussed further in section \ref{secresum:eikonal}, where we discuss the construction of eikonal cross sections.

The elastic limit of the cross section is isolated by introducing an appropriate weight \cite{Kidonakis:1998bk} 
for each final state. The 
weight, which is a linear function of all final state particles if it is possible to resum the observable, 
determines the contribution of each state
to the cross section and vanishes in the elastic limit. In effect the weight parameterises the distance in phase 
space from 
the elastic limit. We shall assume that the contributions to the weight of particles in the soft function, $\omega_s$, 
and from particles in the jet functions, $\omega_i\,(i=1\dots n)$, are
additive,
\begin{equation}
\omega=\sum_i^n\omega_i+\omega_s,
\end{equation}
for a~$n$ jet cross section, and the corresponding statement of factorisation near the elastic limit is 
written as a convolution over the weights \cite{Contopanagos:1996nh},
\begin{eqnarray}
\sigma(\omega)=
H^{(0)}_{IJ}\left(\frac{p_i}{\mu},\xi_i\right)
\int && \frac{\mathrm{d}\omega_s}{\omega_s} \left[\prod_i^n \frac{\mathrm{d}\omega_i}{\omega_i} \right]  
 \nonumber \\
\times && \left[ \prod_i^n J_i^{(0)}\left(\frac{p_i\cdot \xi_i}{\mu},\frac{\omega_i Q}{\mu}\right)\right] \nonumber \\
\times &&S^{(0)}_{JI}\left(\frac{\omega_s Q}{\mu},\nu_i,\xi_i\right)
\delta(\omega-\sum_i^n \omega_i-\omega_s),
\end{eqnarray}
where we have suppressed the functional dependence on $\alpha_s$ and~$i=1\dots n$. Note that the arguments 
of~$H^{(0)}$ and~$S^{(0)}$ depend on all possible variables with $i$ subscripts. For example, for~$n=2$ we
would have~$H^{(0)}(p_1/\mu,p_2/\mu,\xi_1,\xi_2)$.
This expression is our starting point and is the cross section for a hard event $H$, described by 
$H_{IJ}$, with~$n$ light-like jets of particles, 
with momenta $p_i$. The factorisation scale is denoted $\mu$. The 
kinematic details are determined by the precise process 
under consideration; for example in $e^+ e^-$ annhilation ($n=2$), each $p_i$ labels an outgoing jet and we define 
\begin{equation} 
Q^2=(p_1+p_1)^2.
\end{equation}
The jets are described by the functions $J_i$.
The indices $I$ and $J$ label the colour structure of the hard scattering and each matrix has two indices~-~one
for the amplitude and another for its complex conjugate. The vectors $\xi_i$ are used to define a jet in terms
of a matrix element (just like a gauge-fixing vector) and hence are arbitrary, and the vectors $\nu_i$ define the
direction of the appropriate jet.
The function $S_{JI}$ describes the dynamics of partons with energy less than the soft scale $\omega_s Q$; 
we will consider its construction later in the chapter but it is enough to say that the soft 
dynamics of this function exactly match those of the 
full theory. Hence resummation of the soft function is equivalent to the resummation of the soft dynamics of the 
full cross section. We have denoted bare, unrenormalised quantities by a $(0)$ superscript, and deal with the
issue of renormalisation in section~\ref{secresum:renorm}.
The proof of the factorisation statement follows standard proofs of factorisation \cite{Collins:ig}, 
and shall not be discussed
in this thesis. In this chapter we are interested in the consequences of factorisation, and so will consider 
a simplified cross section with only~1 jet. Therefore~$Q^2=p^2$ and we write 
\begin{equation}
\sigma(\omega)=
H^{(0)}_{IJ}\left(\frac{p}{\mu},\xi\right)
\int \frac{\mathrm{d}\omega_J}{\omega_J}
\frac{\mathrm{d}\omega_s}{\omega_s} 
J^{(0)}\left(\frac{p\cdot \xi}{\mu},\frac{\omega_J Q}{\mu}\right)
S^{(0)}_{JI}\left(\frac{\omega_s Q}{\mu},\nu,\xi\right)
\delta(\omega-\omega_J-\omega_s),
\end{equation}
where we have suppressed the functional dependence on $\alpha_s$.
The convolution in weight unravels after the transformation,
\begin{equation}
\sigma(N)=\int_0^{\infty} d\omega e^{-N\omega} \sigma(\omega),
\end{equation}
where the moment variable~$N$ is conjugate to~$\omega$, and the leading behaviour in the small~$\omega$ limit is 
determined by the large~$N$ behaviour. 
%limit $\omega\rightarrow 0$ corresponds to $N$ becoming large. 
We therefore obtain, in moment space,
\begin{equation}
\sigma(N)=H^{(0)}_{IJ}\left(\frac{p}{\mu},\xi\right)
\tilde{J}^{(0)}\left(\frac{p\cdot \xi}{\mu},\frac{Q}{N\mu}\right)
\tilde{S}^{(0)}_{JI}\left(\frac{Q}{N\mu},\nu,\xi\right),
\end{equation}
where we define the transform of the soft function by
\begin{equation}
\tilde{S}^{(0)}_{JI}\left(\frac{Q}{N\mu},\nu,\xi\right)=
\int_0^{\infty} \frac{d\omega_s}{\omega_s} e^{-N\omega_s} S^{(0)}_{JI}\left(\frac{\omega_s Q}{\mu},\nu,\xi\right),
\end{equation}
with a similar definition for the transform of the jet function.
Noting the kinematical relation $p^\mu=Q \nu^\mu$ and writing the soft scale as $Q/N=Q_s$ we arrive at 
\begin{equation}
\sigma(N)=H^{(0)}_{IJ}\left(\frac{Q}{\mu},\nu\cdot\xi \right)
\tilde{J}^{(0)}\left(\frac{Q \nu\cdot \xi}{\mu},\frac{Q_s}{\mu}\right)
\tilde{S}^{(0)}_{JI}\left(\frac{Q_s}{\mu},\nu\cdot \xi\right),
\label{secresum:eq1jetfac}
\end{equation}
where we have slightly changed the notation. Equation~(\ref{secresum:eq1jetfac}) 
is our statement of factorisation, in moment space, for a 
schematic~1~jet cross section.

\section{Renormalisation group properties}

\label{secresum:renorm}

In this section we demonstrate the renormalisation of the contributing 
functions \cite{Botts:nd,Botts:kf,Contopanagos:1996nh,Kidonakis:1998bk} appearing
in the factorised cross section. We start from the factorised expression discussed in the last section, 
written with $n$ jets and in matrix notation,
\begin{equation}
\sigma=\mathrm{Tr}(\uuline{H}^{(0)}\,\uuline{S}^{(0)})\prod_i^n J_i^{(0)},
\end{equation}
in terms of the bare quantities. For the remaining part of this chapter, we shall drop the ``tilde'' notation for 
quantities in moment space; it will be clear from the context what is meant.
We assume that $J_i$ renormalises multiplicatively
\begin{equation}
J_i^{(0)}=Z_i(\mu)J_i(\mu)
\end{equation}
where $Z_i(\mu)$ is an UV renormalisation constant, $J_i(\mu)$ is the finite (renormalised) jet function
and $\mu$ is an arbitrary renormalisation scale (different to the factorisation scale used in the last 
section). By observing that the bare jet function must be
independent of the renormalisation scale,
\begin{equation}
\mu \frac{\partial}{\partial \mu}J_i^{(0)}=0,
\end{equation}
we arrive at the renormalisation group equation for $J_i$,
\begin{eqnarray}
\mu \frac{\partial}{\partial \mu}J_i&=&-Z_1^{-1} \left(\mu\frac{\partial}{\partial \mu}Z_i\right) J_i, \\
&=& -\gamma_i J_i.
\end{eqnarray}
In the last equality we have defined the (so called) jet anomalous dimension.

The renormalisation of $\uuline{S}$ comes from the UV limit of the virtual corrections. We assume that the amplitude 
and its complex conjugate renormalise multiplicatively,
\begin{equation}
\uuline{S}^{(0)}=\uuline{Z_s}(\mu)^{\dagger}\,\uuline{S}(\mu)\,\uuline{Z_s}(\mu).
\end{equation}
Again, $\uuline{Z_s}$ is a UV renormalisation constant, $\uuline{S}$ is the finite (renormalised) soft function
and $\mu$ is an arbitrary renormalisation scale. Exploiting the invariance of the bare soft function
against changes in $\mu$,
\begin{equation}
\mu \frac{\partial}{\partial \mu}\uuline{S}^{(0)}=0,
\end{equation}
means we arrive at the RGE for the soft function
\begin{eqnarray}
\mu \frac{\partial}{\partial \mu}\uuline{S}&=&
-\left[\uuline{Z_s}^{-1\,\dagger}\left(\mu \frac{\partial}{\partial\mu}\uuline{Z_s}^{\dagger}\right)\right]\uuline{S}
-\uuline{S}\left[\left(\mu \frac{\partial}{\partial\mu}\uuline{Z_s}\right)\uuline{Z_s}^{-1}\right], \\
&=& -\uuline{\Gamma_s}^{\dagger}\,\uuline{S}-\uuline{S}\,\uuline{\Gamma_s}.
\end{eqnarray}
In the last equality we have defined the soft anomalous dimensions. The scale-independence of $\sigma$ now allows
us to derive the renormalisation properties of $\uuline{H}$. Inserting the renormalised expressions for the bare
soft and jet functions into the factorised expression for the cross section gives
\begin{eqnarray}
\sigma&=&\mathrm{Tr}\left(\uuline{H}^{(0)}\uuline{Z_s}^{\dagger}\,\uuline{S}\uuline{Z_s}\right)
\prod_i Z_i J_i, \\
&=& \mathrm{Tr}(\uuline{H}\,\uuline{S})\prod_i J_i,
\end{eqnarray}
which implies that $\uuline{H}^{(0)}$ renormalises multiplicatively,
\begin{equation}
\uuline{H}^{(0)}=\prod_i Z_i^{-1}\uuline{Z_s}^{-1}\,\uuline{H}\,\uuline{Z_s}^{-1\,\dagger}.
\end{equation}
We can now write down the hard RGE,
\begin{equation}
\mu \frac{\partial}{\partial \mu}\uuline{H}=\sum_i \gamma_i \uuline{H} + \Gamma_s \uuline{H} 
+ \uuline{H}\uuline{\Gamma_s}^{\dagger},
\end{equation}
by noting that the bare hard function is renormalisation scale independent.

The discussion so far has been for an observable in the threshold region of its final state phase space, where 
there is no extra energy available for gluon emission. This restriction that the gluon emission must be 
very soft means that this factorisation is also valid for rapidity gap observables. In such cross sections the
gluon emission into a measured gap region is restricted, whilst being fully inclusive elsewhere, so the soft 
function depends on radiation into the gap $\Omega$, at scale $Q_s$. To be more general, the soft function also
depends on radiation out of the gap, the region $\bar{\Omega}$, at scale $M$. Therefore we can write
$\uuline{S}=\uuline{S}(Q_s/\mu,M/\mu)$ and trade the $\mu$ derivative for
$Q_s$ and $M$ derivatives to obtain
\begin{equation}
\left(Q_s\frac{\partial}{\partial Q_s} - \beta(g_s)\frac{\partial}{\partial g_s}\right)
\uuline{S}=
-\uuline{\Gamma_s}^{\dagger}\,\uuline{S}-\uuline{S}\,\uuline{\Gamma_s}-M\frac{\partial}{\partial M}\uuline{S}.
\end{equation}
In \cite{Berger:2001ns} the $M$ derivative has been interpreted to produce 
non-global logarithmic enhancements in $\uuline{S}$,
starting at $\alpha_s^2 \log^2(Q_s)$. Configurations producing these logarithmically enhanced terms arise from radiation being
emitted into $\Omega$ from secondary gluons in $\bar{\Omega}$ and as a result resolve the colour structure of the secondary
gluon. Therefore such contributions cannot be absorbed in an anomalous dimension matrix for~$2\rightarrow 2$~scattering in
a simple way, and we drop such non-global terms from our soft RGE. We will include such effects (non-global logarithms)
in a new way in chapter~\ref{ch4}. Therefore our soft RGE (correct to primary leading log level) is
\begin{equation}
\left(Q_s\frac{\partial}{\partial Q_s} - \beta(g_s)\frac{\partial}{\partial g_s}\right)
\uuline{S}=
-\uuline{\Gamma_s}^{\dagger}\,\uuline{S}-\uuline{S}\,\uuline{\Gamma_s}.
\end{equation}
We will solve the soft and jet function RGEs in the next section.

\section{Resummation}

\label{secresum:resum}

In this section we discuss how to use the factorised expression to produce resummed
expressions for the various contributing functions in the cross section. The resummation of the jet and 
soft functions parallels that of \cite{Contopanagos:1996nh}.

\subsection{Resummation of the jet functions}

The expression for the factorised 1 jet cross section (in moment space), introduced in section \ref{secresum:factor}, 
is
\begin{equation}
\sigma(Q,Q_s,\alpha_s)=
H_{IJ}\left(\frac{Q}{\mu},\nu\cdot\xi,\alpha_s(\mu) \right)
J\left(\frac{Q \nu\cdot \xi}{\mu},\frac{Q_s}{\mu},\alpha_s(\mu)\right)
S_{JI}\left(\frac{Q_s}{\mu},\nu\cdot \xi,\alpha_s(\mu)\right).
\end{equation}
In this equation, we have restored the $\alpha_s$ dependence of all the contributing functions.
We will now define the scaled gauge parameters, $r$, such that
\begin{equation}
r=\nu \cdot\xi \Rightarrow p\cdot \xi=Q r.
\end{equation}
This factorisation cannot depend on the exact choice of the quantity $r$,
\begin{equation}
r \frac{\partial}{\partial r} \sigma(Q,Q_s,\alpha_s)=0,
\end{equation}
and any changes in the function $J$ must be compensated by changes in the functions $S$ and $H$. Therefore
(suppressing function arguments)
\begin{equation}
\mathrm{Tr}\{\uuline{H}\,\uuline{S}\}J^{-1}r \frac{\partial}{\partial r}J=
-\mathrm{Tr}\left\{(r \frac{\partial}{\partial r}\uuline{H})\uuline{S}\right\}
-\mathrm{Tr}\left\{\uuline{H}(r \frac{\partial}{\partial r}\uuline{S})\right\}.
\end{equation}
If we assume that this equation is true element-by-element,
\begin{equation}
\uuline{H}\,\uuline{S} J^{-1} r \frac{\partial}{\partial r}J=
-(r \frac{\partial}{\partial r}\uuline{H})\uuline{S}
-\uuline{H}(r \frac{\partial}{\partial r}\uuline{S}),
\end{equation}
we obtain (restoring the function arguments)
\begin{eqnarray}
\uuline{\mathbbm{1}}J^{-1} r \frac{\partial}{\partial r}J\left(\frac{Q r}{\mu},\frac{Q_s}{\mu},\alpha_s(\mu)\right)&=&
-\uuline{H}^{-1} r \frac{\partial}{\partial r} \uuline{H}\left(\frac{Q}{\mu},r,\alpha_s(\mu)\right) \nonumber \\
&-&\left(r  \frac{\partial}{\partial r} \uuline{S}\left(\frac{Q_s}{\mu},r,\alpha_s(\mu)\right)\right)\uuline{S}^{-1}.
\end{eqnarray}
The log-derivative of $J$ may depend on the hard scale $Q$ or the soft scale $Q_s$. This variation can 
be split into two functions: one function which depends on the variables held in common by $J$ and 
$\uuline{H}$, namely $Q r/\mu$ and $\alpha_s(\mu)$, and one function which 
depends on the variables held in common by $J$
and $\uuline{S}$, namely $Q_s/\mu$ and $\alpha_s(\mu)$. Therefore we can write
\begin{equation}
\uuline{\mathbbm{1}}J^{-1} r \frac{\partial}{\partial r}J(\frac{Q r}{\mu},\frac{Q_s}{\mu},\alpha_s(\mu))=
G\left(\frac{Q r}{\mu},\alpha_s(\mu)\right)\uuline{\mathbbm{1}}
+K\left(\frac{Q_s}{\mu},\alpha_s(\mu)\right)\uuline{\mathbbm{1}},
\end{equation}
where we define
\begin{eqnarray}
\uuline{H}^{-1} r \frac{\partial}{\partial r} \uuline{H}\left(\frac{Q}{\mu},r,\alpha_s(\mu)\right)&=&
-G\left(\frac{Q r}{\mu},\alpha_s(\mu)\right)\uuline{\mathbbm{1}},  \\
\left(r  \frac{\partial}{\partial r} \uuline{S}\left(\frac{Q_s}{\mu},r,\alpha_s(\mu)\right)\right)\uuline{S}^{-1}&=&
-K\left(\frac{Q_s}{\mu},\alpha_s(\mu)\right)\uuline{\mathbbm{1}}.
\end{eqnarray}
The auxiliary function $G$ depends on the hard scale $Q$ and contains all the short-distance physics; conversely
the function $K$ depends on the soft scale $Q_s$ and contains all the long-distance physics. Both of these functions
depend on the gauge parameter~$r$.
Applying the RG-invariance operator $\mu \partial_{\mu}$ to this equation, swapping the order of 
the differential operators, using the RGE of $J$ and noting that $\gamma_{J}$ is $r$ independent 
implies that the combination $G+K$ is RG-invariant,
\begin{equation}
\mu \frac{\partial}{\partial\mu}\left[G\left(\frac{Q r}{\mu},\alpha_s(\mu)\right)
+K\left(\frac{Q_s}{\mu},\alpha_s(\mu)\right)\right] =0.
\end{equation}
Separating variables we obtain
\begin{eqnarray}
\mu \frac{\partial}{\partial\mu}G\left(\frac{Q r}{\mu},\alpha_s(\mu)\right)&=&+\gamma_K(\alpha_s(\mu)),  \\
\mu \frac{\partial}{\partial\mu}K\left(\frac{Q_s}{\mu},\alpha_s(\mu)\right)&=&-\gamma_K(\alpha_s(\mu)),
\end{eqnarray}
where the separation constants (Sudakov anomalous dimensions) only depend on the common variable $\alpha_s(\mu)$.
To solve these equations, we postulate the solution (which can be checked by differentiating)
\begin{equation}
K\left(\frac{Q_s}{\mu},\alpha_s(\mu)\right)=K(\quad)+\int_{\mu} \frac{d\mu'}{\mu'} \gamma_K(\alpha_s(\mu')).
\end{equation}
Now we set the upper limit of the integral to the scale $Q_s$ and choose the constant $K(\quad)$ (which denotes 
a $\mu$-independent $K$) to satisfy 
the RG equation. Doing this we obtain
\begin{eqnarray}
K\left(\frac{Q_s}{\mu},\alpha_s(\mu)\right)&=&
K(1,\alpha_s(Q_s))+\int_{\mu}^{Q_s} \frac{d\mu'}{\mu'} \gamma_K(\alpha_s(\mu')), \label{secresum:eqk} \\
G\left(\frac{Q r}{\mu},\alpha_s(\mu)\right)&=&
G(1,\alpha_s(Qr))+\int_{Qr}^{\mu} \frac{d\mu'}{\mu'} \gamma_K(\alpha_s(\mu')).
\end{eqnarray}
In order to combine these two expressions, 
we shift the evaluation on the coupling in equation (\ref{secresum:eqk}) from $Q_s$ to $Q r$,
\begin{eqnarray}
K(1,\alpha_s(Q r))&=&K(1,\alpha_s(Q_s))+\int_{\alpha_s(Q_s)}^{\alpha_s(Q r)} d\alpha_s 
\frac{\partial K(1,\alpha_s)}{\partial \alpha_s},  \\
&=&K(1,\alpha_s(Q_s))+\int_{Q_s}^{Q r} \frac{d\mu'}{\mu'}\beta(\alpha_s(\mu'))
\frac{\partial K(1,\alpha_s)}{\partial \alpha_s}.
\end{eqnarray}
We have used the QCD $\beta$-function in the last line, 
\begin{equation}
\alpha_s=\alpha_s(\mu')\Rightarrow \mathrm{d}\alpha_s=\beta(\alpha_s)\frac{\mathrm{d}\mu'}{\mu'},
\end{equation}
where $\beta(\alpha_s)=\mu' \partial \alpha_s/\partial \mu'$.
Combining the $K$ and $G$ functions, we now obtain
\begin{eqnarray}
G\left(\frac{Q r}{\mu},\alpha_s(\mu)\right)
&+&K\left(\frac{Q_s}{\mu},\alpha_s(\mu)\right)=G(1,\alpha_s(Qr))+K(1,\alpha_s(Qr)) \\
&-&\int_{Q_s}^{Q r} \frac{d\mu'}{\mu'}
\left(\gamma_K(\alpha_s(\mu'))+\beta(\alpha_s(\mu'))\frac{\partial K(1,\alpha_s)}{\partial\alpha_s}\right).
\end{eqnarray}
This expression can be written in the compact form
\begin{equation}
G\left(\frac{Q r}{\mu},\alpha_s(\mu)\right)
+K\left(\frac{Q_s}{\mu},\alpha_s(\mu)\right)
=-\int_{Q_s}^{Q r} \frac{d\mu'}{\mu'}A(\alpha_s(\mu')) - A'(\alpha_s(Qr)),
\label{secresum:eqgkdef}
\end{equation}
where we define
\begin{eqnarray}
A(\alpha_s)&=&\gamma_K(\alpha_s)+\beta(\alpha_s)\frac{\partial K(1,\alpha_s)}{\partial \alpha_s}, \\
A'(\alpha_s)&=&G(1,\alpha_s)+K(1,\alpha_s).
\end{eqnarray}
We now have two evolution equations for $J$, from RG invariance and the invariance of~$J$ against changes
in the gauge parameters $r$,
\begin{eqnarray}
\mu \frac{\partial}{\partial \mu} 
\log J\left(\frac{Q r}{\mu},\frac{Q_s}{\mu},\alpha_s(\mu)\right)&=&-\gamma_J(\alpha_s(\mu)), \\
r \frac{\partial}{\partial r}\log J\left(\frac{Q r}{\mu},\frac{Q_s}{\mu},\alpha_s(\mu)\right)&=&
G\left(\frac{Q r}{\mu},\alpha_s(\mu)\right)
+K\left(\frac{Q_s}{\mu},\alpha_s(\mu)\right).
\end{eqnarray}
The first of these two equations has the solution
\begin{equation}
\log J \left(\frac{Q r}{\mu},\frac{Q_s}{\mu},\alpha_s(\mu)\right)=
\log J \left(\frac{Q r}{Q_s},1,\alpha_s(Q_s)\right)-
\int_{Q_s}^{\mu}\frac{d\mu'}{\mu'} \gamma_J(\alpha_s(\mu')),
\end{equation}
which must also obey the second equation. This gives
\begin{equation}
r \frac{\partial}{\partial r} \log J \left(\frac{Q r}{Q_s},1,\alpha_s(Q_s)\right)
=G\left(\frac{Q r}{Q_s},\alpha_s(Q_s)\right)+K\left(1,\alpha_s(Q_s)\right),
\end{equation}
which has the solution (recall that both $G$ and $K$ depend on the gauge parameter $r$)
\begin{eqnarray}
\log J \left(\frac{Q r}{Q_s},1,\alpha_s(Q_s)\right)
&=&\log J \left(1,1,\alpha_s(Q_s)\right)\\
&+&\int_{Q_s/Q}^{r} \frac{dr'}{r'}
\left[G\left(\frac{Q r'}{\mu},\alpha_s(\mu)\right)+K(1,\alpha_s(Q_s))\right] \\
&=&\log J \left(1,1,\alpha_s(Q_s)\right)\\
&+&\int_{Q_s/Q}^{r} \frac{dr'}{r'}
\left[-\int_{Q_s}^{Q r'}\frac{d\xi}{\xi}A(\alpha_s(\xi))-A'(\alpha_s(Qr'))\right],
\end{eqnarray}
or, after a change of variable ($r'=\mu'/Q$),
\begin{eqnarray}
\log J \left(\frac{Q r}{Q_s},1,\alpha_s(Q_s)\right)
&=&\log J \left(1,1,\alpha_s(Q_s)\right)\\
&+&\int_{Q_s}^{Q r} \frac{d\mu'}{\mu'}
\left[-\int_{Q_s}^{\mu'} \frac{d\xi}{\xi}A(\alpha_s(\xi))-A'(\alpha_s(\mu'))\right].
\label{secresum:eqgkdef2}
\end{eqnarray}
Now we can write down an expression for $J$ which organises both the $Q r$ dependence and the
$N$ dependence (through $Q_s$),
\begin{eqnarray}
\log J \left(\frac{Q r}{\mu},\frac{Q_s}{\mu},\alpha_s(\mu)\right)&=& 
\log J \left(1,1,\alpha_s(Q_s)\right)
-\int_{Q_s}^{\mu}\frac{d\mu'}{\mu'} \gamma_J(\alpha_s(\mu')) \nonumber \\
&-&\int_{Q_s}^{Q r}\frac{d\lambda}{\lambda}
\left\{
\left[\int _{Q_s}^{\lambda} \frac{d\xi}{\xi}A(\alpha_s(\xi))\right]
+A'(\alpha_s(\lambda))
\right\}.
\end{eqnarray}
Extending the $\xi$ integration range using the $\Theta$-function,
\begin{equation}
I=\int_{Q_s}^{Qr} \frac{d\xi}{\xi}A(\alpha_s(\xi))\Theta(\lambda-\xi),
\end{equation}
allows us to switch the order of the $\xi$ and $\lambda$ integrals, and obtain
\begin{equation}
\int_{Q_s}^{Qr}\frac{d\xi}{\xi}A(\alpha_s(\xi))\int_{\xi}^{Qr}\frac{d\lambda}{\lambda}=
\int_{Q_s}^{Qr}\frac{d\xi}{\xi}A(\alpha_s(\xi))\log\left(\frac{Qr}{\xi}\right).
\end{equation}
Therefore we obtain our final form of the resummed expression for $J$, or in general for the
$i^{th}$ jet, of 
\begin{eqnarray}
\log J_i \left(\frac{Q r}{\mu},\frac{Q_s}{\mu},\alpha_s(\mu)\right)&=&
\log J_i \left(1,1,\alpha_s(Q_s)\right) \nonumber \\
&-&\int_{Q_s}^{\mu}\frac{d\lambda}{\lambda} \gamma_i(\alpha_s(\lambda)) \\
&-&\int_{Q_s}^{Qr}\frac{d\xi}{\xi} 
\left\{
A(\alpha_s(\xi))\log\left(\frac{Q r}{\xi}\right)+A'(\alpha_s(\xi))\right\}. \nonumber 
\end{eqnarray}
This can be generalised to an arbitrary number of jets by repeating the reasoning for each 
jet. Doing so we obtain
\begin{equation}
\prod_i J_i \left(\frac{Q r_i}{\mu},\frac{Q_s}{\mu},\alpha_s(\mu)\right)=
\prod_i J_i \left(1,1,\alpha_s(Q_s)\right) e^{-\mathcal{S}_{J_i}},
\label{secresum:eqjetresum}
\end{equation}
where we define the jet Sudakov exponent by
\begin{equation}
\mathcal{S}_{J_i}=
\int_{Q_s}^{\mu}\frac{d\lambda}{\lambda} \left( \gamma_i(\alpha_s(\lambda))\right)
+\int_{Q_s}^{Qr}\frac{d\xi}{\xi} 
\left\{
A(\alpha_s(\xi))\log\left(\frac{Qr}{\xi}\right)+A'(\alpha_s(\xi))\right\}.
\end{equation}
The resummation is controlled by the exponents $A$,$A'$ and the set $\gamma_i$. The latter are known as
anomalous dimensions, which are expressed as series in $\alpha_s$. This completes the resummation, in moment 
space, of the jet function.

\subsection{Resummation of the soft function}

The RGE for the soft function at scale $\mu$ is (suppressing $\alpha_s$ dependence)
\begin{equation}
\mu \frac{\partial}{\partial \mu}\uuline{S}(\mu)=
-\uuline{\Gamma_s^{\dagger}}\,\uuline{S}(\mu)-\uuline{S}(\mu)\,\uuline{\Gamma_s}.
\end{equation}
To solve this equation, we need to use the path-ordered exponential, which allows us to expand 
exponentials of non-commuting objects (in this case the non-commuting objects are matrices)
\begin{eqnarray}
\mathrm{P}\exp\left(\int_{\mu}^{\mu_0} \frac{d\mu'}{\mu'}\uuline{\Gamma_s}(\mu')\right)&=&
\mathbbm{1}+\int_{\mu}^{\mu_0}\frac{d\mu_1}{\mu_1}\uuline{\Gamma_s}(\mu_1)
+\int_{\mu}^{\mu_0}\frac{d\mu_1}{\mu_1}\int_{\mu_1}^{\mu_0}\frac{d\mu_2}{\mu_2} 
\uuline{\Gamma_s}(\mu_2)\uuline{\Gamma_s}(\mu_1) \nonumber \\
&+& \dots \,\,\,+ \nonumber \\
&+& \int_{\mu}^{\mu_0}\frac{d\mu_1}{\mu_1} \dots
\int_{\mu_{n-1}}^{\mu_0}\frac{d\mu_n}{\mu_n}\uuline{\Gamma_s}(\mu_n)\dots\uuline{\Gamma_s}(\mu_1).
\end{eqnarray}
We have introduced a new scale $\mu_0$ ($< \mu$).
In this definition of the path-ordered exponential, we arrange the matrices in order of increasing scale from
left to right (increasing $\mu$). Therefore at a new order of the expansion, we gain a new inner integration of a new 
variable $\mu_n$,
from $\mu_{n-1}$ to $\mu_0$, and place the new $\uuline{\Gamma_s}(\mu_n)$ on the left of the existing matrices. 
The anti-path ordered exponential, denoted $\bar{\mathrm{P}}$, is similar, except that we arrange the matrices in reverse order,
\begin{eqnarray}
\bar{\mathrm{P}}\exp\left(\int_{\mu}^{\mu_0} \frac{d\mu'}{\mu'}\uuline{\Gamma_s^{\dagger}}(\mu')\right)&=&
\mathbbm{1}+\int_{\mu}^{\mu_0}\frac{d\mu_1}{\mu_1}\uuline{\Gamma_s^{\dagger}}(\mu_1)
+\int_{\mu}^{\mu_0}\frac{d\mu_1}{\mu_1}\int_{\mu_1}^{\mu_0}\frac{d\mu_2}{\mu_2} 
\uuline{\Gamma_s}^{\dagger}(\mu_1) \uuline{\Gamma_s}^{\dagger}(\mu_2) \nonumber \\
&+& \dots \,\,\,+ \nonumber \\
&+& \int_{\mu}^{\mu_0}\frac{d\mu_1}{\mu_1} \dots
\int_{\mu_{n-1}}^{\mu_0}\frac{d\mu_n}{\mu_n}\uuline{\Gamma_s}^{\dagger}(\mu_1)\dots\uuline{\Gamma_s}^{\dagger}(\mu_n).
\end{eqnarray}
The effect of applying $\mu \partial_{\mu}$ to a path-ordered object can be seen by inspection to be
\begin{eqnarray}
\mu \partial_{\mu} \mathrm{P}\exp\left(\int_{\mu}^{\mu_0} \frac{d\mu'}{\mu'}\uuline{\Gamma_s}(\mu')\right)&=&
-\mathrm{P}\exp\left(\int_{\mu}^{\mu_0} \frac{d\mu'}{\mu'}\uuline{\Gamma_s}(\mu')\right)
\uuline{\Gamma_s}(\mu), \nonumber \\
\mu \partial_{\mu} \bar{\mathrm{P}}
\exp\left(\int_{\mu}^{\mu_0} \frac{d\mu'}{\mu'}\uuline{\Gamma_s^{\dagger}}(\mu')\right)&=&
-\uuline{\Gamma_s^{\dagger}}(\mu)\bar{\mathrm{P}}
\exp\left(\int_{\mu}^{\mu_0} \frac{d\mu'}{\mu'}\uuline{\Gamma_s}^{\dagger}(\mu')\right).
\end{eqnarray}
Therefore it is easy to see that we can solve the soft RGE by
\begin{equation}
\uuline{S}(\mu)=\bar{\mathrm{P}}
\exp\left(\int_{\mu}^{\mu_0} 
\frac{d\mu'}{\mu'}\uuline{\Gamma_s^{\dagger}}(\mu')
\right)
\uuline{S}(\mu_0)\mathrm{P}
\exp\left(\int_{\mu}^{\mu_0} 
\frac{d\mu'}{\mu'}\uuline{\Gamma_s}(\mu')\right).
\end{equation}
We have picked the scale of $\uuline{S}$ on the right-hand side so that the equation is true when $\mu=\mu_0$.
Note that if $\Gamma_s(\mu)$ commutes with $\Gamma_s(\mu')$ the path-ordering becomes
irrelevant and we recover the usual exponential form. For the soft function defined in this chapter, $\uuline{S}$ is
a function of the soft scale ratio $Q_s/\mu$, and so the RGE is given by
\begin{equation}
\mu \frac{\partial}{\partial \mu}\uuline{S}\left(\frac{Q_s}{\mu}\right)=
-\uuline{\Gamma_s^{\dagger}}\,\uuline{S}\left(\frac{Q_s}{\mu}\right)
-\uuline{S}\left(\frac{Q_s}{\mu}\right)\,\uuline{\Gamma_s}.
\end{equation}
This equation is solved by
\begin{equation}
\uuline{S}\left(\frac{Q_s}{\mu}\right)=\bar{\mathrm{P}}
\exp\left(\int_{\mu}^{Q_s} 
\frac{d\mu'}{\mu'}\uuline{\Gamma_s^{\dagger}}(\mu')
\right)
\uuline{S}(1)\mathrm{P}
\exp\left(\int_{\mu}^{Q_s} 
\frac{d\mu'}{\mu'}\uuline{\Gamma_s}(\mu')\right),
\label{secresum:eqsoftsol}
\end{equation}
where we have picked the arbitrary scale $\mu_0=Q_s$.

\subsection{The resummed expression}

We can now present (in matrix notation) the resummed cross section in moment space for our 1 jet cross section. 
We have chosen $\mu=Q$ throughout, and so the hard function becomes $\uuline{H}(Q/\mu)\rightarrow\uuline{H}(1)$ and
the integrals in the path ordered exponentials of the soft function and the jet Sudakov exponent 
are now between $Q$ and $Q_s$.
We have made the choice of $r=1$ for the gauge parameter. We obtain
\begin{equation}
\sigma(Q,Q_s,\alpha_s)=\mathrm{Tr}
\left\{
\uuline{H}(1)\bar{\mathrm{P}}
\exp\left(\int_{Q}^{Q_s}\frac{d\lambda}{\lambda}\uuline{\Gamma_s^{\dagger}}(\lambda)\right)
\uuline{S}(1)
\exp\left(\int_{Q}^{Q_s}\frac{d\lambda}{\lambda}\uuline{\Gamma_s}(\lambda)\right)
\right\} J\left(1,\frac{Q_s}{Q}\right),
\end{equation}
where we have not written the full expression for the
resummed jet function, which can be read off from equation (\ref{secresum:eqjetresum}).

\section{Diagonalisation}

\label{secresum:diag}

In this section we develop the solution, equation (\ref{secresum:eqsoftsol}), of the soft RGE.
We do this by choosing a basis
in which the $\uuline{\Gamma_s}$ matrices are diagonal \cite{Kidonakis:1998nf,Berger:2001ns}, and hence 
are given by the eigenvalues of the
matrix $\uuline{\Gamma_s}$,
\begin{equation}
\left(\tilde{\Gamma}_s\right)_{\gamma\beta}=\lambda^{(\beta)}\delta_{\gamma\beta},
\end{equation}
where $\lambda_\beta$ denotes the eigenvalues of $\uuline{\Gamma_s}$. We will denote quantities in the diagonal
basis with tildes. If $\uline{e}^{(J)}$ is the $J^{\mathrm{th}}$
eigenvector of $\uuline{\Gamma_s}$ (with eigenvalue $\lambda^{(J)}$) then the matrix whose columns are the eigenvectors,
\begin{equation}
R_{IJ}=e_I^{(J)},
\end{equation}
will diagonalise $\uuline{\Gamma_s}$. We therefore write the anomalous dimension matrices in the new diagonal basis using a
similarity transform,
\begin{equation}
\left(\tilde{\Gamma_s}(\Delta\eta,\Omega)\right)_{\gamma \beta}=\lambda^{(\beta)} \delta_{\gamma \beta}
=R^{-1}_{\gamma I} \left(\Gamma_s(\Delta\eta,\Omega)\right)_{IJ}R_{J \beta}.
\end{equation}
To do this we need to explicitly expand the anomalous dimension matrices to one-loop (work at leading 
order in $\alpha_s$),
\begin{equation}
\uuline{\Gamma_s}(\lambda)=\frac{\alpha_s(\lambda)}{\pi}\uuline{G}.
\end{equation} 
where $\uuline{G}$ is a constant (i.e. scale independent) matrix. This implies that different scales will 
commute with each other (all scales have the same constant matrix $\uuline{G}$) inside of the path-ordering, and the
matrix can come outside of the integral on the left hand side. Hence
\begin{equation}
\mathrm{P}\exp\left(\int_Q^{Q_s}\frac{d\lambda}{\lambda}\uuline{\Gamma_s}(\lambda)\right)
=\exp\left(\uuline{G}\int_Q^{Q_s}\frac{d\lambda}{\lambda}\frac{\alpha_s(\lambda)}{\pi}\right).
\end{equation} 
Now by inserting $\uuline{R}\,\uuline{R}^{-1}=\uuline{I}$ as required, we rewrite the path-ordered exponential as
\begin{equation}
\left(
\uuline{R}^{-1}\mathrm{P}\exp\left(\int_Q^{Q_s} \frac{d\lambda}{\lambda} \uuline{\Gamma_s}(\lambda)\right)
\uuline{R}\right)_{IJ}=\delta_{IJ}\exp\left(\int_Q^{Q_s}\frac{d\lambda}{\lambda}\lambda^{(J)}(\lambda)\right),
\end{equation}
where the eigenvalues have been expanded to one-loop,
\begin{equation}
\lambda^{(I)}=\frac{\alpha_s}{\pi}\lambda^{(I),1},
\label{secresum:eqlambda}
\end{equation}
and we have moved the constant piece of the eigenvalues back inside the integral over~$\lambda$.
Transforming the hard and soft matrices to the new basis,
\begin{eqnarray}
\uuline{\tilde{S}}&=&\uuline{R}^{\dagger}\,\uuline{S}\,\uuline{R}, \\
\uuline{\tilde{H}}&=&\uuline{R}^{-1}\,\uuline{S}\,\uuline{R}^{-1\,\dagger},
\end{eqnarray} 
allows us to write down the resummed expression for the cross section (with $i$ jets) in the diagonal basis,
\begin{equation}
\sigma(Q,Q_s,\alpha_s)=\mathrm{Tr}
\left\{
\uuline{\tilde{H}}(1)
\uuline{\tilde{S}}(1)
\exp\left(\int_Q^{Q_s}\frac{d\lambda}{\lambda}\left(\lambda^{*(J)}+\lambda^{(I)}\right)\right)
\right\}
\prod_i J_i\left(1,\frac{Q_s}{Q}\right).
\end{equation}
By introducing the following combination of the eigenvalues,
\begin{equation}
E^{(IJ)}=\lambda^{(I),1\,*}+\lambda^{(J),1},
\end{equation}
using equation (\ref{secresum:eqlambda}), we write the resummed expression as
\begin{equation}
\sigma(Q,Q_s,\alpha_s)=\mathrm{Tr}
\left\{
\uuline{\tilde{H}}(1)
\uuline{\tilde{S}}(1)
\exp\left(E^{(IJ)} \int_Q^{Q_s}\frac{d\lambda}{\lambda} \frac{\alpha_s(\lambda)}{\pi} \right)
\right\}
\prod_i J_i\left(1,\frac{Q_s}{Q}\right).
\end{equation}
Inserting the one-loop definition of the running
coupling,
\begin{equation}
\alpha_s(Q)=\frac{2\pi}{\beta_0 \log(Q/\Lambda)},
\end{equation}
where $\beta_0=11-\frac{2}{3}n_f$ and $\Lambda\sim0.2$~GeV, allows us to rewrite the soft Sudakov exponential as
\begin{equation}
\exp\left(E^{(IJ)} \int_Q^{Q_s}\frac{d\lambda}{\lambda} \frac{\alpha_s(\lambda)}{\pi} \right)
=\left(\frac{\log(Q_s/\Lambda)}{\log(Q/\Lambda)}\right)^{\frac{2 E^{(IJ)}}{\beta_0}}.
\end{equation}
We have, therefore, written a resummed soft function in moment space in which all leading logarithms
of $Q_s/Q$ have been rearranged (or resummed) in terms of the eigenvalues of the soft anomalous dimensions, which are
matrices in the space of possible colour flows of the system. From the factorisation arguments earlier in this 
chapter, the resummed soft function contains the same soft logarithms to all-orders as the full cross section.

\section{Resummation and colour flow in rapidity gap processes}

\label{secresum:rapgap}

The Collins-Soper-Sterman resummation formalism discussed in this chapter takes place in moment space, and an 
inverse Mellin transform is required to undo the transformation. The two main ways to do this are a fixed
order expansion of the exponents \cite{Kidonakis:1999ze}, where the inverse Mellin transforms are 
trivial, and by a numerical
inversion prescription \cite{Berger:1996ad}. In the latter case the cross section depends on an integral 
in moment space over all scales,
including the Landau pole, and a prescription must be used to avoid this region. 
The disadvantage of this 
procedure is that the resummed cross section becomes prescription
dependent. For a discussion and further references on the inversion procedure see \cite{Berger:1996ad,Kidonakis:2000ui}. 

In this thesis we are interested in the dynamics of wide-angle soft radiation into a restricted
angular region $\Omega$ and the role of colour flow in the hard scattering. The colour-diagonal 
jet functions are incoherent to
the soft function colour structure and only include the dynamics of collinear soft radiation. Therefore in the
study of rapidity gap processes, we only consider the soft function dynamics and the interjet logarithms, 
which arise from soft, wide-angle emission are the leading logarithms.
In general, the differential distribution in the gap energy $Q_s$ is given by
\begin{equation}
\frac{1}{\sigma}\frac{d\sigma}{dQ_s}=
\delta(Q_s)
+ \sum_{n\ge 1}\alpha_s(Q)^n \sum_{0\le m \le n-1}C_{nm}
\left(\frac{1}{Q_s}\log^m\frac{Q_s}{Q}\right)_+,
\label{secresum:eqdiffdist}
\end{equation}
where we see logarithmic enhancements of $Q/Q_s$, $Q_s<Q$, at each order in the perturbation series. We have
only shown the leading behaviour for $Q_s\ll Q$ (there are other non-singular terms at each order in 
perturbation theory) and the coefficients~$C_{nm}$ are numbers. The terms in
the plus distribution 
diverge as $Q_s\rightarrow 0$, singular but integrable in this limit. Here we recall the 
definition of the plus distribution for the function $g(z)$,
\begin{equation}
\left[\frac{g(z)}{1-z}\right]_+=\frac{g(z)}{1-z}-\delta(1-z)\int_0^1 dz' \frac{g(z)}{1-z'},
\end{equation}
where
\begin{equation}
z=1-\frac{Q_s}{Q}.
\end{equation}
This definition of the plus distribution can be expressed as an integral with some smooth function $f(x)$,
\begin{equation}
\int_z^1 dx\,f(x) \left[\frac{g(x)}{1-x}\right]_+=
\int_z^1dx\,\bigg(f(x)-f(1)\bigg)\frac{g(x)}{1-x}-f(1)\int_0^zdx\,\frac{g(x)}{1-x}.
\end{equation}
The case of~$m=n-1$ is the leading logarithmic set. The corresponding integrated distribution 
in momentum space is
\begin{equation}
\Sigma(Q,Q_\mathrm{max})=\int_0^{Q_\mathrm{max}} \frac{1}{\sigma} \frac{d\sigma}{dQ_s} dQ_s
=1+\sum_{n\ge 1} \alpha_s(Q)^n  \sum_{0\le m\le n-1} C_{nm} 
\left(\frac{1}{m+1}\log^{m+1}\frac{Q_{\mathrm{max}}}{Q}\right).
\label{secresum:eqintdist}
\end{equation}
We can take the Mellin moment of the differential cross section, equation (\ref{secresum:eqdiffdist}), using
\begin{equation}
\int_0^1 dz\,z^{N-1}\left[\frac{\log^m(1-z)}{1-z}\right]_+
=\frac{1}{m+1}\log^{m+1}\frac1N +\mathcal{O}\left(\log^{m}N\right),
\end{equation}
which generates terms like
\begin{equation}
\sigma(N) = 1+\sum_{n \ge 1} \alpha_s(Q)^n \sum_{0 \le m \le n-1}
C_{nm}\left(\frac{1}{m+1}\log^{m+1}1/N + \mathcal{O}(\log^{m}N)\right),
\end{equation}
where we only show the leading logarithms in moment space. 
Note that, in the limit of large~$|N|$, the Mellin transform used in this section is equivalent 
to the Laplace transform used in section~\ref{secresum:factor}.
The soft resummation procedure organises terms
like this (at LL) and hence amounts to a knowledge of the LL coefficients to all-orders. 
Comparing this with the integrated
distribution in momentum space, equation (\ref{secresum:eqintdist}), 
we see that the coefficients of the LL term in momentum space correspond to the soft resummed terms in
Mellin space. Therefore, to LL level, our resummation of the soft function in Mellin space corresponds to
a resummation of all leading soft logarithms in momentum space. Therefore we can write down the factorised
cross section for the production of~2 jets at fixed rapidity interval as
\begin{equation} 
\frac{d\sigma}{d\Delta\eta}=
H_{IL}(\hat{s},\hat{t},\mu,\alpha_s)S_{LI}(Q_s,\mu,\alpha_s),
\label{secresum:eqfac}
\end{equation}
where $S_{LI}$ is a soft matrix in momentum space, $\hat{s}$ and $\hat{t}$ are the Mandelstam variables 
and we write the soft scale as $Q_s$. 
The resummation of the soft function in moment space presented in this chapter therefore 
allows logarithms of $Q_s/Q$ to be resummed 
in $d\sigma/d\Delta\eta$, the cross section in momentum space.

\section{Eikonal cross sections}

\label{secresum:eikonal}

In this section we will briefly describe the interpretation of the soft function, which appears in the factorised
expression of the cross section. This topic has been studied in \cite{Botts:kf,Kidonakis:1998nf}. 

This function, which describes the emission of soft radiation in the cross section, is constructed 
by representing the initial and final state partons by Wilson lines. The soft radiation pattern of these Wilson line 
constructions, so-called eikonal cross sections, exactly matches that of the full theory. We represent the initial 
and final 
state particles by an ordered exponential of the gluon field,
\begin{equation}
\Phi_{\beta}^{(f)}(\lambda_1,\lambda_2;x)=\mathrm{P}\exp\left(
-ig \int_{\lambda_1}^{\lambda_2} dl\,\beta\cdot A^{(f)}(l\beta+x)\right),
\end{equation}
where the integration is a straight line path in the direction of the particle 4-velocity~$\beta$. The vector
potential~$A^{(f)}$ is in the appropriate representation to correctly describe a parton of flavour~$f$. Such 
a Wilson, or eikonal, line is a special case of the more general Wilson line which appears many times in 
quantum field theory, where
the path can be any curve~$C$. The path ordering symbol is needed to define the Wilson line in terms of a power
expansion of the exponential with the SU(3) matrices ordered so that higher values of~$l$ stand to the left.
The wide angle, soft radiation pattern which is obtained from replacing each of the partons in the hard event by Wilson
lines exactly matches the wide, angle soft radiation pattern of the full event. Therefore we construct an operator
consisting of products of Wilson lines, tied together by their colour indices. As a transparent example, consider  
two Wilson lines representing partons in a Drell-Yan process,
\begin{equation}
W^{(DY)}_{b2,b2}(x)=\delta_{a_1 a_2} \Phi^{(\bar{q})}_{\beta_2}(0,-\infty;x)_{a_2 b_2}
\Phi^{(q)}_{\beta_1}(0,-\infty;x)_{a_1 b_1},
\end{equation}
where the two Wilson lines extend from negative infinity and meet at the vertex~$x$. 
The initial colour content of the lines 
is represented by the indices~$b$, and the colour exchange at the vertex (in this case singlet) is represented by
the tensor~$\delta_{a_1 a_2}$. This vertex, when dressed with higher order gluons, contains UV divergences; this is
also true of the one-loop external leg self-interaction diagram. The sum of these is renormalised in the following manner,
\begin{equation}
W^{(DY)(0)}(x)=Z_W^{(DY)}\left(\alpha_s,\beta_1,\beta_2\right)W^{(DY)}(x),
\end{equation}
and the invariance of the renormalised operator under changes in the renormalisation scale implies that the renormalised 
quantity obeys the standard RGE,
\begin{equation}
\mu\frac{d}{d\mu}W^{(DY)}=\Gamma_W\left(\alpha_s,\beta_1,\beta_2\right)W^{(DY)},
\end{equation}
where we define the colour singlet anomalous dimension,
\begin{equation}
\Gamma_W=\frac{1}{Z_W} \mu \frac{d}{d\mu} Z_W.
\end{equation}
A one-loop calculation gives
\begin{equation}
\Gamma_W=\frac{\alpha_s}{\pi}C_F \left[\log\left(\frac{-2\beta_1\cdot\beta_2}{\sqrt{\beta_1^2 \beta_2^2}}\right)
-1\right],
\end{equation}
where $\beta_i$ is  the 4-velocity of eikonal line~$i$. The divergences $\beta_i^2\rightarrow 0$ have the 
interpretation of collinear singularities.

We can build eikonal operators for 4-jet processes by tying together four Wilson lines, two for the initial 
state partons and two for the final state partons. The colour structure is now more complicated, and the eikonal 
operator is constructed with a colour structure $c$ connecting the colour indices of the eikonal lines.
We now choose to span the colour space of the process with the set of colour tensors $c_I$, and expand $c$ 
over this basis. Therefore the set of process-dependent tensors $c_I$ link the four Wilson line colour indices 
together. Hence we 
define the eikonal operator,
\begin{eqnarray}
W_I^{(f)}(x)_{r_k}=
&&\sum_{d_i} 
\Phi_{\beta_2}^{(f_2)}(\infty,0;x)_{r_2,d_2} \Phi_{\beta_1}^{(f_1)}(\infty,0;x)_{r_1,d_1}
\left(c_I^{(f)}\right)_{d_2 d_1,d_B d_A} \nonumber \\
&& \times \Phi_{\beta_A}^{(f_A)}(0,-\infty;x)_{d_A,r_A} \Phi_{\beta_B}^{(f_B)}(0,-\infty;x)_{d_B,r_B},
\end{eqnarray}
for the process $f_A(l_A,r_A)+f_B(l_B,r_B)\rightarrow f_1(p_1,r_1)+f_2(p_2,r_2)$, with a 
corresponding notation,~$l$ and~$p$, 
for the particle 4-velocities. The index~$r$ denotes the colour index of the relevant parton. The $c_I$'s form a basis in colour space; an example would be singlet and octet
exchange in the t-channel.
The initial state Wilson lines extend from negative infinity to the vertex at $x$, and the final state Wilson lines
extend from the vertex at $x$ to positive infinity. This operator should reproduce the same wide angle, soft radiation 
as the full theory. In terms of the operators $W_I^{(f)}$ we can define an eikonal soft function for fixed energy 
$Q_{\mathrm{gap}}$ flowing into $\Omega$,
\begin{equation}
S_{LI}^{(f)}\left(\frac{Q_{\mathrm{gap}}}{\mu},\Omega\right)=\sum_{\xi} \delta(Q_{\mathrm{gap}}-Q_s^{(\xi)})
\langle0|\overline{T}\left[(W_L^{(f)}(0))^{\dagger}_{\{b_i\}}\right]|\xi\rangle\langle\xi|
T\left[W_I^{(f)}(0)_{\{b_i\}}\right]|0\rangle,
\label{secresum:eqsoftfuncdiff}
\end{equation}
and the corresponding integrated soft function,
\begin{eqnarray}
S_{LI}^{(f)}\left(\frac{Q_s}{\mu},\Omega\right)&=&
\int_0^{Q_s} dQ'_s
\sum_{\xi} \delta(Q_s^{(\xi)}-Q'_s) \nonumber \\
&&\langle0|\overline{T}\left[(W_L^{(f)}(0))^{\dagger}_{\{b_i\}}\right]|\xi\rangle\langle\xi|
T\left[W_I^{(f)}(0)_{\{b_i\}}\right]|0\rangle,
\label{secresum:eqsoftfuncint}
\end{eqnarray}
which are defined as matrices in the space of the colour tensors~$c_I$.
In these equations~$Q_s^{(\xi)}$ is the energy flowing into the gap region 
of the intermediate state~$\xi$ and 
we sum over all intermediate states contributing to the cross section. 
If we demand the sum is over all intermediate states of 
maximum energy~$Q_s$ into the gap region~$\Omega$ (i.e. away from the eikonal lines),
the eikonal cross section is free from potential collinear singularities.

If we define the soft function using the differential form, equation (\ref{secresum:eqsoftfuncdiff}), in moment space 
then, following the arguments in section \ref{secresum:rapgap}, the resummation of this soft function is 
equivalent to the resummation of the integrated soft function 
in momentum space. Hence we can write the soft function in momentum space as equation (\ref{secresum:eqsoftfuncint}), with
the associated factorisation statement of equation (\ref{secresum:eqfac})
and resum the soft logarithms using the resummation formalism (defined using equation (\ref{secresum:eqsoftfuncint}))
described in this chapter.

If there are no soft gluons, at tree level, then equations (\ref{secresum:eqsoftfuncdiff}) 
and (\ref{secresum:eqsoftfuncint})
reduce to colour traces.  A set of eikonal Feynman rules
have been extracted from the definition of the eikonal line \cite{Botts:kf,Kidonakis:1998nf} to allow the 
computation of the 
UV poles of the soft function and hence the soft anomalous dimension matrices, which were defined in section
\ref{secresum:resum}. The higher order diagrams which contribute to the soft anomalous dimension matrix are 
known as eikonal diagrams. The full set of eikonal Feynman rules are in appendix~\ref{secresum:appeikonal}. 
For a review of these ideas, see the references given above.

\section{Colour mixing under renormalisation}

\label{secresum:mixing}

The resummation in this chapter is driven by the eigenvalues of the soft anomalous dimension matrices, which 
are expressed as matrices in the space of the possible colour flows of the system. The matrices are found from 
the coefficients of the UV poles of higher order contributions to the soft function, which was defined in
section~\ref{secresum:eikonal}; hence the calculation of 
the anomalous dimension matrices is equivalent to the calculation of the renormalisation constants of eikonal 
diagrams, discussed in section~\ref{secresum:eikonal}. 
Once we specify a basis of colour tensors, the higher order gluons which cause the UV divergences result in 
the set of colour tensors mixing into each other~\cite{Berger:2001ns}. In other words, we generate a quantum 
mechanical mixing matrix 
by ``dressing'' a colour tensor~$c_I$~\cite{Botts:nd} with soft gluons and project the result along the 
direction of~$c_J$. This is
just standard mixing of operators under quantum corrections. The calculation of the colour mixing matrices was 
discussed in section~\ref{secintro:mixing} and the full set of matrices appear in 
appendix~\ref{secef:appdecomp}, along with the set of colour bases used in this work in appendix 
\ref{secef:appbases}.

\section{Conclusions}

\label{secresum:conc}

In this chapter we have discussed the factorisation theorems of the cross section used in the rest of this 
thesis. These properties were used to write down resummed soft and jet functions in moment space.  
We then turned our attention to the application of these results to rapidity gap processes, and we argued that,
as we are only interested in wide angle, soft radiation into a gap region, we can neglect the colour-diagonal jets
and focus on the soft function. The inversion of the Mellin transform is a trivial issue in this situation and we
are able to write down a factorised cross section, equation~(\ref{secresum:eqfac}), in momentum space where the soft 
function has the interpretation of an eikonal cross section. The renormalisation properties of the eikonal cross section
 then allow the soft logarithms to be resummed.

We will apply these results to gaps-between-jets processes in chapter~\ref{ch5}, but first we have to deal with a class
of leading logarithms not included in the resummed soft function. We deal with this complication, 
non-global logarithms, in the next chapter.

\chapter{Non-global logarithms in interjet energy flow}
\label{ch4}
% Chapter 4

\section{Introduction}

\label{secngl:intro}

The identification of the inter-jet energy flow as an infrared safe way to study gaps-between-jets processes has produced
extensive interest in the last few years. By considering such observables  we may start to formulate a perturbative approach to
the description of the cross section, as well as probe the interface between perturbative and non-perturbative physics
at energy scales $\sim 1$~GeV. The inter-jet energy distribution was calculated by Sterman {\it et al} 
\cite{Oderda:1998en,Oderda:1999kr,Berger:2001ns} by separating out
the primary emission Bremsstrahlung component and calculating this quantity to all-orders for a 4 jet system. We discussed 
this resummation procedure, and its application to rapidity gap processes, in chapter~\ref{ch3} and will use the 
formalism in chapter~\ref{ch5} to calculate detailed predictions for energy flow processes. 
However Dasgupta and Salam have pointed out \cite{Dasgupta:2001sh,Dasgupta:2002bw} that this procedure does not 
include the effects of so-called non-global logarithms, a set of leading logarithms 
which were shown to be numerically important at the energy scales probed by current colliders. 

Recently the H1 and ZEUS collaborations \cite{Adloff:2002em,zeus:2003} performed improved gaps-between-jets analyses in which the entire event
is clustered into (possibly soft) jets using the inclusive kt algorithm \cite{Catani:1993hr,Ellis:tq,Butterworth:2002xg}. A gap event is then defined by the total minijet
energy in the interjet region, rather than the total hadronic energy as in previous analyses.
The hope was that this would `clean up' the edges of the gap and make this observable less sensitive to 
hadronic uncertainties. 

In this chapter we show that by demanding
the gluonic final state to survive a clustering criterion the effect of non-global logarithms is reduced, 
but they are still numerically important,
at HERA and the Tevatron. The study presented here is particularly interesting in the light of the recent analyses by the H1 and ZEUS
collaborations.

The calculation of the Bremsstrahlung component of the energy flow observable has resulted from extensive work
in the last decade on threshold resummation, for examples
see  \cite{Berger:2001ns,Kidonakis:1998nf,Contopanagos:1996nh,Catani:1996yz,Kidonakis:1997gm}. 
This resummation procedure was the subject of chapter~\ref{ch3}.
It was found 
that by considering the emission of soft, wide angle gluons by light-like quarks one may describe the soft gluon dynamics 
of hadronic processes by an effective theory known as eikonal theory. This subject is now very well developed and the resummation
of such primary logarithms has recently made contact with high-but-fixed order calculations \cite{Sterman:2002qn}. This formalism now allows detailed
calculations of inter-jet energy flow as well as illuminating insights into the topology of colour mixing \cite{Kidonakis:1998nf}.

However, in the study of single-hemisphere observables \cite{Dasgupta:2001sh} Dasgupta and Salam identified an important class of logarithms that 
were missing from the soft gluon calculations. These omitted contributions arise in observables that are sensitive to radiation in a restricted
region of phase space. In \cite{Dasgupta:2002bw}, they showed that the omitted contributions are of leading logarithmic order for interjet energy
flow observables, as confirmed in \cite{Berger:2001ns}, where they were neglected.

\begin{figure}
\begin{center}
\epsfig{figure=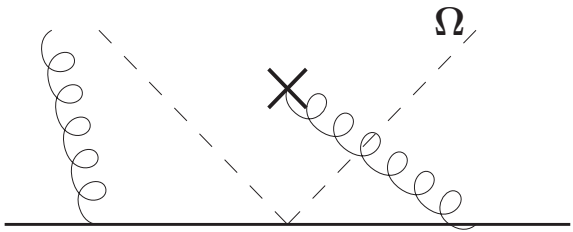,width=3in,height=1.2in}
\caption{A veto of primary emission going into $\Omega$. The kind of LL terms produced by this kind
of diagram are understood from resumming independent gluon emission.
\label{secngl:pic1}}
\end{center}
\end{figure}

\begin{figure}
\begin{center}
\epsfig{figure=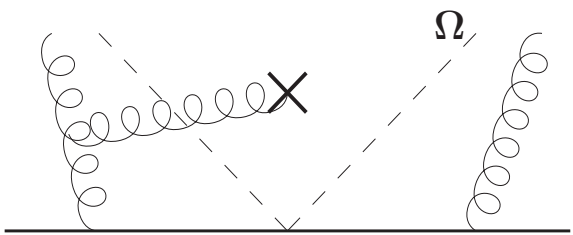,width=3in,height=1.2in}
\caption{A veto of emission into $\Omega$ from a gluon radiated outside of $\Omega$. This configuration will produce
a non-global logarithm.
\label{secngl:pic2}}
\end{center}
\end{figure}

To see the origin of these logarithms more clearly, consider soft gluonic radiation
into a patch of phase space $\Omega$ arising from a 2 jet system, where we restrict the total energy of
radiation into $\Omega$ to be less than $Q_{\Omega}$.
The primary logarithms arise from gluons that are emitted directly into $\Omega$, as shown
in figure \ref{secngl:pic1}, with energy vetoed down to a scale $Q_{\Omega}$, which is denoted
by a crossed gluon line in the figure, and form the single logarithmic (SL) set 
$\left(\alpha_s \log \frac{Q}{Q_{\Omega}}\right)^n$, $n \geq 1$, where $Q$ is the scale of the jet line. 
Now consider a gluon being emitted outside
of~$\Omega$ with intermediate energy~$Q_1$, and then vetoing subsequent emission from this gluon into~$\Omega$ 
down to scale~$Q_{\Omega}$, as shown in figure~\ref{secngl:pic2}. Integrating~$Q_1$ up to~$Q$ then generates 
another SL set of $\left(\alpha_s \log \frac{Q}{Q_{\Omega}} \right)^n$,~$n \geq 2$, 
which are formally the same order
as the primary emission terms. This was studied in detail in~\cite{Dasgupta:2002bw,Banfi:2002hw} and in this work we consider how the conclusions 
found are modified phenomenologically when the kt clustering algorithm is applied to the final state.
This new class of logarithm, which enter at leading log (LL) accuracy in energy flow observables, are known
as non-global logarithms (NGLs), as they arise in non-global observables, or as secondary logarithms, as they
are produced by secondary gluonic radiation.

The organisation of this chapter is as follows. We start in section \ref{secngl:defs} with a definition
of our observable, and a discussion of how we include NGLs in our calculations. Section \ref{secngl:kt} then 
describes the kt 
clustering algorithm used
in the H1 and ZEUS analyses, known as the inclusive jet clustering algorithm. We then develop the algorithm, in order to 
derive a form
suitable for use with a 2-jet system. Section~\ref{secngl:calclo} then describes the order $\alpha_s^2$ calculation of 
the effect of NGLs in our 2-jet system, with the clustering algorithm included in the calculation. We find that the asymptotic 
suppression factor with
clustering is reduced with respect to the non-clustering case. We then proceed to calculate the non-global effect to all orders in section \ref{secngl:calcao} using the combination of the
large $N_c$ limit and a Monte Carlo algorithm and find that the NGLs act as a suppression to the observable, and this
suppression is reduced when we impose clustering on the final state.
In section \ref{secngl:research} we give an overview of some of the current research work 
in the area of non-global observables and finally we conclude with a summary and directions for further work in 
section \ref{secngl:conc}.

To summarise our conclusions, the all-orders treatment used in this chapter allows direct inclusion of the kt 
algorithm, in exactly the same way as is used
experimentally, and we find that the clustering process reduces the magnitude of the non-global corrections to the 
primary suppression factor. However they are still 
phenomenologically relevant at HERA. The research work in this chapter 
is published in \cite{Appleby:2002ke,Appleby:2003ai}.

\section{Non-global logarithms in a 2-jet system}

\label{secngl:defs}

Following Dasgupta and Salam \cite{Dasgupta:2002bw}, the observable we are interested in is the total transverse energy $E_t$ flowing into
a region of phase space $\Omega$ for an event characterised by the hard scale $Q$,
\begin{equation}
E_t=\sum_{i \in \Omega} E_{t,i}.
\end{equation}
We are specifically interested in the cases of $\Omega$ being either a slice in rapidity or a patch,
bounded in rapidity and azimuthal angle. The quantity we shall calculate is called $\Sigma_{\Omega}$ and is 
defined to be the probability that $E_t$ is less than some energy scale $Q_\Omega$,
\begin{equation}
\Sigma_{\Omega}=\frac{1}{\sigma_o}\int_0^{Q_{\Omega}} dE_t \frac{d\sigma}{dE_t}.
\end{equation}
We normalise this equation to the process cross section, for example 2 jet production in $e^+e^-$ annihilation, and
we shall assume
the strong ordering $Q_\Omega \ll Q$.
The aim of this work is to calculate the importance of the non-global contribution to 
$\Sigma_{\Omega}$ and so it is convenient to factorise this expression into a function describing 
primary emission into $\Omega$, $\Sigma_{\Omega,P}(t)$, and a function describing (secondary) emission into $\Omega$ from 
large-angle soft gluons outside of $\Omega$, $S(t)$,
\begin{equation}
\Sigma_{\Omega}(t)=S(t)\Sigma_{\Omega,P}(t).
\end{equation}
This expression is accurate to LL level (it is helpful to think of this as a probabilistic expression, and the
observable can be written as the probability for the region~$\Omega$ to stay empty from primary gluons
\emph{times} the probability for the region~$\Omega$ to stay empty from secondary gluons) and 
the primary emission function $\Sigma_{\Omega,P}$ is the 2 jet analogue of what has been calculated
for 4 jet systems.
We have denoted the following integral of~$\alpha_s$ by $t$,
\begin{eqnarray}
t(Q_{\Omega},Q)&=&\frac{1}{2\pi}\int_{Q_{\Omega}}^{Q/2} \frac{dk_t}{k_t} \alpha_s(k_t), \\
\label{runningt}
&=& \frac{1}{4\pi \beta_0}\log\left(\frac{\alpha_s(Q_{\Omega})}{\alpha_s(Q/2)}\right),  \\
&=& \frac{\alpha_s}{2\pi}\log\frac{Q}{2Q_{\Omega}},
\end{eqnarray}
where the first equality is exact, the second holds at one loop, the third assumes a fixed coupling and $\beta_0=(11 C_A - 2 n_f)/(12\pi)$.
The leading order contribution to $S(t)$ comes in at $\alpha_s^2$ and we shall calculate this for a
2 jet system in the next section but it is useful to first consider the primary emission function at 
first order in $\alpha_s$. If we do not restrict the phase space for gluon emission then, order by order in perturbation 
theory, we expect a complete cancellation of real and virtual soft gluon contributions to the primary emission 
form factor. However the requirement of a gap in a restricted region of phase space results in this cancellation
being spoilt and we are left with an integral over the vetoed region. Hence, to
order $\alpha_s$,
\begin{eqnarray}
\Sigma_{\Omega}^{(1)}(Q_{\Omega},Q)&=&-4C_F\frac{\alpha_s}{2\pi}\int_{Q_{\Omega}}^{Q/2} \frac{dk_t}{k_t} \int_{\Omega}
d\eta \frac{d\phi}{2\pi}, \\
&=& -\frac{4C_F\alpha_s}{2\pi} A_{\Omega} \log \left( \frac{Q}{2Q_{\Omega}} \right),
\end{eqnarray}
where $A_{\Omega}$ denotes the area in $(\eta,\phi)$ space of the region $\Omega$. When the energy scales
are strongly ordered, the logarithm can become large enough to overcome the smallness of $\alpha_s$ and it becomes
necessary to include terms like $\left(\alpha_s \log\frac{Q}{Q_{\Omega}}\right)^n$ to all orders.
The assumption that the primary gluons are emitted independently according to a two-particle
antenna pattern allows one to exponentiate the one loop answer and run the coupling
to the scale $k_t$. We can then write $\Sigma_{\Omega,P}(t)$ to all orders,
\begin{equation}
\Sigma_{\Omega,P}(t)=\exp \left( -4C_F A_{\Omega} t \right).
\end{equation}
This equation only includes contributions from independent primary emission. We also
need to account for secondary gluon emission, which generates NGLs, through
the function $\mathcal{S}(t)$. This has the following expansion in $\alpha_s$,
\begin{equation}
\mathcal{S}(t)=1+S_2\,t^2+S_3\,t^3+ \ldots=1+\sum_{n=2}S_n\,t^n.
\end{equation}
Note that $\mathcal{S}(t)$ has its first non-trivial term at order $\alpha_s^2$ because we need at least 
two gluons to cause a NGL to appear. The function $\mathcal{S}(t)$ has been studied for a 2 jet
system without any final state requirements~\cite{Dasgupta:2002bw} and the purpose of this work is to 
calculate the function, at leading
and at all orders, with the requirement of kt clustering on the final state. Therefore, before we turn
to the details of the calculation of $\mathcal{S}(t)$, let us discuss the kt clustering algorithm.

\section{The \boldmath{kt} clustering algorithm}

\label{secngl:kt}

In this section we shall review the formulation of the kt clustering algorithm which we will
use in our calculations.
 We also provide a formulation that can be directly applied to our fixed 
order calculation in the next section.

The version of the
algorithm we use, which is the one used in the HERA analyses, is known as the inclusive kt algorithm \cite{Catani:1993hr,Ellis:tq,Butterworth:2002xg}.
The main features of importance to the present study are: the clustering procedure starts from the particles of lowest relative transverse 
momenta and iteratively merges them to construct pseudoparticles of higher transverse momentum; the decision of whether a particular 
pair of pseudoparticles are merged depends on their relative opening angle and their transverse 
energy (see below); despite this, it is possible 
for soft particles to be `dragged' through relatively large angles by being merged with harder particles, which are merged with even harder 
particles, and so on. Nevertheless, we do not expect our results to be qualitatively different from other infrared safe 
jet algorithms, such as the improved Legacy Cone algorithm of \cite{Blazey:2000qt}.

In the formulation of the algorithm, we represent the final state of a process by a set of ``protojets'' 
$i$ with momenta $p_i^{\mu}$. The algorithm works in an
iterative way and groups pairs of protojets together to form new ones. The aim is to group almost-parallel protojets together
so that they are part of the same protojet. Once certain criteria are met, a protojet is considered a jet and is not considered
further. The algorithm therefore needs to specify what it means for two objects to be close together and also how to merge two protojets
together. Here we follow the H1 and ZEUS analyses closely and set the radius parameter, $R$, to unity.
The procedure is 
\begin{enumerate}
\item
For each protojet, define 
\begin{equation}
d_i = E_{T,i}^2
\end{equation}
and for each protojet pair define
\begin{equation}
d_{ij}=\mathrm{min}(E_{T,i}^2,E_{T,j}^2) [ (\eta_i - \eta_j)^2 + (\phi_i - \phi_j)^2]/R^2.
\end{equation}
\item
Find the smallest of all the $d_i$ and $d_{ij}$, and label it $d_{\mathrm{min}}$.
\item
If $d_{\mathrm{min}}$ is a $d_{ij}$, merge protojets $i$ and $j$ into the new protojet $k$ with
\begin{equation}
E_{T,k}=E_{T,i}+E_{T,j},
\end{equation}
and
\begin{eqnarray}
\eta_k = [E_{T,i}\eta_i + E_{T,j}\eta_j]/E_{T,k}, \nonumber \\
\phi_k = [E_{T,i}\phi_i + E_{T,j}\phi_j]/E_{T,k}.
\end{eqnarray}
\item
If $d_{\mathrm{min}}$ is a $d_i$, the corresponding protojet $i$ is not ``mergable''. Remove it from the list of protojets and add it to the 
list of jets.
This procedure continues until there are no more protojets. As it proceeds, it produces a list of jets with successively larger 
values of $d_i = E_{T,i}^2$.
\end{enumerate}
This algorithm is implemented in the package KTCLUS \cite{ktclus}, which is used for the all-orders calculation later in this work.

It is necessary to use the full iterative algorithm for experimental analyses and
for Monte Carlo applications, but when we consider a 2 gluon final state in this work we can reduce the algorithm to
a convenient analytic form. We start by considering a hard jet line at scale $Q$ which radiates a gluon with some transverse 
energy $E_{T,1}$ in some direction $(\eta_1,\phi_1)$. This gluon then proceeds to radiate a secondary soft gluon with transverse
energy $E_{T,2}$ in some direction $(\eta_2,\phi_2)$. We assume that the energies of the gluons are strongly ordered, 
\begin{equation}
E_1 \gg E_2,
\end{equation}
which implies the same for the transverse energies,
\begin{equation}
E_{T,1} \gg E_{T,2}.
\end{equation}
Applying the clustering algorithm to the two gluons gives,
\begin{eqnarray}
d_1 &=& E_{T,1}^2, \nonumber \\
d_2 &=& E_{T,2}^2, \nonumber \\
d_{12}&=& E_{T,2}^2 [ (\eta_1 - \eta_2)^2 + (\phi_1 - \phi_2)^2]/R^2. 
\end{eqnarray}
By considering the strong ordering of the transverse momenta, the two gluons will be clustered if $d_{ij} < d_2$, so we
require
\begin{equation}
(\eta_1 - \eta_2)^2 + (\phi_1 - \phi_2)^2 > R^2
\end{equation}
for the two gluons to constitute separate jets and not be merged by the algorithm. Therefore for a 2-gluon final state to pollute the
gap and generate secondary logarithms we require that gluon 1 is outside the gap, gluon 2 is inside the gap and that they be sufficiently
separated in the $(\eta,\phi)$ plane to avoid being merged. The result of this section is that the clustering condition manifests itself
as a $\Theta$-function in our calculation,
\begin{equation}
\Theta( (\eta_1 - \eta_2 )^2 + \phi_2^2 -R^2 ),
\end{equation}
where we have used our freedom to set $\phi_1=0$. We will use this result in the next section.
Therefore the clustering procedure, when applied to a 2 gluon system, will pull the softer gluon out of
the gap and merge it into the harder gluon, provided they are sufficiently close in the $(\eta,\phi)$ plane;  hence
we expect that clustering will reduce the impact of NGLs in such systems.

\section{Controlling the non-global logarithms with \boldmath{kt} clustering}

\label{secngl:calc}

In this section we describe our calculation of the NGLs with clustering
for a 2-jet system. We contrast our work to the unclustered case and find that the clustering
procedure reduces the numerical impact of the NGLs. We then use a Monte Carlo program in the
large $N_c$ limit to find that this conclusion persists to all orders and we make numerical estimates 
of the impact of NGL in energy flow measurements at HERA.

\subsection{The LO result}

\label{secngl:calclo}

\begin{figure}
\begin{center}
\epsfig{figure=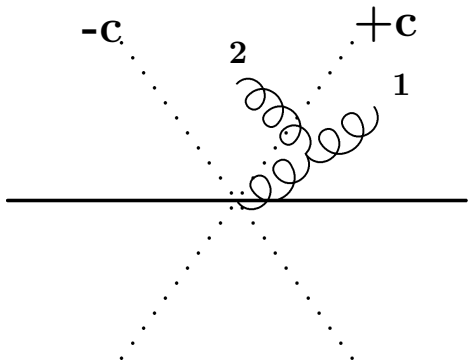,width=3in,height=2in}
\caption{The integration region for the polar angle integrations of the emitted gluons. Gluon 1 is emitted
outside of the gap, and subsequently emits gluon 2 into the gap.
\label{secngl:pic3}}
\end{center}
\end{figure}

\begin{figure}
\begin{center}
\epsfig{figure=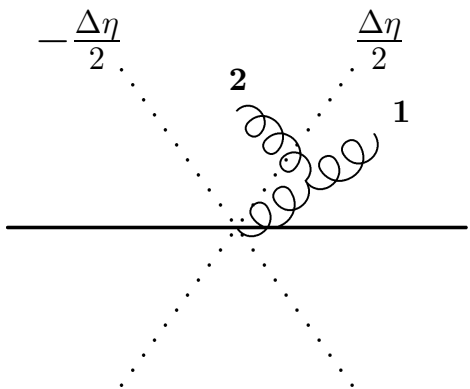,width=3in,height=2in}
\caption{The integration region for the rapidity integrations of the emitted gluons. Gluon 1 is emitted
outside of the gap, and subsequently emits gluon 2 into the gap.
\label{secngl:pic4}}
\end{center}
\end{figure}

\begin{figure}
\begin{center}
\epsfig{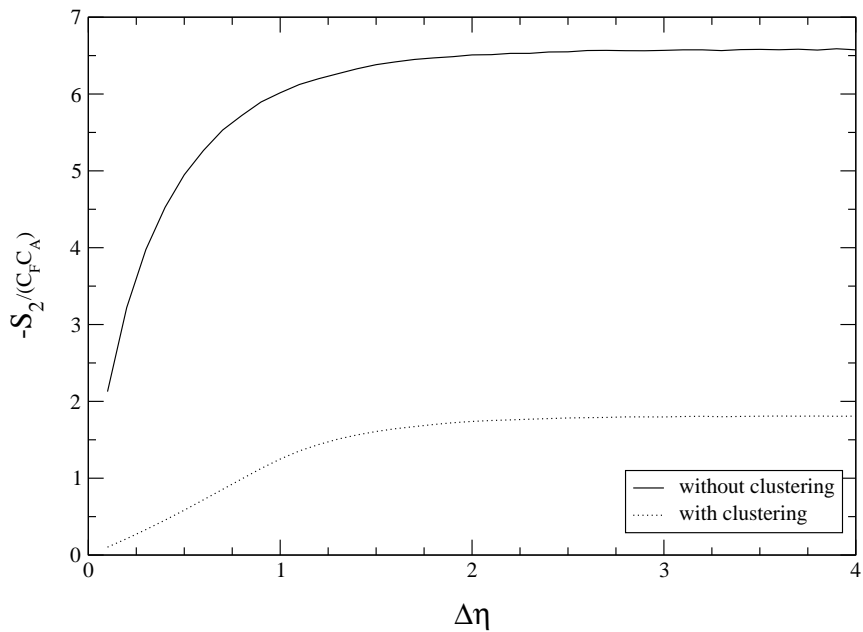}
\caption{$S_2$ as a function of $\Delta\eta$. The clustered curve is obtained using a kt algorithm with
a radius parameter of 1.0.
\label{figngl:s2}}
\end{center}
\end{figure}

\begin{figure}
\begin{center}
\epsfig{figure=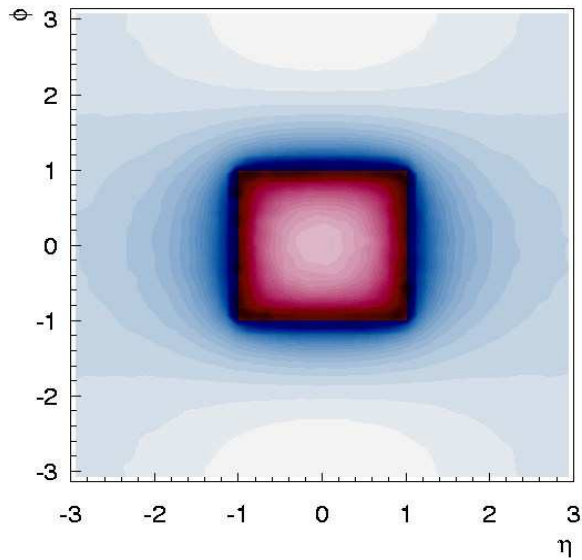,width=3in,height=3in}
\caption{Contributions to the leading order NGLs for a patch in rapidity, as a function of rapidity and azimuthal
angle. Darker shades correspond to a larger contribution and it can be seen that the largest contribution
comes from the edge of the patch. This figure is taken from \cite{Appleby:2003ai}.
\label{figngl:patch2col}}
\end{center}
\end{figure}

In this section we will calculate the leading order contribution to the non-global logarithm
function $\mathcal{S}(t)$. As has already been shown, this has its first non-trivial term at order
$\alpha_s^2$ and has the perturbative expansion,
\begin{equation}
\mathcal{S}(t)=1+S_2\,t^2+S_3\,t^3+ \ldots=1+\sum_{n=2}S_n t^n.
\end{equation}
Our goal is to calculate the $S_2$ piece for $\Omega$ defined as a slice in rapidity of width~$\Delta\eta$ 
with the condition that the topology of the gluon tree satisfies the kt clustering algorithm. 
We shall initially perform the calculation in terms of the 
polar angles of the emitted gluons, and then again
in terms of the rapidities of the emitted gluons because the latter form of the calculation will
turn out to have more convenient properties. A precise definition of $S_2$ can be given by 
following \cite{Dasgupta:2001sh,Dasgupta:2002bw} and using the 2 body phase space of the emitted gluons.
%\begin{equation}
%d(PS)=\frac{1}{(2\pi)^6} \int \frac{d^3 \vec{k}_1}{2k_1^0} \frac{d^3 \vec{k}_2}{2k_2^0}.
%\end{equation}
If we only keep the leading logarithmic piece of the integrals over the momentum fractions, defined by
\begin{equation}
k^0_i=x_i \frac{Q}{2},
\end{equation}
we obtain the following expression for $S_2$,
\begin{eqnarray}
S_2 \log^2\left(\frac{Q}{2 Q_{\Omega}}\right) + \mathcal{O}\left(\log\left(\frac{Q}{2Q_{\Omega}}\right)\right) =
-C_F C_A \int_{k_1 \not\in \Omega} d\cos\theta_1 \frac{d\phi_1}{2\pi} \nonumber \\
\int_{k_2 \in \Omega} d\cos\theta_2 \frac{d\phi_2}{2\pi}
\frac{Q^4}{16} \int_0^1 x_2 dx_2 \int_{x_2}^1 x_1 dx_1 \Theta\left(x_2-\frac{2Q_{\Omega}}{Q}\right)W_S.
\label{secngl:eqs2polar}
\end{eqnarray}
Therefore $S_2$ is the coefficient of the logarithm-squared term in the expansion of $\mathcal{S}(t)$.
The limits on the angular integration ensure that gluon~2 enters region $\Omega$ while gluon~1 does not.
This expression for $S_2$ contains the secondary part, $W_S$, of the well-known squared-matrix element for the energy ordered emission
of two gluons, which can be derived from standard techniques \cite{Dokshitzer:1992ip},
\begin{eqnarray}
W&=&4C_F \frac{(ab)}{(a1)(1b)} \left( \frac{C_A}{2}\frac{(a1)}{(a2)(21)} + \frac{C_A}{2}\frac{(b1)}{(b2)(21)}
+ \left( C_F - \frac{C_A}{2} \right) \frac{(ab)}{(a2)(2b)} \right), \nonumber \\
&=& C_F^2 W_P + C_F C_A W_S,
\label{secngl:eqs2me}
\end{eqnarray}
where the notation $(ij)$  denotes the dot product of the appropriate 4-momenta. This expression contains the
primary emission piece $W_P$, proportional to $C_F^2$, and the piece that interests us, which is the part
proportional to $C_F C_A$ and denoted $W_S$. Note that the last term is the dipole interference term and 
is absent in the large $N_c$ limit.
We shall begin by writing the 4-momenta in terms of the polar angles of the emitted gluons,
\begin{eqnarray}
k_a &=& \frac{Q}{2}(1,0,0,-1), \\
k_b &=& \frac{Q}{2}(1,0,0,1), \\
k_1 &=& x_1 \frac{Q}{2}(1,0,\sin \theta_1,\cos \theta_1), \\
k_2 &=& x_2 \frac{Q}{2}(1,\sin \theta_2 \sin \phi_2, \sin \theta_2 \cos \phi_2, \cos \theta_2),
\end{eqnarray}
where the indices $a$ and $b$ refer to quarks, the indices $1$ and $2$ refer to radiated gluons and
we have set $\phi_1=0$.
If we evaluate the 4-momenta products we get
\begin{eqnarray}
(ab) &=& \frac{Q^2}{2}, \\
(1a) &=& \frac{Q^2 x_1}{4}(1-\cos\theta_1), \\
(1b) &=& \frac{Q^2 x_1}{4}(1+\cos\theta_1), \\
(2a) &=& \frac{Q^2 x_2}{4}(1-\cos\theta_2), \\
(2b) &=& \frac{Q^2 x_2}{4}(1+\cos\theta_2), \\
(12) &=& \frac{Q^2 x_1 x_2}{4}(1-\sin\theta_1\sin\theta_2\cos\phi_2-\cos\theta_1\cos\theta_2).
\end{eqnarray}
Inserting these results into equation (\ref{secngl:eqs2me}) we obtain
\begin{eqnarray}
W_S&=&\frac{32}{Q^4}\frac{1}{(x_1 x_2)^2}
\Bigg( \frac{1}{(1+\cos\theta_1)(1-\cos\theta_2)(1-\sin\theta_1\sin\theta_2\cos\phi_2-\cos\theta_1\cos\theta_2)} 
\nonumber \\
&& + \frac{1}{(1-\cos\theta_1)(1+\cos\theta_2)(1-\sin\theta_1\sin\theta_2\cos\phi_2-\cos\theta_1\cos\theta_2)} \nonumber \\
&& - \frac{2}{(1+\cos\theta_1)(1-\cos\theta_1)(1+\cos\theta_2)(1-\cos\theta_2)} \Bigg).
\end{eqnarray}
We shall write the angular part of $W_s$ (the part without the overall constant) as~$\hat{W}_s$.
Inserting this result into equation (\ref{secngl:eqs2polar}) and then performing the straightforward 
energy fraction integrals (the $\Theta$-function ensures that sufficient energy reaches
the gap region $\Omega$), where we have kept the leading logarithmic piece (dominant when~$Q_{\Omega}/Q$ is small),
\begin{equation}
\int_0^1 \frac{dx_2}{x_2} \int_{x_2}^1 \frac{dx_1}{x_1} \Theta\left(x_2 - \frac{2Q_{\Omega}}{Q}\right)=
\frac{1}{2}\log^2\left(\frac{2Q_{\Omega}}{Q}\right),
\end{equation}
we arrive at an expression for $S_2$
\begin{equation}
S_2=-C_F C_A \int_{angles} \hat{W}_S.
\end{equation}
Specialising to a slice in rapidity of width $\Delta \eta$, delimited by $\cos\theta = \pm c$ 
and centred on $\eta=0$ we can exploit 
the symmetry of the $c_1$ integral and
perform the $\phi_1$ integral to express the integral over angles as
\begin{equation}
\int_{angles}=2 \int_{-1}^{-c} d\cos_1 \int_{-c}^{+c} d\cos_2 \int_0^{2\pi} \frac{d\phi_2}{2\pi}.
\end{equation}
The polar angle integration region is shown in figure \ref{secngl:pic3}.
For the case of no clustering of the gluons it is possible to perform the azimuthal average using the  result,
\begin{eqnarray}
\int_0^{2\pi} \frac{d\phi_2}{2\pi} \frac{1}{1-\sin\theta_1\sin\theta_2\cos\phi_2-\cos\theta_1\cos\theta_2}
&=& \frac{1}{| \cos \theta_2 - \cos \theta_1|}, \nonumber \\
&=& \frac{1}{\cos \theta_2 - \cos \theta_1},
\end{eqnarray}
where we have exploited the angular ordering of the gluons in writing the second equality. This results
in the following, azimuthal-averaged, expression for the matrix element,
\begin{equation}
\langle W_S\rangle_{\phi}=4\left(\frac{1}{(1+\cos\theta_2)(1-\cos\theta_1)(\cos\theta_2-\cos\theta_1)}\right),
\end{equation}
with two remaining $\cos\theta$ integrations. For this case of no clustering of the gluons,
these remaining integrations have been performed, resulting in the following expression for
$S_2$ \cite{Dasgupta:2002bw},
\begin{equation}
S_2=-4C_FC_A \left[ \frac{\pi^2}{12} + (\Delta\eta)^2 - \Delta\eta \log(e^{2\Delta\eta}-1)
- \frac{1}{2}\mathrm{Li}_2 (e^{-2\Delta\eta}) -  \frac{1}{2}\mathrm{Li}_2 (1-e^{2\Delta\eta})\right],
\end{equation}
which, following the literature, we have expressed in terms of $\Delta\eta$, which is related to~$c$ by the kinematical
relation,
\begin{equation}
\Delta\eta=\ln \left( \frac{1+c}{1-c} \right).
\end{equation}
We have used the dilogarithm function, which is defined
\begin{equation}
\mathrm{Li}_2(y)=\int_y^0 \frac{\log(1-x)}{x}\,dx.
\end{equation}
We can now use the following identities (with $x=~\exp(-2\Delta\eta)$ in the first identity
and $x=\exp(2\Delta\eta)$ in the second identity),
\begin{eqnarray}
\mathrm{Li}_2(x)&=&\frac{\pi^2}{6}-\log(x)\log(1-x)-\mathrm{Li}_2(1-x), \\
\mathrm{Li}_2(1-x)&+&\mathrm{Li}_2(1-x^{-1})=-\frac12 \log^2(x),
\end{eqnarray}
to write this result as
\begin{equation}
S_2=-4C_FC_A\left[ \frac{\pi^2}{6}-\mathrm{Li}_2(e^{-2\Delta\eta})\right].
\end{equation}
Note that as $\Delta\eta$ increases, $S_2$ rapidly saturates at its asymptotic value,
\begin{equation}
\lim_{\Delta\eta \rightarrow \infty}=-C_FC_A\frac{2 \pi^2}{3},
\end{equation}
since we note that $\mathrm{Li}_2(0)=0$.
This analytic evaluation of $S_2$ is not possible for the case of clustered gluons because, as discussed 
previously, the requirement that a gluonic final state is in a configuration that will survive the clustering
algorithm can be written as a $\Theta$-function of all three angular integration variables,
\begin{equation}
\Theta( (\eta_1 - \eta_2 )^2 + \phi_2^2 -R^2 ).
\end{equation}
In this situation we have to resort to a numerical solution, in which we integrate over the
three remaining variables. This method has the advantage of being easily 
extendible to any gap geometry of interest.
It is possible, however, to obtain a result for $S_2$ with clustering which is expressible
as a 1D integral by using the rapidity of the emitted gluons. This has the advantage of requiring a
simpler method of numerical solution.
In analogy with equation (\ref{secngl:eqs2polar}), we can express $S_2$ in terms of the gluon rapidities,
\begin{eqnarray}
-C_F C_A &&\int_{k_1 \not\in \Omega} d\eta_1 \frac{d\phi_1}{2\pi}
\int_{k_2 \in \Omega}d\eta_2 \frac{d\phi_2}{2\pi}
\frac{Q^4}{16}
\int_0^1 x_2 dx_2 \int_{x_2}^1 x_1 dx_1  \Theta\left(x_2-\frac{2Q_{\Omega}}{Q}\right) W_S \nonumber \\
&&= S_2  \log^2\left (\frac{Q}{2 Q_{\Omega}}\right) + \mathcal{O}\left(\log\left(\frac{Q}{2Q_{\Omega}}\right)\right).
\end{eqnarray}
Note that in this equation, $x_1$ and $x_2$ are defined as transverse momentum fractions,
\begin{equation}
 P_{t,1} = x_1 \frac{Q}{2},
\end{equation}
and not \emph{energy} fractions as they are in equation~(\ref{secngl:eqs2polar}), which was written in 
terms of polar angles. However, this does not affect the result. The definition of~$Q_{\Omega}$ 
has also correspondingly changed. 
We can relate the phase space in the two coordinate systems by
\begin{equation}
\frac{d^3\vec{k}}{k^0}=k_t\,dk_t\,d\eta\,d\phi.
\end{equation}
We begin
by defining the following 4-momenta,
\begin{eqnarray}
k_a &=& \frac{Q}{2}(1,0,0,-1), \\
k_b &=& \frac{Q}{2}(1,0,0,1), \\
k_1 &=& P_{t,1}(\cosh\eta_1,\sin\phi_1,\cos\phi_1,\sinh\eta_1), \\
k_2 &=& P_{t,2}(\cosh\eta_2,\sin\phi_2,\cos\phi_2,\sinh\eta_2),
\end{eqnarray}
where the indices are the same as those defined for the polar angle calculation.
Evaluating the 4-momenta products we get
\begin{eqnarray}
(ab) &=& \frac{Q^2}{2}, \\
(1a) &=& \frac{Q^2 x_1}{4}\exp(-\eta_1), \\
(1b) &=& \frac{Q^2 x_1}{4}\exp(+\eta_1), \\
(2a) &=& \frac{Q^2 x_2}{4}\exp(-\eta_2), \\
(2b) &=& \frac{Q^2 x_2}{4}\exp(+\eta_2), \\
(12) &=& \frac{Q^2 x_1 x_2}{4}\left( \cosh(\eta_1-\eta_2)-\cos(\phi_1-\phi_2)\right).
\end{eqnarray}
Inserting these results into the expression for $W_S$, equation (\ref{secngl:eqs2me}) and using our freedom to 
set the azimuthal angle
of the first gluon to zero, $\phi_1=0$, we obtain
\begin{equation}
W_S = \frac{128}{Q^4 x_1^2 x_2^2} \left( \frac{\cosh(\eta_1-\eta_2)}{\cosh(\eta_1-\eta_2)-\cos\phi_2}  - 1 \right).
\end{equation}
The transverse momentum fraction integrals are straightforward, as before,
\begin{equation}
\int_0^1 \frac{dx_2}{x_2} \int_{x_2}^1 \frac{dx_1}{x_1} \Theta\left(x_2 - \frac{2Q_{\Omega}}{Q}\right)=
\frac{1}{2}\log^2\left(\frac{2Q_{\Omega}}{Q}\right).
\end{equation}
Specialising to a slice in rapidity of width $\Delta \eta$ and delimited by $|\eta| < \Delta \eta/2$, where
boost invariance allows us to centre the gap on $\eta=0$, we can exploit 
the symmetry of the $\eta_1$ integral and
perform the trivial $\phi_1$ integral of the first gluon to express the final integral over angles as
\begin{equation}
\int_{angles}=2 \int_{-\infty}^{-\frac{\Delta\eta}{2}} 
d\eta_1 \int_{-\frac{\Delta\eta}{2}}^{\frac{\Delta\eta}{2}} d\eta_2 \int_0^{2\pi} \frac{d\phi_2}{2\pi}.
\end{equation}
The rapidity integration region is shown in figure \ref{secngl:pic4}.
For the case of no clustering of the gluons it is possible to perform the azimuthal average using the result,
\begin{equation}
\int_0^{2\pi} \frac{d\phi_2}{2\pi} \frac{1}{\cosh(\eta_1-\eta_2)-\cos\phi_2} = \frac{1}{|\sinh(\eta_1-\eta_2)|}.
\end{equation}
This can be elegantly proven by noting that the following contour integration over the unit circle,
\begin{equation}
I=\frac{i}{\pi}\oint_{u_1} \frac{dz}{(z-\exp(\eta_2-\eta_1))(z-\exp(\eta_1-\eta_2))},
\end{equation}
is equivalent to the integration over the azimuthal angle of the second gluon. 
To see this, expand
the $\cosh(\eta_1-\eta_2)$ and $\cos(\phi_2)$ as exponentials and write~$z=\exp(i\phi_2)$. 
We get poles at 
\begin{equation}
z=e^y,\,e^{-y},
\end{equation}
where $y=\eta_1-\eta_2$. We only have to consider the poles which satisfy $|z|<1$, which is~$z=\exp(-|y|)$, so
\begin{eqnarray}
I&=&2\pi i \left(\frac{i}{\pi}\right) \lim_{z \rightarrow \exp(-|y|)} 
\left\{\frac{z-\exp(-|y|)}{(z-\exp(-|y|))(z-\exp(|y|))}\right\}, \nonumber \\
&=&-\frac{2}{\exp(-|y|)-\exp(|y|)}, \nonumber \\
&=&\frac{1}{\sinh(|y|)},
\end{eqnarray}
and write $\sinh(|y|)=|\sinh(y)|$.
As for the polar angle case, it is possible to do all the integrations for the case of no clustering.
However, evaluation of $S_2$ is not possible for the case of clustered gluons because, as discussed 
previously, the requirement that a gluonic final state is in a configuration that will survive a clustering
algorithm appears as a $\Theta$-function in our calculation, and the argument of the $\Theta$-function is very 
complicated.
However we can readily reduce the three-dimensional integral to a one-dimensional integral using standard
techniques if we consider the region of phase space that is vetoed by the clustering algorithm, denoted $S_2^v$. Doing this 
we obtain,
\begin{eqnarray}
S_2^v&=&\frac{-32 C_F C_A}{\pi} \int_0^R \mathrm{min}(\eta,\Delta\eta)\bigg[ 2 \coth(\eta) \times \\
&&\arctan\left(\frac{\tan(\sqrt{R^2-\eta^2}/2)}{\tanh(\eta/2)}
\right) \nonumber - \sqrt{R^2-\eta^2} \bigg] d\eta,
\label{secngl:eqsub}
\end{eqnarray}
where we define $\eta=\eta_1-\eta_2$. This result is proved in appendix \ref{secngl:app1}. 
Therefore the solution for $S_2$ with the clustering condition imposed is obtained by subtracting $S_2^v$ from the analytic unclustered
result.
We resort to numerical techniques to solve this equation, which has the advantage of being easily 
extendible to any gap geometry of interest. Our numerics were done using Monte Carlo integration methods and the
routine Vegas.
Figure \ref{figngl:s2} shows our result for $S_2$ as a function of the gap width $\Delta\eta$ and with
the radius parameter, $R$, set to 1. The solid line is $S_2$ 
without clustering and the dotted line is $S_2$ with clustering. 
In \cite{Dasgupta:2002bw}, where the unclustered case was first obtained, the observed saturation was 
given a simple physical
explanation: the dominant contribution to~$S_2$ comes from regions where the two gluons are close together (which
means that gluon~1 is just outside the gap and gluon~2 is just inside the gap.) This 
occurs when
\begin{equation}
\eta_1 \simeq \eta_2 \simeq -\Delta\eta/2,
\end{equation}
and the dominant contribution arises from the collinear region of 
the matrix element. Therefore at large $\Delta\eta$,
$S_2$ received no contribution from the centre of the gap, only the edges, and the value of $-S_2/(C_F C_A)$ saturates.
This is the reason that NGLs are often referred to as an `edge effect', and gluons closest to the boundary of
the restricted region of phase space make the largest contribution. This effect is 
illustrated in figure~\ref{figngl:patch2col}.
We note that~$S_2^v$ itself saturates when~$\Delta\eta \simeq R$. The results we have obtained show that when we 
demand the gluonic final state to survive a clustering
algorithm, the saturation of $S_2$ observed in \cite{Dasgupta:2002bw} is still observed but the saturation value is reduced by $70\%$. In other words
the value that $-S_2/(C_F C_A)$ saturates to is reduced from $6.57$ to $1.81$.
The reason for this reduction is that we have removed gluons from the region of collinear enhancement, but gluons that are still
sufficiently separated in the $(\eta,\phi)$ plane to satisfy
\begin{equation}
(\eta_1 - \eta_2 )^2 + \phi_2^2 > R^2,
\end{equation}
will survive the clustering process and contribute to the NGLs.
Therefore we conclude that the saturation of the non-global contribution at fixed order is still seen when we demand clustering 
of the final state.

\subsection{Non-global logarithms to all orders}

\label{secngl:calcao}

The analytic treatment of the 2 gluon system, with or without clustering, is fairly straightforward. The extension
to the many-gluon case presents considerable problems due to the geometry and the colour structure of the
`mexican cactus' gluon ensemble. Therefore we have extended our calculation to all orders by working in 
the large~$N_c$ limit and by employing numerical methods. The complicated colour algebra one obtains is simplified
in the large~$N_c$ limit, because the squared-matrix element for the gluon radiation can be broken down
into a series of independent terms, each associated with a different colour dipole. The complicated
geometry of multi-gluon events is handled by use of a Monte Carlo algorithm, in which the gluon cascade is
built up from successive dipole branching. Therefore we start with a single colour dipole, representing the 
initial quark line, which emits a gluon and splits into two dipoles, which themselves may go
on to emit. The cascade carries on until an emitted gluon enters the gap region. This method also
allows the kt clustering algorithm to be directly implemented, in exactly the same way it is
used experimentally, and before a given gluon configuration can be deemed to have polluted the gap it must
survive the algorithm. If the gluon which has entered the gap is pulled out by the clustering procedure, then
the dipole cascade is allowed to continue.

Note that there are a lot of configurations with
many gluons at high~$t$  and this, coupled with the fact that the speed of the kt algorithm scales 
like~$N^3$ (where~$N$ is the number of gluons), means that the Monte Carlo algorithm can be very slow in this region.
The result is that we have large statistical errors for~$t \ge 0.3$, so we do not show this region.
\begin{figure}
\begin{center}
\epsfig{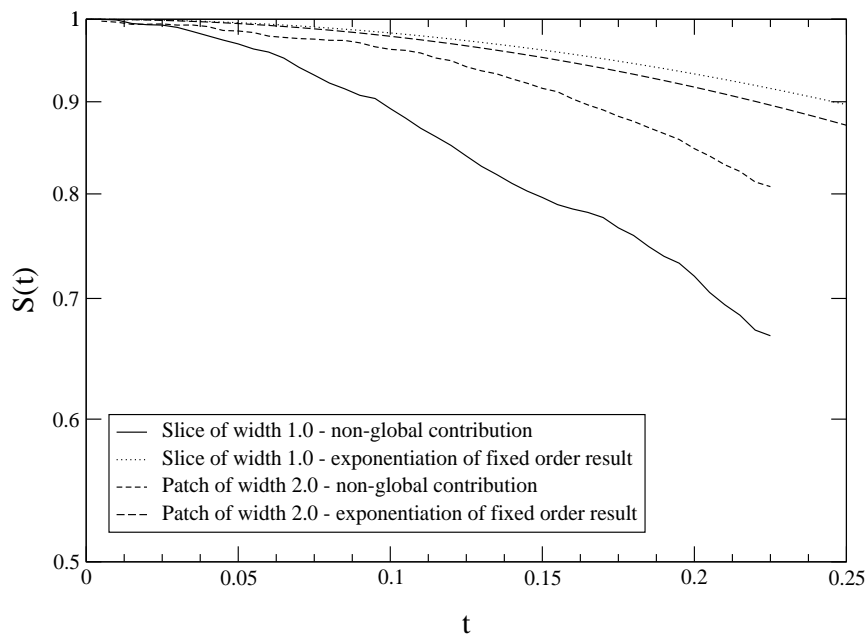}
\caption{The function $S(t)$ for a slice and a patch of phase space, with the condition of kt clustering imposed on the gluons. These curves
were obtained using a Monte Carlo procedure in the large $N_c$ limit. Also shown are the curves for $S(t)$ obtained by exponentiating
the fixed order result. The geometry independence of the $t$ dependence indicates that the buffer zone mechanism identified to exist in previous work
survives the clustering algorithm. Note that the slice shows greater non-global suppression than the patch at high~$t$, in contrast 
to the unclustered case~\cite{Dasgupta:2002bw}.
\label{figngl:st}}
\end{center}
\end{figure}

We have verified the results obtained by Dasgupta and Salam \cite{Dasgupta:2002bw}, which have been calculated with no clustering requirement imposed on the gluons, and written a program, based on their program, to perform the all-orders 
calculation with clustering.
\begin{figure}
\begin{center}
\epsfig{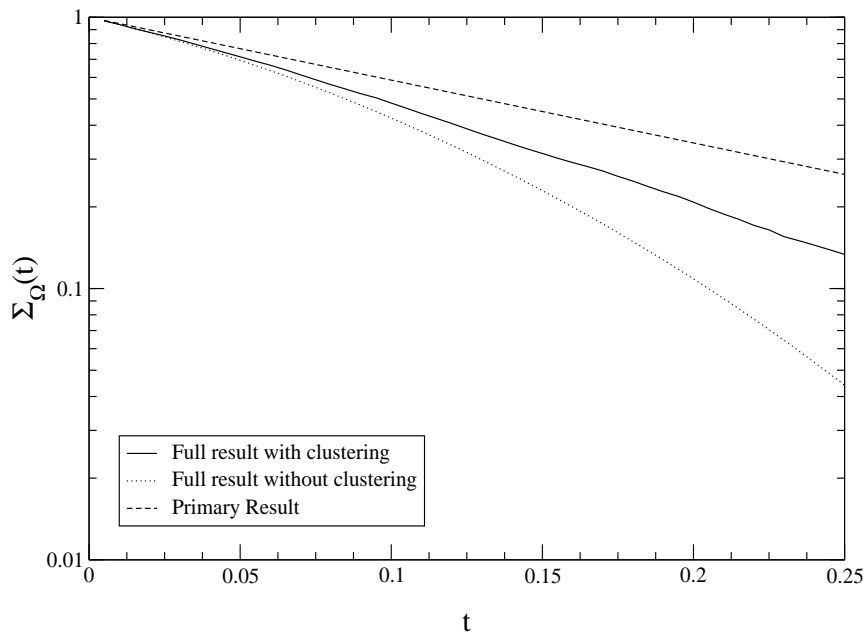}
\caption{The phenomenological implications of clustering for the case of $\Omega$ being a slice of rapidity of width $1.0$. The reduction
of the non-global suppression factor when clustering is included can be seen.
\label{figngl:sigmaslice}}
\end{center}
\end{figure}
Figure \ref{figngl:st} shows the function $S(t)$ for two different geometries for $\Omega$: a slice in rapidity 
of width $\Delta\eta=1.0$ and a square patch in rapidity and azimuthal angle of side length~$\Delta\eta=\Delta\phi=2.0$,
with the requirement of clustering on the final state gluons. The curves labelled `exponentiation of the fixed
order result' are those obtained if there were a simple exponentiation of the $S_2$ term.

Firstly the figure shows that at high $t$ ($t > 0.2$) the suppression increases faster for the full calculation than for the
exponentiation of $S_2$ with clustering and the two curves have different shapes in this region. This implies
a more complex geometry dependence in the full result than is seen by the simple exponentiation of $S_2$.
 Secondly we see that the $t$ dependence of the suppression
is geometry (i.e. the definition of $\Omega$) independent at high $t$; the implication of this is that the clustering process maintains a 
so-called buffer region\footnote{The buffer region is defined
by the absence of any reconstructed jets within it. It may therefore contain gluons, provided that they get pulled
out of the buffer region by the clustering algorithm.}  of suppressed intermediate
radiation around $\Omega$. Such a buffer
mechanism has been postulated to exist in the unclustered case, and be responsible for the lack of gluons 
in the gap region.
 Therefore figure \ref{figngl:st} shows that the all-orders result with clustering is more complicated
than simple exponentiation of the fixed order result. Our work also indicates, due to the
geometry independence at high $t$,  that a buffer region exists around the gap $\Omega$ which is responsible
for the suppression of radiation into the gap.

Figure \ref{figngl:sigmaslice} shows the reduction of the phenomenological significance of the NGLs when we cluster the final state. The figure shows
the function~$\Sigma(t)$ with only primary emissions and the full all-orders treatment with and without clustering effects. 
This is done with 
$\Omega$ defined as a slice in rapidity of size $\Delta\eta = 1.0$. There are several points to note. Firstly the effect of the NGLs is a large 
suppression of the cross section relative to the primary-only result. The value of $t$ which we can consider the realistic upper limit for Tevatron Run II experiments 
is about $0.15$ so we shall take $t=0.15$ as our reference value. We can translate this into an energy scale 
by using the running coupling definition 
of~$t$, equation~(\ref{runningt}), and obtain $Q\sim 100$~GeV for $Q_{\Omega}=1$~GeV. In this region the inclusion of the non-global effects without clustering increases 
the suppression relative to the primary-only result by about a factor of $1.65$. When we include clustering effects, this difference
is reduced to about $1.2$. Therefore, at all orders, the requirement of clustering on the final state reduces the phenomenological significance of the
non-global effects by about 70\%. This reduction in magnitude of the effect can be seen to persist to all orders. Hence when calculating cross sections, if we exclude the effect of NGLs then we will overestimate the cross section by $65\%$ for a non-clustered final state and by $20\%$ for
a clustered final state. At larger $t$ values, the overestimation increases. For comparison, the typical errors on the H1 gaps-between-jets data is
$\sim 30\%$.

\begin{figure}
\begin{center}
\epsfig{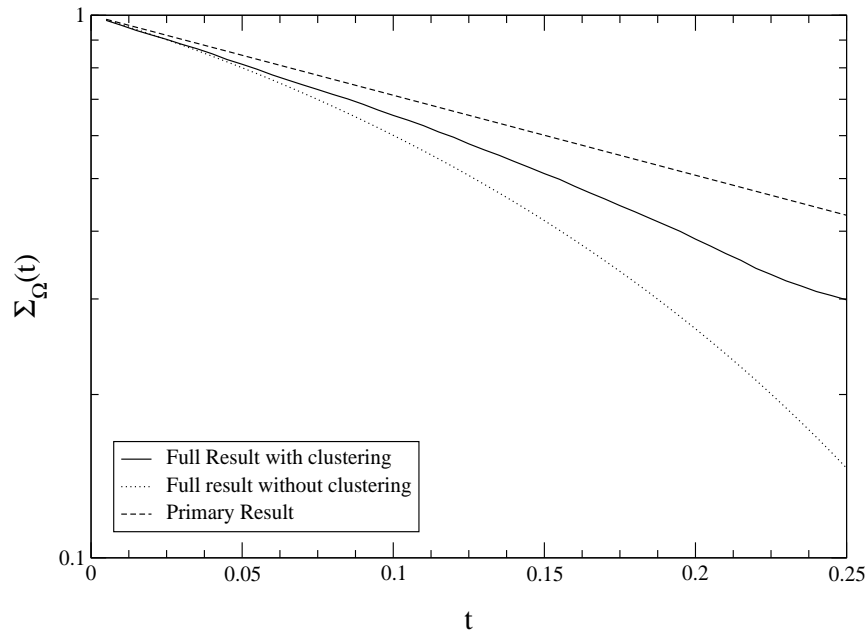}
\caption{The effect of the clustering of the gluons on the function $S(t)$ for a patch of phase space of size  $\Delta\eta = \Delta\phi = 2.0$.
\label{figngl:sigmapatch}}
\end{center}
\end{figure}

Finally, figure \ref{figngl:sigmapatch} shows the same comparison of the full result, with and without clustering, 
for a patch 
of size $\Delta\eta = \Delta\phi = 2.0$. We can see
that the conclusions we made for figure \ref{figngl:sigmaslice} apply and the effect of the NGLs is of similar magnitude.

\begin{figure}
\begin{center}
\epsfig{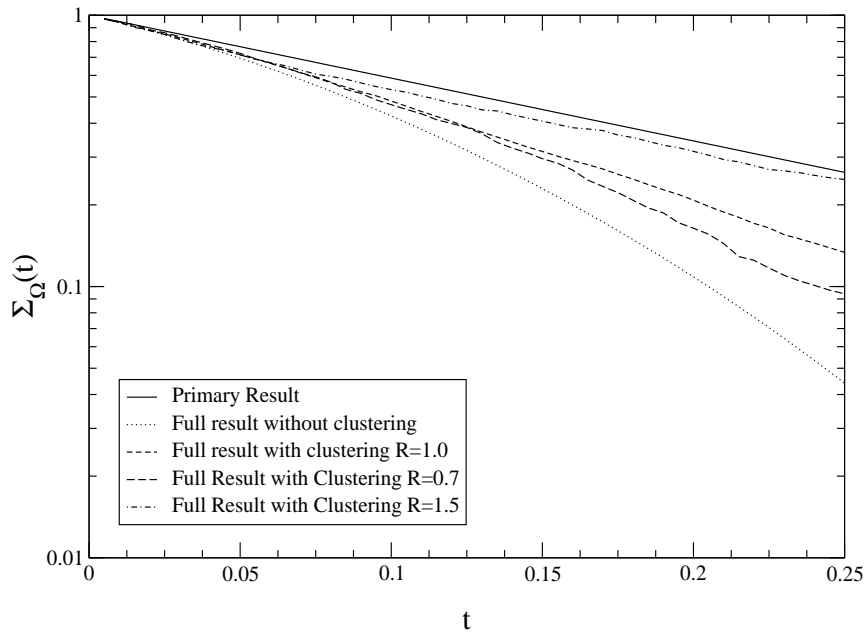}
\caption{The effect of varying radius parameter on the full function $\Sigma_{\Omega}$, for a slice
in rapidity of width $\Delta\eta=1.0$.
\label{figngl:radiusdepend}}
\end{center}
\end{figure}

We have also performed calculations to see how the non-global suppression is affected by varying the radius parameter, $R$, 
of the clustering procedure. Figure \ref{figngl:radiusdepend} shows the full function $\Sigma_{\Omega}$ with
NGLs included and with varying $R$.
By decreasing~$R$ the impact of the clustering algorithm is reduced and the effect of the NGLs is restored
to the non-clustered case.
In fact, we expect the magnitude of the non-global suppression to tend to the non-clustered case as $R \rightarrow 0$.
Similarly, increasing the radius causes more gluons to be included in the clustering and hence 
the magnitude of the suppression to reduce. In fact, we have found that for $R$ close to $1.5$ the full result is almost
identical to the primary result. 

%Finally, The numerical correctness of our results can be seen in figure \ref{figngl:smallt}, which shows the dependence of the full result, the full result
%with clustering and the primary result at small values of $t$. If we think of $t$ as a time variable, then at low $t$, corresponding
%to a low number of gluons or equivilently the low order terms in the perturbative
%expansion, the three results 
%should coalesce. This behaviour can be clearly seen in the figure.
%\begin{figure}
%\begin{center}
%\includegraphics[width=6cm,height=9cm,angle=-90]{../mc.e.plot.ps}
%\end{center}
%\caption{The full result for a patch of phase space of size 2.0}
%\label{mc.e.plot}
%\end{figure}

In summary, non-global logarithms are important not only from the point of view of correctness of the leading logarithm 
series but also result in significant
numerical corrections to cross sections. These corrections are reduced by about a factor of 3 if we cluster the final
 state. 
It is clear, therefore, that while the use of the kt algorithm has aided the control of the NGLs, 
they still have a 
significant numerical effect at HERA.

\section{Ongoing research topics in non-global observables}

\label{secngl:research}

Non-global observables are a new and exciting research area in QCD. In this review section, we will give
an brief overview of some current research directions and the progress made.

\subsection{The 4 jet system}

\label{secngl:4jet}

In this section, we will introduce the calculation of NGLs for a 4 jet system, working in the 
large $N_c$ limit at fixed order. This is physically motivated by the wish to understand
the effect of NGLs for hadrons to jets in, for example, rapidity gap event in photoproduction. 
The main difference between a 2 jet system and a 4 jet system 
is the greater number of dipole radiation sources in the latter case.
We model the 4 jet system in the large $N_c$ limit, using non-interacting and recoilless dipoles. The initial
state jets have 4-momenta,
\begin{eqnarray}
a^\mu &=& \frac{Q}{2}(1,0,0,1), \nonumber \\
b^\mu &=& \frac{Q}{2}(1,0,0,-1),
\end{eqnarray}
and the final state jets have 4-momenta,
\begin{eqnarray} 
c^\mu &=& p_t^c ( \cosh(\Delta\eta/2),1,0,\sinh(\Delta\eta/2)), \nonumber \\
d^\mu &=& p_t^d ( \cosh(\Delta\eta/2),-1,0,-\sinh(\Delta\eta/2)).
\end{eqnarray}
In this work we assume that the jets are produced at fixed~$\Delta\eta=|\eta_1-\eta_2|$.
The matrix element squared for emitting a gluon can be written, in the large $N_c$ limit as
\begin{equation}
W_1=4 C_F \left( \frac{(ad)}{(a1)(1d)} + \frac{(bc)}{(b1)(1c)} \right),
\label{ngleq:4jetme}
\end{equation}
where the first term describes the probability of the gluon being emitted from the~$(ad)$ dipole, and the
second term describes the probability of the gluon being emitted from the~$(bc)$ dipole. There is, of course, other terms
describing $s$ and $t$ channel dipole emission but we initially 
restrict ourselves to equation~(\ref{ngleq:4jetme}) for simplicity. 
In these
equations the notation $(ij)$ means the scalar product of the corresponding 4-momenta and we write
the 4-momentum of the emitted gluon by
\begin{equation}
\mathbbm{1}^\mu=p_t^1(\cosh\eta_1,\cos\phi_1,\sin\phi_1,\sinh\eta_1).
\end{equation}
The matrix element squared for the first gluon to emit a second gluon is, in the same
approximation
\begin{eqnarray}
W_S&=&4C_F \frac{(ad)}{(a1)(1d)} \left( \frac{C_A}{2}\frac{(a1)}{(a2)(21)} + 
\frac{C_A}{2}\frac{(1d)}{(d2)(21)} \right) \nonumber \\
 &+& 4C_F \frac{(bc)}{(b1)(1c)} \left( \frac{C_A}{2}\frac{(b1)}{(b2)(21)} + \frac{C_A}{2}\frac{(1c)}{(c2)(21)} \right),
\end{eqnarray}
where
\begin{equation}
\mathbbm{2}^\mu=p_t^2(\cosh\eta_2,\cos\phi_2,\sin\phi_2,\sinh\eta_2).
\end{equation}
By proceding in analogy to the fixed order calculation for a 2 jet system, we may extract the non-global suppression 
factor for a hadron-hadron process which produces~2~jets at fixed~$\Delta\eta$. The physical expectation is that the
greater quantity of dipole sources will give a larger non-global suppression factor for a 4 jet system, relative to
the~2 jet case. The~2-jet all orders calculation can be extended in a similar way, to extract the all-orders~4-jet 
non-global suppression factor, as a function of the gap width and the time-energy variable~$t$. Such a
study is being performed at present.

\subsection{Event shape/energy flow correlations}

\label{secngl:esef}

Another way of reducing the phenomenological impact of NGLs, in addition to the use of
clustering algorithms, has
been proposed 
\cite{Berger:2003iw}, and is the study of associated
distributions in two variables. In this work, one combines measurement
of a jet shape $V$ in the whole of phase space (for example thrust,
$V=1-T$) and that of the transverse away-from-jets energy flow
$E_{\mathrm{out}}$.  The former is a global measurement and the latter
is a non-global measurement. If the observable $V$ selects 2-jet-like
configurations, one measures the associated distribution,
\begin{equation}
\Sigma_{\mathrm{2ng}}(Q,V,E_{out}),
\end{equation}
where $Q$ is the hard scale. It has been shown that this distribution
factorises \cite{Dokshitzer:2003uw},
\begin{equation}
  \Sigma_{\mathrm{2ng}}(Q,V,E_{out})=\Sigma(Q,V)
  \cdot\Sigma_{\mathrm{out}}(VQ,E_{\mathrm{out}}),
\end{equation}
where $\Sigma(Q,V)$ is the standard global distribution of $V$ and
$\Sigma_{\mathrm{out}}(VQ,E_{\mathrm{out}})$ contains the logarithmic
distribution in $E_{\mathrm{out}}$. This latter distribution,
containing NGLs is evaluated at the reduced scale
$VQ$, and hence the logarithmic terms will be $(\alpha_s
\log(VQ/E_{\mathrm{out}}))^n$. The work of Berger, K\'ucs and
Sterman\cite{Berger:2003iw} considered the region in which $VQ$ and~$E_{\mathrm{out}}$ 
were comparable, so that the NGLs give a negligible
contribution. Thus, for a restricted subset of appropriately selected
events, it is possible, to `tune out' the non-global logarithmically
enhanced terms in associated distributions. Hence the use of event shape/energy flow
correlations is another method, along with clustering algorithms, to control the 
effect of NGLs in non-global observables.

\section{Conclusions}

\label{secngl:conc}

The observation that observables that are sensitive to radiation in only a restricted part of phase space, so-called non-global 
observables, are strongly sensitive to secondary radiation is a new and exciting discovery. For a long time
it was widely thought, now it seems incorrectly, that it was sufficient to consider only primary emission contributions to such observables.
These primary, or Bremsstrahlung, contributions are well understood for a variety of processes. Recent measurements of gaps-between jets 
events at H1 and ZEUS are exactly the class of observable that are sensitive to these effects and to deal with this fact, a study of
NGLs in the context of HERA measurements is required.
In this chapter we considered the NGLs generated from secondary radiation into a restricted region of phase space under the condition of 
the final state surviving a clustering algorithm. Such kt clustering algorithms have been used for the HERA measurements. Our work verified the study of Dasgupta and Salam, who found that inclusion of NGLs resulted
in a strong suppression of the theoretical prediction of the non-global observable.

We have also studied the NGLs for a 2 jet system with clustering imposed and found that the clustering 
process
reduced, but did not eliminate, the non-global logarithm effect.

Our main conclusion is that the final state specified in the H1 and ZEUS analysis will mean that a primary emission calculation in the manner
of Sterman {\it et al} \cite{Oderda:1998en,Oderda:1999kr,Berger:2001ns}, and chapter \ref{ch3}, of energy flow observables will overestimate the observed gaps-between-jets rate by around 20\%. This is to be contrasted with an overestimation of~65\%
that would be found for a non-clustered final state.
This result was calculated to all orders in the large $N_c$ limit for a 2 jet system. The extension of both these results
to beyond the large $N_c$ limit would be a major step forward in the understanding of these effects in
non-global observables and we reserve this for future work.

\chapter{Resummation of energy flow observables}
\label{ch5}
% Chapter 5

\section{Introduction}

\label{secef:intro}

The subject of interjet energy flow \cite{Marchesini:1988} has attracted considerable interest ever since 
it was proposed \cite{dokshitzer,Bjorken:1992er} as a way to study 
rapidity gap processes using the tools of perturbation theory. Rapidity gap processes are defined as processes
containing two high $p_t$ jets with the region of rapidity between the jets containing nothing more than soft
radiation. This region is known generically as the gap. The presence of a range of scales offers a chance to study 
the interface between the soft, non-perturbative scales and the hard, perturbative scales of
QCD. 

In this chapter we will calculate the perturbative contribution to gaps-between-jets cross sections in photoproduction 
at HERA, which have 
been measured by the H1 \cite{Adloff:2002em} and ZEUS \cite{Derrick:1995pb,zeus:2003} collaborations.  
A feature of the recent analyses is the use of a clustering algorithm to define the hadronic final state 
and hence the gap. 
The restriction of transverse radiation in a region of phase space, defined as $\Omega$ and directed away
from the observed jets and the beam directions, produces logarithms at each order of QCD perturbation theory of
the interjet energy flow, $Q_{\Omega}$, over some hard scale, $Q$. The precise definition of the restricted 
region, or gap, is totally free and in this work we are interested in the gap region defined by experimental 
rapidity gap analyses. The source of the large logarithms is twofold. The so-called
primary (or global) logarithms arise from radiation emitted directly into $\Omega$; these wide-angle gluons decouple 
from the dynamics of the colour-diagonal jets and are described by an effective, eikonal theory 
\cite{Berger:2001ns,Oderda:1998en,Oderda:1999kr,Kidonakis:1998nf,Kidonakis:1998bk}.
 The second source of leading
logarithms arise from gluons emitted outside of the gap region, an area of phase space generically denoted
as $\bar{\Omega}$, which subsequently radiate into $\Omega$. These terms are known as non-global (secondary)
logarithms, or NGLs \cite{Dasgupta:2001sh,Dasgupta:2002bw,Appleby:2002ke,Appleby:2003ai}, and were introduced 
in chapter \ref{ch4}.

The primary logarithms are resummed using the formalism of Collins, Soper and Sterman (CSS)
\cite{Collins:ig,Sotiropoulos:1993rd,Contopanagos:1996nh,Kidonakis:1998bk}, which was introduced in 
chapter \ref{ch3}. In this method 
the cross section is factorised into a soft
part describing the emission of soft, wide angle gluons up  to scale $Q_{\Omega}$ and a hard part, 
describing harder quanta. 
A unique feature of QCD
is that the soft and the hard functions are expressed as matrices in the space of possible colour flow of the system.
%The soft function coincides with the QCD eikonal cross section \cite{Kidonakis:1998nf}. 
The scale invariance and factorisation
properties of the cross section are then exploited to resum primary logarithms of $Q_{\Omega}/Q$. 
This resummation is driven by the ultraviolet pole parts of eikonal Feynman graphs and we write the resummed cross section 
in terms of the eigenvalues 
of $\Omega$-dependent soft anomalous dimension matrices. These matrices are known for gap definitions based on the
cone definition of the final state \cite{Oderda:1998en,Oderda:1999kr} and for a gap defined as a square patch in 
rapidity and azimuthal angle
\cite{Berger:2001ns};
here we are interested in gaps defined in terms of the clustering algorithms employed in the recent 
analyses. Hence we are 
required to calculate the corresponding anomalous dimension matrices.

The NGLs \cite{Dasgupta:2002bw,Dasgupta:2001sh} are unable to be incorporated into the resummation of the 
primary logarithms, because the gluon 
emission patterns that produce the NGLs are sensitive to underlying colour flows not included in the formalism. 
The effect of NGLs, which is a suppressive effect, on energy flow processes has been studied using numerical methods 
in the large $N_c$ limit and overall factors describing their effect have been extracted for a two jet system, both
without \cite{Dasgupta:2002bw} and with \cite{Appleby:2002ke} clustering. The latter case was described in 
chapter \ref{ch4}. This factor is not directly applicable to 
the 4 jet systems\footnote{Note that for a two-to-two process the incoming and outgoing partons radiate, so we 
consider the process to be of ``four jet'' type, although only two jets are seen in the final state.} relevant in the 
photoproduction of jets but, in the lack of a four jet formalism, we nevertheless use the 2-jet factor in our predictions.

Our aim is to derive LL resummed predictions for the gap cross section, with primary logarithms correct
to all orders and secondary logarithms correct in the large~$N_c$ limit. The gap cross section will follow the HERA
analyses and demand two hard jets, defined using the kt clustering algorithm 
\cite{Catani:1993hr,Ellis:tq,Butterworth:2002xg}, and we will closely follow the
H1 and ZEUS gap definition. The technical aspects of soft gluon resummation
give a strong dependence on the gluon emission phase space, and hence a considerable part of our work will be 
concerned with the
calculation of soft gluon effects for the specific detector geometry of the H1 and ZEUS experiments.

The organisation of this chapter is as follows. Section \ref{secef:hera} describes, in detail, 
the energy flow analyses of H1 and ZEUS. We 
summarise the experimental cuts employed and the range of measured observables. We also discuss the theoretical
implementation of the inclusive kt algorithm employed to define the hadronic final state and the impact
on soft gluon resummation. Section \ref{secef:fact} describes the theoretical definition of our cross section and we employ the
standard QCD factorisation theorems to write it as the convolution of non-perturbative parton distributions and
a short-distance hard scattering function. We then proceed to refactorise the hard scattering function and exploit this 
factorisation to resum the large interjet logarithms. Section \ref{secef:kt} then derives the 
soft anomalous dimension matrices 
for the kt defined final state and in section \ref{secef:results} we present detailed predictions of 
rapidity gap processes and compare 
to the H1 data. Finally we draw our conclusion in section \ref{secef:conc}. 
We find that our description of the data is good, although the approximate treatment of NGLs results in a
relatively large normalisation uncertainty. The research work in this chapter 
is published in \cite{Appleby:2003sj}.

\section{The HERA energy flow analyses}

\label{secef:hera}

In this section we will outline the experimental analyses of the photoproduction of gaps-between-jets processes and 
summarise the experimental cuts and rapidity gap observables. We will also describe the clustering algorithm used
to define the final hadronic state in the more recent H1 \cite{Adloff:2002em} 
and ZEUS \cite{zeus:2003} analyses.

\begin{figure}
\begin{center}
\epsfig{figure=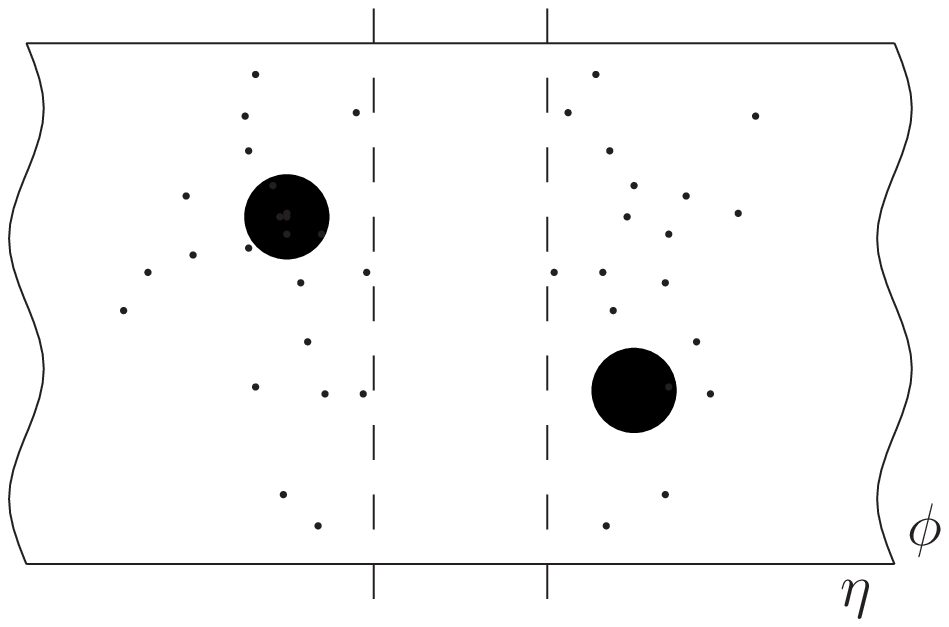,width=3in,height=2.0in}
\caption{The experimental signature of a ``jet-gap-jet'' event. The dark circles denote two jets, and the gap region 
is bounded by the dashed lines.
\label{secef:picjgj}}
\end{center}
\end{figure}

The data for these events were collected when HERA collided $27.6$~GeV positrons\footnote{The positron energy 
varied a negligible amount between the two sets of analyses.}  with $820$~GeV protons, giving 
a centre of mass energy of $\sqrt{s}\simeq 300$~GeV. Following the jet-finding phase, which we will comment on
later, the total transverse energy flow between the two highest $E_T$ jets, denoted $E_T^{\mathrm{GAP}}$, is calculated
by summing the transverse energy of all jets produced by the kt algorithm that are not part of the dijets 
in the pseudorapidity region between
the two highest jets. An event is then defined as a gap event if the energy is less than some energy cut
$E_T^{\mathrm{CUT}} \equiv Q_{\Omega}$. A gap fraction is then calculated by dividing the cross section at fixed $E_T^{\mathrm{CUT}}$ by
the inclusive cross section. The ZEUS collaboration performed a rapidity gap analysis 
several years ago~\cite{Derrick:1995pb} using the
cone algorithm for the jet definition and presented the gap fraction at~$Q_{\Omega}=0.3$~GeV. We consider this 
value of $Q_{\Omega}$ as being too small for our perturbative analysis and will not make any predictions for this
data set. The gaps-between-jets experimental signature is illustrated in figure~\ref{secef:picjgj}.

The more recent H1 and ZEUS analyses used the kt definition of the 
final state and both collaborations presented the gap fraction at four different values of $Q_{\Omega}$, as shown in 
table~\ref{secef:tabheracuts}. We will make predictions and compare to data for the  H1 data sets and, due 
to the fact that the ZEUS data is still preliminary, confine ourselves to making predictions for
the ZEUS analysis. We have summarised the cuts used in table~\ref{secef:tabheracuts}. 

\begin{table}
\begin{center}
\begin{tabular}{|c|c|c|}
 \hline
 & H1 & ZEUS  \\ \hline \hline
$E_T^{\mathrm{jet1}}$ &  $>6.0$~GeV &  $>6.0$~GeV \\ \hline
$E_T^{\mathrm{jet2}}$ &  $> 5.0$~GeV &  $> 5.0$~GeV \\ \hline
$\eta^{\mathrm{jet1}}$ & $< 2.65$ &  $< 2.4$\\ \hline
$\eta^{\mathrm{jet2}}$ & $< 2.65$ & $< 2.4$\\ \hline
$\Delta\eta$ & $ 2.5 < \Delta\eta < 4$ & $2 < \Delta\eta <4$\\ \hline
$\eta_{\mathrm{jj}}$ & N/A & $< 0.75$ \\ \hline
$y$ & $0.3 < y < 0.6$ & $0.2 < y < 0.85$ \\ \hline
$Q^2$ & $< 0.01$~GeV$^2$ & $< 1$~GeV$^2$ \\ \hline
jet def. & kt & kt \\ \hline
gap def. & $\Delta y=\Delta\eta$ & $\Delta y=\Delta\eta$ \\ \hline
$R$ & $1.0$  & $1.0$ \\ \hline
$Q_{\Omega}$ & $0.5,1.0,1.5,2.0$~GeV  & $0.5,1.0,1.5,2.0$~GeV \\ \hline
\end{tabular}
\caption{The experimental cuts used for the HERA analyses.}
\label{secef:tabheracuts}
\end{center}
\end{table}

\subsection{The \boldmath{kt} algorithm}

\label{secef:ktalgor}

Of special interest to those going about soft gluon calculations is the method used to define the hadronic final state, 
the reason being that this jet-finding process determines the phase space
for soft gluon emission; the method used in the H1 and ZEUS data sets 
is the kt algorithm \cite{Catani:1993hr,Ellis:tq,Butterworth:2002xg}. 
Here we follow the inclusive scheme used at H1 and ZEUS which depends on the parameter R, normally set to unity. 
If we assume that any radiation into the gap is much softer than any parent radiation, then this radiation
 with $E_T < E_T^{\mathrm{jet}}$ will be merged into the jet (with kinematical variables 
$(\eta_{\mathrm{jet}},\phi_{\mathrm{jet}})$) if it
satisfies
\begin{equation}
(\eta_r-\eta_{\mathrm{jet}})^2+
(\phi_r-\phi_{\mathrm{jet}})^2 
< R^2,
\end{equation}
where we denote the kinematical variables of the radiated gluon by $(\eta_r,\phi_r)$.
Once merged, a gluon will be pulled out of the gap and can no longer produce a primary or secondary logarithm.
The gap is defined as the interjet region minus the
region of clustered radiation around the jets and may contain soft protojets. The gap transverse energy is then 
defined by the (scalar) sum of the protojets within the gap region, $\eta_1 < \eta < \eta_2$. We have given a 
detailed description of how the kt algorithm works in chapter \ref{ch4}.

The kt gap definition can be contrasted to the older ZEUS analysis \cite{Derrick:1995pb}, which used the well known 
cone definition of the
final state with $R=1.0$. The gap transverse energy is then defined as the scalar sum of the transverse 
energy of the hadrons within it, 
$\eta_1+R < \eta < \eta_2-R$.

\section{Factorisation, refactorisation and resummation of the cross section}

\label{secef:fact}

In this section we will exploit the standard factorisation theorems of QCD to write down the dijet production
cross section from the interaction of a proton and a positron. We will then refactorise the hard scattering function
into the product of two matrices in the space of possible hard scattering colour flows, one matrix describing
soft gluons radiated into the gap region and the other a hard scattering matrix. The renormalisation properties of the
cross section are then used to resum primary interjet logarithms, and write the result in terms of the
eigenvalues of the matrix of counterterms used to renormalise the soft function. In the following section we will 
calculate these matrices and their eigenvalues.

\subsection{Photoproduction cross sections}

The scattering of positrons and protons at HERA proceeds predominantly through the exchange of photons with
very small virtuality and produces a large subset of events with jets of high transverse momentum, $E_T$. 
The presence of this large scale allows the 
application of the perturbative methods of QCD to predict the cross section for multiple jet production.
This process is otherwise 
known as jet photoproduction.

The leading order (LO) QCD contribution can be divided into two types \cite{Oderda:1998en}. The first is 
the direct process in which the photon
interacts directly with a parton from the proton and proceeds through either the Compton process, $\gamma q \rightarrow gq$, or the photon-gluon fusion process, $\gamma g \rightarrow q\bar{q}$. The second contribution is the resolved contribution, in which
the virtual photon fluctuates into a hadronic state that acts as a source of partons, which then scatter off the partonic content of the proton. Therefore
the reaction proceeds through standard QCD $2\rightarrow2$ parton scattering processes. Note that the
precise determination of the partonic content of the photon is a very open question and there is a 
relatively large error associated with the photonic parton densities.
The spectrum of virtual
photons is approximated by the Weiz\"{a}cker--Williams \cite{williams:1934} formula,
\begin{equation}
F_{\gamma/e}(y)=\frac{\alpha}{2\pi}\frac{(1+(1-y)^2)}{y}\log\left( \frac{Q^2_{\mathrm{max}}(1-y)}{m_e^2 y^2}\right),
\end{equation}
where $m_e$ is the electron mass, $y$ is the fraction of the positron's energy that is transfered to the photon, and $Q^2_{\mathrm{max}}$ is the maximum
virtuality of the photon, which is determined by the experimental cuts employed in the analyses. 
Then, by using the equivalent photon approximation, the cross section for the process $e^+p\rightarrow e^{+}X$ is given by
the convolution
\begin{equation}
\mathrm{d}\sigma(e^+p\rightarrow e^{+}X) =\int_{y_{\mathrm{min}}}^{y_{\mathrm{max}}} \mathrm{d}y\,F_{\gamma/e}(y) 
\,\mathrm{d}\sigma(\gamma p \rightarrow X),
\end{equation}
where we write $\mathrm{d}\sigma(\gamma p \rightarrow X)$ for the cross section of $\gamma p \rightarrow X$. The
centre of mass energy squared for the photon-proton system is $W^2=ys$, where $s$ is the centre of mass energy 
squared for the positron-proton system. At HERA,
$s\simeq 90,000$~GeV$^2$ and the values for $y_{\mathrm{min}}$ and $y_{\mathrm{max}}$ are determined by the experimental analyses.
We can now write down the specific expression for the production of two high $E_T$ jets from the photon-proton system, which is written as a sum
of the direct and resolved contributions,
\begin{eqnarray}
d\sigma_{e^+p}(s,\hat{t},\Delta\eta,\alpha_s(\mu_r),Q_{\Omega})
=&& \int_{y_{min}}^{y_{max}} \mathrm{d}y\,F_{\gamma/e}(y) \bigg(
\mathrm{d}\sigma_{\gamma p}^{\mathrm{dir}}
(s_{\gamma p},\hat{t},\Delta\eta,\alpha_s(\mu_r),Q_{\Omega}) \nonumber \\
&&+ \mathrm{d}\sigma_{\gamma p}^{\mathrm{res}}
(s_{\gamma p},\hat{t},\Delta\eta,\alpha_s(\mu_r),Q_{\Omega})\bigg),
\end{eqnarray}
where we denote the 4-momentum transfer squared in the hard scattering as~$\hat{t}$ and have temporarily suppressed the
function arguments on the right-hand side.
We define the rapidity difference and average of the two hard jets by
\begin{eqnarray}
\Delta\eta&=&|\eta_1-\eta_2|, \nonumber \\
\eta_{JJ}&=&\frac{1}{2}(\eta_1+\eta_2).
\end{eqnarray}
At this point we can appeal to the factorisation theorems of QCD and, by working in the $\gamma p$ frame, 
write down factorised forms for the direct and
resolved cross sections. The factorised direct cross section is 
\begin{eqnarray}
\frac{\mathrm{d}\sigma_{\gamma p}^{\mathrm{dir}}}{\mathrm{d}\hat{\eta}}
(s_{\gamma p},\hat{t},\Delta\eta,\alpha_s(\mu_r),Q_{\Omega})=&&
\sum_{f_p,f_1,f_2} 
\int_{\mathrm{R_d}} \mathrm{d}x_p \, \phi_{f_p/p}(x_p,\mu_f) \nonumber \\
&&\times \frac{\mathrm{d}\hat{\sigma}^{(\gamma f)}}{\mathrm{d}\hat{\eta}}
(\hat{s},\hat{t},\Delta\eta,\alpha_s(\mu_r),Q_{\Omega},\mu_f),
\end{eqnarray}
and the factorised resolved cross section is
\begin{eqnarray}
\frac{\mathrm{d}\sigma_{\gamma p}^{\mathrm{res}}}{\mathrm{d}\hat{\eta}}
(s_{\gamma p},\hat{t},\Delta\eta,\alpha_s(\mu_r),Q_{\Omega})=&&
\sum_{f_{\gamma},f_p,f_1,f_2} \int_{\mathrm{R_r}} \mathrm{d}x_{\gamma}\,\mathrm{d}x_p 
\, \phi_{f_{\gamma}/\gamma}(x_{\gamma},\mu_f) \phi_{f_p/p}(x_p,\mu_f) \nonumber \\
&&\times \frac{\mathrm{d}\hat{\sigma}^{(f)}}{\mathrm{d}\hat{\eta}}
(\hat{s},\hat{t},\Delta\eta,\alpha_s(\mu_r),Q_{\Omega},\mu_f)
\end{eqnarray}
which are written in terms of the jet rapidity, $\hat{\eta}$, in the partonic centre-of-mass frame, and we write the 
factorisation scale and the renormalisation scale as $\mu_f$ and $\mu_r$ respectively. Note that
 $\hat{\eta}=\Delta\eta/2$, $\hat{s}=x_p W^2$ for the direct case and $\hat{s}=x_{\gamma}x_p W^2$ 
for the resolved case.
In these equations we denote the integration regions of the direct and resolved convolutions, which are defined by 
the experimental cuts, by ${\mathrm{R_d}}$ and~${\mathrm{R_r}}$. The parton distribution for a parton of flavour $f$ 
in the photon and the proton are denoted 
by $\phi_{f/\gamma}(x_{\gamma},\mu_f)$ and $\phi_{f/p}(x_p,\mu_f)$ respectively and finally 
$\frac{\mathrm{d}\hat{\sigma}^{(\gamma f)}}{\mathrm{d}\hat{\eta}}$ and 
$\frac{\mathrm{d}\hat{\sigma}^{(f)}}{\mathrm{d}\hat{\eta}}$ are the hard scattering functions which, at lowest order, start
from the Born cross sections. These are the functions that will contain the logarithmic enhancements of $Q_{\Omega}/Q$, 
and hence depend on the definition of the gap~$\Omega$ and the gap energy flow~$Q_{\Omega}$. We
assume that~$Q_{\Omega}$ is sufficiently soft that we can ignore the effects of emission on the parent jet, known as 
recoil, but large enough so that~$Q_{\Omega}^2 \gg \Lambda_{QCD}^2$.
The index $f$ denotes the process~$f_{\gamma}+f_p \rightarrow f_1 + f_2$ and the index~$f\gamma$ denotes the 
process~$\gamma+f_p \rightarrow f_1 + f_2$. Since the aim of this chapter is to calculate ratios of cross sections and 
compare with data, we 
will take the renormalisation scale to equal the factorisation scale and set~$\mu_f=\mu_r=p_t$, where $p_t$ is the 
transverse momentum of the produced jets.

\subsection{Refactorisation}

Following \cite{Berger:2001ns,Kidonakis:1998bk}, and the arguments in chapter \ref{ch3},  
we now refactorise the $2\rightarrow 2$ hard scattering 
function into a hard matrix and a soft matrix,
\begin{eqnarray}
\frac{d\hat{\sigma}^{(f)}}{d\hat{\eta}}(\hat{s},\hat{t},\Delta\eta,\alpha_s(\mu_r),Q_{\Omega},\mu_f)=
&& \sum_{L,I} H^{(f)}_{IL}(\hat{s},\hat{t},\Delta\eta,\alpha_s(\mu_r),\mu_f,\mu) \nonumber \\
&& \times S^{(f)}_{LI}(Q_{\Omega},\alpha_s(\mu_r),\mu).
\end{eqnarray}
We introduce a factorisation scale $\mu$, separate to the parton distribution factorisation scale $\mu_f$, and all dynamics
at scales less than $\mu$ are factored into $S_{LI}$. Therefore $H_{IL}$ is $Q_{\Omega}$ independent, and all the 
$Q_{\Omega}$ dependence is included in $S_{LI}$. This latter function describes the soft gluon dynamics. 
The proof of this statement follows standard factorisation arguments \cite{Kidonakis:1998bk}. 
The indices $I$ and $L$ label the basis of colour tensors which describe the possible colour exchange in the hard 
scattering, over which the hard and soft matrices are expressed.
Soft, wide angle 
radiation decouples from the
dynamics of the hard scattering and can be approximated by an effective cross section and in this effective theory
the partons are treated as recoilless sources of gluonic radiation and replaced by eikonal lines, or path ordered
exponentials of the gluon field \cite{Kidonakis:1998nf}. The soft radiation pattern of this effective eikonal theory 
then mimics the radiation
pattern of the partons participating in the hard event, or in other words the effective eikonal theory will
contain the same logarithms of the soft scale as the full theory. The hard scattering function will begin at order
$\alpha_s^2$ for the resolved process and order $\alpha\alpha_s$ for the direct process, and the soft function will 
begin at zeroth order. The lowest order soft function, denoted 
$S^{0}_{LI}$, reduces to a set of colour traces. Note that the definition of the gap, and hence the soft function, depends 
on the jet separation $\Delta\eta$ but we have suppressed this argument of the soft function for clarity.

The construction of the soft function, and in particular its 
renormalisation properties, have been extensively studied 
elsewhere \cite{Kidonakis:1998nf,Kidonakis:1998bk}, and are reviewed in chapter \ref{ch3}. 
A non-local operator is constructed from a product of Wilson lines, which ties four lines (representing 
the four jet process) together with a colour tensor. This operator, which contains ultraviolet divergences and hence 
requires renormalisation, is used to construct a so-called eikonal cross section, which serves as an effective theory
for the soft emission. By summing over intermediate states the eikonal cross section is free of potential collinear singularities.
It is the ultraviolet renormalisation of the eikonal operator that allows colour mixing and the resummation of soft 
interjet logarithms.

\subsection{Factorisation leads to resummation of soft logarithms}

The partonic cross section, which has been factorised into a hard and a soft function, should not depend
on the choice of the factorisation scale $\mu$,
\begin{equation}
\mu \frac{\partial}{\partial\mu}\left(\frac{\mathrm{d}\hat{\sigma}}{\mathrm{d\hat{\eta}}}\right)=0.
\end{equation}
Following chapter \ref{ch3}, this means the soft function obeys
\begin{equation}
\left( \mu \frac{\partial}{\partial\mu} + \beta(g_s) \frac{\partial}{\partial g_s} \right)
\uuline{S}=
-\uuline{\Gamma_s}^{\dagger}(\hat{\eta},\Omega) \uuline{S}-\uuline{S}\uuline{\Gamma_s}(\hat{\eta},\Omega).
\label{secef:rge}
\end{equation}
It is important to point out that we have deliberately ignored the complications of terms in this equation arising from
radiation into~$\bar{\Omega}$, as in~\cite{Berger:2001ns} and chapter~\ref{ch3}, and only include 
radiation emitted by the soft function directly 
into~$\Omega$. 
The implication of ignoring these non-global terms is discussed in section \ref{secef:ngl}, where we also describe how
to include their effect in a different way. Therefore we have never 
included the, technically correct,~$\bar{\Omega}$ argument of the soft function. 
The matrices~$\Gamma_s(\hat{\eta},\Omega)$ are process-dependent 
soft anomalous dimension matrices which depend on the details of the gap definition and the hard scattering. 
This equation is solved by transforming to a basis in which these matrices are
diagonal and hence we require a knowledge of the eigenvectors and eigenvalues of the soft anomalous dimension matrices.
We obtain the entries for~$\Gamma_s(\hat{\eta},\Omega)$ from the coefficients of the ultraviolet poles in the matrix of 
counterterms which
renormalise the soft function; we can write the renormalisation constant as a sum over 
terms from different eikonal lines, each with the
form of a colour factor multiplied by a scaleless integral:
\begin{equation}
(Z_S)_{LI}=\sum_{i,j}(Z_S^{(ij)})_{LI}=\sum_{i,j}\mathcal{C}_{LI}^{(ij)} \omega^{(ij)}.
\end{equation}
The eikonal momentum integrals are process independent, and only depend on~$i$ and~$j$, the eikonal lines that are 
connected by the virtual gluon. The colour factor is found
from consideration of the colour flow for a given process and the basis over which the colour flow is to be decomposed.
The result is a basis- and process-dependent set of colour mixing matrices, which we have listed in 
appendix~\ref{secef:appdecomp}, 
together with our choice of bases in appendix~\ref{secef:appbases}.
The colour mixing matrices were discussed in detail in chapter~\ref{ch1} and 
have been obtained in \cite{Berger:2001ns,Berger:2003zh,Kidonakis:2000gi} for all relevant 
subprocesses, and involves using SU(3) 
colour identities like
\begin{equation}
t^a_{ij}t^a_{kl}=\frac12\delta_{il}\delta_{kj}-\frac{1}{2N_c}\delta_{ij}\delta_{kl},
\end{equation}
for quark processes and
\begin{eqnarray}
d_{abc}&=&2\left[ \mathrm{Tr}\left(t^a t^c t^b\right)+\mathrm{Tr}\left(t^a t^b t^c\right)\right], \\
f_{abc}&=&-2i\left[ \mathrm{Tr}\left(t^a t^b t^c\right)-\mathrm{Tr}\left(t^a t^c t^b\right)\right],
\end{eqnarray}
for gluon processes, to decompose one-loop graphs over a colour basis. A more complete set of 
SU(3) group theoretic identities can be found in appendix \ref{secintro:app1}. We use the fact that the colour
flow for a real graph is the same as the corresponding virtual graph, valid for primary emission.

Therefore we need to calculate the ultraviolet divergent contribution to the momentum function~$\omega^{(ij)}$ from
all contributing eikonal graphs. Working in the Feynman gauge there are two possible sources of divergence. The first 
 is one loop eikonal graphs with a virtual gluon connecting eikonal lines~$i$ and~$j$. From the eikonal Feynman rules
listed in the appendix, these graphs will give a real and an imaginary contribution to~$\Gamma_s$. Note that as
we are working in the Feynman gauge the self energy diagrams~($\omega^{(ii)}$) give no contribution. The second source
of ultraviolet divergences are the real emission diagrams, when the emitted gluon is directed out of the gap. This can
produce an ultraviolet divergence in the eikonal graph as we only measure energy flow into the gap and are fully inclusive
out of the gap. Hence the virtual graphs will only depend on the relative direction of the two eikonal lines and the
real graphs will give a gap (and hence a jet algorithm) dependence. This sum over real and virtual eikonal graphs 
ensures that the soft function remains free of collinear divergences. 
The imaginary (and geometry independent) part of all
our anomalous dimension matrices can be extracted 
from~\cite{Berger:2001ns,Oderda:1999kr,Kidonakis:1998nf,Kidonakis:2000gi}, and the 
calculation for a cone-algorithm defined 
final state has been done in~\cite{Oderda:1999kr}. For the latter case, 
we have re-expressed their results in accordance with our notation in appendix~\ref{secef:appcone}.

\begin{figure}
\begin{center}
\epsfig{figure=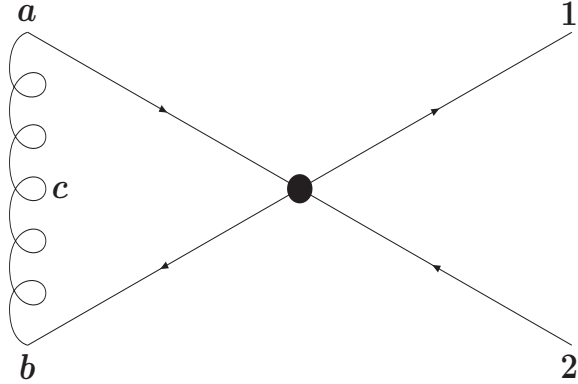,width=3in,height=2in}
\caption{The virtual eikonal graph obtained if the virtual gluon connects eikonal lines $a$ and $b$. 
 In the notation in the text, this 
diagram is denoted $I_v^{(ab)}$.
\label{secef:eikonal}}
\end{center}
\end{figure}
By performing the energy integral of the virtual graphs, we can combine the result with the
corresponding real graph at the integrand level and obtain a partial cancellation. Therefore we start by evaluating
a generic virtual eikonal graph, with the virtual gluon connecting eikonal lines $i$ and $j$, using the eikonal 
Feynman rules in appendix \ref{secresum:appeikonal}. We obtain
\begin{equation}
I_v^{(ij)}=g_s^2 \Delta_i \Delta_j \beta_i\cdot \beta_j
\int \frac{d^dk}{(2\pi)^d} 
\frac{-i}{(k^2+i\epsilon)}\frac{1}{(\delta_i \beta_i \cdot k + i\epsilon)}\frac{1}{(\delta_j \beta_j\cdot k
+i\epsilon)},
\end{equation}
where $\delta_{i,j}=\pm 1$ and we denote the momenta of the virtual gluon by $k$. In this chapter we use the 
dimensional regularisation convention $d=4-2\epsilon$. 
Figure \ref{secef:eikonal} shows the virtual eikonal diagram $I_v^{(ab)}$.
Note that in this section we will use the symbol $\epsilon$ to denote a UV pole and
also as a small parameter when doing contour integration. However there should be no confusion. Writing 
\begin{eqnarray}
d^d k &=& dk^0\,d^{d-1}\vec{k}, \\
k^\mu &=& (k^0,\vec{k}),  \\
\beta_i^\mu&=&(1,\vec{\beta_i}),  \\
k^2+i\epsilon &=& (k^0 - |\vec{k}| + i\epsilon)(k^0 + |\vec{k}| - i\epsilon) \label{secef:k2expand},
\end{eqnarray}
we get
\begin{equation}
I_v^{(ij)}=g_s^2 \Delta_i \Delta_j \beta_i\cdot \beta_j
\int dk^0 \frac{d^{d-1}\vec{k}}{(2\pi)^d} 
\frac{-i}{((k^{0})^2 - \vec{k}^2 + i\epsilon)}
\frac{1}{(\delta_i[k^0-\vec{\beta_i}\cdot\vec{k}] + i\epsilon)}
\frac{1}{(\delta_j[k^0-\vec{\beta_j}\cdot\vec{k}] + i\epsilon)}.
\end{equation}
We need to evaluate the~$k^0$ integral using contour integration, and there are four combinations
of the~$\delta_i's$ to consider; however the integrals only depend on the product~$\delta_i \delta_j$ 
and so it is sufficient to 
only consider two of the four possible combinations of~$\delta_i$ and~$\delta_j$. 
The four poles in the complex~$k^0$ plane are
\begin{eqnarray}
k^0&=&-|\vec{k}|+i\epsilon, \nonumber \\
k^0&=&+|\vec{k}|-i\epsilon, \nonumber \\
k^0&=&(\delta_i\vec{\beta_i}\cdot\vec{k}-i\epsilon)/\delta_i, \nonumber \\
k^0&=&(\delta_j\vec{\beta_j}\cdot\vec{k}-i\epsilon)/\delta_j.
\end{eqnarray}
We consider the case of $\delta_i=\delta_j=-1$, for which the poles are
\begin{eqnarray}
k^0&=&-|\vec{k}|+i\epsilon, \nonumber \\
k^0&=&+|\vec{k}|-i\epsilon, \nonumber \\
k^0&=&\vec{\beta_i}\cdot\vec{k}+i\epsilon, \nonumber \\
k^0&=&\vec{\beta_j}\cdot\vec{k}+i\epsilon.
\end{eqnarray}
We close the contour in the lower-half plane and only pick up the pole $k^0=+|\vec{k}|-i\epsilon$. The residue
of this pole is extracted from the standard result of the Laurent expansion
\begin{equation}
\dots+\frac{a_{-1}}{z-z_o}|_{z=z_0}+\dots
\end{equation}
and we get for the residue (where we have used equation (\ref{secef:k2expand}) to obtain the desired form)
\begin{equation}
a_{-1}=\frac{-i}{(2|\vec{k}|-i\epsilon)(-|\vec{k}|+\vec{\beta_i}\cdot \vec{k} + i\epsilon)
(-|\vec{k}|+\vec{\beta_j}\cdot \vec{k} + i\epsilon)}.	
\end{equation}
Therefore, multiplying by $2\pi i$ and by $-1$ (for going clockwise around the contour), we get
\begin{eqnarray}
I^{(\delta_i=\delta_j=-1)}&=&(-2\pi i) g_s^2 \Delta_i \Delta_j \beta_i\cdot \beta_j
 \int \frac{d^{d-1}k}{(2\pi)^d} \frac{-i}{2|\vec{k}|(-\beta_i\cdot k)(-\beta_j\cdot k)}, \nonumber \\
&=& - g_s^2 \Delta_i \Delta_j \beta_i\cdot \beta_j 
\int \frac{d^{d-1}k}{(2\pi)^{d-1}} \frac{1}{2|\vec{k}|(-\beta_i\cdot k)(-\beta_j\cdot k)},
\end{eqnarray}
as $\epsilon\rightarrow0$. We note that this expression 
is the {\bf same} as the real diagram expression integrated over all of phase space, and with an overall
minus sign. This real diagram expression is readily obtained from the evaluation of an eikonal diagram with real gluon
emission using the eikonal Feynman rules in appendix \ref{secresum:appeikonal}. We can also see that, if we have no 
restrictions on the real gluon emission energy, we expect
a complete cancellation of real and virtual diagrams.
Now consider the case of~$\delta_i=+1$ and~$\delta_j=-1$. This configuration corresponds to virtual diagrams where
the virtual gluon connects the incoming eikonal lines. Now we have the following arrangement of poles,
\begin{eqnarray}
k^0&=&-|\vec{k}|+i\epsilon, \nonumber \\
k^0&=&+|\vec{k}|-i\epsilon, \nonumber \\
k^0&=&\vec{\beta_i}\cdot\vec{k}-i\epsilon, \nonumber \\
k^0&=&\vec{\beta_j}\cdot\vec{k}+i\epsilon.
\end{eqnarray}
with two in the upper half plane, and two in the lower-half plane. We choose to close the contour in the 
lower-half plane again and pick up two poles. The residues are computed in exactly the same manner as the first case
and we obtain, for the $k^0=+|\vec{k}|-i\epsilon$ pole
\begin{equation}
a_{-1}=\frac{-i}{(2|\vec{k}|-i\epsilon)}
\frac{1}{(|\vec{k}|-\vec{\beta_i}\cdot \vec{k})}
\frac{1}{(-|\vec{k}|+\vec{\beta_j}\cdot \vec{k} + i\epsilon)},
\end{equation}
and for the $k^0=\vec{\beta_i}\cdot\vec{k}-i\epsilon$ pole
\begin{equation}
a_{-1}=\frac{-i}{((\vec{\beta_i}\cdot \vec{k} - i\epsilon)^2-\vec{k}^2+i\epsilon)
(\vec{\beta_j}\cdot \vec{k} - \vec{\beta_i}\cdot \vec{k} + i\epsilon)}.
\end{equation}
Note that to get the correct form of the first residue, we need to use the form of the integrand containing 
equation (\ref{secef:k2expand}), and to get the correct form for the second residue we need to use 
the $k^2=(k^{0})^2 - \vec{k}^2$ form of the integrand.
Therefore the integral for this arrangement of poles has two contributions - one from the $k^2$ pole
and another from the $\beta\cdot k$ pole. The first gives 
\begin{equation}
I^{(\delta_i=+1,\delta_j=-1)}=-g_s^2 \Delta_i \Delta_j 
\beta_i\cdot \beta_j \int \frac{d^{d-1}k}{(2\pi)^{d-1}} \frac{1}{2|\vec{k}|(\beta_i\cdot k)(-\beta_j\cdot k)},
\end{equation}
and the latter gives
\begin{equation}
I_s=-g_s^2 \beta_i\cdot \beta_j \int \frac{d^{d-1}k}{(2\pi)^{d-1}} 
\frac{1}{((\vec{\beta_i}\cdot\vec{k})^2 - \vec{k}^2 \pm i\epsilon)
(\vec{\beta_j}\cdot \vec{k}-\vec{\beta_i}\cdot \vec{k}+i\epsilon)}.
\label{secef:eqimagbit}
\end{equation}
We will give an explanation for the $s$ subscript later in this section.
The $\pm$ piece comes from the unknown sign of $\vec{\beta_i}\cdot k$. Note that, once again, the piece arising
from the $k^2$ poles gives a contribution looking like the negative of the real contribution. At first sight, it 
looks like the real diagrams should give a $\beta_i\cdot k$ pole if we write the Lorentz invariant phase space in 
4 dimensions
using
\begin{equation}
\int \frac{d^3 p}{(2\pi)^3}=\int \frac{d^4 p}{(2\pi)^4} (2\pi)\delta^+(p^2),
\end{equation}
where we express the on-shell condition of real gluons as
\begin{equation}
\delta^+(p^2)=\delta(p^2)\Theta(p^0).
\end{equation}
The $\beta_i\cdot k$ pole is not picked up by the real diagram energy integral, due to this condition.
We have determined that if we perform the energy integral of a virtual eikonal graph, we obtain the same form
as the corresponding (same $i$ and $j$) real graph. We shall not consider the analysis of the remaining 
pole configurations, for which we will draw the same conclusions, and describe how
everything fits together. We are interested in the ultraviolet pole part of the real and virtual one-loop 
eikonal diagrams, in
order to calculate the soft anomalous dimension matrix. The virtual loop diagrams contribute 
through their loop integrals and
the real diagrams contribute in all regions where we are completely inclusive of gluonic radiation (everywhere
except any ``gap'' region.) Therefore
\begin{eqnarray}
\omega^{(ij)}&=&I_v^{(ij)} + I_r^{(ij)}, \nonumber \\
&=& g_s^2 \Delta_i \Delta_j \beta_i\cdot \beta_j \int \frac{d^d k}{(2\pi)^d} \frac{-i}{(k^2+i\epsilon)}
\frac{1}{(\delta_i \beta_i \cdot k + i\epsilon)}\frac{1}{(\delta_j \beta_j\cdot k
+i\epsilon)} \nonumber \\
&+& g_s^2 \Delta_i \Delta_j \beta_i\cdot \beta_j \int \frac{d^{d-1}k}{(2\pi)^{d-1}} (1-\Theta(\vec{k}))
\frac{1}{2|\vec{k}|(\delta_i\beta_i\cdot k)(\delta_j\beta_j\cdot k)}.
\end{eqnarray}
In writing this line we have used the explicit forms of the virtual and real diagrams, with the
virtual gluon joining eikonal lines $i$ and $j$. The function $\Theta(\vec{k})=1$ when the real gluon is
directed into the gap, and is zero otherwise. We can now do the virtual~$k^0$ integral,
\begin{eqnarray}
\omega^{(ij)}&=&-g_s^2 \Delta_i \Delta_j \beta_i\cdot \beta_j \int \frac{d^{d-1}k}{(2\pi)^{d-1}} 
\frac{1}{2|\vec{k}|(-\beta_i\cdot k)(-\beta_j\cdot k)}, \nonumber \\
&+& g_s^2 \Delta_i \Delta_j \beta_i\cdot \beta_j \int \frac{d^{d-1}k}{(2\pi)^{d-1}} (1-\Theta(\vec{k}))
\frac{1}{2|\vec{k}|(\delta_i\beta_i\cdot k)(\delta_j\beta_j\cdot k)}, \nonumber \\
&+& \frac12 I_s(1-\delta_i\delta_j) \nonumber \\
&=& -g_s^2 \Delta_i \Delta_k \beta_i\cdot \beta_j \int \frac{d^{d-1}k}{(2\pi)^{d-1}} \Theta(\vec{k})
\frac{1}{2|\vec{k}|(\delta_i\beta_i\cdot k)(\delta_j\beta_j\cdot k)} \nonumber \\
&+& \frac12 I_s(1-\delta_i\delta_j),
\end{eqnarray}
and find that its real part can be written as $-1 \,\times$ the real contribution, but with no phase space
restrictions. This part is the bit left over from picking up the $k^2$ pole. The function $I_s$ is the result 
of picking up the $\beta_i\cdot k$ pole, and is only present
when $\delta_i=-1\times\delta_j$, or when the virtual gluon connects two initial or final state eikonal lines
(s-channel diagrams). This is the reason for the~$s$ subscript for the function $I_s$.
We have also performed a real/virtual cancellation and have been left with an integral over the 
vetoed gap region, plus an additional function~$I_s$, equation~(\ref{secef:eqimagbit}).  
To evaluate this latter piece we 
use polar coordinates in~$(d-1)$ dimensions,
\begin{equation}
\mathrm{d}^{d-1}k=k^{d-2}\mathrm{d}k\,\,\mathrm{d}\cos{\theta}\,\,\mathrm{d}\phi\,\,\mathrm{d}^{d-4}\Omega,
\end{equation}
where we have noted that this integral only appears in diagrams in which $\vec{\beta_i}$ and $\vec{\beta_j}$ are 
back-to-back (and so $\beta_i\cdot \beta_j=2$), and write
\begin{eqnarray}
\vec{\beta_i}\cdot\vec{k}&=&k\cos\theta, \\
\vec{\beta_j}\cdot\vec{k}&=&-k\cos\theta.
\end{eqnarray}
Note that the a general polar d-vector for $k$ is
\begin{equation}
k^\mu=k(1;\dots,\sin\theta\sin\phi,\sin\theta\cos\phi,\cos\theta).
\end{equation}
Doing this we obtain the integrand
\begin{eqnarray}
\frac{1}{(\cos^2(\theta)-1\pm i\epsilon)(-2\cos(\theta)+i\epsilon)} 
&=& \frac{1}{(2\cos(\theta)-i\epsilon)(1\mp i\epsilon + 1/4\epsilon^2)}
\nonumber \\ 
&+& \frac{(2\cos(\theta)+i\epsilon)}{(1-\cos^2(\theta)\mp i\epsilon)(4\mp 4i\epsilon + \epsilon^2)},
\end{eqnarray}
where we have used partial fractions to write the second form. We can safely set~$\epsilon=0$ in the
second term, since the only poles present are at $\cos\theta=\pm 1$ and these are protected
by dimensional regularisation. Therefore the second term is an odd function of $\cos\theta$ and
integrates (over the range $-1$ to $+1$) to zero. Therefore we are left with
\begin{equation}
\int_{-1}^{+1} \frac{d\cos(\theta)}{2\cos(\theta)-i\epsilon}= + \frac{\pi}{2} i - \mathcal{O}(\epsilon),
\end{equation}
Hence we find that the imaginary piece is 
\begin{eqnarray}
I_s&=&-\frac{g_s^2}{(2\pi)^2} \frac{1}{2\epsilon} 2\Delta_i \Delta_j \frac{\pi}{2}{i}, \\
&=&\Delta_i\Delta_j\delta_i\delta_j \frac{\alpha_s}{\pi} \frac{i\pi}{2\epsilon},
\end{eqnarray}
where we have used the fact that
\begin{equation}
\int_{\mathrm{PP}} \frac{k^{d-2} dk}{k^3}=\frac{1}{2\epsilon},
\end{equation}
for $d=4-2\epsilon$ and an integration from some fixed energy to infinity.
Therefore the momentum integrals take the form,
\begin{eqnarray}
\omega^{(ij)}&=&I_{v}^{(ij)}+I_{r}^{(ij)},\nonumber \\
&=& -g_s^2 \Delta_i \Delta_k \beta_i\cdot \beta_j \int \frac{d^{d-1}k}{2|\vec{k}|(2\pi)^{d-1}} \Theta(\vec{k})
\frac{1}{(\delta_i\beta_i\cdot k)(\delta_j\beta_j\cdot k)}, \nonumber \\ &+& 
\delta_i \delta_j \Delta_i \Delta_j \frac{\alpha_s}{2\pi}\frac{i\pi}{2\epsilon}(1-\delta_i \delta_j),
\end{eqnarray}
which we can write in terms of the rapidity and transverse 
energy of the emitted gluon,
\begin{equation}
\omega^{(ij)}=-\frac{g_s^2}{2} \Delta_i \Delta_j \delta_i \delta_j \frac{1}{(2\pi)^2} \int
k_t\,dk_t\,d\eta\,\frac{d\phi}{2\pi} \Theta(\vec{k})
\frac{\beta_i\cdot \beta_j}{(\beta_i\cdot k)(\beta_j\cdot k)} + I.P..
\end{equation} 
We have written the geometry independent imaginary part as I.P..
Once we have obtained the momentum integrals for the kt defined
final state we can construct the anomalous dimension matrices using the colour mixing matrices in 
appendix \ref{secef:appdecomp}.
Consideration of the eigenvalues and eigenvectors of these matrices, together with the process-dependent hard and 
soft matrices (the full set of hard and soft matrices is shown in appendix~\ref{secef:hardsoft}
) allows the resummed cross section to be written down,
\begin{equation}
\frac{d\hat{\sigma}^{(f)}}{d\hat{\eta}}=
\sum_{L,I} \bar{H}^{0,(f)}_{IL}\bar{S}^{0,(f)}_{LI} 
\exp\left\{ \frac{1}{\beta_0}(\hat{\lambda}_L^*(\hat{\eta},\Omega)+\hat{\lambda}_I(\hat{\eta},\Omega))
\int_{p_t}^{Q_{\Omega}} 
\frac{d\mu}{\mu}\beta_0\alpha_s(\mu)\right\},
\label{secef:eqresum}
\end{equation}
which follows from the diagonalisation of the soft RGE, equation~(\ref{secef:rge}). This was discussed in 
chapter~\ref{ch3}, where we developed this resummed expression from the factorisation properties of the cross section. 
We denote matrices in the diagonal basis by barred
matrices, the eigenvalues of the anomalous dimension matrices by $\lambda_i=\alpha_s\hat{\lambda}_i$ 
and we write the lowest-order piece of the QCD beta function as $\beta_0=(11N_c-2n_f)/(6\pi)$. Note that our 
normalisation of the eigenvalues differs by a factor of~$\pi$ from that used in chapter~\ref{ch3}. 
We will observe that, in 
agreement with~\cite{Oderda:1999kr},~$\mathrm{Re}(\lambda)>0$ for all physical channels 
and hence the resummed cross sections are suppressed relative to the 
fully inclusive cross section. 

\subsection{Non-global effects}

\label{secef:ngl}

As we have discussed in the last section, we have deliberately ignored terms arising from secondary radiation into
$\Omega$, or non-global logarithms (NGLs) \cite{Appleby:2003ai,Appleby:2002ke,Dasgupta:2002bw,Dasgupta:2001sh}. Such
terms arise from radiation at some intermediate scale,~$M$, being emitted outside of~$\Omega$, i.e.~into $\bar{\Omega}$,
and then subsequently radiating into $\Omega$. In energy flow observables such effects give rise to leading
logarithms. Inclusion of NGLs in the formalism of the last section would result in an explicit $M$ dependence of the
soft function and a sensitivity to more complicated, $2\rightarrow n$, colour flows for all $n>2$. For further details
see \cite{Berger:2001ns}.  NG effects have been studied for a two-jet system by Dasgupta and Salam
\cite{Dasgupta:2002bw,Dasgupta:2001sh}, by Appleby and Seymour with the complication of clustering
\cite{Appleby:2002ke}, and in the context of energy flow/event shape correlations by Dokshitzer and Marchesini
\cite{Dokshitzer:2003uw} and Berger, K\'ucs and Sterman \cite{Berger:2003iw}.

The effect of NGLs for four-jet kinematics has not been explicitly calculated to date.  The best that has been managed
is a two-jet calculation in the large-$N_c$ limit.  The NG contributions to the gap cross section factorize into an
overall suppression factor $S^{NG}$, making it smaller than would be predicted by the resummation of primary logarithms
alone.  In the absence of a complete calculation, we include the NGLs approximately, by using our all-order results in
the large-$N_c$ limit for $S^{NG}$ in a two-jet system \cite{Appleby:2002ke}.  Since four-jet configurations are
dominated, in the large-$N_c$ limit and for large $\Delta\eta$, by colour flows in which two colour dipoles stretch
across the gap region, we approximate the four-jet NG suppression factor by the square of the two-jet one.

We have reperformed our previous calculation for the kinematic range relevant to HERA and find that the variation of
$S^{NG}$ with $\Delta\eta$ is very weak, so we neglect it.  The variation with $Q_\Omega$ is very strong on the other
hand.  $S^{NG}$~is a function of $t$,
\begin{equation}
t=\frac{1}{2\pi \beta_0}\log\left(\frac{\alpha_s(Q_{\Omega})}{\alpha_s(Q)}\right),
\end{equation}
where $\beta_0=(11 C_A-2 n_f)/(6\pi)$, and is well-approximated by a Gaussian in $t$.  Thus if~$Q_\Omega$ is too close
to $\Lambda_{QCD}$, $t$ varies rapidly with it and $S^{NG}$ varies very rapidly.

\begin{table}
\begin{center}
\begin{tabular}{|c|c|}
 \hline
$Q_{\Omega}$ [GeV] & $S^{NG}(t)^2$ \\ \hline\hline
0.5 & 0.10$^{+0.30}_{-0.10}$ \\ \hline 
1.0 & 0.47$^{+0.16}_{-0.22}$ \\ \hline
1.5 & 0.65$^{+0.10}_{-0.13}$ \\ \hline
2.0 & 0.74$^{+0.07}_{-0.08}$ \\ \hline
\end{tabular}
\caption{The non-global emission suppression factors for the 4-jet system, obtained from an all-orders
calculation for $Q=6$~GeV.}
\label{secef:nglsupp}
\end{center}
\end{table}
It is impossible to quantify the uncertainties in this approximation, without a more detailed understanding of the
underlying physics.  To get an idea however, we estimate the possible size of higher order corrections, by varying the
hard scale at which~$\alpha_s$ is evaluated.  To leading logarithmic accuracy, this is equivalent to varying the value
of $\alpha_s(Q)$ by an amount proportional to its value. Therefore we estimate the uncertainty by
\begin{eqnarray}
\alpha_s^{\mathrm{up}}&=&\alpha_s(1+\alpha_s), \\
\alpha_s^{\mathrm{down}}&=&\alpha_s(1-\alpha_s).
\end{eqnarray}
We therefore evaluate $t$, and hence $S^{NG}(t)^2$, using our
central value of $\alpha_s(M_z)=0.116$, which implies $\alpha_s(Q=\mbox{6~GeV})=0.196$, and with raised and lowered
values $\alpha_s^{\mathrm{up}}(\mbox{6~GeV})=0.234$ and $\alpha_s^{\mathrm{down}}(\mbox{6~GeV})=0.158$.  For
$Q_\Omega=1.0$~GeV, for example, these values result in $t=0.097^{+0.056}_{-0.032}$ and hence
$S^{NG}(t)^2=0.47^{+0.16}_{-0.22}$.  We show the results for all relevant values of $Q_\Omega$ in table~\ref{secef:nglsupp}.
Note that $Q_\Omega=0.5$~GeV is so low that the range of uncertainty in $t$ extends beyond~$\Lambda_{QCD}$ and hence the
estimate of $S^{NG}$ extends to zero.  We have not shown any results for the 1995 cone-based ZEUS energy flow analysis
\cite{Derrick:1995pb} because the low value of~$Q_{\Omega}=0.3$~GeV means that the central value of the NG suppression
is already zero, indicating a breakdown of our perturbative approach.

The uncertainty on the secondary emission probability estimated in this way should be added to that on the primary
emission probability, described in section~\ref{secef:results}. However, we will see that the secondary uncertainty
generally dominates the two. This is therefore clearly an area that needs more work if more precise quantitative
predictions are to be made.

\section{Soft gluon dynamics for a \boldmath{kt} defined final state}

\label{secef:kt}

\begin{figure}
\begin{minipage}{0.9\textwidth}
\begin{center}
\includegraphics*[width=7cm,height=6cm]{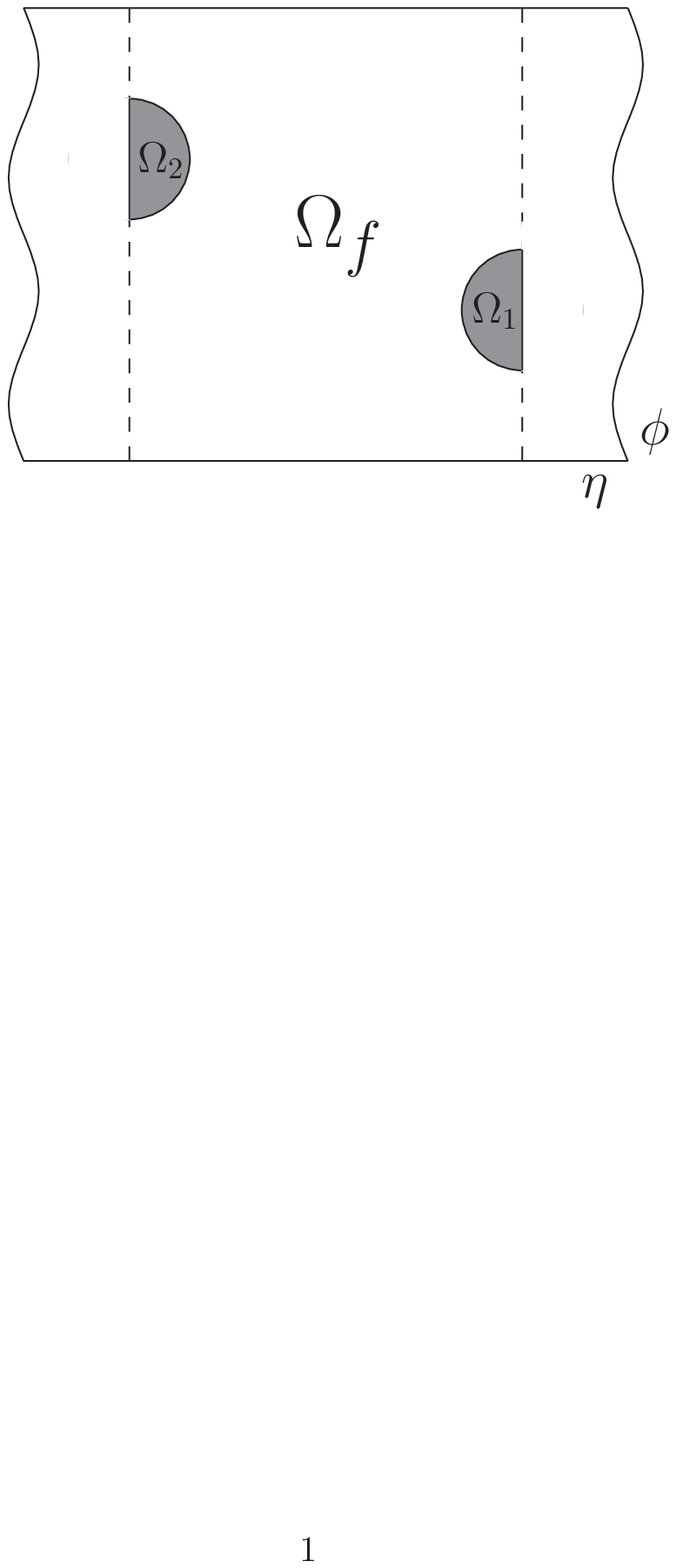}
\end{center}
\end{minipage}
\caption{The phase space regions for a kt defined final state. The shading denotes the regions 
vetoed by the algorithm, which are subtracted from the $\Omega_f$ piece. Note that we have dropped 
the ${(ij)}$ superscripts in this figure.}
\label{secef:figktphasespace}
\end{figure}

We now evaluate the momentum integral, $\omega^{(ij)}$, over the gap region $\Omega$. 
The region of integration is determined by the experimental geometry, in which the final state is defined
by the kt algorithm, and we shall work with the quantity
\begin{equation}
\Omega_{kt}^{(ij)}=\int_{kt} \mathrm{d}\eta \int_{kt} \frac{\mathrm{d}\phi}{2\pi}
\frac{\beta_i\cdot \beta_j}{(\beta_i\cdot \bar{k})(\beta_j\cdot \bar{k})},
\end{equation}
where we define $\bar{k}=k/k_t$. Therefore
\begin{equation}
\omega^{(ij)}=-\frac{\alpha_s}{2\pi} \Delta_i \Delta_j \delta_i \delta_j \frac{1}{2\epsilon}\Omega_{kt}^{(ij)}+I.P..
\end{equation}  
We denote the geometry independent imaginary part by $I.P.$, and we define the finite piece $\Gamma^{(ij)}$ by
\begin{equation}
\omega^{(ij)}=-\mathcal{S}_{ij}\frac{\Gamma^{(ij)}}{2\epsilon}.
\end{equation}
We have extracted the sign function from $\Gamma^{(ij)}$,
\begin{equation}
\mathcal{S}_{ij}=\Delta_i \Delta_j \delta_i \delta_j,
\label{secef:eqsign}
\end{equation}
so that 
\begin{equation}
\Gamma^{(ij)}=\frac{\alpha_s}{2\pi}\Omega^{(ij)}_{kt}+I.P..
\end{equation}
In this work we denote the rapidity separation of the jets by 
$\Delta\eta$ and the width of an azimuthally symmetric rapidity gap by $\Delta y$ ($<\Delta\eta$). Therefore the available 
phase space for soft gluon emission for a kt defined final state is given by
\begin{equation}
\Omega_{kt}^{(ij)}=\lim_{\Delta y \rightarrow \Delta\eta} \left(
\Omega^{(ij)}_f(\Delta y,\Delta\eta) - \Omega^{(ij)}_1(\Delta y,\Delta\eta,R) - 
\Omega^{(ij)}_2(\Delta y,\Delta\eta,R)\right),
\end{equation}
where the first term arises from an azimuthally symmetric gap of width $\Delta y$, and we subtract the region around 
each jet which is vetoed by the kt algorithm. The regions of this equation are shown in figure \ref{secef:figktphasespace}. 
In these regions any soft radiation is clustered into the jet, and cannot 
form part of $\Omega$. In the first term we take $\Delta y$ approaching~$\Delta\eta$, 
and hence it contains 
a collinear divergence when the emitted gluon is collinear to one of the jets. The two subtracted pieces then
 remove the regions of phase space defined by
\begin{equation}
(\eta_k-\eta_i)^2+(\phi_k-\phi_i)^2<R^2,
\end{equation}
where the index $i$ labels final state jets and $k$ labels the emitted gluon. 
The collinear divergences in the subtracted pieces exactly match the 
divergences in the first piece and hence the function $\Omega^{(ij)}_{kt}(\Delta\eta)$ is
collinear safe. Explicit definitions of the $\Omega$ functions are
\begin{eqnarray}
\Omega^{(ij)}_f&=&\int_{-\Delta y/2}^{+\Delta_y/2}\mathrm{d}\eta \int_0^{2\pi} \frac{\mathrm{d}\phi}{2\pi} 
\frac{\beta_i\cdot \beta_j}{(\beta_i\cdot \bar{k})(\beta_j\cdot \bar{k})}, \nonumber \\
\Omega^{(ij)}_1&=&\int_{\Delta\eta/2-R}^{+\Delta y/2}\mathrm{d}\eta 
\int_{-\phi_{\mathrm{lim}}}^{+\phi_{\mathrm{lim}}} \mathrm{d}\phi 
\frac{\beta_i\cdot \beta_j}{(\beta_i\cdot \bar{k})(\beta_j\cdot \bar{k})},
\end{eqnarray}
where we write $\phi_{\mathrm{lim}}=\sqrt{R^2-(\eta-\Delta\eta/2)^2}$ and obtain $\Omega^{(ij)}_2$ by
the symmetry $\Omega^{(ij)}_2 =
\Omega^{(\bar\imath\bar\jmath)}_1$, where the mapping $i\to\bar\imath$
is given by $\{a,b,1,2\}\to\{b,a,2,1\}$. If we define the following combinations of momentum integrals,
\begin{eqnarray}
\alpha&=&\mathcal{S}_{ab}\Gamma^{(ab)}+\mathcal{S}_{12}\Gamma^{(12)}, \nonumber \\
\beta&=&\mathcal{S}_{a1}\Gamma^{(a1)}+\mathcal{S}_{b2}\Gamma^{(b2)}, \nonumber \\
\gamma&=&\mathcal{S}_{b1}\Gamma^{(b1)}+\mathcal{S}_{a2}\Gamma^{(a2)},
\end{eqnarray}
where we have combined classes of diagram with the same colour structure, we obtain the following closed form 
for the positive gap contributions, in the limit $\Delta y \rightarrow \Delta\eta$,
\begin{eqnarray}
\alpha &=&
\frac{\alpha_s}{\pi}\Bigl(\phantom{-}2\Delta\eta+
\log\bigl(1-e^{-2\Delta\eta}\bigr)+\log\frac1{\Delta\eta-\Delta y}
-2i\pi\Bigr), \phantom{(10)} \label{secef:eqalpha} \\
\beta &=&
\frac{\alpha_s}{\pi}\Bigl(\phantom{-2\Delta\eta+{}}
\log\bigl(1-e^{-2\Delta\eta}\bigr)+\log\frac1{\Delta\eta-\Delta y}
\Bigr), \\
\gamma &=&
\frac{\alpha_s}{\pi}\Bigl(-2\Delta\eta-
\log\bigl(1-e^{-2\Delta\eta}\bigr)-\log\frac1{\Delta\eta-\Delta y}
\Bigr).
\label{secef:eqgamma}
\end{eqnarray}
The subtraction pieces are straightforward to 
express as power series in $R$ and~$e^{-\Delta\eta}$~and we shall illustrate the calculation of the momentum 
integrals with an example.

\subsection{Calculation of \boldmath{$\Omega_{kt}^{(a1)}$}}

We can write the matrix element in terms of the rapidity of the emitted gluon and obtain the 
following matrix element
\begin{equation}
\frac{\beta_a\cdot \beta_1}{(\beta_a\cdot \bar{k})(\beta_1\cdot \bar{k})}=
\frac{e^{-\Delta\eta/2}}{e^{-\eta}(\cosh(\Delta\eta/2-\eta)-\cos\phi)},
\end{equation}
using the 4-momenta products,
\begin{eqnarray}
\beta_a\cdot\beta_1&=&\cosh(\Delta\eta/2)-\sinh(\Delta\eta/2), \\
\beta_a\cdot \bar{k}&=&\cosh(\eta/2)-\sinh(\eta/2), \\
\beta_1\cdot\bar{k}&=&\cosh(\Delta\eta/2)\cosh\eta-\sinh(\Delta\eta/2)\sinh\eta-\cos\phi.
\end{eqnarray}
The integrations for the function $\Omega_f^{(a1)}$ are straightforward, and we obtain
\begin{equation}
\Omega_f^{(a1)}=-\Delta y + \log\left(\frac{\sinh(\Delta\eta/2+\Delta y/2)}{\sinh(\Delta\eta/2-\Delta y/2)}\right).
\end{equation}
The expression for $\Omega_1^{(a1)}$ is
\begin{eqnarray}
\Omega^{(a1)}_1&=&\int_{\Delta\eta/2-R}^{+\Delta y/2}\mathrm{d}\eta 
\int_{-\phi_{\mathrm{lim}}}^{+\phi_{\mathrm{lim}}} \frac{\mathrm{d}\phi}{2\pi} 
\frac{e^{-\Delta\eta/2}}{e^{-\eta}(\cosh(\Delta\eta/2-\eta)-\cos\phi)}, \nonumber \\
&=&\int_{\Delta\eta/2-R}^{+\Delta y/2}\mathrm{d}\eta \, f(\eta,\Delta\eta,R), \nonumber \\
&=&\int_{\Delta\eta/2-\Delta y/2}^{R}\mathrm{d}\eta' \, f(\eta',\Delta\eta,R),
\end{eqnarray}
where $\phi_{\mathrm{lim}}$ is defined in the previous section, we have performed the azimuthal 
integration in the second step and changed variable to $\eta'=\Delta\eta/2-\eta$ in the third step. 
The function $f$ can be easily obtained, but it is rather lengthy so we do not reproduce it here. 
We now note that this expression for $\Omega^{(ij)}_1$ only involves jet 1 and hence, by Lorentz
invariance, cannot depend on the other jet and so may not be a function of the jet separation 
$\Delta\eta$. Therefore we write
\begin{equation}
\Omega^{(a1)}_1=\int_{\Delta\eta/2-\Delta y/2}^{R}\mathrm{d}\eta' \, f(\eta',R).
\end{equation}
This function $f(\eta',R)$ has a divergence as $\eta'\rightarrow 0$, so we add and subtract this 
divergence to obtain
\begin{equation}
\Omega^{(a1)}_1=\int_{\Delta\eta/2-\Delta y/2}^R \mathrm{d}\eta' \left(f(\eta',R)-\frac{1}{\eta'}\right) 
+ \int_{\Delta\eta/2-\Delta y/2}^R \frac{\mathrm{d}\eta'}{\eta'}.
\end{equation}
We can rewrite the lower limit of the first, divergence free, integral as 0, and the collinear
divergence is now contained in the second term. Therefore we have used $\Delta y$ as a cut-off for
the divergence, and we can write
\begin{equation}
\Omega^{(a1)}_1=\bar{\Omega}^{(a1)}_1+\log{2R}-\log(\Delta\eta-\Delta y)+\mathcal{O}(\Delta\eta-\Delta y).
\end{equation}
We will always denote the divergence free angular integrals, which result
from such subtractions, as barred quantities. We can now rescale the $\bar{\Omega}^{(a1)}_1$ integral,
using
\begin{equation}
\bar{\eta}=\eta'/R, 
\end{equation}
to obtain
\begin{equation}
\bar{\Omega}^{(a1)}_1=\int_0^1 \mathrm{d}\bar{\eta} \left(R\cdot g(\bar{\eta},R)-\frac{1}{\bar{\eta}}\right). 
\end{equation}
We have denoted the rescaled version of~$f$ by~$g$. 
This quantity, which is only a function of $R$, can now be expressed as a power series in $R$ and the
integrals done on a term-by-term basis. Doing this we obtain the rapidly converging series,
\begin{equation}
\bar{\Omega}^{(a1)}_1=-\log(2)-\frac{2R}{\pi}+\frac{R^2}{8}-\frac{R^3}{18\pi}+\frac{R^4}{576}
-\frac{R^5}{5400\pi}-\frac{R^7}{529200\pi}+\frac{R^8}{4147200}+\dots.
\end{equation}
To calculate $\Omega^{(a1)}_2$ we use the parity symmetry mapping, $\{a,b,1,2\}\to\{b,a,2,1\}$, 
and obtain the expression,
\begin{eqnarray}
\Omega^{(a1)}_2&=&\Omega^{(b2)}_1 \nonumber \\
&=&\int_{\Delta\eta/2-R}^{+\Delta y/2}\mathrm{d}\eta 
\int_{-\phi_{\mathrm{lim}}}^{+\phi_{\mathrm{lim}}} \frac{\mathrm{d}\phi}{2\pi} 
\frac{e^{-\Delta\eta/2}}{e^{\eta}(\cosh(\Delta\eta/2+\eta)+\cos\phi)}.
\end{eqnarray}
We now perform similar manipulations to the case of $\Omega^{(a1)}_1$. However, as $\Omega^{(a1)}_2$
is a function of both final state jets, the resulting expression must be a function of $\Delta\eta$ and
we also note that $\Omega^{(a1)}_2$ is not divergent. We hence obtain the expression
\begin{equation}
\bar{\Omega}_2^{(a1)}=\int_0^1 d\bar{\eta}\left(
R\cdot f(\bar{\eta},\Delta\eta,R)\right),
\end{equation}
which we can expand as a power series in the variables $R$ and $z=\exp(-\Delta\eta)$, and perform
the remaining integrations term-by-term.

The pole arising in the subtraction term $\Omega_1^{(a1)}$ now cancels against an equivalent pole
in the function $\Omega_f^{(a1)}$, when we expand the latter in $\Delta y$ around the point $\Delta\eta$,
\begin{equation}
\lim_{\Delta y\rightarrow \Delta\eta}\Omega_f^{(a1)}
\sim-\Delta \eta -\log(\Delta\eta-\Delta y) + \log(2\sinh\Delta\eta).
\end{equation}
Therefore we find the final, divergence free, form of $\Omega_{kt}^{(a1)}$ as
\begin{equation}
\Omega_{kt}^{(a1)}=-\Delta\eta+\log(2\sinh\Delta\eta)-\log(2R)-\bar{\Omega}_1^{(a1)}-\bar{\Omega}_2^{(a1)}.
\end{equation}
We have presented the full set of series expansions in appendix \ref{secef:appgamma} and these, together with 
equations (\ref{secef:eqalpha}--\ref{secef:eqgamma}), are sufficient to compute the set of kt defined momentum 
integrals and hence the
corresponding anomalous dimension matrix. 
It is worth noting that, although the off-diagonal terms for the kt anomalous dimension matrices are no longer
pure imaginary, as in the cone case, their real parts still vanish for large~$\Delta\eta$.
Indeed for~$\Delta\eta=2$, the real part is more than two orders
of magnitude smaller than the imaginary part. We have listed the
 closed-form momentum integrals for the cone defined final state using our notation 
in appendix~\ref{secef:appcone}.

\section{Results}

\label{secef:results}

We now have the tools we need to calculate resummed cross sections at HERA, which correctly include primary emission 
to all orders and secondary emission approximately in the large $N_c$ limit. The colour bases
used for the contributing partonic cross sections are presented in the appendix, along with 
the decomposed hard and soft matrices. We also present the complete colour mixing matrices and the correct 
sign structure for the three classes of diagram. Therefore we can use the eigenvectors and eigenvalues
of the soft anomalous dimension matrices, together with the hard and soft matrices, to calculate the primary 
resummed cross section using equation (\ref{secef:eqresum}), for either a kt or a cone defined final state. 
The differential cross section, in $\Delta\eta$, 
can then be computed using the cuts given in section \ref{secef:hera}, both for the totally inclusive cross section
 (no gap) and for the gap cross section at fixed $Q_{\Omega}$. The gap fraction is then found by dividing the latter
quantity by the former. All our results are computed using GRV-G (LO) photon parton densities \cite{Gluck:1991jc} and 
the (HERWIG \cite{Corcella:2002jc} default) MRST-LO proton parton densities \cite{Martin:1998np}. We have included an 
estimate of the theoretical
uncertainty in the primary resummation by varying the hard scale in the evaluation of $\alpha_s$, while keeping the 
ratio of the hard and soft scales fixed.

\subsection{Totally inclusive \boldmath{ep} cross section and the gap cross section}

\begin{figure}
\begin{center}
 \begin{minipage}{0.48\textwidth}
\includegraphics*[width=8cm,height=7cm]{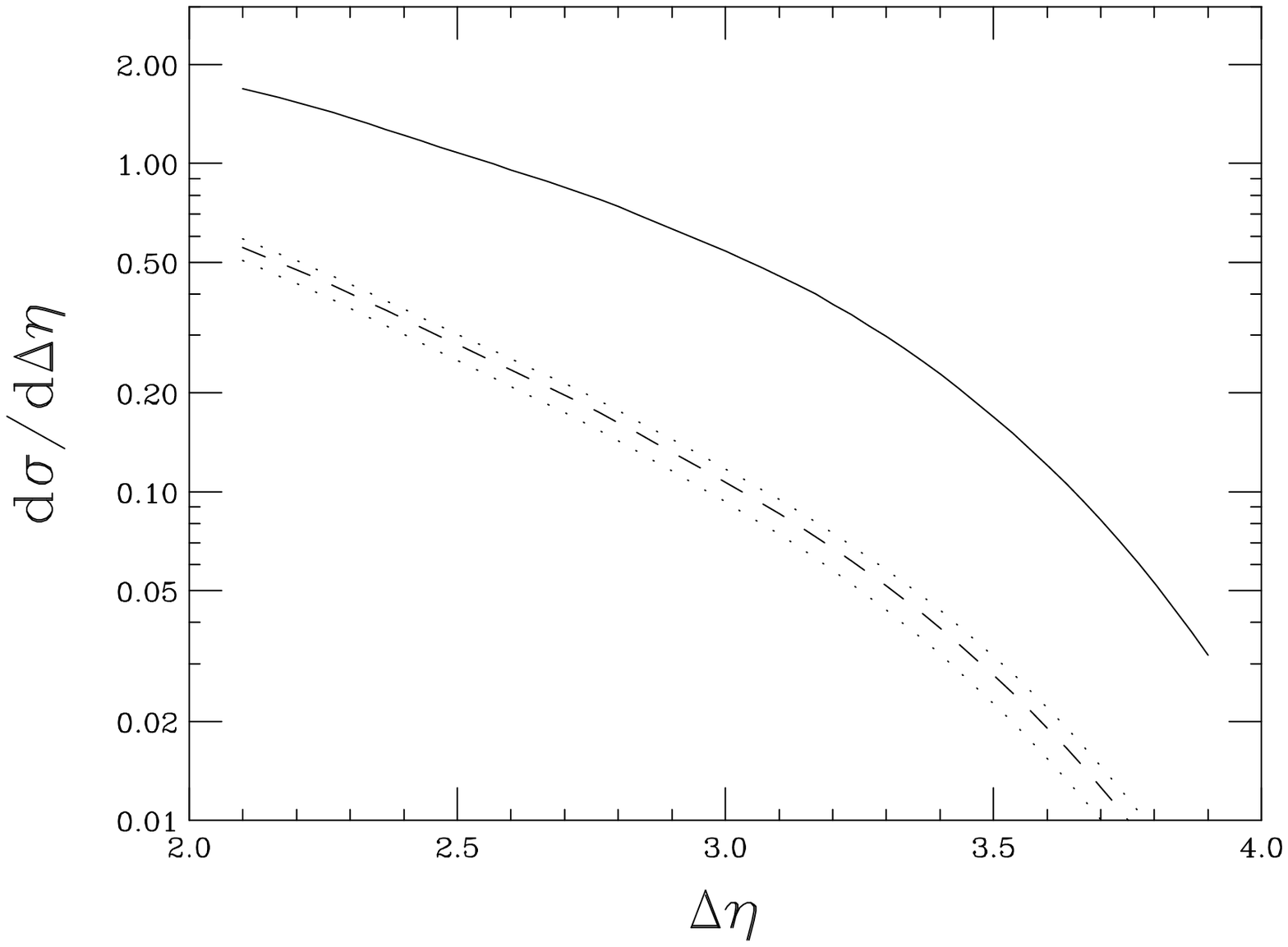}
\end{minipage}
\hfill
\begin{minipage}{0.48\textwidth}
\includegraphics*[width=8cm,height=7cm]{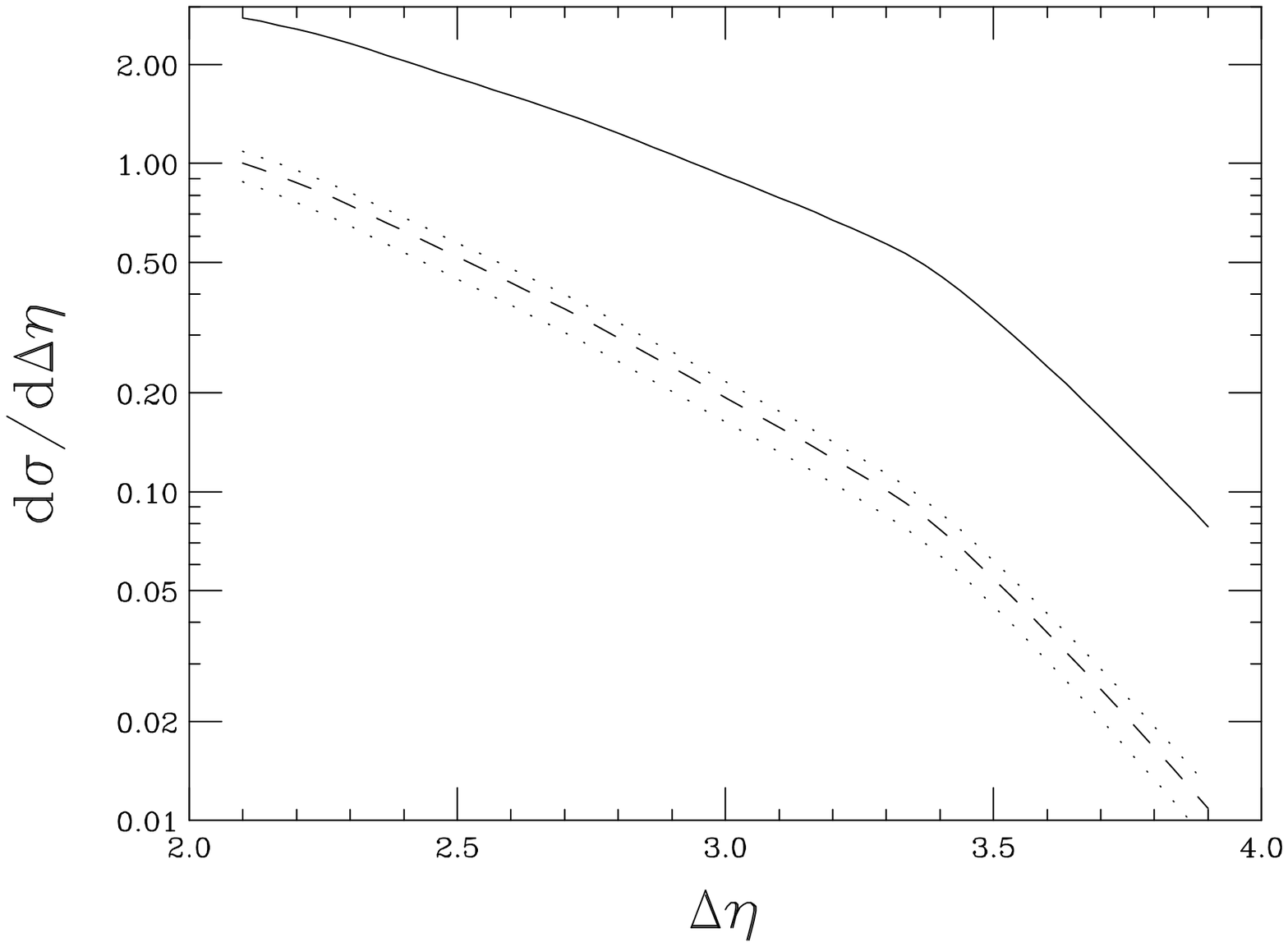}
\end{minipage}
\caption{The cross sections for the H1 data (left) and the ZEUS data (right), which was defined using the kt
algorithm with $R=1.0$. On both plots the solid line is the total 
inclusive cross section, the dashed line is the gap cross section
for $Q_{\Omega}=1$~GeV with only primary emission included, and the dotted lines indicate the range of 
theoretical uncertainty in the prediction.}
\label{secef:figgap}
\end{center}
\end{figure}

\begin{figure}
\begin{center}
  \begin{minipage}{0.48\textwidth}
    \includegraphics*[width=8cm,height=7cm]{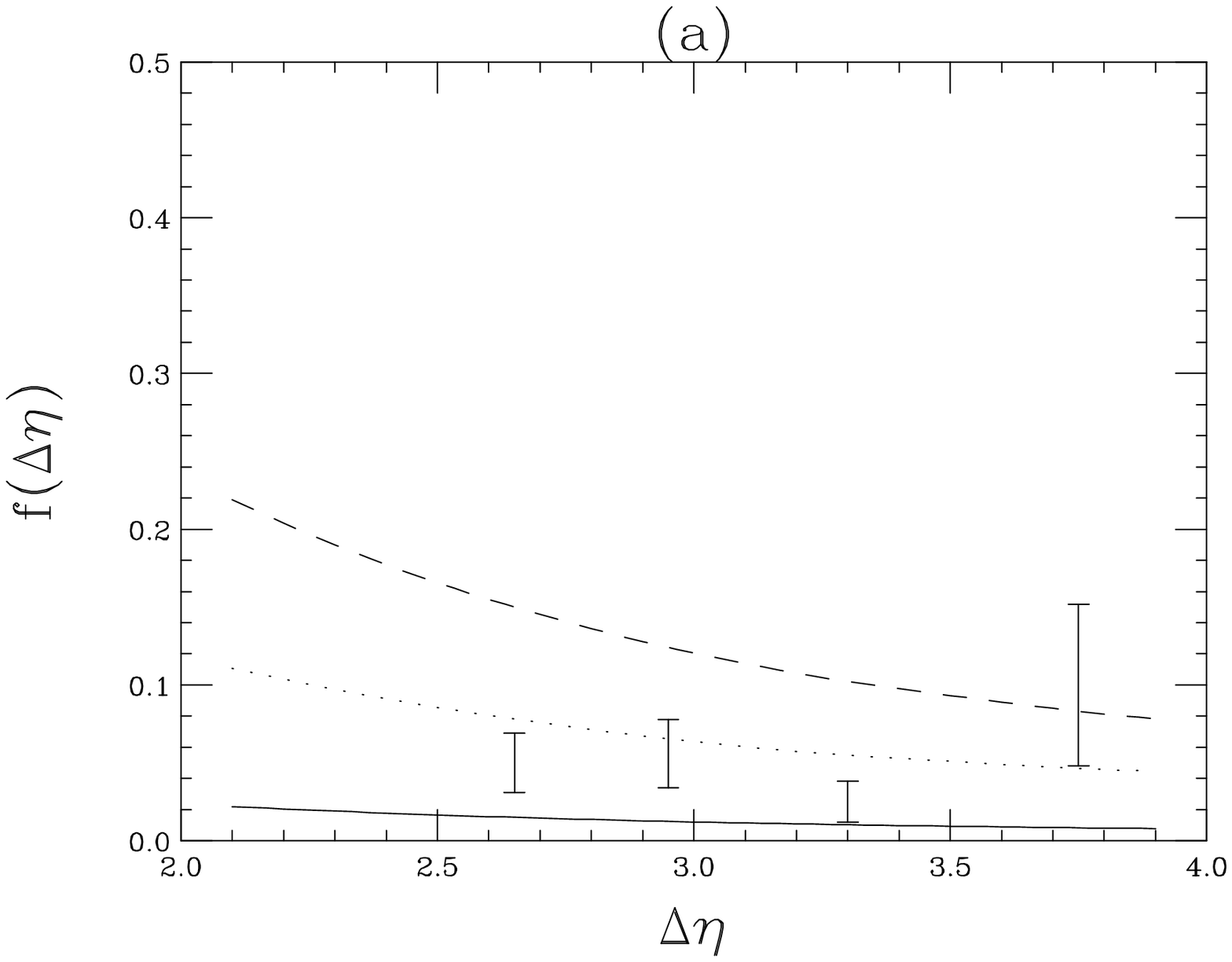}	
  \end{minipage}
  \hfill
  \begin{minipage}{0.48\textwidth}\vspace{-5pt}
\includegraphics*[width=8cm,height=7cm]{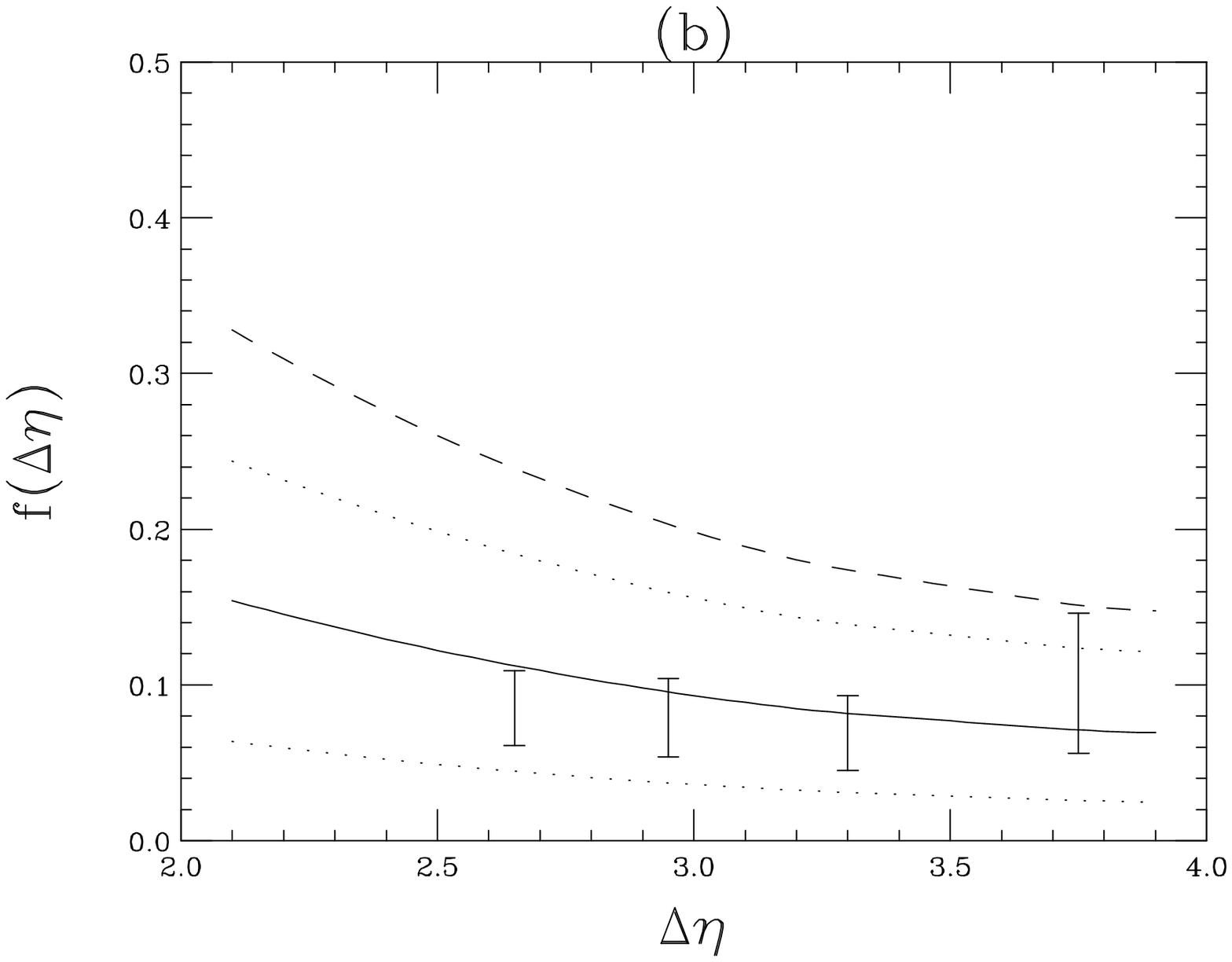}
    %\bigskip
  \end{minipage} \\
   \begin{minipage}{0.48\textwidth}
    \hfill\includegraphics*[width=8cm,height=7cm]{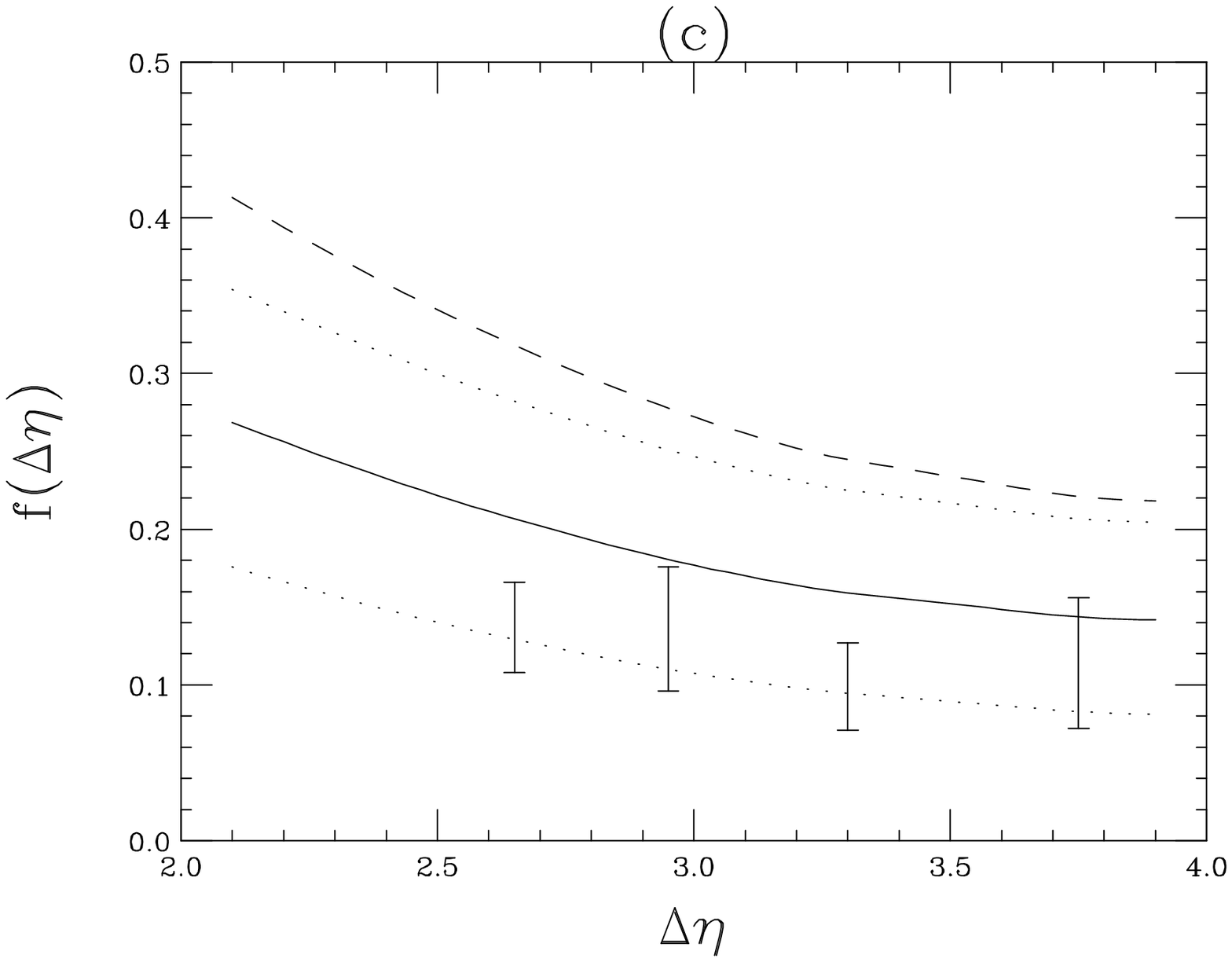}	
  \end{minipage}
  \hfill\hfill
  \begin{minipage}{0.48\textwidth}\vspace{-5pt}
\includegraphics*[width=8cm,height=7cm]{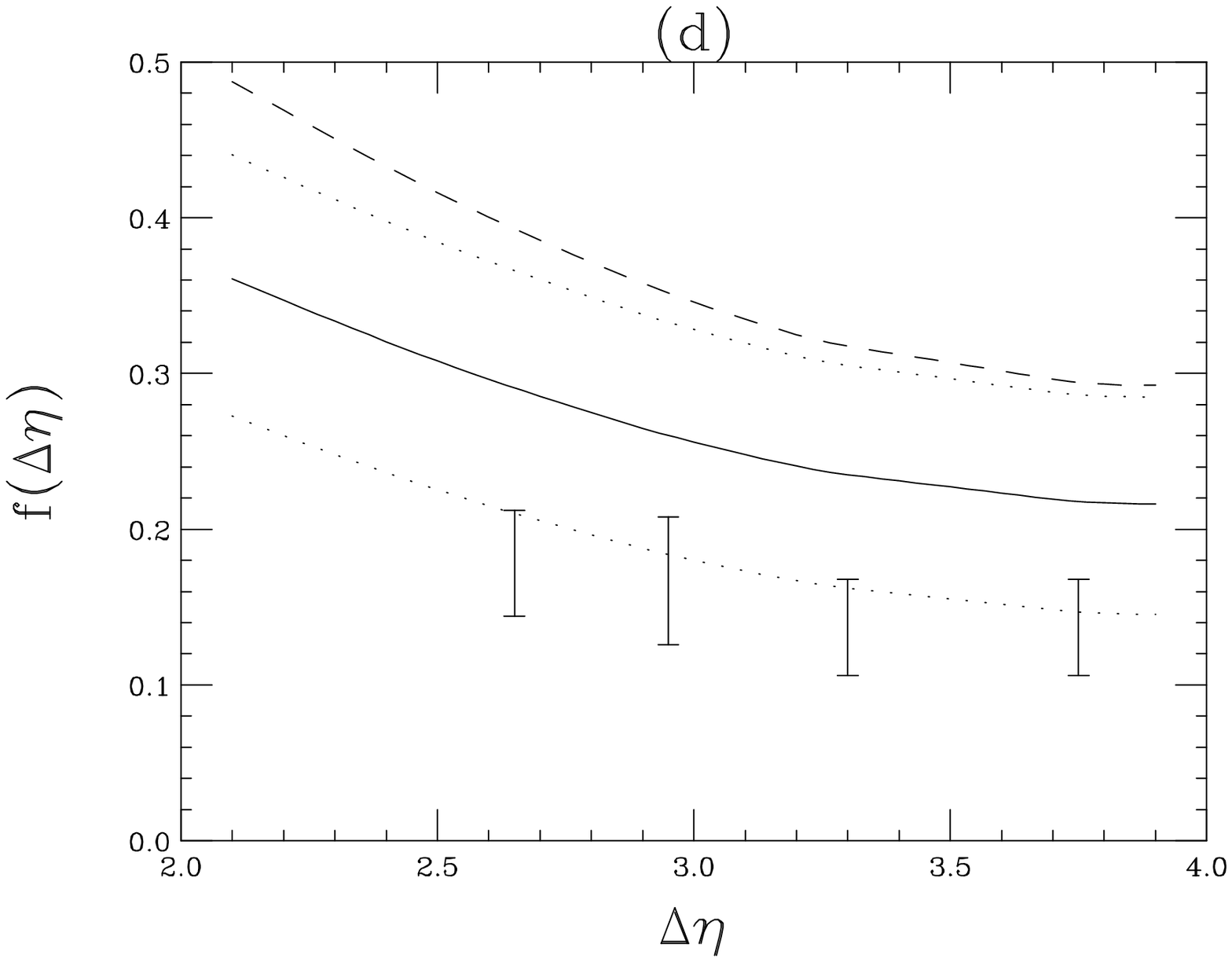}
    %\bigskip
  \end{minipage}
\caption{The gap fractions for the H1 analysis with a kt defined final state~($R=1.0$), 
at varying $Q_{\Omega}$. $Q_{\Omega}=0.5,\,1.0,\,1.5,\,2.0$~GeV for plots (a), (b), (c) and (d) respectively.
The H1 data is also shown. The solid line includes the effects of primary emission and 
the secondary emission suppression factor. The overall theoretical
uncertainty, including the primary uncertainty and the secondary uncertainty, is shown by the
dotted lines. The dashed line indicates the gap fraction obtained by including only primary emission.}
\label{secef:figh1frac}
\end{center}
\end{figure}

\begin{figure}
\begin{center}
  \begin{minipage}{0.48\textwidth}
    \includegraphics*[width=8cm,height=7cm]{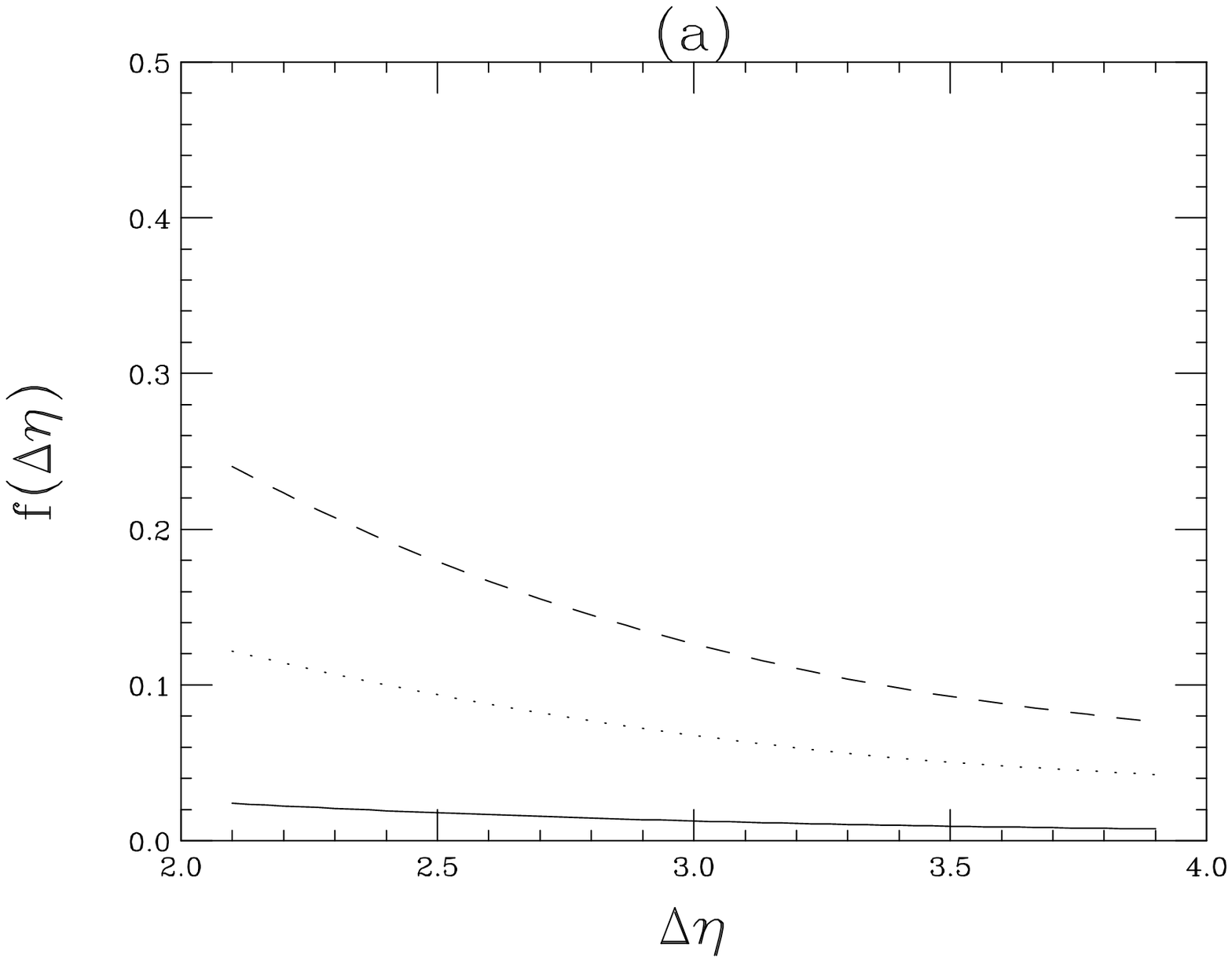}	
  \end{minipage}
  \hfill
  \begin{minipage}{0.48\textwidth}\vspace{-5pt}
\includegraphics*[width=8cm,height=7cm]{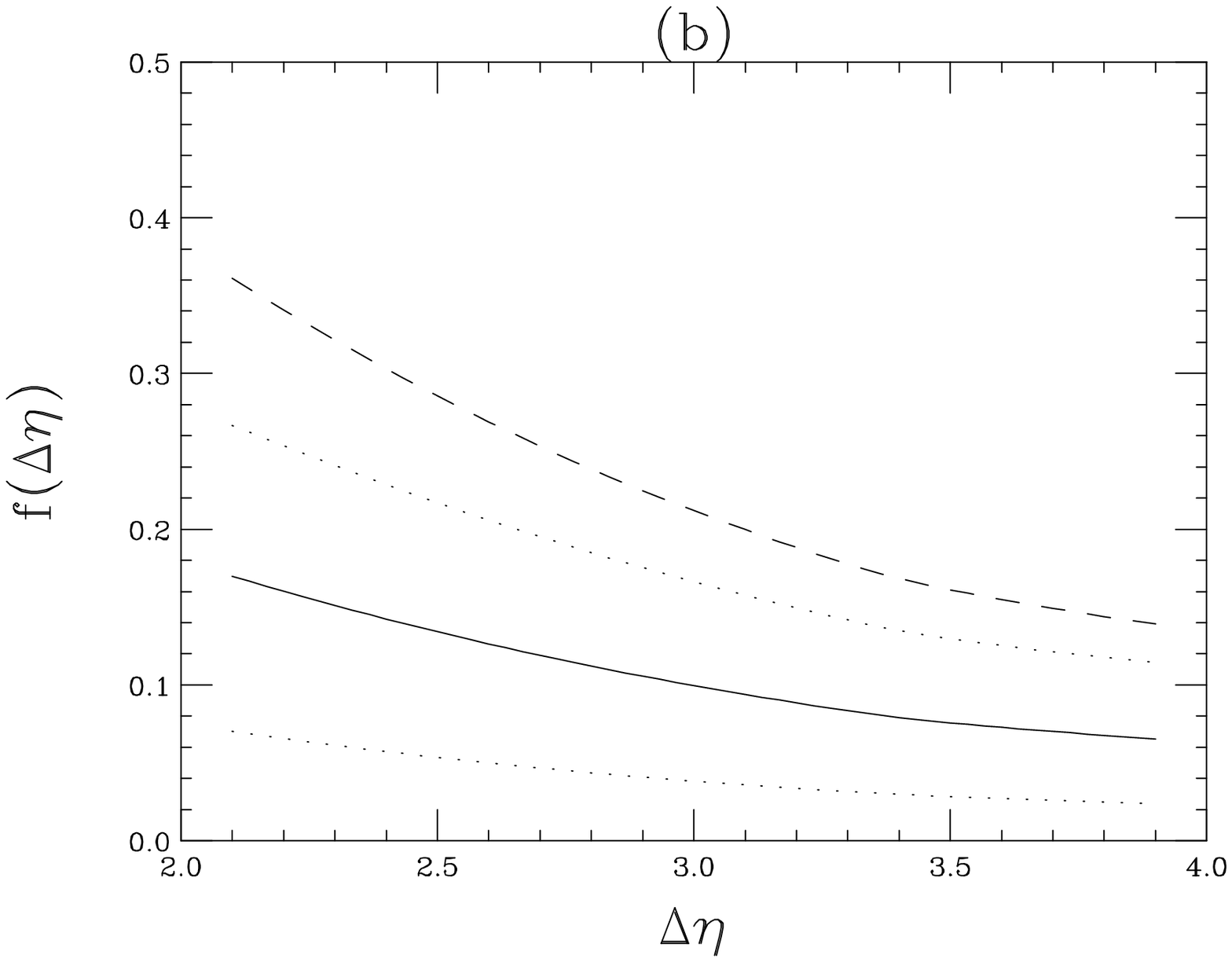}
    %\bigskip
  \end{minipage} \\
   \begin{minipage}{0.48\textwidth}
    \hfill\includegraphics*[width=8cm,height=7cm]{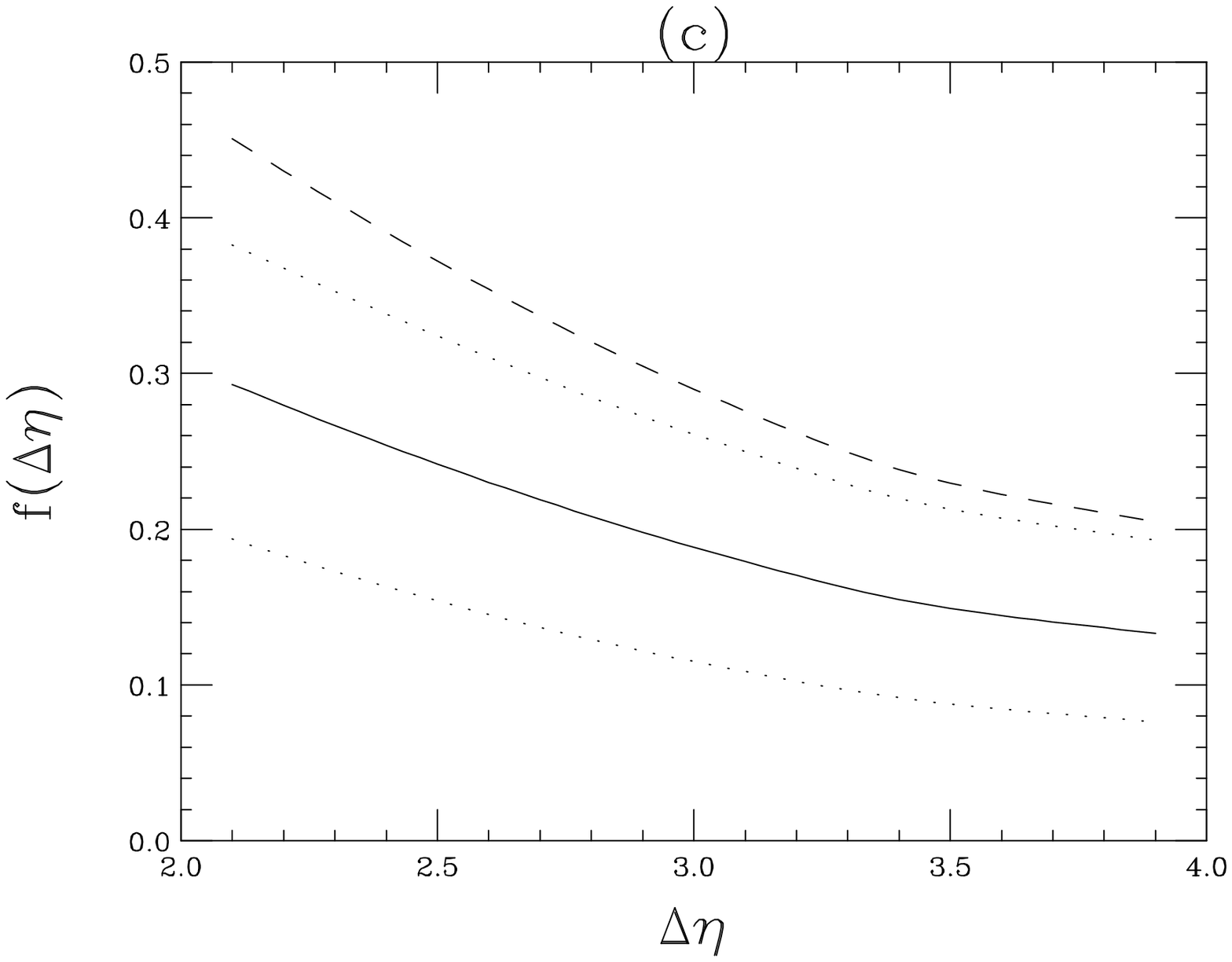}	
  \end{minipage}
  \hfill
  \begin{minipage}{0.48\textwidth}\vspace{-5pt}
\includegraphics*[width=8cm,height=7cm]{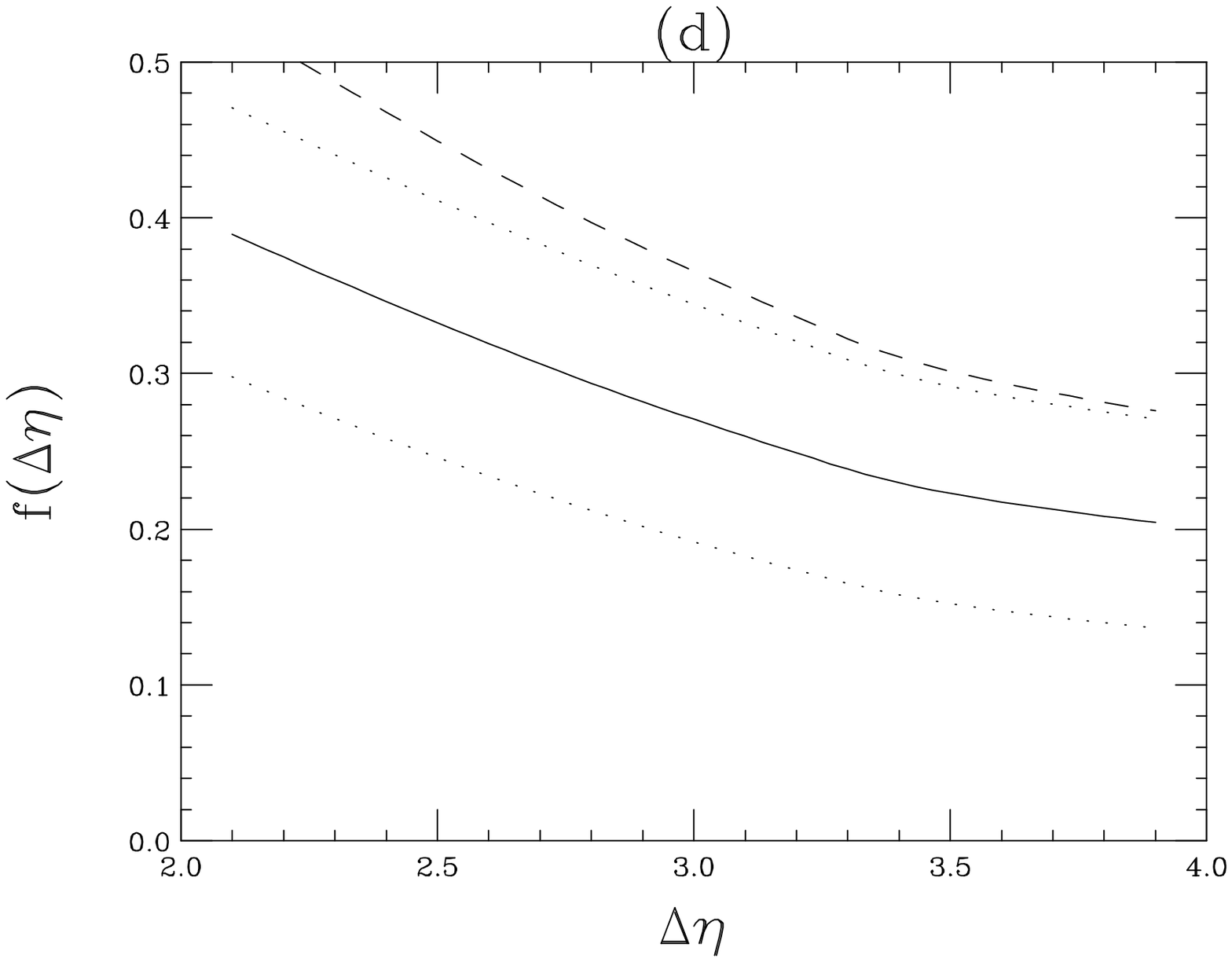}
    %\bigskip
  \end{minipage}
\caption{The gap fractions for the ZEUS analysis with a kt defined final state~($R=1.0$), 
at varying $Q_{\Omega}$. $Q_{\Omega}=0.5,\,1.0,\,1.5,\,2.0$~GeV for plots (a), (b), (c) and (d) respectively.
The solid line includes the effects of primary emission and 
the secondary emission suppression factor. The overall theoretical
uncertainty, including the primary uncertainty and the secondary uncertainty, is shown by the
dotted lines. The dashed line indicates the gap fraction obtained by including only primary emission.}
\label{secef:figz2frac}
\end{center}
\end{figure}
The left hand side of figure \ref{secef:figgap} shows the totally inclusive dijet cross section for the H1 analysis and the 
gap cross section
 for $Q_{\Omega}=1.0$~GeV. We have not shown further values of $Q_{\Omega}$ as all the plots show qualitatively the same
behaviour. 
We have cross-checked our total inclusive cross section against the Monte Carlo event generator HERWIG 
\cite{Corcella:2000bw,Corcella:2002jc} and we 
obtained complete agreement for the H1 and both the ZEUS sets of cuts. 
In figure \ref{secef:figgap} the solid curve is
the total inclusive cross section, the dashed line is the cross section with the primary interjet logarithms resummed 
and the
dotted lines show the theoretical uncertainty of the primary resummation, estimated by varying $\alpha_s$ as described 
above. The inclusion of the primary gap
logarithms gives a substantial suppression of the cross section; our analysis confirms simple ``area of phase space''
arguments which say that the kt defined final state will have greater soft gluon suppression than a cone defined
final state due to the increased gap area in the $(\eta,\phi)$ plane. This plot for the ZEUS analysis is shown 
in the right hand side of figure~\ref{secef:figgap}.

\subsection{Gap fractions}

The gap fraction is defined as the gap cross section, at fixed $Q_{\Omega}$, divided by the total inclusive cross section.
Figure \ref{secef:figh1frac} shows the gap fraction for the H1 cuts at the four experimentally measured values of 
$Q_{\Omega}$ and figure \ref{secef:figz2frac} shows the gap fractions for the ZEUS analysis. 
The solid line is the gap fraction curve obtained by including the primary emission and the 
NG suppression factors of table~\ref{secef:nglsupp} in the prediction for the gap cross section. The dotted lines show
the theoretical uncertainty of both the primary and secondary emission probabilities, and the dashed line shows the 
gap fraction obtained by including only the primary emission contribution. 
We find that our gap fraction is consistent with the H1 values for the
measured $Q_{\Omega}$. The large uncertainty in the gap fraction predictions comes from an approximate 
treatment of the NG suppression and from using perturbation theory at $\sim$1 GeV. Nonetheless, this uncertainty is 
principally in the normalisation of the curves and we expect our resummation to describe accurately 
the shape of the gap fraction curves. 

\section{Conclusions}

\label{secef:conc}

In this chapter we have computed resummed predictions for rapidity gap processes at HERA. We include primary logarithms 
using the soft gluon techniques of CSS, as surveyed in chapter \ref{ch3}, and include the effects of NGLs using an overall 
suppression factor computed from an extension of our earlier work in chapter \ref{ch4}. 
The kt definition of a hadronic final state 
determines the
phase space available for soft primary emission and we have computed a set of anomalous dimension matrices specific 
to the geometry of the H1 and ZEUS analyses. Of course this method can be used for any definition of the gap, 
provided $\Omega$ is directed away
from all hard jets. 
We then compared our predictions with gaps-between-jets data from the H1 collaboration and found a
consistent agreement. The theoretical uncertainty of our predictions is relatively large, and generally 
dominated by the secondary emission
uncertainty. However our resummed predictions correctly predict the shape of the H1 data, and the normalisation 
agrees within errors. There is a suggestion that the $Q_{\Omega}$ dependence is not quite right, with our central 
$Q_{\Omega}=0.5$~GeV prediction below data and our central $Q_{\Omega}=2.0$~GeV prediction above data, although all 
are within our uncertainty. It is possible that a more complete treatment of the perturbative/non-perturbative interface 
would improve this. We expect that calculation of primary emission will be correct if $Q_{\Omega}$ is not too large, so that we
can neglect jet recoil. However our
 calculation is of sufficient accuracy in the region of phase space probed at HERA.

Our treatment of the NGLs is very approximate. For a fuller treatment, it is necessary to extend the 
extraction of the suppression factor to beyond the large $N_c$ limit and overcome the inherent disadvantages of
the numerical methods used. For the current application, consideration of the four jet system described in 
chapter \ref{ch4} is also necessary. We reserve 
the latter extension, in the large $N_c$ limit, for future work.
Our calculation has not included power corrections \cite{Korchemsky:1999kt}. The inclusion of such 
non-perturbative effects is required for a full and correct comparison to the experimental data. Again, we reserve 
this for future work.

Our calculation involves a numerical integration over all kinematic variables, so it would be straightforward 
to calculate the dependence of the gap fraction on, for example, the fraction of the photon's momentum 
participating in the hard process,~$x_{\gamma}$. 

In conclusion, in this chapter we have shown that the calculation of primary and secondary gluon emission patterns, 
using the
tools of chapters \ref{ch3} and \ref{ch4}, can give a good description 
of rapidity gap data at HERA. A fuller treatment would refine our approximation of NGLs and include power corrections.

\chapter{Conclusions}
\label{conc}
% Conclusion

In this thesis we have studied rapidity gap processes at modern colliders. Using Quantum 
Chromodynamics, we have made detailed calculations and made comparisons to the 
newest experimental analyses; we have found that the predictions of the theory agree well with the
available data.

In the study of gap-jet-gap processes, we were able to show that the factorised
model of Ingelman and Schlein is able to describe the diffractive observations at the Tevatron. To achieve
this we used pomeron parton densities obtained from HERA experiments and a gap survival factor which is
consistent with theoretical estimations. This issue of gap survival is not very well understood at the present time
but it is encouraging that we are able to use the HERA parton densities at the Tevatron and have a framework
in which to understand hadronic diffractive processes.

We then turned our attention to the total gap energy flow in rapidity gap processes, which is 
an exciting way study these processes in a perturbative way. The
contribution to these cross sections from primary emission, meaning gluons that are emitted directly 
into the gap, can be
resummed if we replace the partons in the hard scattering by ordered exponentials, or eikonal lines. This exponentiation is
driven by a set of anomalous dimension matrices, which are sensitive to the geometry of the soft gluon emission 
phase space. We were led to compute a set of these matrices for the geometric definition of the gap cross section used
in recent H1 and ZEUS analyses, which we then used to calculate resummed primary emission gap cross sections.

However, the leading logarithmic behaviour in such events also includes emission into the gap by harder gluons, which 
are emitted out of the gap. Such secondary, or non-global, logarithms have a significant numerical importance in rapidity 
gap cross sections. We were unable to include the effects of secondary emission in our primary emission calculations, and 
we performed numerical calculations in the large~$N_c$ limit to extract an overall scale and gap dependent 
non-global suppressive factor.

The cross sections we computed, which included both primary and secondary emission, were then 
compared to H1 gaps-between-jets 
data. The predictions were consistent with the data within errors, although the theoretical uncertainty was 
relatively large. The description of the shape of the data was good. The inclusion of power corrections, and a refined 
treatment of non-global logarithms, would improve the fit to data. This is reserved for future work.

In summary, the research work in this thesis is based on the calculation of rapidity gap processes using QCD and 
the successful comparison to experimental data. Further study of these topics will shed light on the mechanisms 
of QCD and the interplay between short and long distance physics. 

\begin{appendix}

\chapter{Useful identities from group theory}
\label{secintro:app1}
There are many SU(3) group theoretic identities used in this thesis. In this appendix, 
based on \cite{MacFarlane:1968vc}, we will
summarise the most frequently occuring identities.
The fundamental quark indentity is
\begin{equation}
t_{ij}^a t^a_{kl}=\frac12 \left( \delta_{il}\delta_{jk}-\frac{1}{N_c}\delta_{ij}\delta_{kl}\right),
\end{equation}
which allows the expansion of quark-gluon vertices. For gluons we have
\begin{eqnarray}
d_{abc}&=&2\left[ \mathrm{Tr}\left(t^a t^c t^b\right)+\mathrm{Tr}\left(t^a t^b t^c\right)\right], \\
f_{abc}&=&-2i\left[ \mathrm{Tr}\left(t^a t^b t^c\right)-\mathrm{Tr}\left(t^a t^c t^b\right)\right],
\end{eqnarray}
which allow the expansion of triple gluon vertices into their colour content. The first one is the symmetric form 
and the second is the antisymmetric form. Only the latter appears in the QCD Feynman rules. We can also expand 
a closed quark-gluon loop using
\begin{equation}
\left( t^a t^a \right)_{ij}=C_F \delta_{ij},
\end{equation}
where $C_F=4/3$, and multiply two SU(3) matrices using
\begin{equation}
t^a_{ij} t^b_{jk}=\frac{1}{2}\left[\frac{1}{N_c}\delta_{ab}\delta_{ik}+\left(d_{abc}+i f_{abc}\right)t^c_{ik}\right].
\end{equation}
The antisymmetric structure constants are antisymmetric under the interchange of two indices 
and obey the Jacobi relation,
\begin{equation}
f_{abe}f_{ecd}+f_{cbe}f_{aed}+f_{dbe}f_{ace}=0.
\end{equation}
The summetric structure constants are symmetric 
under the exchange of any two indices, obey $d_{abb}=0$ and satisfy
\begin{equation}
f_{abe}d_{ecd}+f_{cbe}d_{aed}+f_{dbe}d_{ace}=0.
\end{equation}
These constants also satisfy
\begin{equation}
d_{acd}d_{bcd}=\frac{N_c^2-4}{N_c}\delta_{ab},
\end{equation}
and the antisymmetric and the symmetric structure constants are related
\begin{equation}
\{t^a,t^b\}=\frac{1}{N_c}\delta^{ab}\mathbbm{1}+d^{abc}t^c.
\end{equation}
The following identies are also useful,
\begin{eqnarray}
d_{ijk}f_{ljk}&=&0, \label{secintroappen:eqdfstart} \\
f_{ijk}f_{ljk}&=&N_c \delta_{il}, \\
d_{ijk}d_{ljk}&=&\frac{N_c-4}{N_c}\delta_{il}, \\
f_{mig}f_{gjn}f_{nkm}&=&-\frac{N_c}{2}f_{ijk}, \\
d_{mig}f_{gjn}f_{nkm}&=&-\frac{N_c}{2}d_{ijk}, \\
d_{mig}d_{gjn}f_{nkm}&=&\frac{N_c^2-4}{2N_c}f_{ijk}, \\
d_{mig}d_{gjn}d_{nkm}&=&\frac{N_c^2-12}{2N_c}d_{ijk}.
\label{secintroappen:eqdfend}
\end{eqnarray}

\chapter{Eikonal Feynman rules}

\label{secresum:appeikonal}

The Feynman rules for the eikonal graphs can be extracted from the definition of the 
eikonal cross section.  These rules have appeared several times in the literature and we list them
here for completeness. A propagator for a quark, antiquark or gluon eikonal line is,
\begin{equation}
\frac{i}{\delta v\cdot q+i\epsilon},
\end{equation}
where $\delta=+1(-1)$ for the loop momentum flowing in the same (opposite) direction as the vector $v^{\mu}$.
A propagator for a gluon is given by
\begin{equation}
\frac{-i}{k^2+i\epsilon}.
\end{equation}
The interaction vertex for a virtual gluon (with colour index c) with a quark or an antiquark eikonal line is
\begin{equation}
-ig_s t^c_{ba}v^{\mu} \Delta,
\end{equation}
where $t^c$ is an SU(3) matrix in the fundamental representation and $\Delta$ is +1(-1) for a quark 
(antiquark) eikonal line. 
The interaction vertex for a virtual gluon with a gluon eikonal line is
\begin{equation}
-g_sf^{abc}v^{\mu}\Delta,
\end{equation}
where $f^{abc}$ is a SU(3) structure constant and $\Delta$ is +1(-1) for a gluon located above (below)
the eikonal line. We agree to read the gluon colour indices in an anticlockwise direction. This last convention
is necessary because of a minus-sign ambiguity in the gluon/gluon vertices.
\chapter{The derivation of the subtraction formula (\ref{secngl:eqsub})}
\label{secngl:app1}

The goal of this appendix is to derive the subtraction formula (\ref{secngl:eqsub}), for $S_2$ with clustering. Therefore
we wish to integrate over the phase space which is vetoed by the clustering procedure, where the two
gluons are merged by the algorithm. The full result for $S_2$ with clustering is then found by subtracting
this result from the unclustered result. We begin by writing down the general integral,
\begin{equation}
I^v=\int_{-\infty}^{\infty} d\eta_1 \int_{-\infty}^{\infty} d\eta_2 \int_{-\pi}^{\pi}\frac{d\phi}{2\pi}
f(\eta_1-\eta_2,\phi),
\end{equation}
where we integrate over all possible phase space, and the function $f(\eta_1-\eta_2,\phi)$ is obtained
from the squared-matrix element, equation (\ref{secngl:eqs2me}). We can use the fact that the integrand is
only a function of $\eta_1-\eta_2$ (boost invariant) to centre the edge of the gap on $\eta=0$. Therefore, ensuring
that gluon one is outside the gap and gluon two is inside the gap, we
obtain the following integration region, expressed as $\Theta$-functions,
\begin{equation}
\Theta(\eta_1)\Theta(-\eta_2)\Theta(\eta_2+\Delta\eta)\Theta(R^2-(\eta_1+\eta_2)^2-\phi^2).
\end{equation}
The final $\Theta$-function ensures we only integrate over a region where the two gluons would 
be clustered i.e. within $R^2$ in the $(\eta,\phi)$ plane. Now we make a variable change
\begin{eqnarray}
\eta&=&\eta_1-\eta_2, \nonumber \\
\bar{\eta}&=&\eta_1+\eta_2,
\end{eqnarray}
which has the Jacobian
\begin{equation}
d\eta_1\,d\eta_2=\frac{1}{2}d\eta\,d\bar{\eta}.
\end{equation}
The $\Theta$-functions become
\begin{equation}
\Theta(\frac{1}{2}(\eta+\bar{\eta}))
\Theta(-\frac{1}{2}(\bar{\eta}-\eta))
\Theta(\frac{1}{2}(\bar{\eta}-\eta)+\Delta\eta)
\Theta(R^2-\eta^2-\phi^2),
\end{equation}
which can be trivially rewritten
\begin{equation}
\Theta(\bar{\eta}>-\eta)
\Theta(\eta>\bar{\eta})
\Theta(\bar{\eta}>\eta-2\Delta\eta)
\Theta(R^2>\eta^2-\phi^2),
\end{equation}
where we have adopted a non-standard notation which is better to use in this 
appendix. We now expand the first and the
third $\Theta$-functions into a sum of two terms with definite orderings: the first with the ordering 
$\bar{\eta}>-\eta>\eta-2\Delta\eta$ and
the second with the ordering $\bar{\eta}>\eta-2\Delta\eta>-\eta$,
\begin{eqnarray}
\Theta(\bar{\eta}>-\eta)
\Theta(\bar{\eta}>\eta-2\Delta\eta)
&=&
\Theta(\bar{\eta}>-\eta)\Theta(-\eta>\eta-2\Delta\eta) \nonumber \\
&+&
\Theta(\bar{\eta}>\eta-2\Delta\eta)\Theta(\eta-2\Delta\eta>-\eta).
\end{eqnarray}
We can now deduce that
\begin{eqnarray}
\Theta(\bar{\eta}>-\eta)\Theta(\bar{\eta}<\eta)&\Rightarrow&\Theta(\eta>0), \nonumber \\
\Theta(-\eta>\eta-2\Delta\eta)&\Rightarrow&\Theta(\eta<\Delta\eta), \nonumber \\
\Theta(\bar{\eta}>\eta-2\Delta\eta)\Theta(\bar{\eta}<\eta)&\Rightarrow&\Theta(\Delta\eta>0), \nonumber \\
\Theta(\eta-2\Delta\eta>-\eta)&\Rightarrow&\Theta(\eta>\Delta\eta).
\end{eqnarray}
Therefore we find that
\begin{eqnarray}
I^v&=&\frac{1}{2}
\int_{-\infty}^{\infty} d\eta \int_{-\pi}^{\pi}\frac{d\phi}{2\pi}  \int_{-\infty}^{\infty} d\bar{\eta}
f(\eta,\phi) [\Theta(\Delta\eta-\eta)\Theta(\eta)\Theta(\eta-\bar{\eta})\Theta(\bar{\eta}+\eta) \nonumber \\
&&+ \Theta(\eta-\Delta\eta)\Theta(\eta-\bar{\eta})\Theta(\bar{\eta}-\eta+2\Delta\eta)] \Theta(R^2-\eta^2-\phi^2).
\end{eqnarray}
The integrand is not a function of $\bar{\eta}$ and so we can do that integral, with different limits for the two
terms. The $\Theta$-functions for the first term give 
\begin{equation}
\int_{-\eta}^{\eta}d\bar{\eta}=2\eta,
\end{equation}
and the second term gives
\begin{equation}
\int_{\eta-2\Delta\eta}^{\eta}d\bar{\eta}=2\Delta\eta.
\end{equation}
The remaining $\Theta$-functions now allow us to cover the whole of the integration region (the first
term goes from $0$ to $\Delta\eta$ and the second term goes from $\Delta\eta$ to $\infty$) and
we get 
\begin{equation}
I^v=\int_0^{\infty} d\eta \int_{-\pi}^{\pi} \frac{d\phi}{2\pi} f(\eta,\phi) \mathrm{min}(\eta,\Delta\eta)
\Theta(R^2-\eta^2-\phi^2),
\end{equation}
The expression for $f(\eta,\phi)$ is derived in section \ref{secngl:calclo},
\begin{equation}
f(\eta,\phi)=8\left(\frac{\cosh(\eta)}{\cosh(\eta)-\cos(\phi)}-1\right),
\end{equation}
and we note it is even in $\phi$. We arrive at
\begin{eqnarray}
I^v&=&16 \int_0^{\infty} d\eta \int_{0}^{\pi} \frac{d\phi}{2\pi} 
\left(\frac{\cosh(\eta)}{\cosh(\eta)-\cos(\phi)}-1\right)\,\mathrm{min}(\eta,\Delta\eta)
\Theta(R^2-\eta^2-\phi^2), \nonumber \\
&=&16 \int_0^{R} d\eta \, \mathrm{min}(\eta,\Delta\eta) 
\int_{0}^{\sqrt{R^2-\eta^2}} \frac{d\phi}{2\pi} 
\left(\frac{\cosh(\eta)}{\cosh(\eta)-\cos(\phi)}-1\right),
\end{eqnarray}
where we have changed the upper limit of the $\eta$ integration because of the $\Theta$-function and the 
requirement that $\phi$ be a real number.
We can do the final integration to obtain
\begin{equation}
I^v=\frac{8}{\pi}\int_0^Rd\eta\,\mathrm{min}(\eta,\Delta\eta)
\left( 2\coth\eta\,\arctan\left(\frac{\tan(\frac{\sqrt{R^2-\eta^2}}{2})}{\tanh(\frac{\eta}{2})}\right)
-\sqrt{R^2-\eta^2}\right),
\end{equation}
and we obtain our final result from
\begin{equation}
S_2^v=-4C_F C_A I^v.
\end{equation}

\chapter{Colour bases}

\label{secef:appbases}

In this section we present the colour bases used in chapter \ref{ch5} of this thesis. 
All the bases in this section have appeared in \cite{Oderda:1999kr,Berger:2001ns,Kidonakis:2000gi,Berger:2003zh}.

\section*{The process \boldmath{$q\bar{q}\rightarrow q\bar{q}$}}
\begin{eqnarray}
c_1&=&\delta_{a1}\delta_{b2}, \nonumber \\
c_2&=&-\frac{1}{2N_c}\delta_{a1}\delta_{b2}+\frac12\delta_{ab}\delta_{12}.
\label{qqbarbasis}
\end{eqnarray}

\section*{The process \boldmath{$qq\rightarrow qq$}}
\begin{eqnarray}
c_1&=&\delta_{a1}\delta_{b2}, \nonumber \\
c_2&=&-\frac{1}{2N_c}\delta_{a1}\delta_{b2}+\frac12\delta_{a2}\delta_{b1}.
\label{qqbasis}
\end{eqnarray}

\section*{The process \boldmath{$qg\rightarrow qg$}}
\begin{eqnarray}
c_1&=&\delta_{a1}\delta_{b2}, \nonumber \\
c_2&=&d_{b2c}(T_F^c)_{1a}, \nonumber \\
c_3&=&if_{b2c}(T^c_F)_{1a}.
\label{qgbasis}
\end{eqnarray}

\section*{The processes \boldmath{$q\bar{q}\rightarrow gg$} and \boldmath{$gg\rightarrow q\bar{q}$}}
The process $gg\rightarrow q\bar{q}$ has the basis,
\begin{eqnarray}
c_1&=&\delta_{ab}\delta_{12}, \nonumber \\
c_2&=&d_{abc}(T_F^c)_{12}, \nonumber \\
c_3&=&if_{abc}(T_F^c)_{12}.
\label{qqbarggbasis}
\end{eqnarray}
To find the basis for $q\bar{q}\rightarrow gg$, we interchange $a\leftrightarrow 2$ and $b\leftrightarrow 1$.

\section*{The process \boldmath{$gg\rightarrow gg$}}
The complete basis is
\begin{equation}
\left\{
c_1,c_2,c_3,P_1,P_{8_S},P_{8_A},P_{10\bigoplus \overline{10}},P_{27}\right\},
\label{ggbasis}
\end{equation}
where
\begin{eqnarray}
c_1&=&\frac{i}{4}\left[f_{abc}d_{12c}-d_{abc}f_{12c}\right], \nonumber \\
c_2&=&\frac{i}{4}\left[f_{abc}d_{12c}+d_{abc}f_{12c}\right], \nonumber \\
c_3&=&\frac{i}{4}\left[f_{a1c}d_{b2c}+d_{a1c}f_{b2c}\right], \nonumber \\
P_1&=&\frac{1}{8}\delta_{a1}\delta_{b2}, \nonumber \\
P_{8_S}&=&\frac{3}{5}d_{a1c}d_{b2c}, \nonumber \\
P_{8_A}&=&\frac{1}{3}f_{a1c}f_{b2c}, \nonumber \\
P_{10\bigoplus\overline{10}}&=&\frac12\left(\delta_{ab}\delta_{12}-\delta_{a2}\delta_{b1}\right)-
\frac13 f_{a1c}f_{b2c}, \nonumber \\
P_{27}&=&\frac12 \left(\delta_{ab}\delta_{12}+\delta_{a2}\delta_{b1}\right)-\frac18\delta_{a1}\delta_{b2}
-\frac35 d_{a1c} d_{b2c}.
\end{eqnarray}

\section*{The direct processes}

Since there is only one colour structure, these are basis independent.

\chapter{The hard and soft matrices}

\label{secef:hardsoft}

We now show the complete set of hard and soft matrices used in \ref{ch5}. These matrices have appeared 
in a variety of forms in \cite{Oderda:1999kr,Berger:2001ns,Kidonakis:2000gi,Berger:2003zh}. In all these equations we
have set $N_c=3$ and have written the coupling scale as $\mu$. Note that all our hard matrices differ from 
the normalisation used in~\cite{Oderda:1999kr,Kidonakis:2000gi} by 
a factor of~$2\hat{s}/\pi$ and from that used in~\cite{Berger:2001ns} by a factor of $4\hat{t}\hat{u}/\hat{s}^2$, while
they agree with that used in~\cite{Berger:2003zh}.

\section*{The process \boldmath{$q\bar{q}\rightarrow q\bar{q}$}}

The hard matrix has, in the basis \ref{qqbarbasis}, the form
\begin{equation}
H^{(1)}=\frac{1}{9}\frac{\alpha_s^2(\mu)\pi}{\hat{s}}
\left(\begin{array}[c]{cc}
\frac{16}{81} \chi_1 & \frac{4}{27}\chi_2 \\
\frac{4}{27} \chi_2 & \chi_3 
\end{array}\right),
\end{equation}
where we define
\begin{eqnarray}
\chi_1&=&\frac{\hat{t}^2+\hat{u}^2}{\hat{s}^2}, \nonumber \\
\chi_2&=&3\frac{\hat{u}^2}{\hat{s}\hat{t}}-\frac{\hat{t}^2+\hat{u}^2}{\hat{s}^2}, \nonumber \\
\chi_3&=&\frac{\hat{s}^2+\hat{u}^2}{\hat{t}^2}+\frac19\frac{\hat{t}^2+\hat{u}^2}{\hat{s}^2}-\frac23\frac{\hat{u}^2}
{\hat{s}\hat{t}}.
\end{eqnarray}
The unequal flavour process $q\bar{q}'\rightarrow q\bar{q}'$ is found by dropping the $s$-channel terms from 
these equations, and the unequal flavour process $q\bar{q}\rightarrow q'\bar{q}'$ is found by dropping the
$t$-channel terms. The hard matrix for $q\bar{q}\rightarrow \bar{q}q$ is found using the
transformation $\hat{t} \leftrightarrow \hat{u}$.
The corresponding soft matrix for all these processes is
\begin{equation}
S^{(0)}=
\left(\begin{array}[c]{cc}
N_c^2 & 0 \\
0 & \frac{1}{4}(N_c^2-1)
\end{array}\right).
\end{equation}

\section*{The process \boldmath{$qq\rightarrow qq$}}

The hard matrix has, in the basis \ref{qqbasis}, the form
\begin{equation}
H^{(1)}=\frac{1}{9}\frac{\alpha_s^2(\mu)\pi}{\hat{s}}
\left(\begin{array}[c]{cc}
\frac{16}{81} \chi_1 & \frac{4}{27}\chi_2 \\
\frac{4}{27} \chi_2 & \chi_3 
\end{array}\right),
\end{equation}
where we define
\begin{eqnarray}
\chi_1&=&\frac{\hat{t}^2+\hat{s}^2}{\hat{u}^2}, \nonumber \\
\chi_2&=&3\frac{\hat{s}^2}{\hat{u}\hat{t}}-\frac{\hat{t}^2+\hat{s}^2}{\hat{u}^2}, \nonumber \\
\chi_3&=&\frac{\hat{u}^2+\hat{s}^2}{\hat{t}^2}+\frac19\frac{\hat{t}^2+\hat{s}^2}{\hat{u}^2}-\frac23\frac{\hat{s}^2}
{\hat{u}\hat{t}}.
\end{eqnarray}
For the process $qq'\rightarrow qq'$ only keep the $t$-channel terms.
The corresponding soft matrix is
\begin{equation}
S^{(0)}=
\left(\begin{array}[c]{cc}
N_c^2 & 0 \\
0 & \frac{1}{4}(N_c^2-1)
\end{array}\right).
\end{equation}

\section*{The process \boldmath{$qg\rightarrow qg$}}

The hard matrix has, in the basis \ref{qgbasis}, the form
\begin{equation}
H^{(1)}=\frac{1}{24}\frac{\alpha_s^2(\mu)\pi}{2\hat{s}}
\left(\begin{array}[c]{ccc}
\frac{1}{18} \chi_1 & \frac{1}{6}\chi_1 & \frac13\chi_2 \\
\frac{1}{6} \chi_1 & \frac12\chi_1 & \chi_2 \\
\frac13\chi_2 &  \chi_2 & \chi_3
\end{array}\right),
\end{equation}
where we define
\begin{eqnarray}
\chi_1&=&2-\frac{\hat{t}^2}{\hat{s}\hat{u}}, \nonumber \\
\chi_2&=&1-\frac12 \frac{\hat{t}^2}{\hat{s}\hat{u}}-\frac{\hat{u}^2}{\hat{s}\hat{t}}-\frac{\hat{s}}{\hat{t}}, \nonumber \\
%\chi_2&=&-1-\frac{2\hat{s}}{\hat{t}}+\frac{\hat{u}}{2\hat{s}}-\frac{\hat{s}}{2\hat{u}}, \nonumber \\
\chi_3&=&3-4\frac{\hat{s}\hat{u}}{\hat{t}^2}-\frac12 \frac{\hat{t}^2}{\hat{s}\hat{u}}.
\end{eqnarray}
The hard matrix for the process $qg\rightarrow gq$ is found by the transformation $\hat{t}\leftrightarrow \hat{u}$.
The corresponding soft matrix is
\begin{equation}
S^{(0)}=
\left(\begin{array}[c]{ccc}
N_c(N_c^2-1) & 0 & 0\\
0 & \frac{1}{2N_c}(N_c^2-4)(N_c^2-1) & 0 \\
0 & 0 & \frac12 N_c (N_c^2-1)
\end{array}\right).
\end{equation}

\section*{The processes \boldmath{$q\bar{q}\rightarrow gg$} and \boldmath{$gg\rightarrow q\bar{q}$}}
In the basis \ref{qqbarggbasis} the hard matrix for these processes has the form
\begin{equation}
H^{(1)}=\frac{1}{\Delta}\frac{\alpha_s^2(\mu)\pi}{2\hat{s}}
\left(\begin{array}[c]{ccc}
\frac{1}{18} \chi_1 & \frac{1}{6}\chi_1 & \frac16\chi_2 \\
\frac{1}{6} \chi_1 & \frac12\chi_1 & \frac12\chi_2 \\
\frac16\chi_2 &  \frac12\chi_2 & \frac12\chi_3
\end{array}\right),
\end{equation}
where we define
\begin{eqnarray}
\chi_1&=&\frac{\hat{t}^2+\hat{u}^2}{\hat{t}\hat{u}}, \nonumber \\
\chi_2&=&\left(1+\frac{2\hat{t}}{\hat{s}}\right)\chi_1, \nonumber \\
\chi_3&=&\left(1-\frac{4\hat{t}\hat{u}}{\hat{s}^2}\right)\chi_1.
\end{eqnarray}
The constant $\Delta=9$ for the process $q\bar{q}\rightarrow gg$ and 
$\Delta=64$ for the process $gg\rightarrow q\bar{q}$. The matrix for the process 
$gg\rightarrow \bar{q}q$ is found from the transformation $\hat{t}\leftrightarrow \hat{u}$.
The soft matrix is
\begin{equation}
S^{(0)}=\frac{N_c^2-1}{2N_c}
\left(\begin{array}[c]{ccc}
2N_c^2 & 0 & 0 \\
0 & N_c^2-4 & 0 \\
0 & 0 & N_c^2 
\end{array}\right).
\end{equation}

\section*{The process \boldmath{$gg \rightarrow gg$}}
The hard matrix, in the basis \ref{ggbasis} has the block-diagonal form
\begin{equation}
H^{(1)}=
\left(\begin{array}[c]{cc}
0_{3\times3} & 0_{3\times5} \\
0_{5\times3} & H^{(1)}_{5\times5} 
\end{array}\right),
\end{equation}
where the matrix $H^{(1)}_{5\times5}$ has the form
\begin{equation}
H^{(1)}_{5\times5}=\frac{1}{16}\frac{\alpha_s^2(\mu)\pi}{2\hat{s}}
\left(\begin{array}[c]{ccccc}
9\chi_1 & \frac92\chi_1 & \frac92\chi_2 & 0 & -3\chi_1 \\
\frac92\chi_1 & \frac94\chi_1 & \frac94\chi_2  & 0 & -\frac32\chi_1 \\
\frac92\chi_2 & \frac94\chi_2 & \chi_3 & 0 & -\frac32\chi_2 \\
0 & 0 & 0 & 0 & 0 \\
-3\chi_1 & -\frac32\chi_1 & -\frac32\chi_2 & 0 & \chi_1
\end{array}\right),
\end{equation}
and we write
\begin{eqnarray}
\chi_1&=&1-\frac{\hat{t}\hat{u}}{\hat{s}^2}-\frac{\hat{s}\hat{t}}{\hat{u}^2}+\frac{\hat{t}^2}{\hat{s}\hat{u}}, \nonumber \\
\chi_2&=&\frac{\hat{s}\hat{t}}{\hat{u}^2}-\frac{\hat{t}\hat{u}}{\hat{s}^2}+\frac{\hat{u}^2}{\hat{s}\hat{t}}
-\frac{\hat{s}^2}{\hat{t}\hat{u}}, \nonumber \\
\chi_3&=&\frac{27}{4}
-9\left(\frac{\hat{s}\hat{u}}{\hat{t}^2}+\frac14\frac{\hat{t}\hat{u}}{\hat{s}}+
\frac14\frac{\hat{s}\hat{t}}{\hat{u}^2}\right)
+\frac92\left(\frac{\hat{u}^2}{\hat{s}\hat{t}}+\frac{\hat{s}^2}{\hat{t}\hat{u}}-
\frac12\frac{\hat{t}^2}{\hat{s}\hat{u}}\right).
\end{eqnarray}
For this process the soft matrix is
\begin{equation}
S^{(0)}=
\left(\begin{array}[c]{cccccccc}
5 & 0 & 0 & 0 & 0 & 0 & 0 & 0 \\
0 & 5 & 0 & 0 & 0 & 0 & 0 & 0 \\
0 & 0 & 5 & 0 & 0 & 0 & 0 & 0 \\
0 & 0 & 0 & 1 & 0 & 0 & 0 & 0 \\
0 & 0 & 0 & 0 & 8 & 0 & 0 & 0 \\
0 & 0 & 0 & 0 & 0 & 8 & 0 & 0 \\
0 & 0 & 0 & 0 & 0 & 0 & 20 & 0 \\
0 & 0 & 0 & 0 & 0 & 0 & 0 & 27 
\end{array}\right).
\end{equation}

\section*{The direct processes}
For both these processes the zeroth order soft factor is unity and the 
hard functions are
\begin{eqnarray}
H^{(1)}(\gamma g \rightarrow q\bar{q})&=&
\Bigl(\sum_qe_q^2\Bigr)
\frac{\alpha_s \alpha_{\mathrm{em}}\pi}{2\hat{s}}
\;\frac{4\hat{t}\hat{u}}{\hat{s}^2}
\left(\frac{\hat{u}}{\hat{t}}+\frac{\hat{t}}{\hat{u}}\right), \nonumber \\
H^{(1)}(\gamma q(\bar{q}) \rightarrow g q(\bar{q}))&=&\frac83e_q^2\frac{\alpha_s\alpha_{\mathrm{em}}\pi}{2\hat{s}}
\;\frac{4\hat{t}\hat{u}}{\hat{s}^2}
\left(\frac{-\hat{u}}{\hat{s}}+\frac{\hat{s}}{-\hat{u}}\right),
\end{eqnarray}
where $e_q$ is the electric charge of quark flavour $q$, in units of the electron charge.  Note that if the
sum for $\gamma g \rightarrow q\bar{q}$ is taken to be over four flavours, then this gives a factor of~$10/9$.

\chapter{Colour decomposition matrices}

\label{secef:appdecomp}

We now give the full set of colour decomposition matrices used in chapter \ref{ch5} of this thesis, and also the 
sign function $\mathcal{S}$, defined by equation (\ref{secef:eqsign}), for $\alpha$, $\beta$ and $\gamma$, defined 
by 
\begin{eqnarray}
\alpha&=&\mathcal{S}_{ab}\Gamma^{(ab)}+\mathcal{S}_{12}\Gamma^{(12)}, \nonumber \\
\beta&=&\mathcal{S}_{a1}\Gamma^{(a1)}+\mathcal{S}_{b2}\Gamma^{(b2)}, \nonumber \\
\gamma&=&\mathcal{S}_{b1}\Gamma^{(b1)}+\mathcal{S}_{a2}\Gamma^{(a2)}.
\end{eqnarray}

\section*{The process \boldmath{$q\bar{q}\rightarrow q\bar{q}$}}
\begin{equation}
\mathcal{C}^{q\bar{q}\rightarrow q\bar{q}}=
\left(\begin{array}[c]{cc}
C_F \beta \,\,\,& \frac{C_F}{2N_c}(\alpha+\gamma) \\
\alpha+\gamma \,\,\,&  C_F\alpha-\frac{1}{2N_c}(\alpha+\beta+2\gamma)
\end{array}\right).
\end{equation}
The signs are
\begin{eqnarray}
\mathcal{S}_\alpha&=&+1, \\
\mathcal{S}_\beta&=&+1, \\
\mathcal{S}_\gamma&=&-1.
\end{eqnarray}

\section*{The process \boldmath{$qq\rightarrow qq$}}
\begin{equation}
\mathcal{C}^{qq\rightarrow qq}=
\left(\begin{array}[c]{cc}
C_F \beta \,\,\,& \frac{C_F}{2N_c}(\alpha+\gamma) \\
\alpha+\gamma \,\,\,&  C_F\gamma-\frac{1}{2N_c}(2\alpha+\beta+\gamma)
\end{array}\right).
\end{equation}
The signs are
\begin{eqnarray}
\mathcal{S}_\alpha&=&-1, \\
\mathcal{S}_\beta&=&+1, \\
\mathcal{S}_\gamma&=&+1.
\end{eqnarray}

\section*{The process \boldmath{$qg\rightarrow qg$}}
\begin{equation}
\mathcal{C}^{qg\rightarrow qg}=
\left(\begin{array}[c]{ccc}
C_F \Gamma^{(a1)}+C_A\Gamma^{(b2)} & 0 &  -\frac12(\alpha+\gamma) \\
0 & \chi & -\frac{N_c}{4}(\alpha+\gamma) \\
-(\alpha+\gamma) & -\frac{N_c^2-4}{4N_c}(\alpha+\gamma) & \chi
\end{array}\right).
\end{equation}
The signs are
\begin{eqnarray}
\mathcal{S}_\alpha&=&+1, \\
\mathcal{S}_\beta&=&+1, \\
\mathcal{S}_\gamma&=&-1,
\end{eqnarray}
and we define
\begin{equation}
\chi=\frac{N_c}{4}(\alpha-\gamma)-\frac{1}{2N_c}\Gamma^{(a1)}+\frac{N_c}{2}\Gamma^{(b2)}.
\end{equation}

\section*{The processes \boldmath{$q\bar{q}\rightarrow gg$} and \boldmath{$gg \rightarrow q\bar{q}$}}

For $q\bar{q}\rightarrow gg$ we have
\begin{equation}
\mathcal{C}^{q\bar{q}\rightarrow gg}=
\left(\begin{array}[c]{ccc}
C_F \Gamma^{(ab)}+C_A\Gamma^{(12)} & 0 &  \frac12(\beta+\gamma) \\
0 & \chi' & \frac{N_c}{4}(\beta+\gamma) \\
(\beta+\gamma) & \frac{N_c^2-4}{4N_c}(\beta+\gamma) & \chi'
\end{array}\right).
\end{equation}
The signs are
\begin{eqnarray}
\mathcal{S}_\alpha&=&+1, \\
\mathcal{S}_\beta&=&+1, \\
\mathcal{S}_\gamma&=&-1,
\end{eqnarray}
and we define
\begin{equation}
\chi'=\frac{N_c}{4}(\beta-\gamma)-\frac{1}{2N_c}\Gamma^{(ab)}+\frac{N_c}{2}\Gamma^{(12)}.
\end{equation}

For $gg\rightarrow q\bar{q}$ we have
\begin{equation}
\mathcal{C}^{gg\rightarrow q\bar{q}}=
\left(\begin{array}[c]{ccc}
C_F \Gamma^{12}+C_A\Gamma^{(ab)} & 0 &  \frac12(\beta+\gamma) \\
0 & \chi'' & \frac{N_c}{4}(\beta+\gamma) \\
(\beta+\gamma) & \frac{N_c^2-4}{4N_c}(\beta+\gamma) & \chi''
\end{array}\right).
\end{equation}
The signs are
\begin{eqnarray}
\mathcal{S}_\alpha&=&+1, \\
\mathcal{S}_\beta&=&+1, \\
\mathcal{S}_\gamma&=&-1,
\end{eqnarray}
and we define
\begin{equation}
\chi''=\frac{N_c}{4}(\beta-\gamma)-\frac{1}{2N_c}\Gamma^{(12)}+\frac{N_c}{2}\Gamma^{(ab)}.
\end{equation}

\section*{The process \boldmath{$gg\rightarrow gg$}}
\begin{equation}
\mathcal{C}^{gg\rightarrow gg}=
\left(\begin{array}[c]{cc}
\mathcal{M}_{3\times3} & 0_{3\times5} \\
0_{5\times3} & \mathcal{M}_{5\times5}
\end{array}\right),
\end{equation}
where the matrix $\mathcal{M}_{3\times3}$ is 
\begin{equation}
\mathcal{M}_{3\times3}=
\left(\begin{array}[c]{ccc}
\frac{N_c}{2}(\alpha+\beta) & 0 & 0 \\
0 & \frac{N_c}{2}(\alpha-\gamma) & 0 \\
0 & 0 & \frac{N_c}{2}(\beta-\gamma),
\end{array}\right),
\end{equation}
and the matrix $\mathcal{M}_{5\times5}$ is 
\begin{equation}
\mathcal{M}_{5\times5}=
\left(\begin{array}[c]{ccccc}
3\beta & 0 & 3(\alpha+\gamma) & 0 & 0 \\
0 & \frac{3}{4}(\alpha+2\beta-\gamma) & \frac{3}{4}(\alpha+\gamma) & \frac{3}{2}(\alpha+\gamma) & 0 \\
\frac{3}{8}(\alpha+\gamma)&\frac{3}{4}(\alpha+\gamma)&\frac{3}{4}(\alpha+2\beta-\gamma) & 0 & \frac{9}{8}(\alpha+\gamma) \\
0 & \frac{3}{5}(\alpha+\gamma) & 0 & \frac{3}{2}(\alpha-\gamma) & \frac{9}{10}(\alpha+\gamma) \\
0 & 0 & \frac{1}{3}(\alpha+\gamma) & \frac{2}{3}(\alpha+\gamma) & 2\alpha-\beta-2\gamma
\end{array}\right),
\end{equation}
for $N_c=3$.
The signs are
\begin{eqnarray}
\mathcal{S}_\alpha&=&+1, \\
\mathcal{S}_\beta&=&+1, \\
\mathcal{S}_\gamma&=&-1.
\end{eqnarray}

\section*{The direct processes}
This processes has no matrix structure, and so we present colour decomposition functions.
\begin{eqnarray}
\mathcal{C}^{\gamma g \rightarrow q\bar{q}}&=&
-\frac{1}{2N_c}\Gamma^{(12)}+\frac{N_c}{2}\left(\Gamma^{(b1)}+\Gamma^{(b2)}\right), \nonumber \\
\mathcal{C}^{\gamma q \rightarrow gq}&=&
-\frac{1}{2N_c}\Gamma^{(b2)}+\frac{N_c}{2}\left(\Gamma^{(b2)}+\Gamma^{(12)}\right).
\end{eqnarray}

\chapter{The \boldmath{$\Gamma^{(ij)}$} series expansions}

\label{secef:appgamma}

We have not found a closed form for these integrals, but they are
straightforward to express as power series in $R$ and $e^{-\Delta\eta}$ (by Lorentz invariance, 
only the contributions from dipoles containing jet 2 are $\Delta\eta$-dependent),

\begin{eqnarray}
\Omega^{(ab)}_1 &=& \frac{\alpha_s}{\pi}\Bigl(
\frac14R^2\Bigr), \\
\Omega^{(a1)}_1 &=& \frac{\alpha_s}{\pi}\Bigl(
\frac12\log R+\frac12\log\frac1{\Delta\eta-\Delta y} \nonumber 
\\&&{}
 - 0.31831\,R + 0.06250\,R^2 - 0.00884\,R^3 + 0.00087\,R^4 \nonumber \\
&& - 0.00003\,R^5
\Bigr), \\
\Omega^{(b1)}_1 &=& \frac{\alpha_s}{\pi}\Bigl(
-\frac12\log R-\frac12\log\frac1{\Delta\eta-\Delta y} - 0.31831\,R - 0.06250\,R^2
\\&&{}
  - 0.00884\,R^3 - 0.00087\,R^4 -
   0.00003\,R^5
\Bigr),
\end{eqnarray}

\begin{eqnarray}
\Omega^{(12)}_1 &=& \frac{\alpha_s}{\pi}\Bigl(
\frac12\log R+\frac12\log\frac1{\Delta\eta-\Delta y}+
\\&&
(+ 0.31831\,R + 0.06250\,R^2 + 0.00884\,R^3 + 0.00087\,R^4 +
   0.00003\,R^5)
\hspace*{-1cm}\nonumber\\&&
+(- 0.08616\,R - 0.03383\,R^2 -
   0.01197\,R^3 - 0.00282\,R^4
\nonumber\\&&
 - 0.00039\,R^5 +
   0.00001\,R^7)e^{-(\Delta\eta-2)}+
\nonumber\\&&
(+ 0.01166\,R + 0.00916\,R^2 +
   0.00551\,R^3 + 0.00305\,R^4
\nonumber\\&&
 + 0.00122\,R^5 +
   0.00038\,R^6 + 0.00011\,R^7 + 0.00003\,R^8)e^{-2(\Delta\eta-2)}
\nonumber\\&&
+(- 0.00158\,R - 0.00186\,R^2 - 0.00162\,R^3 -
   0.00139\,R^4
\nonumber\\&&
 - 0.00088\,R^5 - 0.00041\,R^6 -
   0.00017\,R^7 - 0.00007\,R^8 
\nonumber\\&&
- 0.00002\,R^9)e^{-3(\Delta\eta-2)}+
\hspace*{-3cm}\nonumber\\&&
(+ 0.00021\,R +
   0.00034\,R^2 + 0.00039\,R^3 + 0.00045\,R^4
\nonumber\\&&
 + 0.00039\,R^5 + 0.00024\,R^6 + 0.00013\,R^7 +
   0.00007\,R^8
\nonumber\\&&
 + 0.00003\,R^9 +
   0.00001\,R^{10})e^{-4(\Delta\eta-2)}+
\nonumber\\&&
(- 0.00003\,R -
   0.00006\,R^2 - 0.00008\,R^3 - 0.00012\,R^4
\nonumber\\&&
 - 0.00013\,R^5 - 0.00010\,R^6 - 0.00007\,R^7 -
   0.00004\,R^8
\nonumber\\&&
 - 0.00003\,R^9 -
   0.00001\,R^{10})e^{-5(\Delta\eta-2)}+
\nonumber\\&&
(+ 0.00001\,R^2 + 0.00002\,R^3 +
   0.00003\,R^4 + 0.00004\,R^5
\nonumber\\&&
 + 0.00003\,R^6 +
   0.00003\,R^7 + 0.00002\,R^8 + 0.00001\,R^9)e^{-6(\Delta\eta-2)}
\Bigr),
\end{eqnarray}

\begin{eqnarray}
\Omega^{(b2)}_1 &=& \frac{\alpha_s}{\pi}\Bigl(
(+ 0.00458\,R^2 + 0.00389\,R^3 + 0.00229\,R^4 +
   0.00104\,R^5
\nonumber\\&&
 + 0.00038\,R^6 + 0.00012\,R^7 +
   0.00003\,R^8)e^{-2(\Delta\eta-2)}+
\nonumber\\&&
(- 0.00124\,R^2 -
   0.00158\,R^3 - 0.00124\,R^4 - 0.00079\,R^5
\nonumber\\&&
 - 0.00041\,R^6 - 0.00018\,R^7 - 0.00007\,R^8 -
   0.00002\,R^9)e^{-3(\Delta\eta-2)}
\nonumber\\&&
+(+ 0.00025\,R^2 + 0.00043\,R^3 + 0.00042\,R^4 +
   0.00034\,R^5
\nonumber\\&&
 + 0.00024\,R^6 + 0.00014\,R^7 +
   0.00007\,R^8 + 0.00003\,R^9 \nonumber \\
&& + 0.00001\,R^{10})e^{-4(\Delta\eta-2)}+
\hspace*{-3cm}\nonumber\\&&
(- 0.00005\,R^2 -
   0.00010\,R^3 - 0.00011\,R^4 - 0.00011\,R^5
\nonumber\\&&
 - 0.00010\,R^6 - 0.00007\,R^7 - 0.00004\,R^8 -
   0.00002\,R^9 \nonumber \\
&& - 0.00001\,R^{10})e^{-5(\Delta\eta-2)}+
\hspace*{-3cm}\nonumber\\&&
(+ 0.00002\,R^3 +
   0.00003\,R^4 + 0.00003\,R^5 + 0.00003\,R^6
\nonumber\\&&
 + 0.00003\,R^7 + 0.00002\,R^8 + 0.00001\,R^9)e^{-6(\Delta\eta-2)}
\Bigr),
\end{eqnarray}

\begin{eqnarray}
\Omega^{(a2)}_1 &=& \frac{\alpha_s}{\pi}\Bigl(
-\frac14R^2+
\\&&{}
(+ 0.06767\,R^2 + 0.02872\,R^3 - 0.00096\,R^5 +
   0.00002\,R^7)e^{-(\Delta\eta-2)}
\nonumber\\&&
+(- 0.01374\,R^2 -
   0.01166\,R^3 - 0.00229\,R^4 - 0.00038\,R^6
\nonumber\\&&
 - 0.00021\,R^7 - 0.00003\,R^8)e^{-2(\Delta\eta-2)}+
\nonumber\\&&
(+ 0.00248\,R^2 + 0.00316\,R^3 + 0.00124\,R^4 +
   0.00032\,R^5
\nonumber\\&&
 + 0.00041\,R^6 + 0.00027\,R^7 +
   0.00007\,R^8 + 0.00001\,R^9)e^{-3(\Delta\eta-2)}+
\nonumber\\&&
(- 0.00042\,R^2 -
   0.00071\,R^3 - 0.00042\,R^4 - 0.00019\,R^5
\nonumber\\&&
 - 0.00024\,R^6 - 0.00019\,R^7 - 0.00007\,R^8 -
   0.00002\,R^9 \nonumber \\
&& - 0.00001\,R^{10})e^{-4(\Delta\eta-2)}+
\hspace*{-3cm}\nonumber\\&&
(+ 0.00007\,R^2 + 0.00014\,R^3 + 0.00011\,R^4 +
   0.00007\,R^5
\nonumber\\&&
 + 0.00010\,R^6 + 0.00009\,R^7 +
   0.00004\,R^8 + 0.00002\,R^9 \nonumber \\
&& +
   0.00001\,R^{10})e^{-5(\Delta\eta-2)}+
\hspace*{-3cm}\nonumber\\&&
(- 0.00001\,R^2 -
   0.00003\,R^3 - 0.00003\,R^4 - 0.00002\,R^5 - 0.00003\,R^6
\nonumber\\&&
  - 0.00004\,R^7 - 0.00002\,R^8 -
   0.00001\,R^9)e^{-6(\Delta\eta-2)}
\Bigr),
\end{eqnarray}
where all coefficients larger than $10^{-5}$ are shown (recall that we
are mainly interested in the case $R=1$, $\Delta\eta>2$). By symmetry, we have $\Omega^{(ij)}_2 =
\Omega^{(\bar\imath\bar\jmath)}_1$, where the mapping $i\to\bar\imath$
is given by $\{a,b,1,2\}\to\{b,a,2,1\}$.

\chapter{The \boldmath{$\Omega_f^{(ij)}$} angular integrals for a cone geometry}
\label{secef:appcone}
We present these results as they have not appeared previously in this form.
They have the expression,
\begin{equation}
\Omega^{(ij)}_f=\int_{-\Delta y/2}^{+\Delta_y/2}d\eta \int_0^{2\pi} \frac{d\phi}{2\pi} 
\frac{\beta_i\cdot \beta_j}{(\beta_i\cdot \bar{k})(\beta_j\cdot \bar{k})},
\end{equation}
where the integrand is found from the appropriate 4-momenta, and the phase space is taken to
be of width $\Delta y$ and azimuthally symmetric. Note that these expressions do not include
the sign factors. We obtain
\begin{eqnarray}
\Omega_f^{(ab)}&=&2\Delta y, \nonumber \\
\Omega_f^{(12)}&=&2\log\left(\frac{\sinh(\Delta\eta/2+\Delta y/2)}{\sinh(\Delta\eta/2-\Delta y/2)}\right), \nonumber \\ 
\Omega_f^{(a1)}&=&-\Delta y+
\log\left(\frac{\sinh(\Delta\eta/2+\Delta y/2)}{\sinh(\Delta\eta/2-\Delta y/2)}\right), \nonumber \\ 
\Omega_f^{(b2)}&=&-\Delta y+
\log\left(\frac{\sinh(\Delta\eta/2+\Delta y/2)}{\sinh(\Delta\eta/2-\Delta y/2)}\right), \nonumber \\ 
\Omega_f^{(a2)}&=&\Delta y+
\log\left(\frac{\sinh(\Delta\eta/2+\Delta y/2)}{\sinh(\Delta\eta/2-\Delta y/2)}\right), \nonumber \\ 
\Omega_f^{(b1)}&=&\Delta y+
\log\left(\frac{\sinh(\Delta\eta/2+\Delta y/2)}{\sinh(\Delta\eta/2-\Delta y/2)}\right).
\end{eqnarray}

\chapter{The ZEUS energy flow measurements}
\label{secef:appzeusdata}

\begin{figure}
\begin{center}
\epsfig{figure=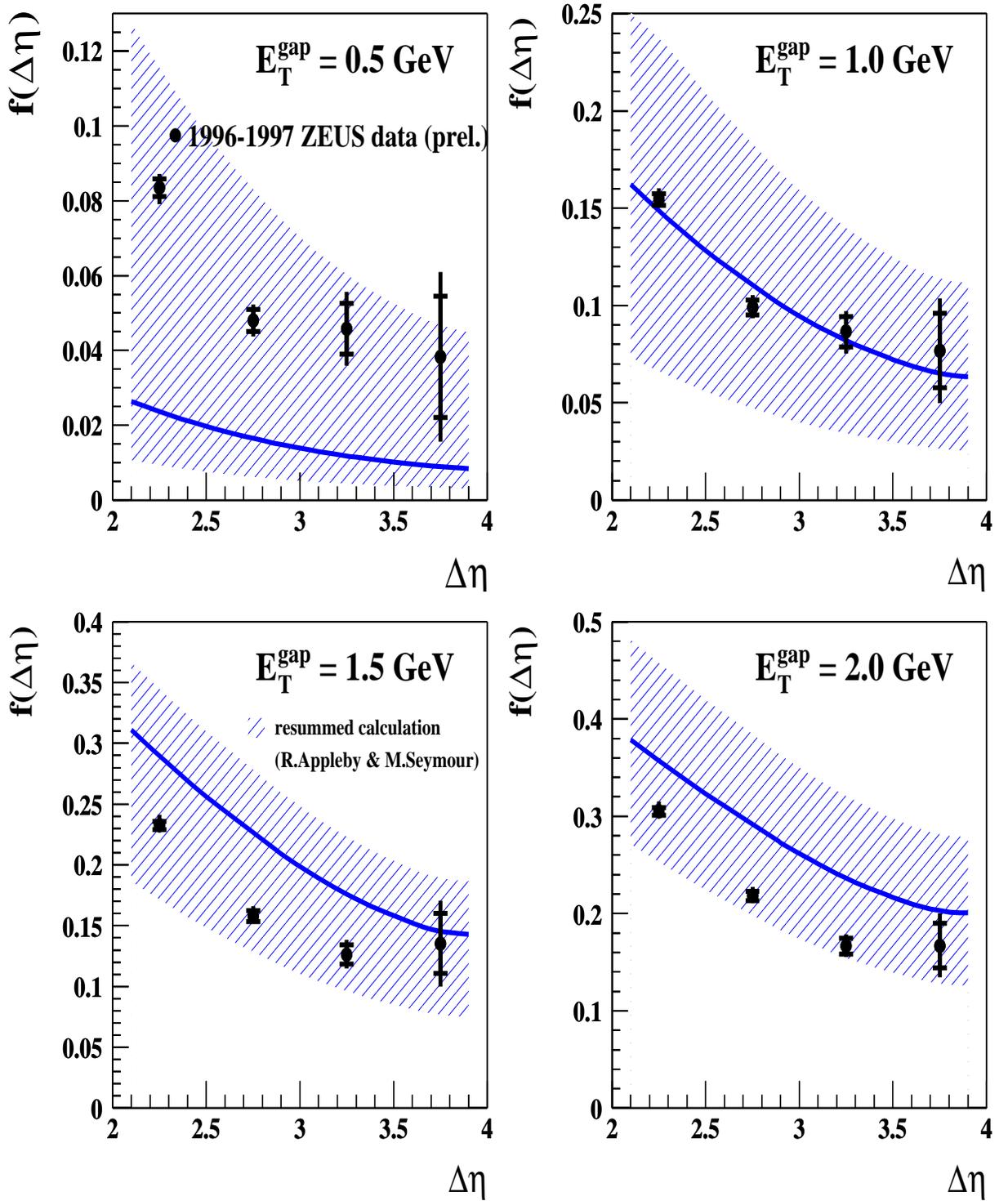,width=6.4in,height=7.8in}
\caption{The gap fractions for the ZEUS analysis with a kt defined final state, 
at varying~$E_T^{\mathrm{gap}}(=Q_{\Omega})$. The 
(preliminary) ZEUS data is also shown. The solid line includes the effects of primary emission and 
the secondary emission suppression factor. The overall theoretical
uncertainty, including the primary uncertainty and the secondary uncertainty, is shown by the
shaded region.}
\label{figef:zeusplot}
\end{center}
\end{figure}

In this appendix, we present our theoretical gap fraction predictions for the ZEUS analysis, figure~\ref{secef:figz2frac}, 
together with a comparison to the (brand new) preliminary ZEUS data points. This comparison was not possible
in chapter~\ref{ch5} due to the unavailability of data when this thesis was written, and we perform the comparison in 
this extra appendix.
Figure~\ref{figef:zeusplot} shows the gap fraction curves and the ZEUS data, together with the theoretical uncertainty. 
This uncertainty, in agreement with figure~\ref{secef:figz2frac}, includes both the primary and secondary emission 
uncertainty.

We find that our gap fraction predictions are consistent with the ZEUS values for the
measured $E_T^{\mathrm{gap}}$ although, as for the H1 data, there is a large theoretical uncertainty. 
However, this uncertainty is 
principally in the normalisation of the curves and we can see that our resummation describes accurately 
the shape of the gap fraction curves. 

\end{appendix}

\label{references}
% my references

\end{document}